\documentclass{iopjournal}

\usepackage[utf8]{inputenc} 
\usepackage[T1]{fontenc} 

\usepackage{graphicx}
\usepackage{subcaption}
\usepackage{tikz}
\usetikzlibrary{positioning}
\usepackage{changepage}
\usepackage{pdfpages}
\usepackage{subcaption}
\usepackage{titletoc}
\usepackage{multirow}
\usepackage{titlesec}


\graphicspath{{./ImagesNew/Methods}{./ImagesNew/bipartite_examples_main}{./ImagesNew/didactical_intro}{./ImagesNew/sector_lengths}{./ImagesNew/threshold}{./ImagesNew/ent_quantifier}{./images/}{./images/product_states}{./images/class_class_states}{./images/entangled_states}{./images/optimization_results}{./images/stacked_plot}{./images/separable_states}{./images/bloch_analysis_bipartite}{./images/hubertigam3}{./images/hansjoerg}{./images/quantum_without_class}{./images/ghz_w_mixed}{./images/ghz_plusminus_mixed}{./images/ghz_w_pure}{./images/ghz_w_coherent}{./images/entanglement_distillation}{./images/noise_reopt/dephasing}{./images/noise_reopt/depolarization}{./images/noise_reopt/amplitude_damping}{./images/tripartite_only}{./images/noisyghzclass}{./images/noisyghzwhite}}

\newcommand{\beginsupplement}{%
    \counterwithout{figure}{section}%
    \setcounter{table}{0}%
    \renewcommand{\thetable}{S\arabic{table}}%
    \setcounter{figure}{0}%
    \renewcommand{\thefigure}{S\arabic{figure}}%
    \setcounter{section}{0}%
    \renewcommand{\thesection}{S\arabic{section}}%
    \renewcommand{\theHsection}{supplementary.\arabic{section}}%
    \setcounter{equation}{0}%
    \renewcommand{\theequation}{S\arabic{equation}}%
}

\usepackage{calc} 
\usepackage{tabto} 
\usepackage{parskip}
\usepackage{tabularx}
\usepackage{wrapfig}
\usepackage{float}
\usepackage{bm}
\usepackage{placeins}

\usepackage{siunitx}
\usepackage{mhchem}

	\usepackage{verbatim}
	\usepackage{amsfonts}
	\usepackage {amsmath}
	\usepackage{bbm}
	\usepackage{braket}
	\usepackage[capitalise,nameinlink,noabbrev]{cleveref}
	\crefname{equation}{Eq.}{Eqs.}
	\crefrangelabelformat{equation}{#3#1#4--#5#2#6}
	
	\usepackage[nohyperlinks]{acronym}
	\usepackage{color}
	\usepackage{enumitem}
	
	\usepackage{booktabs}
	\usepackage{soul}
	
	\newcommand{\centerlabel}[1]{%
    	\makebox[2em][c]{#1}%
	}

	\newcommand*\emptycirc[1][1ex]{\raisebox{-.8ex}{\tikz\draw[line width=.75pt] (0, 0) circle (#1);}}

	\newcommand*\emptytri[1][1ex]{%
		\raisebox{-.0ex}{%
			\tikz[
			x=1pt, y=1pt, line width=.75pt, baseline
			]{
				\pgfmathsetmacro{\a}{2*sqrt(3)} 
				\def\radius{#1}
				
				\draw 
				(0,{2*\radius}) -- 
				({-\a/2*\radius}, {-\radius}) -- 
				({\a/2*\radius}, {-\radius}) -- 
				cycle;
			}%
		}%
	}
	
	\newcommand*\threestar[1][1ex]{%
		\raisebox{.8ex}{%
			\tikz[
			x=1pt, y=1pt, line width=.75pt, baseline
			]{
				\pgfmathsetmacro{\a}{sqrt(3)} 
				\def\radius{#1}
				
				\fill
				(0,{-1*\radius}) -- 
				(-0.3*\a/2*\radius,{-(0.5*0.3)*\radius}) -- 
				({-\a/2*\radius}, {0.5*\radius}) -- 
				(0,{.3*\radius}) -- 
				({\a/2*\radius}, {0.5*\radius}) -- 
				(0.3*\a/2*\radius,{-(0.5*0.3)*\radius}) -- 
				cycle;
			}%
		}%
	}

	\newcommand*\circstar[1][1ex]{%
		\raisebox{-.8ex}{%
			\tikz[ x=1pt, y=1pt, line width=.75pt]{%
				\def\radius{#1}
				\pgfmathsetmacro{\a}{2*sqrt(3)} 
				
				\draw (0,0) circle[radius=\radius];
				
				\fill
				(0,{-1*\radius}) -- 
				(-0.3*\a/4*\radius,{-(0.5*0.3)*\radius}) -- 
				({-\a/4*\radius}, {0.5*\radius}) -- 
				(0,{.3*\radius}) -- 
				({\a/4*\radius}, {0.5*\radius}) -- 
				(0.3*\a/4*\radius,{-(0.5*0.3)*\radius}) -- 
				cycle;
			}%
		}%
	}
	
	\newcommand*\emptycirctristar[1][1ex]{%
		\raisebox{-.0ex}{%
			\tikz[ x=1pt, y=1pt, line width=.75pt, baseline]{%
				\def\radius{#1}
				\pgfmathsetmacro{\a}{2*sqrt(3)} 
				
				\draw (0,0) circle[radius=\radius];
				
				\draw 
				(0,{2*\radius}) -- 
				({-\a/2*\radius}, {-\radius}) -- 
				({\a/2*\radius}, {-\radius}) -- 
				cycle;
				
				\fill
				(0,{-1*\radius}) -- 
				(-0.3*\a/4*\radius,{-(0.5*0.3)*\radius}) -- 
				({-\a/4*\radius}, {0.5*\radius}) -- 
				(0,{.3*\radius}) -- 
				({\a/4*\radius}, {0.5*\radius}) -- 
				(0.3*\a/4*\radius,{-(0.5*0.3)*\radius}) -- 
				cycle;
			}%
		}%
	}
	
	\newcommand*\emptycirctri[1][1ex]{%
		\raisebox{-.0ex}{%
			\tikz[ x=1pt, y=1pt, line width=.75pt, baseline]{%
				\def\radius{#1}
				\pgfmathsetmacro{\a}{2*sqrt(3)} 
				
				\draw (0,0) circle[radius=\radius];
				
				\draw 
				(0,{2*\radius}) -- 
				({-\a/2*\radius}, {-\radius}) -- 
				({\a/2*\radius}, {-\radius}) -- 
				cycle;
				
			}%
		}%
	}
	
	\newcommand*\emptycircquadtristar[1][1ex]{%
		\raisebox{-.0ex}{%
			\tikz[ x=1pt, y=1pt, line width=.75pt, baseline]{%
				\def\radius{#1}
				\pgfmathsetmacro{\a}{2*sqrt(3)} 
				
				\draw[fill=black!10]
				({-\a/2*\radius}, {-\radius}) -- 
				({\a/2*\radius}, {-\radius}) -- 
				({\a/2*\radius}, {2*\radius}) -- 
				({-\a/2*\radius}, {2*\radius}) -- 
				cycle;
				
				\draw[fill=white!10]
				(0,{2*\radius}) -- 
				({-\a/2*\radius}, {-\radius}) -- 
				({\a/2*\radius}, {-\radius}) -- 
				cycle;
				
				\draw[fill=white!10] (0,0) circle[radius=\radius];
				
				\fill
				(0,{-1*\radius}) -- 
				(-0.3*\a/4*\radius,{-(0.5*0.3)*\radius}) -- 
				({-\a/4*\radius}, {0.5*\radius}) -- 
				(0,{.3*\radius}) -- 
				({\a/4*\radius}, {0.5*\radius}) -- 
				(0.3*\a/4*\radius,{-(0.5*0.3)*\radius}) -- 
				cycle;
			}%
		}%
	}
	
	\newcommand*\emptycircquadtri[1][1ex]{%
		\raisebox{-.0ex}{%
			\tikz[ x=1pt, y=1pt, line width=.75pt, baseline]{%
				\def\radius{#1}
				\pgfmathsetmacro{\a}{2*sqrt(3)} 
				
				\draw[fill=black!10] 
				({-\a/2*\radius}, {-\radius}) -- 
				({\a/2*\radius}, {-\radius}) -- 
				({\a/2*\radius}, {2*\radius}) -- 
				({-\a/2*\radius}, {2*\radius}) -- 
				cycle;
				
				\draw[fill=white!10] 
				(0,{2*\radius}) -- 
				({-\a/2*\radius}, {-\radius}) -- 
				({\a/2*\radius}, {-\radius}) -- 
				cycle;
				
				\draw[fill=white!10]  (0,0) circle[radius=\radius];       
				
			}%
		}%
	}
	
	\newcommand*\emptyquad[1][1ex]{%
		\raisebox{-.0ex}{%
			\tikz[ x=1pt, y=1pt, line width=.75pt, baseline]{%
				\def\radius{#1}
				\pgfmathsetmacro{\a}{2*sqrt(3)} 
				
				\draw[fill=black!10] 
				({-\a/2*\radius}, {-\radius}) -- 
				({\a/2*\radius}, {-\radius}) -- 
				({\a/2*\radius}, {2*\radius}) -- 
				({-\a/2*\radius}, {2*\radius}) -- 
				cycle;
			}%
		}%
	}
	
	\newcommand*\trismallquad[1][1ex]{%
		\raisebox{-.0ex}{%
			\tikz[
			x=1pt, y=1pt, line width=.75pt, baseline
			]{
				\pgfmathsetmacro{\a}{2*sqrt(3)} 
				\def\radius{#1}
				\pgfmathsetmacro{\s}{4*\radius / (1 + sqrt(3))} 
				
				\draw[fill=black!10] 
				({-\a/2*\radius}, {-\radius}) -- 
				({\a/2*\radius}, {-\radius}) -- 
				({\a/2*\radius}, {2*\radius}) -- 
				({-\a/2*\radius}, {2*\radius}) -- 
				cycle;
				
				\draw[fill=white!10] 
				(0,{2*\radius}) -- 
				({-\a/2*\radius}, {-\radius}) -- 
				({\a/2*\radius}, {-\radius}) -- 
				cycle;
				
				\pgfmathsetmacro{\xleft}{-\s/2}
				\pgfmathsetmacro{\ybase}{-\radius}
				\draw[dashed, dash pattern=on 1.2pt off 1.2pt, fill=white!10] 
				(\xleft, \ybase) rectangle ({\xleft+\s}, {\ybase+\s});
			}%
		}%
	}
	
	\newcommand*\circsmallquad[1][1ex]{%
		\raisebox{-.0ex}{%
			\tikz[
			x=1pt, y=1pt, line width=.75pt, baseline
			]{
				\pgfmathsetmacro{\a}{2*sqrt(3)} 
				\def\radius{#1}
				\pgfmathsetmacro{\s}{4*\radius / (1 + sqrt(3))} 
				
				\draw[fill=white!10]  (0,0) circle[radius=\radius]; 
				
				\draw [fill=black!10]
				({-0.5*sqrt(2)*\radius},{-0.5*sqrt(2)*\radius}) -- 
				({-0.5*sqrt(2)*\radius},{0.5*sqrt(2)*\radius}) -- 
				({0.5*sqrt(2)*\radius},{0.5*sqrt(2)*\radius}) -- 
				({0.5*sqrt(2)*\radius},{-0.5*sqrt(2)*\radius}) -- 
				cycle;
			}%
		}%
	}
	
	\newcommand*\smallquad[1][1ex]{%
		\raisebox{-.0ex}{%
			\tikz[
			x=1pt, y=1pt, line width=.75pt, baseline
			]{
				\pgfmathsetmacro{\a}{2*sqrt(3)} 
				\def\radius{#1}
				\pgfmathsetmacro{\s}{4*\radius / (1 + sqrt(3))} 
				
				\draw [fill=black!10]
				({-0.5*sqrt(2)*\radius},{-0.5*sqrt(2)*\radius}) -- 
				({-0.5*sqrt(2)*\radius},{0.5*sqrt(2)*\radius}) -- 
				({0.5*sqrt(2)*\radius},{0.5*sqrt(2)*\radius}) -- 
				({0.5*sqrt(2)*\radius},{-0.5*sqrt(2)*\radius}) -- 
				cycle;
			}%
		}%
	}
	
	\newcommand*\bigandsmallquad[1][1ex]{%
		\raisebox{-.0ex}{%
			\tikz[
			x=1pt, y=1pt, line width=.75pt, baseline
			]{
				\pgfmathsetmacro{\a}{2*sqrt(3)} 
				\def\radius{#1}
				\pgfmathsetmacro{\s}{4*\radius / (1 + sqrt(3))} 
				
				\draw[fill=black!10] 
				({-\a/2*\radius}, {-\radius}) -- 
				({\a/2*\radius}, {-\radius}) -- 
				({\a/2*\radius}, {2*\radius}) -- 
				({-\a/2*\radius}, {2*\radius}) -- 
				cycle;
				
				\pgfmathsetmacro{\xleft}{-\s/2}
				\pgfmathsetmacro{\ybase}{-\radius}
				\draw[dashed, dash pattern=on 1.2pt off 1.2pt, fill=white!10] 
				(\xleft, \ybase) rectangle ({\xleft+\s}, {\ybase+\s});
			}%
		}%
	}
	
	\usepackage{pifont}
	\newcommand{\cmark}{\ding{51}}%
	\newcommand{\xmark}{\ding{55}}%

	\usepackage[pass]{geometry}
\begin{document}
\articletype{Paper} 

\title{The Generalized BEADS Representation: An Informationally Complete Visualization of Mixed Quantum States}

\author{Celina Bräutigam$^1$\orcid{0009-0000-1250-9814}, Steffen J. Glaser$^{2,3}$\orcid{0000-0003-4099-3177} and Dennis Huber$^{2,3,*}$\orcid{0000-0002-5182-6078}}

\affil{$^1$ Technical University of Munich, TUM School of Computation, Information and Technology, Department of Computer Science, Boltzmannstrasse 3, 85748 Garching, Germany}\\
\affil{$^2$ Technical University of Munich, TUM School of Natural Sciences, Department of Chemistry, Lichtenbergstrasse 4, 85748 Garching, Germany}\\
\affil{$^3$ Munich Center for Quantum Science and Technology (MCQST), Schellingstrasse 4, 80799 München, Germany}\\
\affil{$^*$ Author to whom any correspondence should be addressed.}

\email{dennis.huber@tum.de}


\begin{abstract}
A deep understanding of quantum information theory is essential for quantum science and technology. However, quantum mechanics and particularly quantum information theory suffer from a lack of intuitive images. Visualization tools offer novel perspectives on many problems in quantum physics by revealing important properties of quantum states and processes that are otherwise obscured by abstract mathematical formalisms. Here, we extend the previously established BEADS representation of pure states to the so-called generalized BEADS representation of mixed quantum states. We introduce spherical functions, so-called beads to visualize correlation structures of mixed quantum states. A recursive optimization scheme is used to obtain minimum-entanglement pure state decompositions of all subsystems of a given $N$-qubit state. We show how correlation function components can be decomposed into individual contributions that directly correspond to different types of correlations of mixed states. This allows us to visually characterize the total correlation
, genuine $n$-partite entanglement
, nonclassical quantum-discord-type correlations 
as well as classical correlations 
. After outlining the mathematical details relevant for constructing the representation, we examine the generalized BEADS representations for various classes of mixed states including product, classically correlated, separable but nonclassically correlated, as well as entangled states. As exemplary applications of the generalized BEADS representation, we showcase entanglement distillation and the effects of noise channels.
\end{abstract}

\section{Introduction}
A deep understanding of quantum information theory is key for a wide range of research disciplines. The notion of correlation, and in particular, the concepts of entanglement and total correlation play a central role in quantum computing, quantum communication, and quantum simulation \cite{horodecki_quantum_2009, jozsa_role_2003, wolf_arealaws_2008}. Entanglement is used as a central resource in quantum information processing \cite{horodecki_quantum_2009,nielsen_quantum_2010} and plays a fundamental role in the characterization of quantum many-body states \cite{amico_entmany_2008, boguslawski_entanglement_2012}. In quantum chemistry, the relation of entanglement and the forming or dissociation of chemical bonds is actively investigated \cite{boguslawski_orbital_2013,szalay_correlation_2017, ding_correlation_2020, ding_chemicalbondig_2026}, whereas in physical systems, the connection of entanglement to quantum critical phenomena such as quantum phase transitions is studied \cite{vidal_entanglement_2003,osborne_entanglement_2002,gu_entanglement_2004}.

Moreover, in the currently ongoing noisy intermediate-scale (NISQ) quantum era, it is essential to understand noise, often referred to as decoherence processes on a fundamental level in order to be able to address them adequately via specifically tailored error correction, error mitigation, or error suppression protocols \cite{nielsen_quantum_2010, terhal_error_2015, cai_mitigation_2023}. Decoherence processes are generally non-unitary, that is, they transform pure into mixed quantum states. This is exactly where a deep understanding of quantum information theory is beneficial as it enables to assess how the correlation and especially the entanglement properties of a quantum state have changed after undergoing a noise process and how this constrains the available computational resources. 

However, a common problem occurring in quantum mechanics and especially quantum information theory is the lack of intuitive memorable images, which prevents students and scientists from developing a deep understanding. Visualization tools offer new perspectives onto quantum mechanics. They are able to reveal numerous important properties of quantum states that are otherwise obscured by the abstract mathematical formalism of quantum physics. Many visualizations of pure quantum states have been introduced over the last decades with prominent examples being the Q-Sphere \cite{javadi_qiskit_2024} or the dimensional circle notation \cite{bley_circle_2024}. On the other hand, visualizations of mixed quantum states such as finite-dimensional phase-space representations, e.g., the Husimi \cite{husimi_1940}, the Wigner \cite{wigner_1932}, and the Glauber-Sudarshan representations \cite{glauber_1969}, or generalizations of the Majorana stellar representation \cite{majorana_1932, braun_majorana_2020}, are commonly restricted to fully permutation-symmetric subspaces of a density operator and thus, are informationally incomplete. They also fail to intuitively visualize the correlation structure of the underlying states. Previously, we have introduced the BEADS representation \cite{huber_beads_2025}, which provides complete visualizations of arbitrary quantum states and is in particular useful for displaying pure states of coupled, distinguishable, and individually addressable quantum mechanical two-level systems (qubits). These systems, which fulfill DiVincenzo's criteria for quantum computing \cite{divincenzo_topics_1996}, can be used for quantum information processing and quantum computing tasks.

In the following, we extend the original BEADS representation \cite{huber_beads_2025} to the so-called \textit{generalized BEADS representation} of mixed quantum states. This is motivated by the fact that mixed quantum states commonly exhibit richer correlation structures than pure states. While the original BEADS representation is in principle also applicable to mixed states, it cannot provide intuitive visualizations of their correlation structures. In the case of bipartite quantum states, correlation types and measures are well-defined. Total correlation measured by the quantum mutual information \cite{ollivier_quantum_2001}, quantifies the total amount of statistical dependence of two parties. Classical correlations as assessed by the classical mutual information \cite{ollivier_quantum_2001} correspond to information accessible via local measurements. Nonclassical or quantum correlations, as measured by quantum discord \cite{ollivier_quantum_2001}, can be attributed to entanglement-based and separable properties of a density operator.  In fact, the only source of genuine correlations found in pure states is entanglement, in which case it is equivalent to the total correlation of the state. However, even in pure states, a comprehensive characterization of multiparty entanglement ($n>2$) remains a challenging question for which several entanglement measures have been proposed, yet no general solution has been established \cite{plenio_introduction_2006, schlienz_description_1995}. Therefore, the scientific questions which have to be answered in order to construct a meaningful visualization of mixed quantum states are the following
\begin{enumerate}[label=\alph*)]
\item How can we reliably differentiate between and visualize pairwise and genuine $n$-partite entanglement of mixed quantum states?
\item How can classical correlation be assessed, isolated and visualized?
\item How can nonclassical correlations as measured by quantum discord be visualized?
\end{enumerate}
In contrast to the original BEADS representation \cite{huber_beads_2025}, we apply a recursive optimization scheme for reduced subsystems of an $N$-qubit state. This enables us to accurately visualize the potential mixedness and the associated correlation structure of reduced subsystems of $n$-qubits, where $n<N$. With this we specifically address the potential ambiguity arising when interpreting $n$-partite connected correlation functions and $n$-partite entanglement of corresponding reduced marginals of a globally pure state in the original BEADS representation.

The article is structured in the following way. First, we introduce the generalized BEADS representation methodologically. We then showcase visualizations of different classes of two-qubit quantum states ranging from product to entangled states. We identify characteristic features of the visualizations and use these to deduce classification rules. Next, we examine tripartite systems and discuss properties of $N$-partite systems, e.g., genuine $n$-partite entanglement and classical correlation, and explain how these are approached in the generalized BEADS representation. Finally, we showcase possible applications of the generalized BEADS representation by presenting visualizations of the BBPSSW entanglement distillation protocol as well as different kinds of decoherence processes and analyze how changes in the underlying correlation structure of the states manifest in their generalized BEADS representation. For readers who are unfamiliar with the quantum information theory of mixed quantum states, we include a didactic and phenomenological introduction to this class of states that is based on the generalized BEADS representation in supplementary \cref{sec:didactic_intro}.
\clearpage
\section{Methods}
In the following, we introduce the methods required for constructing the generalized BEADS representation. We build upon our previous publication \cite{huber_beads_2025} that introduced the BEADS representation of pure states and adapt the underlying theory and construction to match the information theory of mixed quantum states.

\subsection{General visualization approach}
\label{sec:visapproach}
In the BEADS representation, an arbitrary operator $A$ -- for the remainder of this article we will consider the density operator $\rho$ describing an arbitrary quantum state -- is expanded in a symmetry-adapted spherical tensor operator basis, the so-called LISA basis. Here, the name LISA reflects the properties captured by the basis by design, which are \underline{li}nearity, \underline{s}ubsystems, and \underline{a}uxiliary criteria \cite{garon_visualizing_2015,leiner_symmetry-adapted_2020}.

Components of the operator of interest labeled $A^{(\ell')}$ can then be visualized by spherical functions called beads $b_{j}^{(\ell')}$ by mapping the LISA-basis spherical tensor operators $T_{j,m}^{(\ell')}$ of rank $j$ and order $m$ onto corresponding spherical harmonics $Y_{j,m}$ \cite{huber_beads_2025}:
    \begin{equation}
        A^{(\ell')} = \sum_{j\in J(\ell')}\sum_{m=-j}^{j} c_{j,m}^{(\ell')} T_{j,m}^{(\ell')} \longleftrightarrow b_{j}^{(\ell')} = \sum_{j\in J(\ell')}\sum_{m=-j}^{j} {c'}_{j,m}^{(\ell')} Y_{j,m} = \sum_{j\in J(\ell')}\sum_{m=-j}^{j} s_j^{(\ell')}(N,g){c}_{j,m}^{(\ell')} Y_{j,m}.
    \end{equation}
Here, $\ell'$ labels the point symmetry (even or odd) and the qubits involved in a bead, and $c_{j,m}^{(\ell')}$ denote the LISA-basis expansion coefficients, which are applied to the spherical harmonics with additional non-trivial scaling factors $s_j^{(\ell')}(N,g)$ \cite{huber_beads_2025} depending on $N$, the number of qubits in the system and the linearity $g$. An overview of all beads and the associated symmetries for systems consisting of up to three qubits is provided in \cref{sup:beads_overview}.

The scaling factors $s_j^{(\ell')}(N,g)$ enable the spherical function values represented by specific beads to directly match experimentally measurable expectation values of (joint) local measurements or to provide a direct correspondence to the global unitary bound of the underlying LISA tensor operator $T_{j,m}^{(\ell')}$. We revisit the numerical values of these scaling factors for systems of up to three qubits, which we have previously published in \cite{huber_beads_2025}, in supplementary \cref{sup:scaling_factors}.
    
The BEADS representation of a quantum state at least comprises a set of $N$ beads of linearity $g = 1$, so-called $Q$-Beads. $Q$-Beads visualize single-qubit reduced density operators and are thus equivalent to Bloch vectors. In addition, the BEADS representation features correlation function beads of linearity $g > 1$ when $N > 1$. As will be detailed in the following sections, correlation function beads directly represent $n$-partite ($1 < n \leq N$) correlation functions or specific components thereof depending on the type of visualized information. As the simplest possible example, the BEADS representation of a two-qubit quantum state contains two $Q$-Beads $Q_1$ and $Q_2$ and up to two correlation function beads of even and odd point symmetry.

\begin{figure}[htbp]
    \centering

    \begin{subfigure}[c]{0.38\textwidth}
    	\vspace{.7cm}
        \centering
        \textbf{Expectation value arising from separable structure}
        \includegraphics[width=\textwidth]{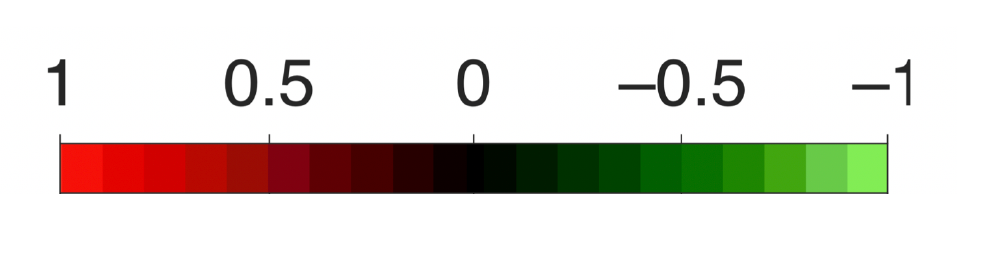}
        \vfill
		\vspace{.6cm}
        \centering
        \textbf{Expectation value arising from genuine $\mathbf{n}$-partite entanglement}
        \includegraphics[width=\textwidth]{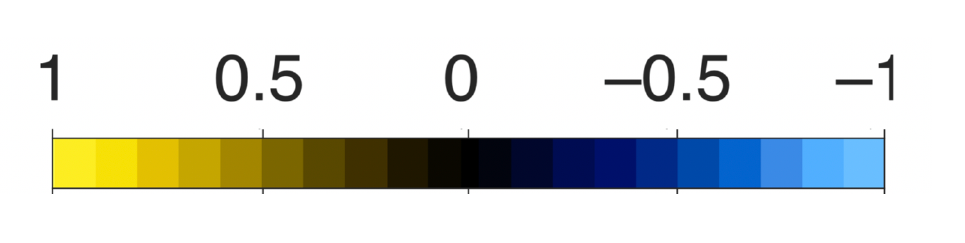}
        \vspace{.1cm}
        \caption{In the generalized BEADS representation, red-green (top) and yellow-blue (bottom) color schemes are used to visually differentiate the separable and entanglement structure of a quantum state.}
    \end{subfigure}
    \hfill
    \begin{subfigure}[c]{0.48\textwidth}
        \centering
        \includegraphics[width=\textwidth]{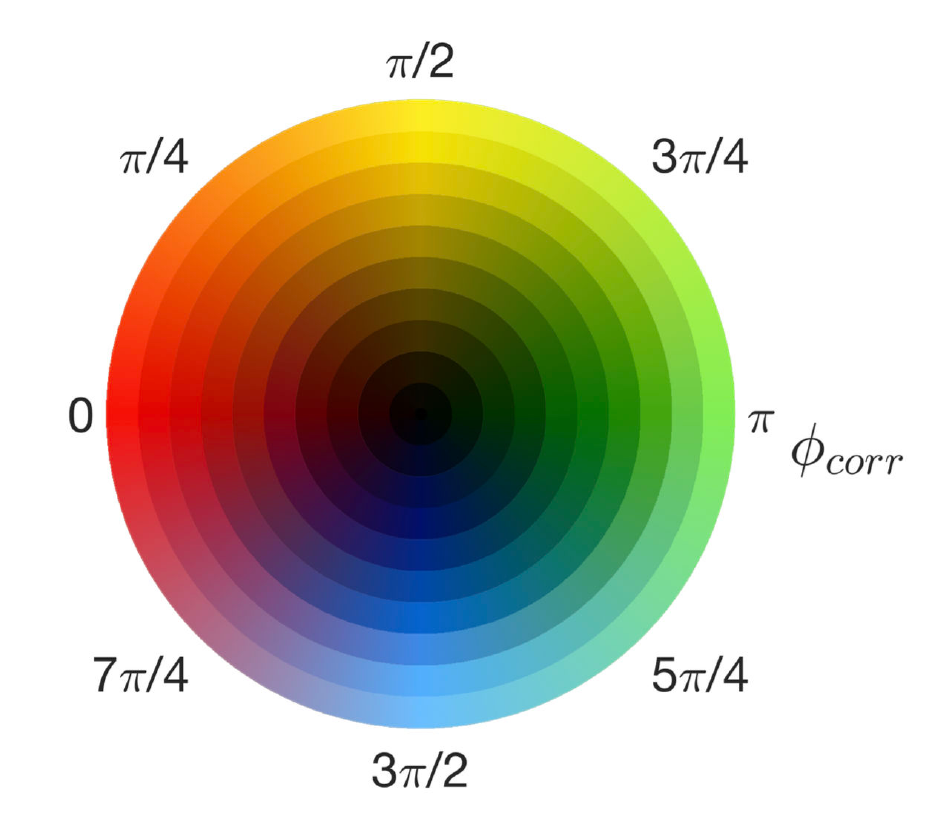}
        \caption{Using a combined color scheme allows to visualize expectation values $\langle O \rangle$ that result from both, separable and entangled properties of a state. Here, the color mixing angle $\phi_{corr} = \text{atan2}(\langle O \rangle_{\text{ent}}, \langle O \rangle_{\text{sep}})$ quantifies the ratio of the underlying properties, while the distance to the color wheel origin corresponds to the magnitude $|\langle O \rangle| \in [0;1]$.}
        \label{fig:third}
    \end{subfigure}

    \caption{Standard color schemes used in the generalized BEADS representation for representing spherical function values, which can be interpreted in terms of expectation values \cite{huber_beads_2025}.}
    \label{fig:ColorSchemes}
\end{figure}

\color{black}
As in the original BEADS representation \cite{huber_beads_2025}, the spherical function values of beads along specific spatial directions are represented by corresponding colors. In the generalized BEADS representation we adopt the color schemes from the original visualization variant, but we slightly alter the way they are employed and interpreted. Red and green colors now generally represent spherical function values associated with the separable structure (including redundant information; see \cref{sec:methods_corrfunctions}) of an arbitrary density operator, while yellow and blue colors are strictly reserved for visualizing $n$-partite correlation function components that directly result from $n$-partite entanglement and thus, represent the $n$-partite entangling structure of a quantum state. Once again a combined color scheme can be constructed that directly quantifies the underlying properties in all spatial directions when visualizing correlation types that manifest in both, the separable and the entangling structure of a state. An overview of the standard color schemes used in the generalized BEADS representation and their interpretation is given in \cref{fig:ColorSchemes}. We refer the interested reader to consult the original publication \cite{huber_beads_2025} for further mathematical details on the construction of the representation.

\FloatBarrier

\subsection{Correlation functions}
\label{sec:methods_corrfunctions}

Correlation functions quantify the expectation values of joint local measurements for arbitrary choices of local measurement bases, that is, measurement directions. In other words, correlation functions measure how the outcomes of joint local measurements are correlated. This is not to be confused with the notion of correlation of a quantum state in quantum information theory which describes the phenomenon that a joint $N$-partite state contains statistical structures that cannot be inferred by observing the individual parties or arbitrary reduced marginals. In a system consisting of $N$ qubits, the \textit{total} $n$-partite correlation function $\Gamma_{T, \mathcal{G}}$ of an $n$-partite set $\mathcal{G}$ (where $n \leq N$) of measured particles is defined as \cite{tran_genuine_2017}
\begin{equation}
	\Gamma_{T,\mathcal{G}}= \text{Tr}(\rho \, O_\mathcal{G}) = \braket{O_\mathcal{G}},
	\label{eq:n_partite_correlation_function}
\end{equation}
which corresponds to the expectation value of an observable
\begin{equation}
	O_\mathcal{G} = \bigotimes_{k=1}^N 
	\left\{
	\begin{array}{ll}
		\vec{r}_k \, \cdot \, \vec{\sigma} & \text{if } k \in \mathcal{G} \\
		\mathbbm{1} & \text{otherwise}.
	\end{array}
	\right.
	\label{eq:O_G}
\end{equation}

Here, the measurement directions are specified by 
$\vec{r}_k$, $\vec{\sigma} = (\sigma_x, \sigma_y, \sigma_z)^T$ is the vector of Pauli matrices, $\mathbbm{1}$ is the single-qubit identity operator, and $k$ is the qubit index \cite{huber_beads_2025}.

In pure states $\ket{\psi}$, entanglement represents the only source of genuine correlation. Schlienz and Mahler found that when an $n$-partite system is in a pure state, the state is genuinely $n$-partite entangled if and only if it exhibits an $n$-partite \textit{connected} correlation function $\Gamma_{E, \mathcal{G}}$ \cite{schlienz_description_1995}. $\Gamma_{E, \mathcal{G}}$ corresponds to the component of the total correlation function $\Gamma_{T, \mathcal{G}}$ for which the correlation coefficient, i.e., the expectation value of $O_\mathcal{G}$, does not factorize into products of expectation values of single-qubit observables $\langle\sigma_{k\vec{r}_k}\rangle$ and/or lower-order connected correlation coefficients. Conversely, expectation values which factorize into such products represent redundant information and are denoted \textit{compound} correlation functions $\Gamma_{C, \mathcal{G}}$. Therefore, the total correlation function $\Gamma_{T, \mathcal{G}}$ is the sum of the connected correlation function $\Gamma_{E, \mathcal{G}}$ and the compound correlation function contributions $\Gamma_{C, \mathcal{G}}$. $\Gamma_{E, \mathcal{G}}$ is obtained by subtracting all compound contributions, that is, all recursively constructed products representing redundant information, from the total correlation function $\Gamma_{T, \mathcal{G}}$. For the bi- and tripartite case one thus arrives at:
\begin{subequations}
\begin{align}
\Gamma_{E,12}&=\Gamma_{T,12}-\Gamma_{C,12}=\left\langle\sigma_{1\vec{r}_1}\sigma_{2\vec{r}_2}\right\rangle-\left\langle\sigma_{1\vec{r}_1}\right\rangle\left\langle\left.\sigma_{2\vec{r}_2}\right\rangle\right.,\label{eq:bipartite_E_correlationfunction}\\[10pt]
\Gamma_{E,123}&= \Gamma_{T,123} -\Gamma_{C,123} \nonumber\\
&= \Gamma_{T,123} -\sum_{\substack{\{j\}\sqcup\{k,l\}=\{1,2,3\}\\
\text{partitions}}}\Gamma_{C,(j)(kl)} \ \ - \Gamma_{C,(1)(2)(3)}\nonumber\\
&= \Gamma_{T,123} -\Gamma_{C,(1)(23)}-\Gamma_{C,(2)(13)}-\Gamma_{C,(3)(12)} - \Gamma_{C,(1)(2)(3)}\nonumber\\
&= \Gamma_{T,123} - \langle\sigma_{1\vec{r}_1}\rangle\Gamma_{E,23} - \langle\sigma_{2\vec{r}_2}\rangle\Gamma_{E,13} - \langle\sigma_{3\vec{r}_3}\rangle\Gamma_{E,12} - \Gamma_{C,(1)(2)(3)} \nonumber\\
&= \left\langle\sigma_{1\vec{r}_1}\sigma_{2\vec{r}_2}\sigma_{3\vec{r}_3}\right\rangle - \langle\sigma_{1\vec{r}_1}\rangle\left[\left\langle\sigma_{2\vec{r}_2}\sigma_{3\vec{r}_3}\right\rangle-\left\langle\sigma_{2\vec{r}_2}\right\rangle\left\langle\sigma_{3\vec{r}_3}\right\rangle\right] - \langle\sigma_{2\vec{r}_2}\rangle\left[\left\langle\sigma_{1\vec{r}_1}\sigma_{3\vec{r}_3}\right\rangle-\left\langle\sigma_{1\vec{r}_1}\right\rangle\left\langle\sigma_{3\vec{r}_3}\right\rangle\right] \nonumber\\
&\phantom{=}- \langle\sigma_{3\vec{r}_3}\rangle\left[\left\langle\sigma_{1\vec{r}_1}\sigma_{2\vec{r}_2}\right\rangle-\left\langle\sigma_{1\vec{r}_1}\right\rangle\left\langle\sigma_{2\vec{r}_2}\right\rangle\right] - \left\langle\sigma_{1\vec{r}_1}\right\rangle\left\langle\sigma_{2\vec{r}_2}\right\rangle\left\langle\sigma_{3\vec{r}_3}\right\rangle \nonumber\\
&=\left\langle\sigma_{1\vec{r}_1}\sigma_{2\vec{r}_2}\sigma_{3\vec{r}_3}\right\rangle-\left\langle\sigma_{1\vec{r}_1}\right\rangle\left\langle\sigma_{2\vec{r}_2}\sigma_{3\vec{r}_3}\right\rangle-\left\langle\sigma_{2\vec{r}_2}\right\rangle\left\langle\sigma_{1\vec{r}_1}\sigma_{3\vec{r}_3}\right\rangle\nonumber\\
&\phantom{=}-\left\langle\sigma_{3\vec{r}_3}\right\rangle\left\langle\sigma_{1\vec{r}_1}\sigma_{2\vec{r}_2}\right\rangle+2\left\langle\sigma_{1\vec{r}_1}\right\rangle\left\langle\sigma_{2\vec{r}_2}\right\rangle\left\langle\sigma_{3\vec{r}_3}\right\rangle.
\label{eq:tripartite_E_correlationfunction}
\end{align}
\end{subequations}

The four-partite connected correlation function is detailed in the supplementary \cref{sup:4partconn}. While, if the underlying $n$-partite state is pure, a nonzero $n$-partite connected correlation function serves as a necessary condition for genuine $n$-partite entanglement, this does not generalize directly to mixed states. In fact, the presence of a connected correlation function merely indicates genuine correlation but not the underlying type of correlation of a mixed state. We will establish a suitable generalization to mixed states in \cref{sec:MixedStateCorrFunc}, which allows to distinguish between all relevant types of correlations, in particular entanglement, of mixed quantum states.

It is also possible to directly extract the density operator component $\rho_{E,\mathcal{G}}$ corresponding to the connected correlation function $\Gamma_{E,\mathcal{G}}$ of a set of qubits $\mathcal{G}$. This is achieved by subtracting the unwanted $n$-body terms corresponding to subsystems $\mathcal{S} \neq \mathcal{G}$ of no interest, all compound correlation function density operator components $\rho_{C,\mathcal{G}}$, the single-qubit density operator components $\rho_{k}$, and the identity operator $\mathbbm{1}$ term from the full density operator $\rho$. This can be conveniently achieved by expanding $\rho$ in a basis of choice, here, the Pauli basis \cite{huber_beads_2025}. In the bipartite case this gives:
\begin{align}
	\rho_{E,12}&=\rho_{12}- \left[ \rho_{C,12}+\rho_{1}+\rho_{2}+\frac{\mathbbm{1}_4}{4} \right] \nonumber\\
	&= \rho- \frac{1}{4} \left[\sum_{\alpha,\beta=x,y,z} \braket{\sigma_{1\alpha}}\braket{\sigma_{2_\beta}} \sigma_{1\alpha} \sigma_{2_\beta} + \sum_{\alpha=x,y,z} \braket{\sigma_{1\alpha}} \sigma_{1\alpha} + \sum_{\beta=x,y,z} \braket{\sigma_{2\beta}} \sigma_{2\beta} + \mathbbm{1}_4 \right].
	\label{eq:rho_E_pure_bipartite}
\end{align}
Importantly, for an arbitrary \textit{pure} density operator $\rho_\mathcal{G}$ describing a system of qubits $\mathcal{G}$, $\rho_{E,\mathcal{G}}$ can be directly viewed as the entangling structure, whereas $\rho_{C,\mathcal{G}}$ can be interpreted as the product structure of $\rho_\mathcal{G}$. Note however, that from a mathematical-physical point of view $\rho_{E,\mathcal{G}}$ and $\rho_{C,\mathcal{G}}$ generally need not be physical density matrices themselves, i.e., they are in general not semidefinite-positive and do not have trace 1. A generalization of \cref{eq:rho_E_pure_bipartite} to tri- and four-partite connected correlation function density operator components is exemplified in \cref{sec:tripartite_connected}.

\subsection{Minimally entangled pure state decompositions of mixed quantum states}
\label{sec:methods_opt}

A central objective of the generalized BEADS representation is to visualize the entanglement of mixed quantum states, which is defined based on a mixed state's minimally entangled pure state decomposition \cite{hughston_complete_1993}. Finding such a decomposition is in general a non-trivial task, which requires nonlinear optimization using a suitable entanglement measure as a cost function. It should also be stressed that there commonly exists no unique solution to this optimization problem, i.e., up to infinitely many different least-entangled pure state ensembles exist that realize the same mixed state.

In the two-qubit case, a suitable cost function for optimization is given by the entanglement of formation \cite{bennett_eof_1996} of a quantum state described by a density operator $\rho$. The entanglement of formation is defined as the minimum average pure-state entanglement, that is, the convex sum of the entropies of entanglement $E(\psi_i)$ (i.e., the von Neumann entropy of either reduced system) of the contributing pure states, where the infimum $E(\rho)$ is taken over all possible pure-state ensembles $\{\ket{\psi_i}\}$ with individual weights $p_i$ constituting $\rho$ \cite{bennett_eof_1996}
\begin{equation}
	E(\rho)= \inf_{\{\ket{\psi_i}\}} \sum_i p_i \, E(\psi_i).
	\label{eq:entanglement_of_formation}
\end{equation}
For bipartite systems, Wootters showed that there also exists an analytic expression for the entanglement of formation \cite{wootters_entanglement_1998}:
\begin{equation}
	E(\rho)= h
	\left(
	\frac{1+\sqrt{1-C(\rho)^2}}{2}
	\right),
	\label{eq:analytical_ent}
\end{equation}
where $h$ is the Shannon entropy function defined as 
\begin{equation}
	h(x)= -x\log_2 x -(1-x) \log_2 (1-x),
	\label{eq:analytical_ent_h}
\end{equation}
and $C$ denotes the concurrence \cite{hill_concurrence_1997}. The analytical expression \cref{eq:analytical_ent} serves as a convenient tool for verifying numerical optimization results.

To find a least-entangled pure state decomposition by numerical optimization, Hughston et al. suggested a procedure called unitary remixing \cite{hughston_complete_1993}, which allows to construct all possible pure state ensembles for any given density matrix $\rho$. Starting with the eigenensemble $\{\ket{e_i}\}$ of the density operator of interest as a canonical choice, the weighted eigenvectors -- that is, the probability amplitude of each eigenvector is directly multiplied to the dedicated vector -- are stacked in a list of state vectors, where each vector represents a single element of the list. By applying a unitary $\mathcal{U}$ to this list of states, a new list of states is generated corresponding to a different ensemble of pure states which constitutes the same $\rho$:
\begin{equation}
\renewcommand{\arraystretch}{1.7}
\left(
\begin{array}{c}
    \ket{\psi_1} \\
    \ket{\psi_2} \\
    \vdots \\
    \vdots \\
    \vdots \\
    \vdots \\
    \ket{\psi_k}
\end{array}
\right)
=
\mathcal{U}
\left(
\begin{array}{c}
    \ket{e_1} \\
    \ket{e_2} \\
    \vdots \\
    \ket{e_r} \\
    \ket{0} \\
    \vdots \\
    \ket{0}
\end{array}
\right).
	\label{eq:unitary_remixing}
\end{equation}
The eigenensemble of the density matrix $\rho$ contains $r$ eigenvectors corresponding to the rank $r$ of $\rho$. However, according to the Carathéodory theorem \cite{caratheodory_1907, caratheodory_uber_1911} it is necessary to allow for a maximum number of up to $k = r^2+1$ vectors to find an optimum least-entangled decomposition of $\rho$. This can be achieved by padding the eigenensemble with $k-r$ zero vectors $\ket{0}=(0\,0\,... \, 0)^\text{T}$ as shown on the right side of \cref{eq:unitary_remixing} \cite{hughston_complete_1993}.

Whereas entanglement measures are universally accepted and well defined for bipartite systems, the same does not apply to higher systems. Indeed, there exists no broadly accepted generalization of the entanglement of formation to $n$-partite systems which permits a usage in ensemble optimizations as described before. Entanglement measures, which have been introduced for $n$-partite systems include generalizations of the concurrence and, in particular, the so-called tangle measures \cite{coffman_distributed_2000}. Despite being well-suited for identifying and quantifying particular kinds of $n$-partite entanglement, such measures are unsuited for pure state ensemble optimizations of mixed quantum states, as they commonly do not detect all relevant types of entanglement simultaneously. A brief discussion on this topic is provided in \cref{sec:app_tangles}.

Thus, in ensemble optimizations for the generalized BEADS representation, we use a lesser known entanglement quantifier which is nonzero for any type of genuine $n$-partite entanglement and which can be easily generalized to larger systems: The so-called connected correlation function norm $\mathcal{N}_{\mathcal{G}}$, which for a pure state can be obtained as \cite{schlienz_description_1995}
\begin{equation}
	\mathcal{N}_{\mathcal{G}}=\|\rho_{E,\mathcal{G}}\|_{_F} = \sqrt{\text{Tr}\left(\rho_{E,\mathcal{G}}^2\right)},
	\label{eq:connected_corr_function_norm}
\end{equation}
using the Frobenius norm of $\rho_{E,\mathcal{G}}$, the density operator component corresponding to the connected correlation function of the system composed of the set of qubits $\mathcal{G}$ as introduced in section~\ref{sec:methods_corrfunctions}.

For mixed quantum states, $\mathcal{N}_{\mathcal{G}}$ is conveniently given by the convex roof extension over all possible pure states ensembles $\{\ket{\psi_{i,\mathcal{G}}}\}$ with weights $p_i$:
\begin{equation}
	\mathcal{N}_{\mathcal{G}}=\inf_{\{\ket{\psi_{i,\mathcal{G}}}\}}\left(\sum_i p_i \|\rho_{i, E,\mathcal{G}}\|_{_F}\right).
	\label{eq:connected_corr_function_norm_cre}
\end{equation}
The fact that $\mathcal{N}_{\mathcal{G}}$ is nonzero for GHZ- and W-entangled pure states \cite{dur_three_2000} or any other class of genuine $n$-partite entanglement in higher systems \cite{verstraete_fourqubit_2001}, despite these families of entangled states corresponding to different SLOCC classes, makes it an excellent choice to serve as a cost function for optimizing the pure state ensemble with respect to a least $n$-partite entangled decomposition for any $n > 2$. The overall cost is then obtained as described in \cref{eq:connected_corr_function_norm_cre} by considering all contributing pure states $\rho_i = \ket{\psi_i}\bra{\psi_i}$ in the ensemble, and minimizing $\mathcal{N}_{\mathcal{G}}$ automatically corresponds to maximally reducing the amount of genuine $n$-partite entanglement.

\subsection{Visualizing correlations of mixed states: Bead types and generalized separation of correlation function components}
\label{sec:MixedStateCorrFunc}

Mixed quantum states generally exhibit rich correlations structures. In the following we will detail how to isolate correlation function components which directly relate to entanglement as well as total, classical, and nonclassical correlations. These correlation function components are directly visualized by corresponding correlation function beads in the generalized BEADS representation.

\subsubsection*{Visualizing entanglement, separable structures, and full density operators: E-, C-, and T-Beads of mixed quantum states} 
\label{sec:ECTBeads}
In \cref{sec:methods_corrfunctions}, the separation of pure state \textit{\underline{t}otal} correlation functions $\Gamma_{T}$ into connected \textit{\underline{e}ntanglement-based} correlation functions $\Gamma_{E}$, and \textit{\underline{c}ompound} correlation function components $\Gamma_{C}$ was introduced in \cref{eq:bipartite_E_correlationfunction,eq:tripartite_E_correlationfunction}. In the mixed state case, we use convex roof extensions of pure state connected and compound correlation functions, or equivalently the corresponding density operator components. 

Thus, the mixed state density operator components corresponding to the connected and compound correlation functions are obtained as convex sums over the corresponding correlation function density operator components of the pure states $\rho_{i,\mathcal{G}} = \ket{\psi_{i,\mathcal{G}}}\bra{\psi_{i,\mathcal{G}}}$ with weights $p_i$ of an optimum least-entangled ensemble $\{\ket{\psi_i}\}$ realizing the density operator $\rho_{\mathcal{G}}$ of a (sub)system $\mathcal{G}$ of interest:
\begin{align}
	\rho_{E, \mathcal{G}}&=\sum_i p_i \,\rho_{E,i,\mathcal{G}}, \text{ where } \{\ket{\psi_{i,\mathcal{G}}}\} \text{ attains } \mathcal{N}_\mathcal{G} = \inf_{\{\ket{\psi_{k, \mathcal{G}}}\}}  \left(\sum_k p_k || \rho_{E,k,\mathcal{G}} ||_F\right),\label{eq:connmixed}\\
	\rho_{C, \mathcal{G}}&=\sum_i p_i \, \rho_{C,i,\mathcal{G}}, \text{ where } \{\ket{\psi_{i,\mathcal{G}}}\} \text{ attains } \mathcal{C}_\mathcal{G} = \sup_{\{\ket{\psi_{k,\mathcal{G}}}\}}  \left(\sum_k p_k || \rho_{C,k,\mathcal{G}} ||_F\right).\label{eq:compmixed}
\end{align}
Here, $\{\ket{\psi_{k,\mathcal{G}}}\}$ denotes all possible pure state ensembles with individual pure state weights $p_k$ realizing the density operator of interest.
Analogously, the total correlation function density operator components obey the following relations:
\begin{equation}
\rho_{T, \mathcal{G}}=\sum_i p_i \, \rho_{T,i,\mathcal{G}} = \rho_{E, \mathcal{G}} + \rho_{C, \mathcal{G}} = \sum_i p_i \,( \rho_{E,i,\mathcal{G}} + \rho_{C,i,\mathcal{G}}).
\end{equation}
Crucially, this approach assumes that the pure state ensembles $\{\ket{\psi_{k,\mathcal{G}}}\}$ constituting the density operators of all possible $n$-partite marginals $\mathcal{G}$ of an $N$-partite quantum state, where again $n \leq N$, are \textit{optimized individually} with respect to least-entangled decompositions (cf. \cref{sec:methods_opt}). That is, we start with the overall $N$-partite density operator, optimize its constituting ensemble, and determine the $N$-partite connected and compound correlation functions according to \cref{eq:connmixed} and \cref{eq:compmixed}. We then repeat this procedure for any of the possible reduced marginals by tracing out distinct qubits to determine the connected and compound $n$-partite correlation function components of all reduced subsystem density operators. As a consequence of this recursive approach, any $n$-partite connected correlation function then strictly corresponds to the genuine $n$-partite entangling structure of the underlying density operator, or put into simpler terms, $n$-partite entanglement of the corresponding subsystem. 
   
Finally, the density operator components $\rho_T$, $\rho_E$, and $\rho_C$ can then be visualized by corresponding $T$-, $E$- and $C$-beads, respectively, following the visualization technique introduced in \cref{sec:visapproach}.

\subsubsection*{Correlations arising from the separable structure of a state: $\Sigma$-Beads}
\label{sec:SigmaBeads}

Unlike in the case of pure states, mixed quantum states exhibit correlations beyond entanglement. In the preceding section we have shown how to isolate the density operator components corresponding to the connected correlation functions of a mixed state that are a direct manifestation of its $n$-partite entanglement (see \cref{eq:connmixed}). Compound correlation function density operator components (\cref{eq:compmixed}) result as a byproduct in the course of this procedure. As pointed out in \cref{sec:methods_corrfunctions}, on the level of the density operator, the compound correlation functions of pure states strictly correspond to the product structure, i.e., redundant information, which does not relate to genuine correlations of the state and merely captures trivial correlations of joint local measurement outcomes.

Whereas these correlation function components are therefore irrelevant for the analysis of genuine correlations of pure states, this is not the case for mixed states. Indeed, it turns out that the resulting mixed state compound correlation functions (cf. \cref{eq:compmixed}) do not strictly represent redundant information that can be recovered from lower-order marginals but may generally encode information arising from genuine correlations of the non-entangling, i.e., \textit{separable structure} of a mixed state density operator. This stems from the fact that forming the convex sum over an optimized set of pure states to realize a mixed state of interest, that is, $\rho = \sum_i p_i \ket{\psi_i}\bra{\psi_i}$ can lead to a cancellation of information encoded in the reduced marginals, for instance in the reduced single-qubit density operators (visualized by $Q$-Beads). At the same time a build-up of information can occur for higher-order subsystems. Consequently, the information encoded in the resulting higher-order correlation function exceeds what can be recovered from the lower-order subsystems.

Thus, we can extract the genuinely redundant information with respect to the mixed state density operator $\rho$ and use it to isolate the genuinely correlation-based component $\rho_\Sigma$ of the mixed state density operator components $\rho_C$. This is again a form of connected correlation function where the expectation values are computed with respect to $\rho$ instead of the individual contributing pure states.
In the bipartite case, this corresponds to:
\begin{equation}
	\rho_{\Sigma,12}= \rho_{C,12} - \frac{1}{4} \sum_{\alpha,\beta=x,y,z} \braket{\sigma_{1\alpha}}_{\rho} \braket{\sigma_{2_\beta}}_{\rho} \sigma_{1\alpha} \sigma_{2_\beta}.
	\label{eq:rho_CD_bipartite}
\end{equation} 
The approach can again be generalized to arbitrary larger multiqubit systems (see \cref{eq:tripartite_E_correlationfunction}, or supplementary \cref{sup:4partconn} and \cref{sec:tripartite_connected} for further details) by computing the corresponding $n$-partite compound correlation function density operator components and eliminating the redundant information with respect to the overall mixed state density operator $\rho$ as shown in \cref{eq:rho_CD_bipartite}. In the generalized BEADS representation, operator components $\rho_{\Sigma}$ are visualized by $\Sigma$-Beads which represent manifestations of genuine correlations arising from the separable structure of $\rho$ which may involve both, classical or nonclassical (quantum-discord-type) correlations.

\subsubsection*{Total correlation: M-Beads}

The ability to isolate correlation function density operator components that correspond to $n$-partite entanglement (see \cref{sec:ECTBeads}) and genuine $n$-partite correlations emerging from the separable structure of a mixed quantum state (\cref{sec:SigmaBeads}) further motivates a combined representation corresponding to the state's \textit{total correlation}, sometimes referred to as \textit{\underline{m}utual information}.
As it turns out, the $n$-partite connected correlation function of a mixed quantum state $\rho$, where all expectation values are computed with respect to $\rho$ itself, can be attributed to genuine $n$-partite correlation without distinguishing the underlying types of correlation. 

For a bipartite density operator $\rho$, we obtain this genuinely correlation-based correlation function density operator component $\rho_{M,12}$  according to:
\begin{align}
	\rho_{M,12} &= \rho_{E,12} + \rho_{\Sigma,12} \nonumber \\
	&= \rho_{E,12} + \rho_{C,12} - \frac{1}{4} \sum_{\alpha,\beta=x,y,z} \braket{\sigma_{1\alpha}}_{\rho} \braket{\sigma_{2_\beta}}_{\rho} \sigma_{1\alpha} \sigma_{2_\beta}\nonumber\\
	&= \rho_{T,12} - \frac{1}{4} \sum_{\alpha,\beta=x,y,z} \braket{\sigma_{1\alpha}}_{\rho} \braket{\sigma_{2_\beta}}_{\rho} \sigma_{1\alpha} \sigma_{2_\beta}\nonumber\\
	&=\rho- \frac{1}{4} \left[\sum_{\alpha,\beta=x,y,z} \braket{\sigma_{1\alpha}}_{\rho} \braket{\sigma_{2_\beta}}_{\rho} \sigma_{1\alpha} \sigma_{2_\beta} + \sum_{\alpha=x,y,z} \braket{\sigma_{1\alpha}}_{\rho} \sigma_{1\alpha} + \sum_{\beta=x,y,z} \braket{\sigma_{2\beta}}_{\rho} \sigma_{2\beta} + \mathbbm{1}_4 \right].
	\label{eq:rho_M_bipartite}
\end{align}
In the generalized BEADS representation, density operator components $\rho_{M}$ are visualized by corresponding $M$-Beads. Note that the approach implies useful additivity relations, which are a direct consequence of the natural properties of correlation functions. A generalization to arbitrarily large systems is again possible using the corresponding connected correlation function relations.

As is particularly obvious from the third line of \cref{eq:rho_M_bipartite}, genuinely correlation-based density operator components $\rho_{M}$ encode the part of information of the full correlation function $\Gamma_T$, or equivalently $\rho_{T}$, which cannot be recovered by observing the reduced marginals (here, the reduced single-qubit operators). In the bipartite case, this is conceptually akin to the commonly used total correlation measure \textit{quantum mutual information} \cite{ollivier_quantum_2001}, which in principle applies the same approach at the level of von Neumann entropies:
\begin{equation}
	I(\rho)= S(\rho_A)+ S(\rho_B)-S(\rho),
	\label{eq:total_correlation}
\end{equation}
where the difference of the sum of von Neumann entropies of the individual reduced subsystems $A$ and $B$, and the composite system $AB$ is taken. $I(\rho)$ quantifies how much information is missing when only adding up the locally accessible information of the individual subsystems. 

\subsubsection*{Classical correlation: $M_{Cl}$-Beads}
\label{sec:genuine_tripartite_$M_{Cl}$}

Having obtained the density operator component corresponding to the total correlation $\rho_M$, it is now possible to determine the contribution $\rho_{M_{Cl}}$ corresponding to the classically accessible part of $\rho_M$, which thus, can be attributed to \textit{\underline{cl}assical correlation}.

We establish the construction of operator components $\rho_{M_{Cl}}$ based on the definition of classical mutual information using conditional entropies and its generalization to quantum systems \cite{ollivier_quantum_2001,fanchini_lectures_2017}. The quantity $M_{Cl}(B:A)_{\{\Pi_l^{A}\}}$ is called the classically accessible mutual information, or in short, the \textit{\underline{cl}assical \underline{m}utual information} or classical correlation of a (bipartite) quantum system. The classical mutual information is defined as the maximum difference between the von Neumann entropy of a subsystem $B$ and the conditional entropy $S(B|\{\Pi_l^{A}\})$ with respect to all possible sets of projectors $\{\Pi_l^A\}$ \cite{ollivier_quantum_2001}:
\begin{equation}
	M_{Cl}(B:A)_{\{ \Pi_l^{A}\}} = \max_{\{\Pi_l^{A}\}} \Big[ S(B) - S(B|\{\Pi_l^{A}\}) \Big],
	\label{eq:classical_mutual}
\end{equation}
where the conditional entropy is obtained as \cite{ollivier_quantum_2001}
\begin{equation}
	S\left(B|\left\{\Pi_l^{A} \right\}\right)= \sum_l p_l S\big(\rho_{B|\Pi_l^{A}}\big),
	\label{eq:conditional_entropy_quantum}
\end{equation} 
using the density operator of subsystem $B$ corresponding to the outcome $\Pi_l^A$:
\begin{equation}
	\rho_{B|\Pi_l^{A}}=\text{Tr}_{A}\left[ \frac{\Pi_l^{A} \rho_{AB}  \Pi_l^{A}} {\text{Tr}[\Pi_l^{A} \rho_{AB}  \Pi_l^{A}]}\right].
	\label{eq:projected_subsystemB}
\end{equation}
Thus, the term $S\big(\rho_{B|\Pi_l^{A}}\big)$ describes the von Neumann entropy of the reduced subsystem $B$ under the condition that subsystem $A$ was measured and the outcome corresponding to the projector $\Pi_l^A$ was obtained. As a consequence, $M_{Cl}(B:A)_{\{ \Pi_l^{A}\}}$ is an asymmetric property and needs to be assessed with respect to local measurements of both parties $A$ and $B$ in order to obtain an informationally complete picture of a bipartite quantum state.

As determining the classical mutual information involves a maximization (see \cref{eq:classical_mutual}), or equivalently, a minimization of the conditional entropy shown in \cref{eq:conditional_entropy_quantum}, a set of local projectors $\{\Pi_j^{(k)}\}$ needs to be optimized using \cref{eq:conditional_entropy_quantum} as a cost function.
We can then extract the classically accessible part of the genuinely correlation-based operator component $\rho_{M,\mathcal{G}}$ corresponding to the classical correlation of (sub)system $\mathcal{G}$ by a projection using the optimized set of projectors:
\begin{equation}
	\rho_{M_{Cl},\mathcal{G}}^{(k)}= \sum_l \Pi_l^{(k)} \rho_{M,\mathcal{G}} \, \Pi_l^{(k)},
	\label{eq:rho_CM_projected}
\end{equation}
where $k$ specifies the projected qubit(s).
In the generalized BEADS representation, density operator components $\rho_{M_{Cl},\mathcal{G}}^{(k)}$ are represented by corresponding $M_{Cl}^{(k)}$-Beads.

Note that for the purpose of visualization, the approach can be conveniently generalized to $n$-partite systems by reformulating the conditional entropy from \cref{eq:conditional_entropy_quantum} for arbitrary sets of locally projected and non-measured qubits. For instance, in a tripartite system it is then possible to assess which parts of the individual $n$-partite operator components $\rho_M$ can be classically accessed, when one or two out of the three qubits in the system are measured locally. Further details are provided in \cref{app:class_corr}.

\subsubsection*{Classically accessible parts of the separable structure of a density operator: $\Sigma_{Cl}$-Beads}
By applying the optimized projectors $\{\Pi_l^{(k)}\}$ to $\rho_{\Sigma,\mathcal{G}}$ instead of $\rho_{M,\mathcal{G}}$, density operator components $\rho_{\Sigma_{Cl},\mathcal{G}}^{(k)}$ -- which correspond to the components of the \textit{\underline{cl}assically} accessible genuinely correlation-based part of the density operator corresponding to a (sub)system $\mathcal{G}$ that arises from its separable structure -- can be isolated:
\begin{equation}
	\rho_{\Sigma_{Cl},\mathcal{G}}^{(k)}= \sum_l \Pi_l^{(k)} \rho_{\Sigma,\mathcal{G}} \, \Pi_l^{(k)}.
	\label{eq:rho_CL_projected}
\end{equation}
Operator components $\rho_{\Sigma_{Cl},\mathcal{G}}^{(k)}$ are visualized by so-called $\Sigma_{Cl}^{(k)}$-Beads.

$\Sigma_{Cl}^{(k)}$-Beads provide a useful additional perspective as optimizing the classically accessible mutual information as introduced in the previous chapter does not query the source of the accessed information. That is, as will be shown later, $M_{Cl}^{(k)}$-Beads can unintuitively capture structures of a density operator that originate from entanglement. On the other hand, some parts of the separable structure of the mixed state density operator of interest may be classically inaccessible. Comparing $\Sigma_{Cl}^{(k)}$-Beads with $M_{Cl}^{(k)}$-Beads, $M$-Beads, and $\Sigma$-Beads allows to easily identify genuinely classically correlated states and distinguish them from other classes of correlated states.

\subsubsection*{Nonclassical structures: D-Beads}
The classical mutual information as defined in \cref{eq:classical_mutual} is closely associated with the concept of quantum discord, which measures the entirety of nonclassical correlations of a bipartite quantum state \cite{ollivier_quantum_2001}.

The definition of quantum discord builds upon the two definitions of mutual information (see \cref{eq:total_correlation} and \cref{eq:classical_mutual}), which are not identical in the quantum case. For bipartite states, the quantum discord $\delta$ is defined as the difference between the two quantities \cite{ollivier_quantum_2001}:
\begin{equation}
	\delta(A:B)_{\{\Pi_l^{A}\}}= I(B:A)-J(B:A)_{\{\Pi_l^{A}\}}= S(A)-S(B,A)+	S(B|\left\{\Pi_l^{A}\right\}).
	\label{eq:quantum_discord}
\end{equation} 
$\delta$ is a non-negative and asymmetric quantity, as it again depends on the measured party. 

For the generalized BEADS representation, we adopt \cref{eq:quantum_discord} on the level of density operator components to identify the nonclassical genuinely correlation-based structure of a density operator, or in other words, the classically inaccessible parts $\rho_{D,\mathcal{G}}^{(k)}$  of the operator components $\rho_{M,\mathcal{G}}$:
\begin{equation}
	\rho_{D,\mathcal{G}}^{(k)}= \rho_{M,\mathcal{G}}-\rho_{M_{Cl},\mathcal{G}}^{(k)},
	\label{eq:rho_D}
\end{equation}
where $k$ again denotes the set of projected qubits and $\mathcal{G}$ is the (sub)system of interest.
The corresponding type of beads is called $D$-Beads marking the conceptual relation to \textit{quantum \underline{d}iscord}.

\color{black}
\subsubsection*{Nonclassical separable structures: $\Sigma_D$-Beads}
The classically inaccessible separable structure of a mixed density operator can be obtained as
\begin{equation}
	\rho_{\Sigma_D,\mathcal{G}}^{(k)}= \rho_{\Sigma,\mathcal{G}}-\rho_{\Sigma_{Cl},\mathcal{G}}^{(k)}.
	\label{eq:rho_SigmaD}
\end{equation}
Operator components $\rho_{\Sigma_D,\mathcal{G}}^{(k)}$ are visualized by $\Sigma_{D}^{(k)}$-Beads and are particularly useful for the identification of nonclassically correlated separable states.

\subsubsection*{Information loss on reduced single-qubit density operators: $\Delta Q$-Beads}
In rare cases, optimized projections $\{\Pi_l^{(k)}\}$ for classically accessing a state $\rho$ of interest can restore the entire genuinely correlation-based structure of a state, i.e., $\rho_M = \rho_{M_{Cl}}^{(k)}$, even if the state is nonclassically correlated. In such cases the nonclassical character of the correlation commonly manifests in a loss of information on the reduced single-qubit density operators induced by the optimized measurement. In order to still be able to identify such unusual cases on a visual level in the generalized BEADS representation, we use $\Delta Q^{(k)}$-Beads which are direct representations of the difference between the pre- and post-measurement single-qubit reduced density operators:
\begin{equation}
	\rho_{\Delta Q_j}^{(k)}= \text{Tr}_{\bar{\mathcal{J}}} \left(\rho -  \sum_l \Pi_l^{(k)} \rho \,\Pi_l^{(k)}\right),
	\label{eq:deltaQ}
\end{equation}
where the operation $\text{Tr}_{\bar{\mathcal{J}}}$ denotes taking the partial trace over all except the $j$-th qubit. Note that the order of terms in the difference is chosen deliberately to avoid unintuitive changes of sign in the visualization.

\newpage
\subsubsection*{Overview of correlation function bead types}
\begin{figure}[H]
\centering
\includegraphics[width=.8\linewidth]{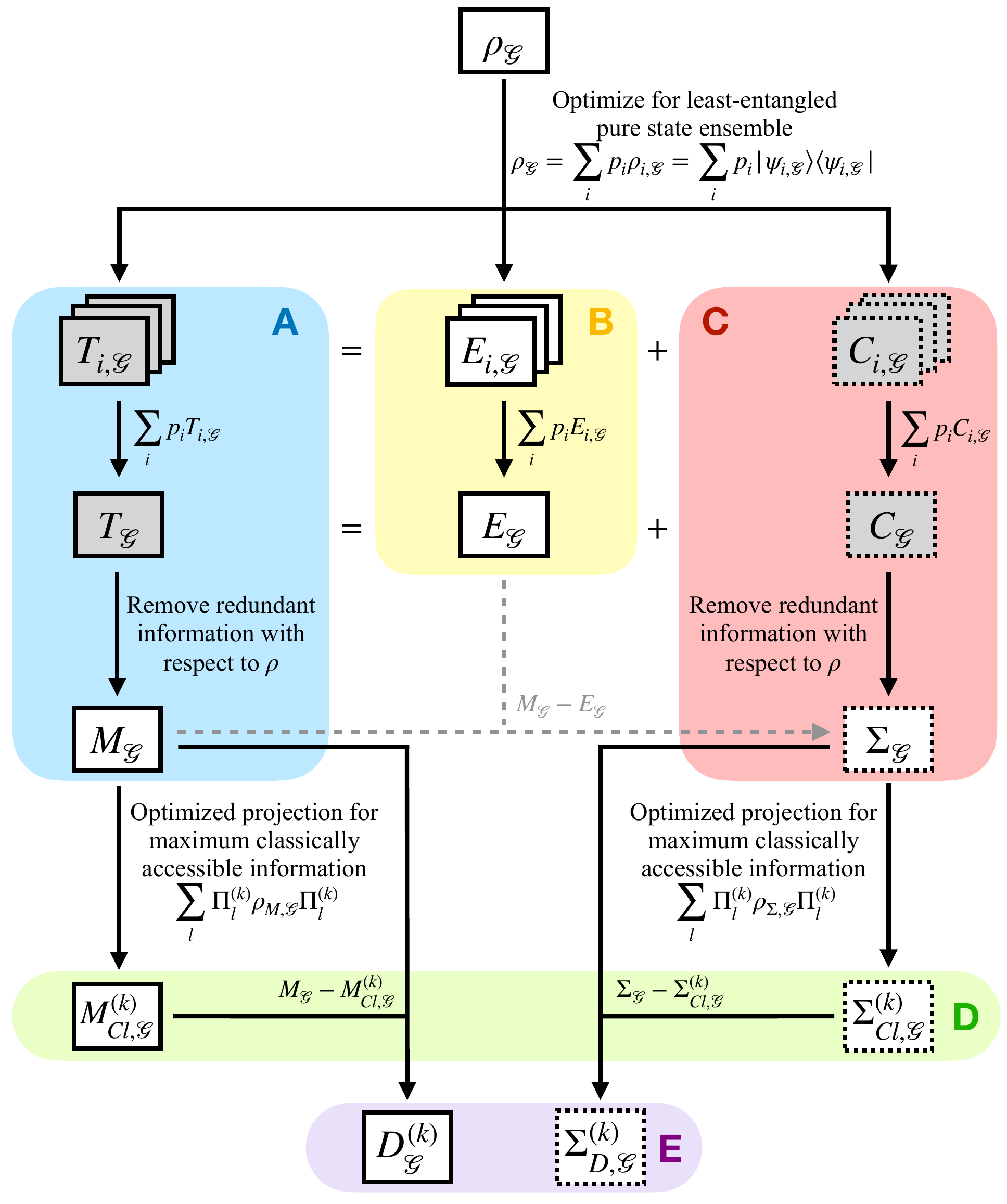}
\caption{Flow chart summarizing the construction of correlation function beads corresponding to arbitrary subsystems $\mathcal{G}$ in the generalized BEADS representation of mixed quantum states. The bead types are grouped in colored boxes indicating the correlation types (\textbf{A}) total correlation,  (\textbf{B}) entanglement,  (\textbf{C}) genuine correlations which arise from the separable structure of the density operator $\rho$,  (\textbf{D}) classical correlation, and  (\textbf{E}) nonclassical correlation. Bead types which contain redundant (irrelevant information) are displayed with a grey background. Boxes of bead types which do only show parts of the information required to assess commonly discussed correlation types (that is, total, classical, and nonclassical correlation as well as entanglement; see \cref{tab:correlation_visualization} for comparison) have dashed lines. Required steps in the workflow are indicated by black solid arrows, whereas optional pathways are sketched using greyed out dashed arrows. The applied colors are used to improve the readability of the figure and do not relate to the color schemes used in the representation.}
\label{fig:flowchart}
\vspace{-.5cm}
\end{figure}

The workflow to generate the different types of correlation function beads used in the generalized BEADS representation of mixed states is illustrated in \cref{fig:flowchart}. 
An overview of all correlation function bead types used in the generalized BEADS representation is provided in \cref{tab:correlation_visualization}. Note that the table needs to be read in the following way: First, the type of visualized correlation for a distinct class of beads can be directly inferred from the \textit{Correlation type} columns. By consulting the columns titled \textit{Visualized structure of $\rho$}, it is then possible to assess which underlying structure of the density operator $\rho$ is actually captured by the bead type (and equivalently, by the corresponding type of correlation). For instance, we find that $M_{Cl}$-Beads correspond to the classical correlation (classically accessible mutual information) of a state $\rho$. However, on the level of density operator structures, this classical access by an optimized measurement (see previous subsections) can actually extract operator components that are in part responsible for the entanglement of $\rho$ or parts of its separable structure.

\begin{table}[H]
	\caption{Overview of bead types, the correlation types they are specifically designed to visualize, whether the bead type captures (parts of) the entangling- or separable structure of a density operator, and the applied color schemes (cf. \cref{fig:ColorSchemes}). The symbols \checkmark, (\checkmark), and $-$ indicate that a criterion is fully met, partially met (i.e., the criterion can but need not necessarily be fully met), or not met at all, respectively.}
	\hspace{-1cm}
	\renewcommand{\arraystretch}{1.2}
	\begin{tabular}{p{0.06\textwidth}<{\centering} p{0.08\textwidth}<{\centering} p{0.09\textwidth}<{\centering} p{0.08\textwidth}<{\centering} p{0.07\textwidth}<{\centering} p{0.001\textwidth}<{\centering} p{0.09\textwidth}<{\centering} p{0.09\textwidth}<{\centering} p{0.125\textwidth}<{\centering} p{0.09\textwidth}<{\centering}}
		\hline
		\hline
		\textbf{Type} & \multicolumn{4}{c}{\textbf{Correlation type}} & & \multicolumn{2}{c}{\textbf{\parbox[t]{0.18\textwidth}{\centering Visualized structures of $\rho$}}} & \textbf{Redundant information w.r.t. $\rho$} & \textbf{Color Scheme} \\
		\cline{2-5} \cline{7-8}
		& \textbf{Total} & \textbf{Entangle-ment} & \textbf{\parbox[t]{0.08\textwidth}{\centering{Non-class.}}} & \textbf{\parbox[t]{0.07\textwidth}{\centering{Class.}}}& & \textbf{\parbox[t]{0.09\textwidth}{\centering{Entan-\\gling}}} & \textbf{\parbox[t]{0.09\textwidth}{\centering{Separa-ble}}} & & \\
		\hline\\[-5pt]
		\textbf{T}  & \checkmark &  $-$ &  $-$ &  $-$ & & \checkmark & \checkmark & yes & color wheel \\
		\textbf{E}  & $-$ & \checkmark & $-$ & $-$    &    &  \checkmark & $-$        & no  & yellow-blue \\
		\textbf{C}  & (\checkmark) & $-$ & $-$  & $-$ & & $-$        & \checkmark & yes & red-green \\
		\bm{$\Sigma$} &  (\checkmark) & $-$  & $-$ & $-$ & & $-$        & \checkmark & no  & red-green \\
		\textbf{M}  & \checkmark & $-$ &  $-$ & $-$ & & \checkmark & \checkmark & no  & color wheel \\
		\bm{$M_{Cl}$} & $-$ & $-$        & $-$ & \checkmark & & (\checkmark) & (\checkmark) & no  & color wheel \\
		\bm{$\Sigma_{Cl}$} & $-$ &  $-$        & $-$        & \checkmark & & $-$        & (\checkmark) & no  & red-green \\
		\textbf{D}  & $-$ &  $-$ & \checkmark       & $-$ & & (\checkmark     ) & (\checkmark) &  no  & color wheel \\
		\bm{$\Sigma_D$}  & $-$ &  $-$ & (\checkmark)        & $-$      & & $-$        & (\checkmark) & no  & red-green \\
		\hline
		\hline
	\end{tabular}
	\label{tab:correlation_visualization}
\end{table}

\subsection{Auxiliary elements of the generalized BEADS representation: Entanglement arcs}\label{sec:entarcs}
In addition to the discussed types of beads, we augment generalized BEADS representations of entangled quantum states by so-called entanglement arcs. These arcs connect the $Q$-Beads corresponding to some subsystem $\mathcal{G}$ and are scaled in diameter with respect to a suitable entanglement quantifier.

In case of bipartite subsystems $\mathcal{G} = \{jk\}$, the corresponding arcs are scaled by means of the two-tangle, i.e., the squared concurrence $\tau_{2, jk} = C^2_{jk}$.

Analogously, for $n$-partite systems, we primarily use the $n$-tangle \cite{coffman_distributed_2000, wong_ntangle_2001} to scale corresponding $n$-partite subsystem entanglement arcs. However, these types of arcs are only useful for visually indicating GHZ-like entanglement, whereas they commonly have zero diameter for W-type entangled states. All tangle arcs are drawn in black color.

Thus, in tripartite subsystems, we explicitly complement the visualization by a second type of arc which is scaled by the quantifier
\begin{equation}
\mathcal{W}_3 = |\overline{\mathcal{N}}_3 - \tau_3|.
\label{eq:Wquant}
\end{equation}
Here, $\overline{\mathcal{N}}_3$ is the modified connected correlation function norm, which is mapped such that it adopts the same values as the three-tangle $\tau_3$ for structurally ideally GHZ-type entangled states and different values otherwise. By applying the absolute difference as shown in \cref{eq:Wquant} for scaling the dedicated entanglement arc, it is possible to directly visually indicate W-type entanglement. To be able to distinguish these W-type arcs from three-tangle arcs, the former are colored green or violet depending on the sign of the difference $\overline{\mathcal{N}}_3 - \tau_3$. The full derivation of $\mathcal{W}_3$ is given in \cref{sec:EntQuant} and further details are presented in supplementary \cref{sec:SuppMathDetailsEntQuant}.
\newpage

\section{Results and Discussion}\label{sec:Results}
\subsection{Mixed bipartite states}\label{sec:mixed_bipartite}

In this section, we will introduce different classes of mixed quantum states -- product states, classically correlated, separable, and entangled states -- and will derive general rules for unambiguously identifying these classes of states using the generalized BEADS representation. We forward readers who are unfamiliar with mixed quantum states to \cref{sec:didactic_intro} of the supplementary material, where a more in-depth didactic introduction to mixed quantum states is provided using the generalized BEADS representation.
\subsubsection{Product states}
\label{sec:product_state}
 
\begin{figure}[htbp]
\centering
\includegraphics[width=\linewidth]{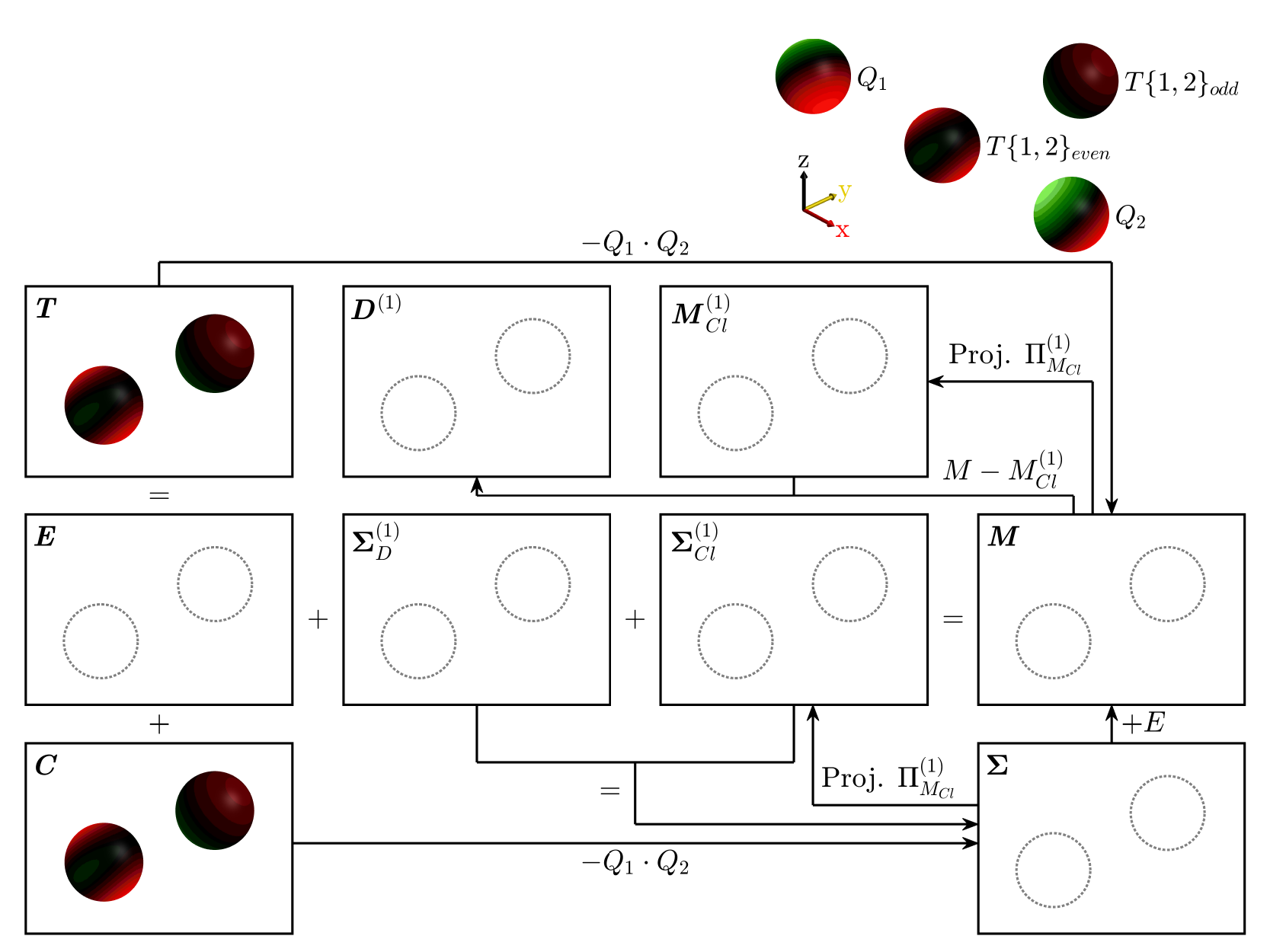}
\caption{Overview of the entirety of bead types for $\rho_\text{product}$ (see \cref{eq:rho_prod}). In the generalized BEADS picture, it is directly clear that the state is not genuinely correlated as there are no $M$-Beads. The $T$- and $C$-Beads are identical, but only encode redundant information, that is the products of the single-qubit reduced density operators $Q_1$ and $Q_2$.}
\label{fig:product_state_all}
\end{figure}

In the state space of mixed quantum states, the family of completely uncorrelated states is called product states
\begin{equation}
	\pi = \bigotimes_{k=1}^{N} \pi_k = \pi_1 \otimes \pi_2 \otimes ... \otimes \pi_N.
	\label{eq:product_state}
\end{equation}
For product states, the density operators of the reduced subsystems $\pi_k$ factorize by means of a tensor product, just like the expectation value of Pauli matrices factorizes into the product of single-qubit expectation values. As an example, consider the mixed product state
\begin{equation}
	\rho_\text{product}=
	\begin{pmatrix}
		0.24 & 0.39 + 0.13i \\
		0.39 - 0.13i & 0.76
	\end{pmatrix}
	\otimes
	\begin{pmatrix}
		0.16 & 0.30 - 0.21i \\
		0.30 + 0.21i & 0.84
	\end{pmatrix},
	\label{eq:rho_prod}
\end{equation}
which is visualized in \cref{fig:product_state_all}. The figure displays all types of beads. Correlation function beads which have a zero spherical function value in all spatial directions are omitted and replaced by a dashed-line placeholder.

In the upper right corner of \cref{fig:product_state_all}, an informationally complete visualization of $\rho_\text{product}$ is provided in terms of $Q$-, and $T$-Beads. The lower part of the figure displays all different types of correlation function beads (see \cref{sec:MixedStateCorrFunc} for details on the construction) which are useful for characterizing the correlation structure of $\rho_\text{product}$. The $Q$-Beads $Q_1$ and $Q_2$ show colors other than black. Hence, the corresponding reduced density operators are not maximally mixed. 

\begin{figure}[htbp]
\centering
\includegraphics[width=.9\linewidth]{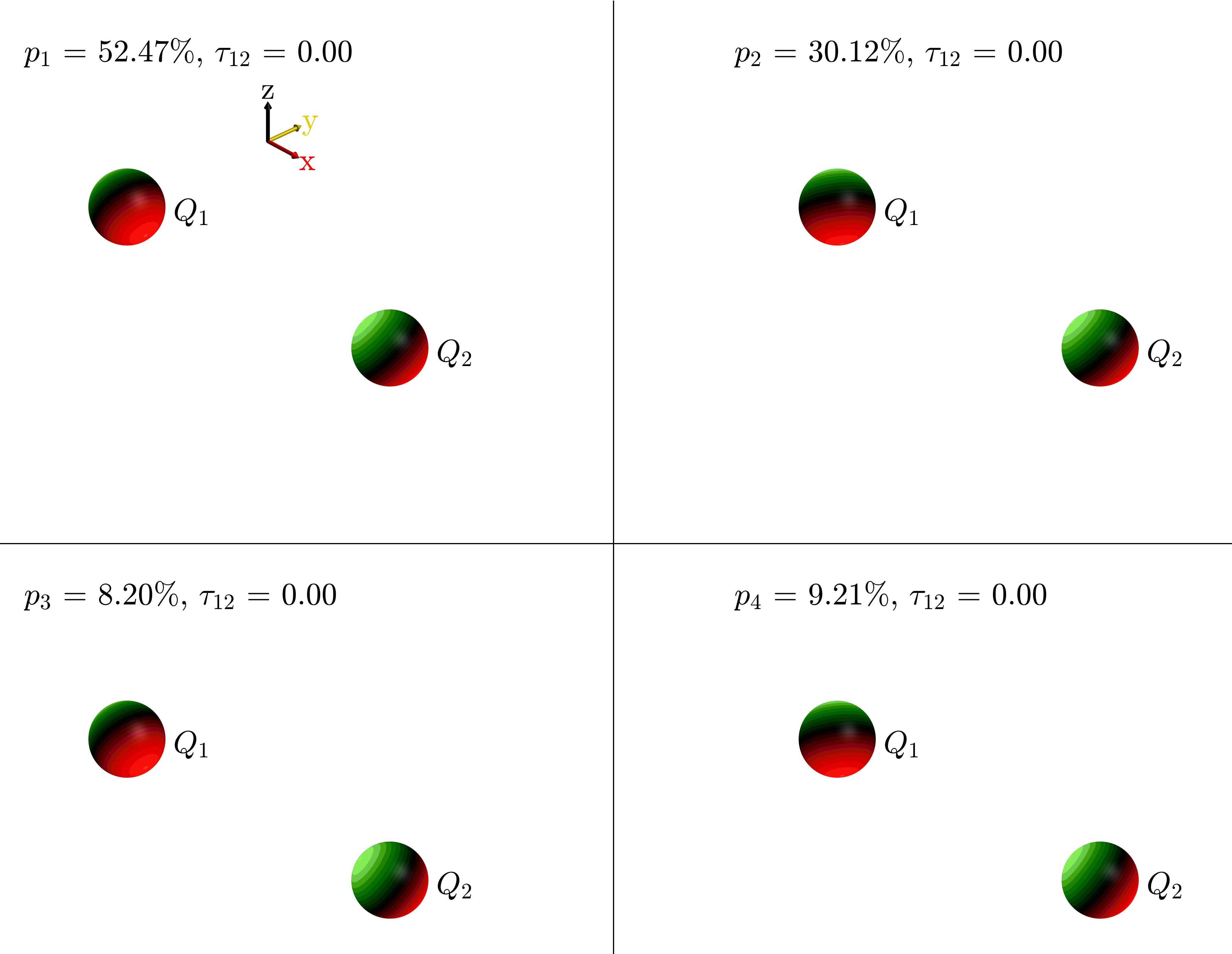}
\caption{$E$-Beads representation of an optimized pure state decomposition of $\rho_\text{product}$. Indeed, the pure states do not show any $E$-Beads for which they are uncorrelated product states. As a direct consequence, the corresponding mixed state does also not show any $E$-Beads as these result as convex sums of the pure state $E$-Beads, which here are zero in all spatial directions. The weights and two-tangles of each pure state are plotted in the top left corner of every state's subfigure.}
\label{fig:product_state_decomposition}
\end{figure}

Mixed product states can always be realized by an ensemble of pure product states \cite{ding_quantum_2023}, which do not show entanglement-based beads ($E$-Beads) in their corresponding pure state BEADS visualization. Such an ensemble is visualized for $\rho_\text{product}$ in \cref{fig:product_state_decomposition}. Hence, following \cref{eq:connmixed}, $\rho_\text{product}$ does not show mixed state $E$-Beads, and it is immediately clear that the state is not entangled. 

As a direct consequence, $\rho_\text{product}$ only has a separable uncorrelated structure and the total correlation function ($T$-Beads) and the compound correlation function components ($C$-Beads) become identical.
Indeed, in this case all nonzero spherical function values can be attributed to products of information encoded in the reduced marginals $Q_1$ and $Q_2$. This does not correspond to genuine correlations and consequently, $\rho_\text{product}$ does not show any $M$-Beads or $\Sigma$-Beads in its generalized BEADS representation and all dependent bead-types, i.e., $M_{Cl}$, $D$, $\Sigma_{Cl}$, and $\Sigma_D$, automatically vanish as well.

\begin{enumerate}[start = 1, label=\textbf{Result \arabic*:}, itemindent=.95cm, labelindent=1cm]
	\item \label{res:BEADS1} Unless the state is the maximally mixed state, product states in general have nonzero and identical $T$- and $C$-Beads despite being uncorrelated.
	\item \label{res:BEADS2} Mixed product states can always be decomposed into a set of pure product states which are fully separable. Hence, there are neither $E$-Beads in the pure state nor in the corresponding mixed state BEADS representations.
	\item \label{res:BEADS3} Product states do not show bead types that are associated with genuine correlations of the state, that is, no $E$-, $M$-, $\Sigma$-, $M_{Cl}$-, $D$-, $\Sigma_{Cl}$-, or $\Sigma_D$-Beads.
	\item \label{res:BEADS4} If a state does not show any $M$-Bead, it is a product state. 
\end{enumerate}

\newpage
\subsubsection{Classically correlated states}
\label{sec:classical_state}

\begin{figure}[H]
	\centering
	\includegraphics[width=\linewidth]{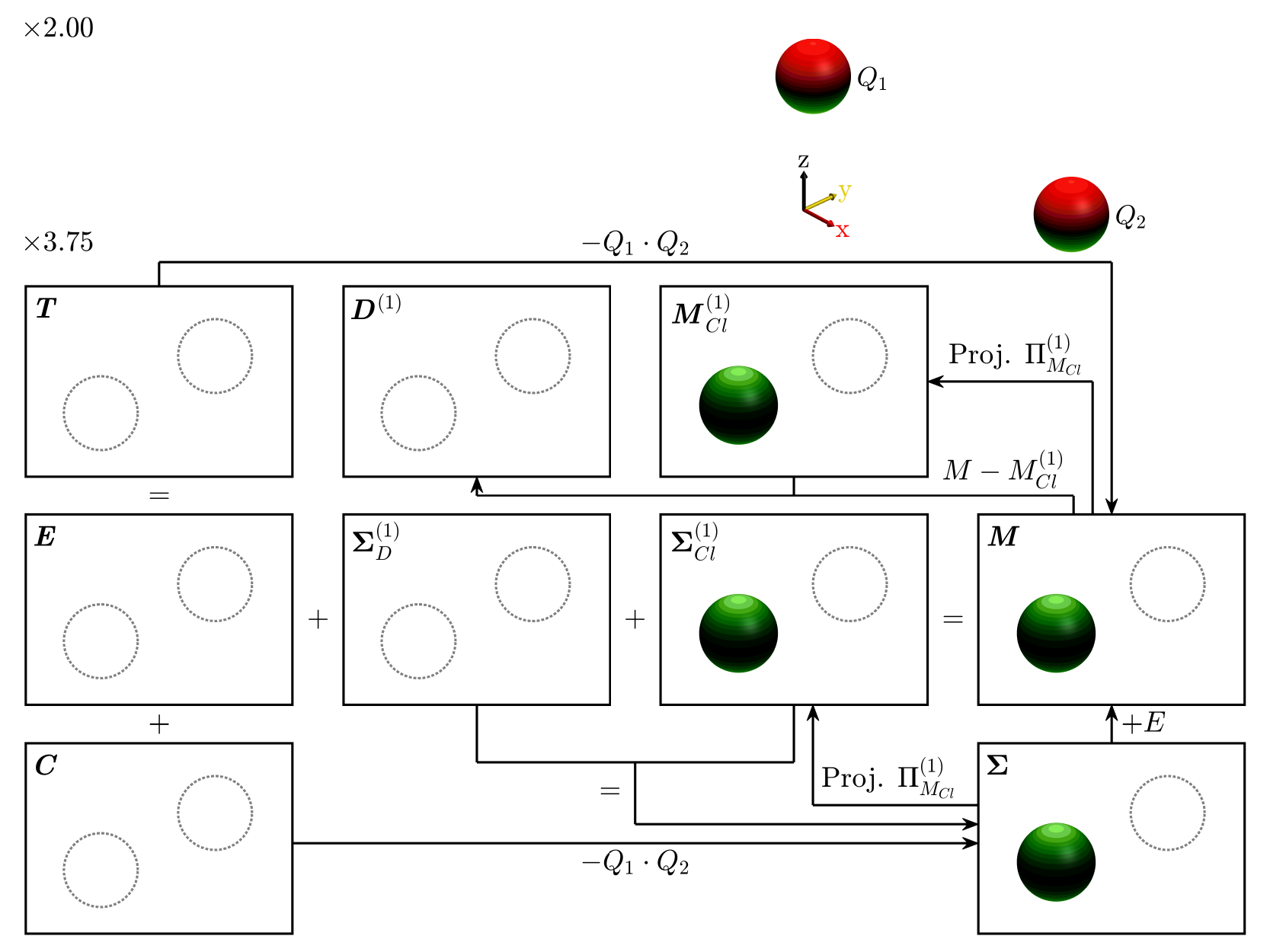}
	\caption{Complete generalized BEADS representation of the classical-classical state $\rho_\text{class-class}$ (\cref{eq:classclass}). Although the state has no total correlation function ($T$-Beads), it is still genuinely correlated ($M$-Bead). Since there are no $E$-Beads, all genuine correlations arise from the separable structure of the state, which is reflected in identical $M$- and $\Sigma$-Beads. Coinciding $M_{Cl}^{(1)}$-, $\Sigma_{Cl}^{(1)}$-, $\Sigma$- and $M$-Beads allow to unambiguously identify the state as being classically correlated with respect to an optimized local measurement of qubit~1. For better visibility, the spherical function values are scaled by the scaling factors shown on the left margin. Note that different scaling factors are used for the $Q$-Bead picture (upper right corner) and the correlation function beads visualizations.}
	\label{fig:classclass_all}
\end{figure}

For classically correlated states $\chi$ each qubit can be expressed in a single orthonormal basis:
\begin{equation}
	\chi = \sum_{r,s,...w} p_{rs...w} \, \ket{r}\bra{r} \otimes \ket{s}\bra{s} \otimes ... \otimes \ket{w}\bra{w}, p_{rs...w} >0,\,\,\, \sum_{r,s,...w} p_{rs...w} =1.
	\label{eq:classically_correlated}
\end{equation}
In other words, is the state diagonal in a single product basis with the local basis chosen independently for each subsystem such that it is possible to find local projectors which project qubits $1, 2, ...\, N$ onto the local eigenstates {$\ket{r}$},{$\ket{s}$} and {$\ket{w}$}. When a joint measurement is performed using these sets of projectors  $\{P_1^{(r)} \otimes P_2^{(s)} \otimes ... \otimes P_N^{(w)}\}$, no disturbance is introduced on the state $\chi$, that is, the post-measurement is identical to the pre-measurement state, and it is therefore categorized as a classically correlated state \cite{luo_using_2008}.
In the following, we examine the classical-classical state
\begin{equation}
	\rho_\text{class-class}= \frac{1}{2} \ket{00}\bra{00}+ \frac{1}{4} \left(\ket{01} \bra{01}+ \ket{10} \bra{10}\right).
	\label{eq:classclass}
\end{equation}
As can be seen in \cref{fig:classclass_all}, this state atypically does neither show total correlation function beads ($T$-Beads), nor entanglement-based beads ($E$-Beads), nor compound correlation function beads ($C$-Beads) in its generalized BEADS representation. The absence of $T$- and $C$-Beads might lead to the wrong conclusion that this state is uncorrelated. However, evaluating the information encoded in the reduced density operators of the overall mixed state, genuinely correlation-based  $M$- and $\Sigma$-Beads are obtained, which unambiguously unveil that the state is genuinely correlated. 

Interestingly, the chosen state is an example of a special case where the effects of redundant information and genuine correlations cancel such that outcomes of joint local measurements are strictly uncorrelated as all $T$-Beads have spherical function value, i.e., expectation value zero in all spatial directions. 

The lack of $E$-Beads again allows to tell that the state is fully separable, whereas the presence of $M$- and $\Sigma$-Beads signals genuine correlation. Performing an optimized local measurement on one qubit, here, qubit~1 to maximize the classically accessible information, yields $M_{Cl}^{(1)}$- and $\Sigma_{Cl}^{(1)}$-Beads according to \cref{eq:rho_CM_projected} and \cref{eq:rho_CL_projected}, respectively. It is directly clear that these projections fully recover the $M$- and $\Sigma$-Beads, for which they result in being identical to the $M_{Cl}^{(1)}$- and $\Sigma_{Cl}^{(1)}$-Beads. There are no classically inaccessible components, which can be deduced from the lack of $D^{(1)}$-and $\Sigma_D^{(1)}$-Beads. Thus, the state is classically correlated with respect to a local measurement of qubit 1. Naturally, the correlation is exclusively encoded in the separable structure of the underlying density operator which coincides with classically correlated states being a special class of separable states and which is reflected in coinciding $M$- and $\Sigma$-Beads.

The exact same observations can be made when performing an optimized local measurement of qubit~2. A corresponding visualization is omitted to avoid redundancy. 

\begin{enumerate}[start = 5, label=\textbf{Result \arabic*:}, itemindent=.95cm, labelindent=1cm]
	\item \label{res:BEADS5} Classically correlated states with respect to local measurements of qubit(s) $k$ have nonzero and identical $M$-, $M_{Cl}^{(k)}$-, $\Sigma$-, and $\Sigma_{Cl}^{(k)}$-Beads. They do not show $E$-, $D^{(k)}$-, or $\Sigma_D^{(k)}$-Beads.
	\item \label{res:BEADS6} If $M$- and $M_{Cl}$-Beads of a quantum state are identical in all and nonzero in at least one spatial direction, the state is strictly classically correlated.
	\item \label{res:BEADS7} The absence of $T$- or $C$-Beads does not relate to a state being uncorrelated. As a consequence, correlated states may show no correlation of measurement outcomes despite being genuinely correlated.
\end{enumerate}

\subsubsection{Separable states}\label{sec:sep_states}
Separable states $\sigma$ exhibit the structure of a convex sum over product states:
\begin{equation}
	\sigma = \sum_i p_i \, \pi_1^{(i)} \otimes \pi_2^{(i)}  \otimes ... \otimes  \pi_N^{(i)},\, p_i >0,\,\,\, \sum_i p_i =1.
	\label{eq:separable_state}
\end{equation}
Werner states \cite{werner_state_1989} are a canonical example of separable states for suitable parameter choices. Consider the Werner state centered on the Bell state $\ket{\Phi^+}=\left(\ket{00}+\ket{11}\right)/\sqrt{2}$:
\begin{equation}
	\rho(z)= \frac{1-z}{4}\mathbbm{1} + z\ket{\Phi^+}\bra{\Phi^+},
	\label{eq:zurek_example_werner}
\end{equation}
which is separable for $z\leq\frac{1}{3}$ and entangled for {$z>\frac{1}{3}$} \cite{werner_state_1989, ollivier_quantum_2001}. As this subsection focuses on the set of separable states, $\rho(z)$ will be visualized for {$z=0.3$}. A central property of Werner states is the invariance of the classical mutual information under unitary rotation of the applied projector, that is, the classical mutual information is independent of the choice of local projectors. Here, $\rho_\text{separable}\,(z=0.3)$ serves as an illustrative example for a separable state that possesses nonclassical correlations as measured by quantum discord despite not being entangled. 

\begin{figure}[H]
\centering
\includegraphics[width=\linewidth]{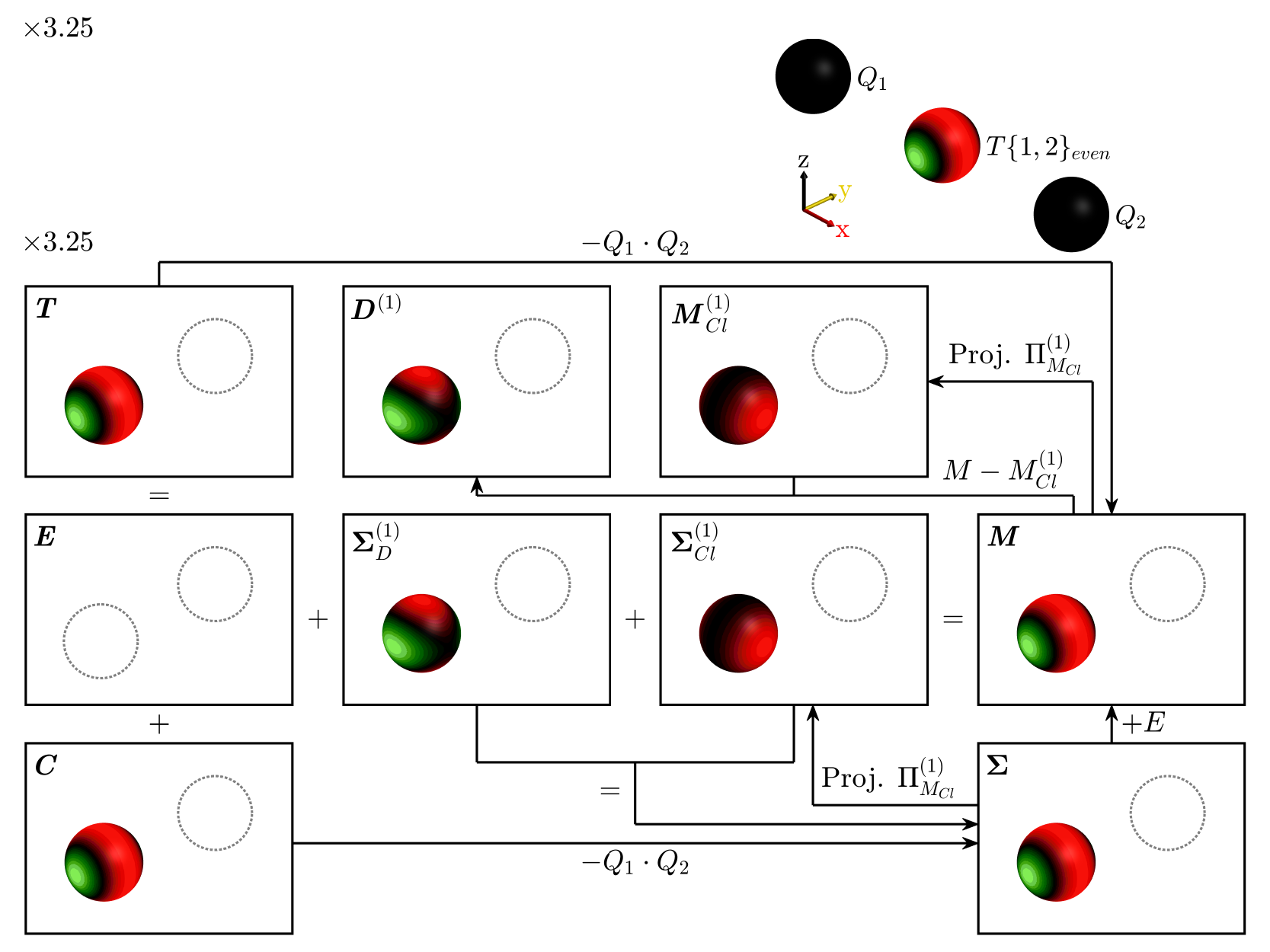}		
	\caption{Full generalized BEADS representation of the Werner state $\rho_\text{separable}\, (z=0.3)$ (\cref{eq:zurek_example_werner}) showing all relevant types of correlation function beads. As for all separable states, there are no entanglement-based $E$-Beads and thus, the $T$- and $C$-Beads coincide. The state is nonclassically correlated, which can be directly inferred from the occurrence of nonzero $D^{(1)}$- and $\Sigma_D^{(1)}$-Beads, which visualize the classically inaccessible part of $\rho_\text{separable}\, (z=0.3)$ for the chosen classical access by a local projection of qubit~1. Complementarily, the $M_{Cl}^{(1)}$- and $\Sigma_{Cl}^{(1)}$-Beads visualize the classically accessible information of the state. As $\rho_\text{separable}\, (z=0.3)$ is a Werner state, the choice of the local measurement basis for classically accessing the state is arbitrary.}
	\label{fig:zurek_werner}
\end{figure}

\cref{fig:zurek_werner} provides the full generalized BEADS representation of the discussed state. Again, as the state is separable, all $E$-Beads vanish and $T$- and the $C$-Beads result in being identical. In this case the $Q$-Beads are all-black, which implies that the reduced single-qubit density operators are maximally mixed and do not encode any information. Thus, the entire information encoded on the correlation function beads corresponds to genuine correlations of the state and $T$-, $C$-, $M$-, and $\Sigma$-Beads all coincide consequently.

\begin{figure}[htbp]          
	\centering
	\includegraphics[width=10cm, trim= 2cm 2cm 2cm 2cm, clip]{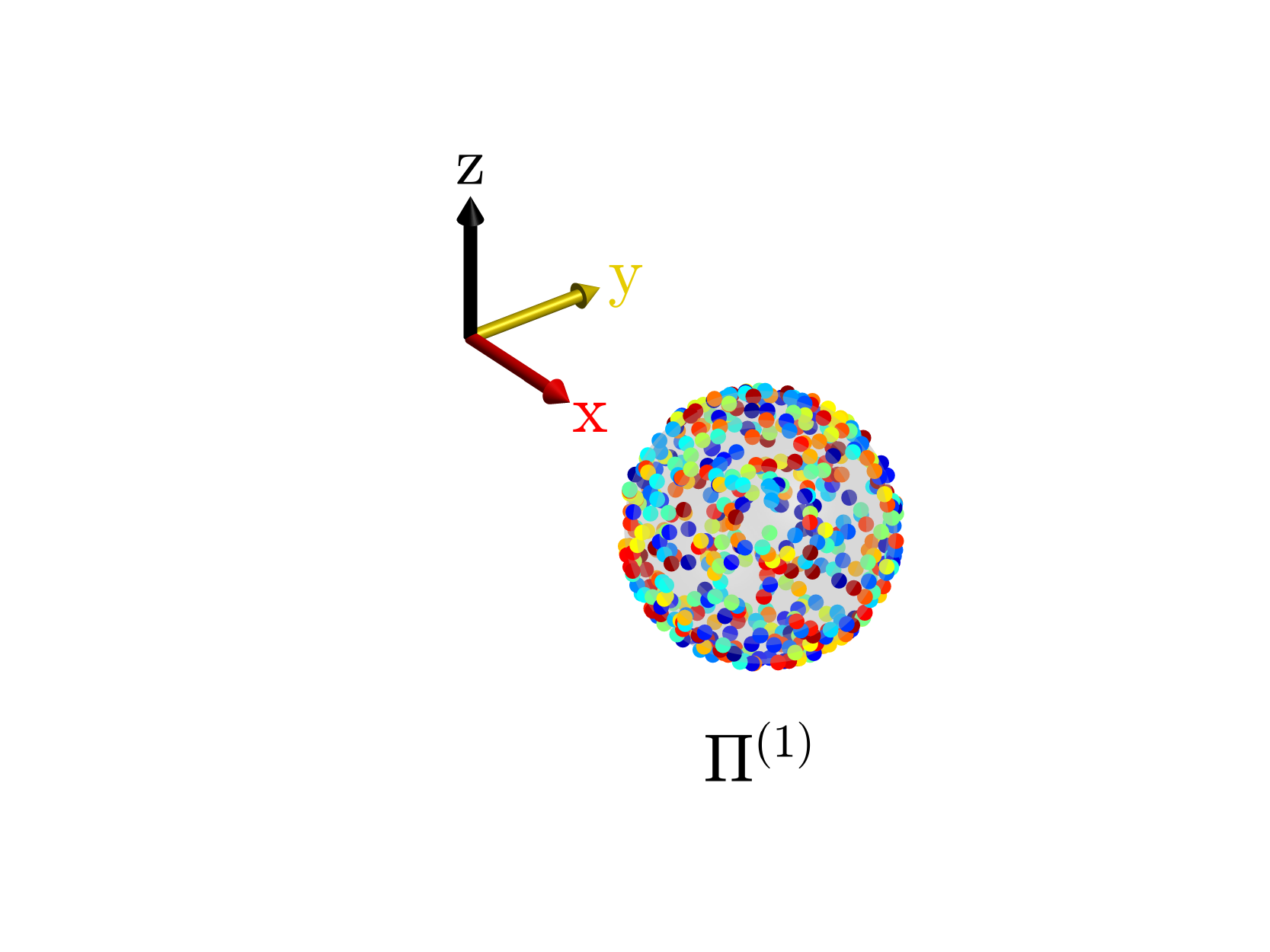}
	\caption{Classical access bloch vector analysis for $\rho_\text{separable}\,(z=0.3)$ of \cref{eq:zurek_example_werner}. The figure shows the local measurement directions that optimize the classical mutual information.  500 Haar-distributed initial unitary seeds were used. The optimum projector orientations are visualized as points on the Bloch sphere. As the state is a Werner state, the choice of the projector $\Pi^{(1)}$ is arbitrary which is reflected in the uniform distribution of points on the Bloch sphere.}
	\label{fig:blochvectors_zurek_werner_500_haar}
\end{figure}

For general nonclassically correlated separable states, the classically accessible mutual information does not coincide with the total mutual information of the state. Unlike in the case of classically correlated states (compare \cref{sec:classical_state}) this results in $M$- and $\Sigma$-Beads differing from $M_{Cl}^{(k)}$- and $\Sigma_{Cl}^{(k)}$-Beads. The classically inaccessible structure of the state is then visualized by $D^{(k)}$ and $\Sigma_D^{(k)}$-Beads, which are obtained as described in \cref{eq:rho_SigmaD} (cf. \cref{sec:methods_corrfunctions}) and which coincide for a specific choice of local projectors in case of separable states. Recall that for the given exemplary state $\rho_\text{separable}\, (z=0.3)$, the local measurement basis can be chosen arbitrarily. To further illustrate this property, the outcome of 500 measurement basis optimizations are shown \cref{fig:blochvectors_zurek_werner_500_haar}. Evidently, there is no preferred measurement basis and the obtained measurement directions are distributed uniformally across the Bloch sphere. Thus, it is clear that $M_{Cl}^{(k)}$-, $\Sigma_{Cl}^{(k)}$-, $D^{(k)}$-, and $\Sigma_D^{(k)}$-Beads must generally be non-unique if the state admits a degree of freedom in the choice of local projectors as different projections capture different components of the density operator of interest.

\begin{enumerate}[start = 8, label=\textbf{Result \arabic*:}, itemindent=.95cm, labelindent=1cm]
	\item \label{res:BEADS8} For separable states, $M$- and $\Sigma$-Beads, $M_{Cl}^{(k)}$- and $\Sigma_{Cl}^{(k)}$-Beads, as well as $D^{(k)}$- and $\Sigma_{D}^{(k)}$-Beads are pairwise identical for a specific optimum choice of local measurement.
	\item \label{res:BEADS9} In contrast to classically correlated states, $\Sigma$-Beads of nonclassically correlated separable states contain additional classically inaccessible components -- the nonclassical structure of the state -- which are visualized by $D^{(k)}$ and $\Sigma_D^{(k)}$-Beads.
\end{enumerate}
\begin{enumerate}[start = 10, label=\textbf{Result \arabic*:}, itemindent=1.15cm, labelindent=1cm]
	\item \label{res:BEADS10} For a given nonclassically correlated state $\rho$, $M_{Cl}^{(k)}$-, $\Sigma_{Cl}^{(k)}$-Beads, $D^{(k)}$- and $\Sigma_{D}^{(k)}$-Beads are generally non-unique unless there exists a single preferred local measurement for classically accessing the state.
\end{enumerate}

\subsubsection{Entangled states}\label{sec:entbipartite}
\begin{figure}[H] 
	\centering
	\includegraphics[width=\linewidth]{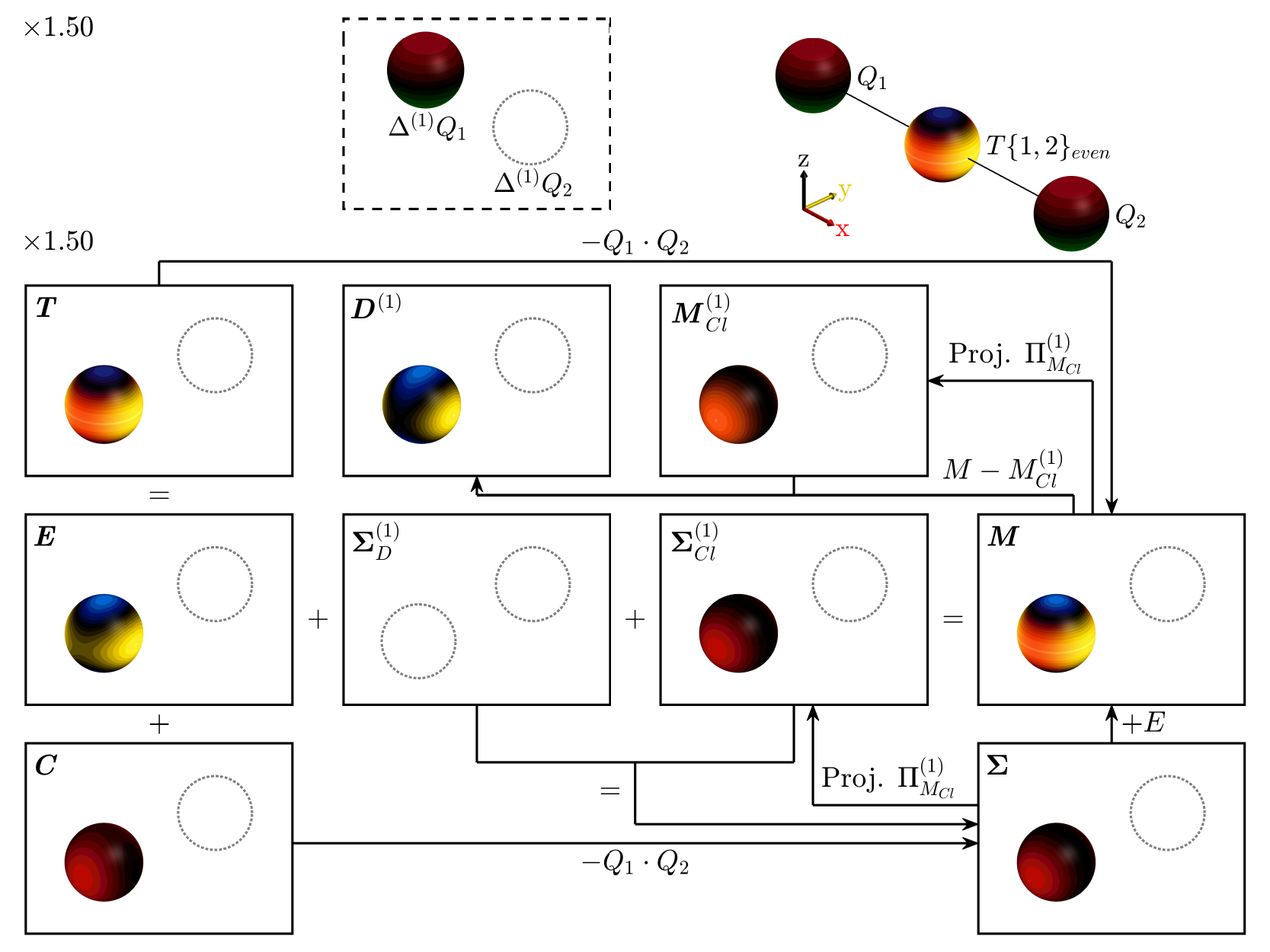}
	\caption{Complete generalized BEADS representation of the entangled state $\rho_\text{entangled}$ (cf. \cref{eq:hubertigam}) corresponding to a reduced bipartite marginal of the tripartite W state. It is clear from the single-qubit reduced density operators visualized by colored $Q$-Beads that the state is neither maximally correlated nor maximally entangled. Classically accessing this state by an optimum local measurement of one qubit always captures parts of its entangling structure ($M_{Cl}^{(1)} \neq \Sigma_{Cl}^{(1)}$ and $D^{(1)} \neq E$) which is directly clear from the shown case where the projection captures the full genuinely correlation-based separable structure of $\rho_\text{entangled}$ ($\Sigma = \Sigma_{Cl}^{(1)}$).}
	\label{fig:hubertigam_all}
\end{figure}

The last bipartite state that shall be analyzed is the reduced bipartite marginal of the tripartite W state $\ket{\text{W}} = 1/\sqrt{3}(\ket{001}+\ket{010}+\ket{100})$ (see \cref{sec:ghz_w_pure}), which is obtained by taking the partial trace over one of the three qubits (here, e.g., qubit 1):
\begin{equation}
	\rho_\text{entangled}=\text{Tr}_1\big(\ket{\text{W}}\bra{\text{W}}\big) = \frac{1}{3}\begin{pmatrix}
	1 & 0 & 0 & 0 \\
	0 & 1 & 1 & 0 \\
	0 & 1 & 1 & 0 \\
	0 & 0 & 0 & 0
	\end{pmatrix}.
	\label{eq:hubertigam}
\end{equation}
In the generalized BEADS visualization of this state shown in \cref{fig:hubertigam_all}, the $T$-Beads exhibit the full color palette as shown in \cref{fig:third} due to contributions from entanglement ($E$-Beads) and the separable structure of the state ($C$-Beads). Clearly, this state is entangled which can be directly inferred from the occurrence of an $E$-Bead. Moreover, the $T$-Bead being the sum of $C$- and $E$-Beads causes the former two to not coincide in this case. 

The reduced single-qubit density operators ($Q$-Beads) are colored and thus, not maximally mixed. Due to the encoded local information, the $M$- and $\Sigma$-Beads again do not coincide with the $T$- and $C$-Beads, respectively, as the latter contain the redundant product $Q_1 \cdot Q_2$.

This is also the first example of a state, where a black auxiliary connector line appears between the two $Q$-Beads, running through the $T\{1,2\}_{even}$-Bead in the full visualization in the top right corner of \cref{fig:hubertigam_all}. As discussed in \cref{sec:entarcs}, this line is called an entanglement arc and is scaled in diameter proportionally to the two-tangle of the state. The arc serves as a simple supporting tool for identifying the state as being bipartite entangled in one glance.

Assessing the classically accessible mutual information of the state reveals a degree of freedom in the choice of local measurements. Here, this degree of freedom is reflected in the magnitude of spherical function values, i.e., the color brightness neglecting the color hues, which is axially symmetric with respect to the $z$-axis. Indeed, any local measurement in the transverse $xy$-plane of either one of the qubits maximizes the classical mutual information of $\rho_\text{entangled}$. In this case, we exemplarily visualize the $M_{Cl}^{(1)}$-, $\Sigma_{Cl}^{(1)}$-, $D^{(1)}$ and $\Sigma_D^{(1)}$-Beads for the projection that captures the smallest contribution from the entangling structure of $\rho_\text{entangled}$, which here simultaneously covers the entire genuinely correlation-based structure of the state. This is directly clear from coinciding  $\Sigma$- and  $\Sigma_{Cl}^{(1)}$-Beads while all $\Sigma_D^{(1)}$-Beads vanish.

Notably, in \cref{fig:hubertigam_all} the $M_{Cl}^{(1)}$- and $\Sigma_{Cl}^{(1)}$-Beads are also not identical. This is due to the $M_{Cl}^{(1)}$-Bead capturing parts of the entangling structure. Using the blended color scheme shown in \cref{fig:third}, we find that the orange color corresponds to simultaneous contributions of the entangling and the separable structures of $\rho_\text{entangled}$. As stated in \cref{sec:genuine_tripartite_$M_{Cl}$}, parts of the entangling structure of a quantum state being classically accessible is a perfectly valid phenomenon. As a consequence, the $D^{(1)}$-Bead, which visualizes the classically inaccessible structure of $\rho_\text{entangled}$ encodes less information for the given local measurement compared to the $E$-Bead. Naturally, the missing information on the $D^{(1)}$-Bead compared to the $E$-Bead is the classically accessible component contained in the $M_{Cl}^{(1)}$-Bead. 

Moreover, it is clear from the generalized BEADS representation that any optimum measurement which maximizes the classical mutual information does not recover the local information encoded on the reduced density operator of the measured qubit. Evidently, for $\rho_\text{entangled}$ this local information is arranged along the $z$-axis. As the optimum measurement is conducted in an arbitrary orthogonal direction, the information is fully lost. This loss is directly visualized by the $\Delta^{(k)}Q$, here, $\Delta^{(1)} Q_1$.

As a last aspect, it is important to point out that the same axial symmetry which leads to the discussed behaviour under classical access of $\rho_\text{entangled}$ also causes the $T$-, $E$-, $C$-, $M$-, and $\Sigma$-Beads to be non-unique. Indeed the rotational degree of freedom induced by the axial symmetry admits that there exist infinitely many variants of the mentioned beads which still correspond to the same density operator. For instance, any version of the shown $T$- and $M$-Beads obtained by a rotation of these beads around the $z$-axis such that the yellow and orange regions are oriented differently, are still valid $T$- and $M$-Beads of $\rho_\text{entangled}$. At the same time, this implies that, while it is true for any entangled mixed state, that there can exist infinitely many different least-entangled pure state ensembles \cite{hughston_complete_1993}, these only manifest in different $T$-, $E$-, $C$-, $M$-, and $\Sigma$-Beads if the state has some intrinsic geometric symmetry which induces a corresponding degree of freedom.

\pagebreak

\begin{enumerate}[start = 11, label=\textbf{Result \arabic*:}, itemindent=1.15cm, labelindent=1cm]
	\item \label{res:BEADS11} If a state is entangled, its generalized BEADS representation has nonzero $E$-Beads at least for some contributing pure state in an optimized least-entangled ensemble. As a consequence, $T$- and $C$-Beads are not identical for this type of states.
\end{enumerate}

\begin{enumerate}[start = 12, label=\textbf{Result \arabic*:}, itemindent=1.15cm, labelindent=1cm]
	\item \label{res:BEADS12} Different least-entangled pure state ensembles realizing the same entangled mixed state manifest in the same $T$-, $E$-, $C$-, $M$-, and $\Sigma$-Beads unless the mixed state has a spatial symmetry which admits a rotational degree of freedom in the generalized BEADS representation. In such cases the mentioned beads are equivalent up to a rotation around the symmetry axis. Conversely, a set of $T$-, $E$-, $C$-, $M$-, and $\Sigma$-Beads is always unique for a given least-entangled pure state ensemble of a mixed density operator of interest.
\end{enumerate}

\begin{enumerate}[start = 13, label=\textbf{Result \arabic*:}, itemindent=1.15cm, labelindent=1cm]
	\item \label{res:BEADS13} Accessing mixed entangled states classically by optimum local measurements can capture separable and entangling structures when viewed on the level of the underlying density operator, which are directly visualized by  $M_{Cl}^{(k)}$-Beads for a given optimal local measurement. As a result, $M_{Cl}^{(k)}$- and $\Sigma_{Cl}^{(k)}$-Beads do not necessarily coincide for entangled states.
\end{enumerate}

\subsubsection{Summary and overview of correlation function beads for different types of mixed quantum states}

To summarize the most important results and findings obtained for the discussed examples of mixed bipartite quantum states, \cref{tab:correlation_visualization2} provides a graphical overview of the encoded information in each type of bead for the different types of mixed states. For instance, for $M_{Cl}$-Beads of entangled states, the grey shaded square symbol \raisebox{.1cm}{\scalebox{0.5}{\emptyquad}} inside a circle \raisebox{.1cm}{\scalebox{0.8}{\emptycirc}} indicates that both, parts of the entangling and separable structures of the given state can contribute to the genuinely correlation-based classically accessible information. 

On the other hand, $C$-Beads of entangled states can contain redundant information (three-forked star \raisebox{.04cm}{\scalebox{0.9}{\threestar}}) but also visualize the genuinely correlation-based separable structure of the state of which some components can be classically accessed (circle \raisebox{.1cm}{\scalebox{0.8}{\emptycirc}}) whereas others may remain classically inaccessible (triangle \raisebox{.1cm}{\scalebox{0.5}{\emptytri}}).

\begin{table}[H]
	\caption{Graphical overview of information encoded in different types of correlation function beads for different sets of mixed quantum states. The same symbol reoccurring for multiple beads indicates that the beads encode the same information. If for a given set of quantum states different beads share the same symbols, they are identical. The symbol $-$ indicates that beads of this type are strictly zero in all spatial directions for the given set of mixed states. That is, these beads do not occur in the generalized BEADS representation of these states.}
	\centering
	\renewcommand{\arraystretch}{1.5}
	\begin{tabular}{p{0.19\textwidth}<{\centering} p{0.06\textwidth}<{\centering} p{0.06\textwidth}<{\centering} p{0.06\textwidth}<{\centering} p{0.06\textwidth}<{\centering} p{0.06\textwidth}<{\centering} p{0.06\textwidth}<{\centering} p{0.06\textwidth}<{\centering} p{0.06\textwidth}<{\centering} p{0.06\textwidth}<{\centering}}
		\hline\hline
		\textbf{Set of states} & \textbf{T} & \textbf{E} & \textbf{C} & \textbf{M} & \textbf{$\Sigma$} & \textbf{$M_{Cl}$} & \textbf{$\Sigma_{Cl}$} & $D$ & \textbf{$\Sigma_D$} \\\hline
		
		Product & \threestar & -- & \threestar & -- & -- & -- & -- & -- & -- \\[10pt]
		
		Classical & \circstar & -- & \circstar & \emptycirc & \emptycirc & \emptycirc & \emptycirc & -- & -- \\[10pt]
		
		Separable & \emptycirctristar & -- & \emptycirctristar & \emptycirctri & \emptycirctri & \emptycirc & \emptycirc & \emptytri & \emptytri \\[10pt]
		
		Entangled & \emptycircquadtristar & \emptyquad & \emptycirctristar & \emptycircquadtri & \emptycirctri & \circsmallquad & \emptycirc & \trismallquad &\emptytri \\[6pt]
		\hline\hline
	\end{tabular}
	\label{tab:correlation_visualization2}
\end{table}
\vspace{-.5cm}
\underline{Legend}\\[5pt]
\begin{minipage}{.49\textwidth}
\begin{itemize}[labelwidth=2em, itemsep=0.5em]
\item[\centerlabel{\threestar}] Redundant information
\item[\centerlabel{\emptycirc}] Classically accessible genuinely correlation-based information arising from separable structure
\item[\centerlabel{\emptytri}] Nonclassical genuinely correlation-based information arising from separable structure
\end{itemize}
\end{minipage}\hspace{.02\textwidth}%
\begin{minipage}{.49\textwidth}
\begin{itemize}[itemsep=0.5em]
\item[\centerlabel{\emptyquad}] Entanglement-based information (entangling structure)
\item[\centerlabel{\smallquad}] Classically accessible information arising from entangling structure
\item[\centerlabel{\bigandsmallquad}] Nonclassical information arising from entangling structure
\end{itemize}
\end{minipage}\\[5pt]

All discussed types of quantum states are further summarized in Table~\ref{tab:state_types_characteristics} which also provides recipes for quickly and unambiguously identifying distinct types of (un)correlated quantum states based on their generalized BEADS representation and the features discussed in the preceding subsections.

\begin{table}[H]
\centering
\caption{Overview of different types of quantum states and how they can easily be identified based on their generalized BEADS representation.}
\begin{tabular}{
    >{\centering\arraybackslash}m{0.3\textwidth}
    >{\centering\arraybackslash}m{0.7\textwidth}
}
\hline
\hline\\[-6pt]
\textbf{Type of states} & \textbf{Distinguishing characteristics} \\[4pt]
\hline\\[-6pt]
\textit{Product} states &
\begin{minipage}{\linewidth}
    \begin{itemize}[nosep]
    	\item No other type of correlation function beads other than $T$- and $C$-Beads
        \item $T$-Beads = $C$-Beads
        \item Pure product states: $Q$-Beads always have full color brightness
        \item Mixed product states: $Q$-Beads do not have full color brightness
    \end{itemize}
\end{minipage}
\\\\[15pt]
\textit{Classically} correlated states & 
\begin{minipage}{\linewidth}
    \begin{itemize}[nosep]
        \item $T$-Beads = $C$-Beads
        \item $M$-Beads = $\Sigma$-Beads = $M_{Cl}$-Beads = $\Sigma_{Cl}$-Beads
        \item No $E$-, $D$, $\Sigma_D$, and $\Delta Q$-Beads
    \end{itemize}
\end{minipage}
\\\\[15pt]
\textit{Nonclassically} (quantum) correlated \textit{separable} states & 
\begin{minipage}{\linewidth}
    \begin{itemize}[nosep]
        \item $T$-Beads = $C$-Beads
        \item $M$-Beads = $\Sigma$-Beads
        \item $M_{Cl}$-Beads = $\Sigma_{Cl}$-Beads
        \item At least one of the following bead types is nonzero (visible): $D$-, $\Sigma_D$-, or $\Delta Q$-Beads
        \item If they exist, $D$-Beads = $\Sigma_D$-Beads
        \item No $E$-Beads
        \item $M_{Cl}$-, $\Sigma_{Cl}$-, $D$-, $\Sigma_D$-, and $\Delta Q$-Beads can be non-unique if the state admits a degree of freedom in the choice of local measurements to maximize the classically accessible information
    \end{itemize}
\end{minipage}
\\\\[15pt]
\textit{Entangled} states & 
\begin{minipage}{\linewidth}
    \begin{itemize}[nosep]
        \item Only type of states that shows $E$-Beads
        \item $T$-Beads = $E + C$-Beads
        \item $M$-Beads = $E + \Sigma$-Beads
        \item $T$-, $E$-, $C$-, $M$-, and $\Sigma$-Beads can be non-unique if the state has a geometric degree of freedom (e.g., an axial symmetry), which admits infinitely many different least-entangled pure state ensembles which manifest in different entangling and separable structures while still realizing the same density operator.
        \item $M_{Cl}$-, $\Sigma_{Cl}$-, $D$-, $\Sigma_D$-, and $\Delta Q$-Beads can be non-unique if the state admits a degree of freedom in the choice of local measurements to maximize the classically accessible information
    \end{itemize}
\end{minipage}
\\\\[-6pt]
\hline
\hline
\end{tabular}
\label{tab:state_types_characteristics}
\end{table}

\subsection{Generalized BEADS representations of tripartite quantum states}
In the following, we will examine generalized BEADS representations of tripartite states. With this we illustrate how the visualization generalizes to multipartite systems and which additional features are required to fully visualize the corresponding quantum states. Further examples are provided in supplementary \cref{sup:tripartite}.

\subsubsection{The GHZ and W states}
\label{sec:ghz_w_pure}

We start with the two inequivalent representatives of genuine tripartite entanglement: The GHZ state $\ket{\text{GHZ}} = 1/\sqrt{2}(\ket{000}+\ket{111})$ and the W state $\ket{\text{W}} = 1/\sqrt{3}(\ket{001}+\ket{010}+\ket{100})$. Note that even if these states are pure from a global perspective, their reduced bipartite marginals are mixed. Hence, it is advisable to visualize them using the generalized BEADS representation to get an informationally complete view of the correlation structures of the reduced subsystems.

\begin{figure}[htbp] 
	\centering
	\makebox[\linewidth][c]{
		\begin{minipage}{1\textwidth}
			\begin{subfigure}[t]{0.5\textwidth}
				\centering
				\includegraphics[width=\textwidth]{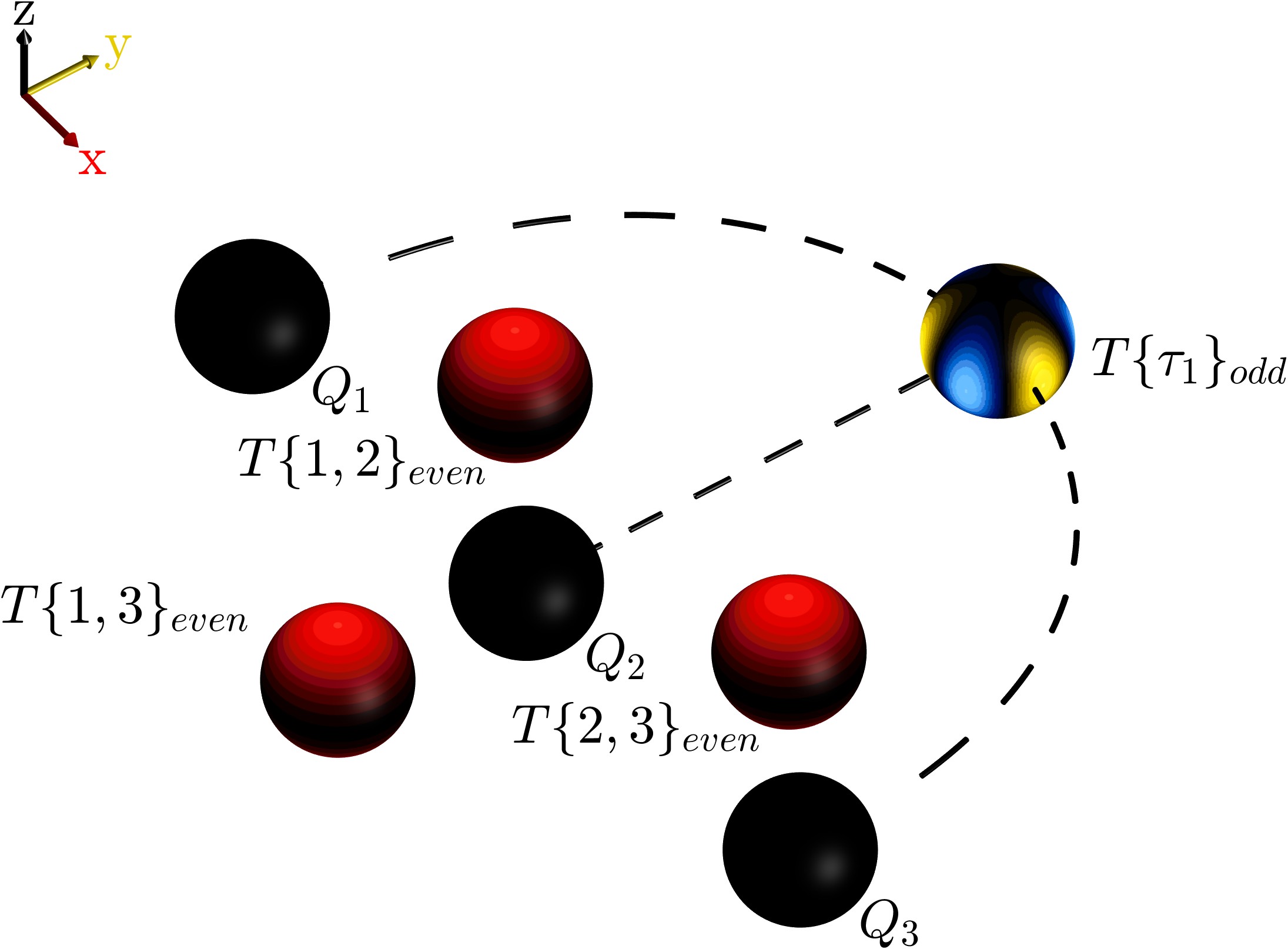}
				\label{fig:ghz_pure}
			\end{subfigure}
			\begin{subfigure}[t]{0.5\textwidth}
				\centering
				\includegraphics[width=\textwidth]{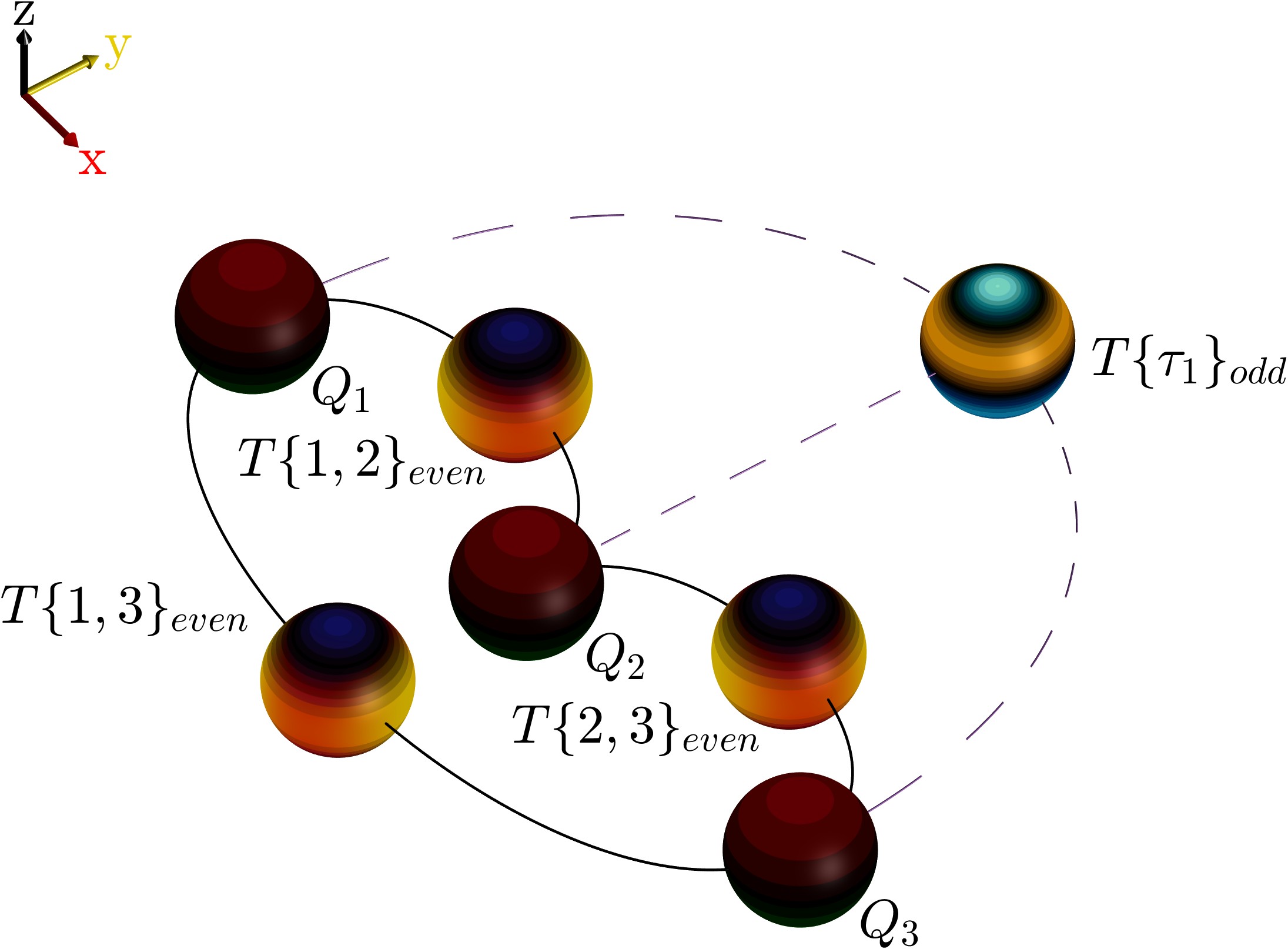}
				\label{fig:w_pure}
			\end{subfigure}
		\end{minipage}
	}
	\caption{Total correlation function (T-Beads) visualization of the GHZ (left) and the W state (right). Despite being globally pure, both states have mixed bipartite reduced marginals of which the correlation structure is directly evident in the generalized BEADS picture.}
	\label{fig:ghz_w_pure}
\end{figure}

\cref{fig:ghz_w_pure} shows the $T$-Beads representations of the GHZ and the W state. The GHZ state exhibits a characteristic surface pattern on the tripartite $\tau_1$-bead, which -- given that the bead has only yellow, blue and black colors -- can be fully attributed to genuine tripartite entanglement. As explained in \cref{sec:entarcs}, GHZ-type entanglement is further indicated by the black dashed entanglement arcs connecting the $Q$-Beads with the tripartite $T$-Bead, which are scaled in diameter proportionally to the three-tangle. 

In contrast to the standard BEADS representation \cite{huber_beads_2025}, in the generalized BEADS representation, visualizations of multipartite qubit states are based on a recursive optimization scheme. Here, this means that the reduced density operators corresponding to the bipartite marginals of the GHZ state are also optimized with respect to least-entangled pure state decompositions. Indeed, it turns out that the bipartite subsystems of the GHZ state share the same surface pattern and are not entangled as is directly clear, inter alia, from the strictly red and black colors and the absence of entanglement arcs. Upon further analysis (see \cref{sec:mixed_bipartite}), all bipartite marginals can be identified to be classically correlated based on their generalized BEADS representation.

The W state shows an axially symmetric surface pattern of the trilinear $\tau_1$-bead with respect to the $z$-axis. Clearly, the state is not maximally tripartite entangled for which the resulting tripartite bead has an underlying residual product structure leading to orange and turquoise colors. As expected, the state is strictly W-type entangled, which is directly evident from missing black tangle arcs and visible violet W-entanglement arcs as introduced in \cref{sec:entarcs}. 

Clearly, the bipartite marginals of the W state are non-maximally entangled which is reflected in the $T$-Bead colors and solid black two-tangle arcs. Note that the generalized BEADS representation of the bipartite {W state} subsystem was analyzed in detail in \cref{sec:entbipartite}.

\subsubsection{Parametrized mixture of the W and anti-W state}\label{sec:Wmixtures}

\begin{figure}[H] 
	\centering
	\makebox[\linewidth][c]{
		\begin{minipage}{1\textwidth}
			\begin{subfigure}[t]{0.5\textwidth}
				\centering
				\includegraphics[width=\textwidth]{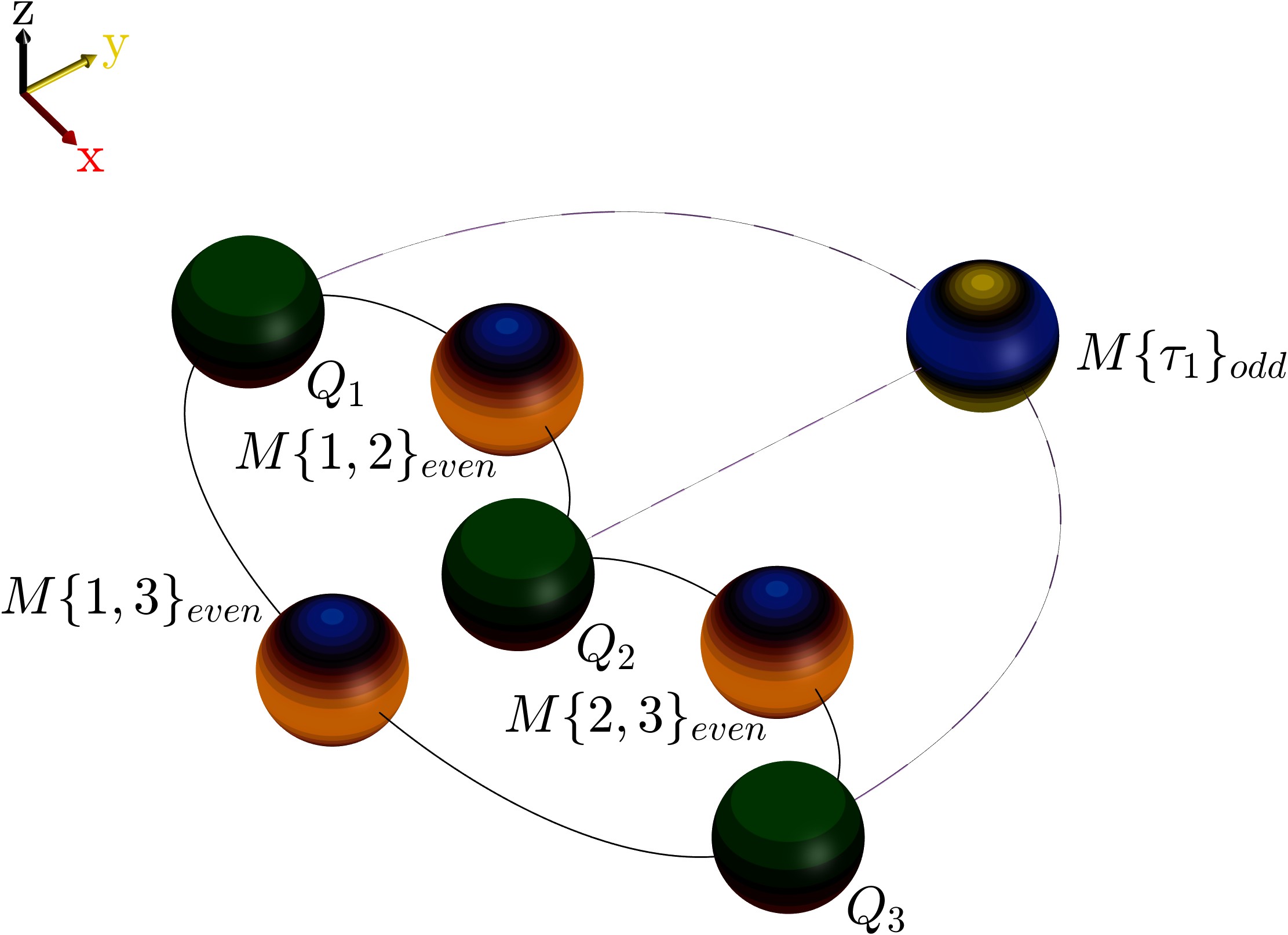}
				\subcaption{$p = 0.05$}
				\label{fig:quantum_without_class_0.05_M}
			\end{subfigure}
			\begin{subfigure}[t]{0.5\textwidth}
				\centering
				\includegraphics[width=\textwidth]{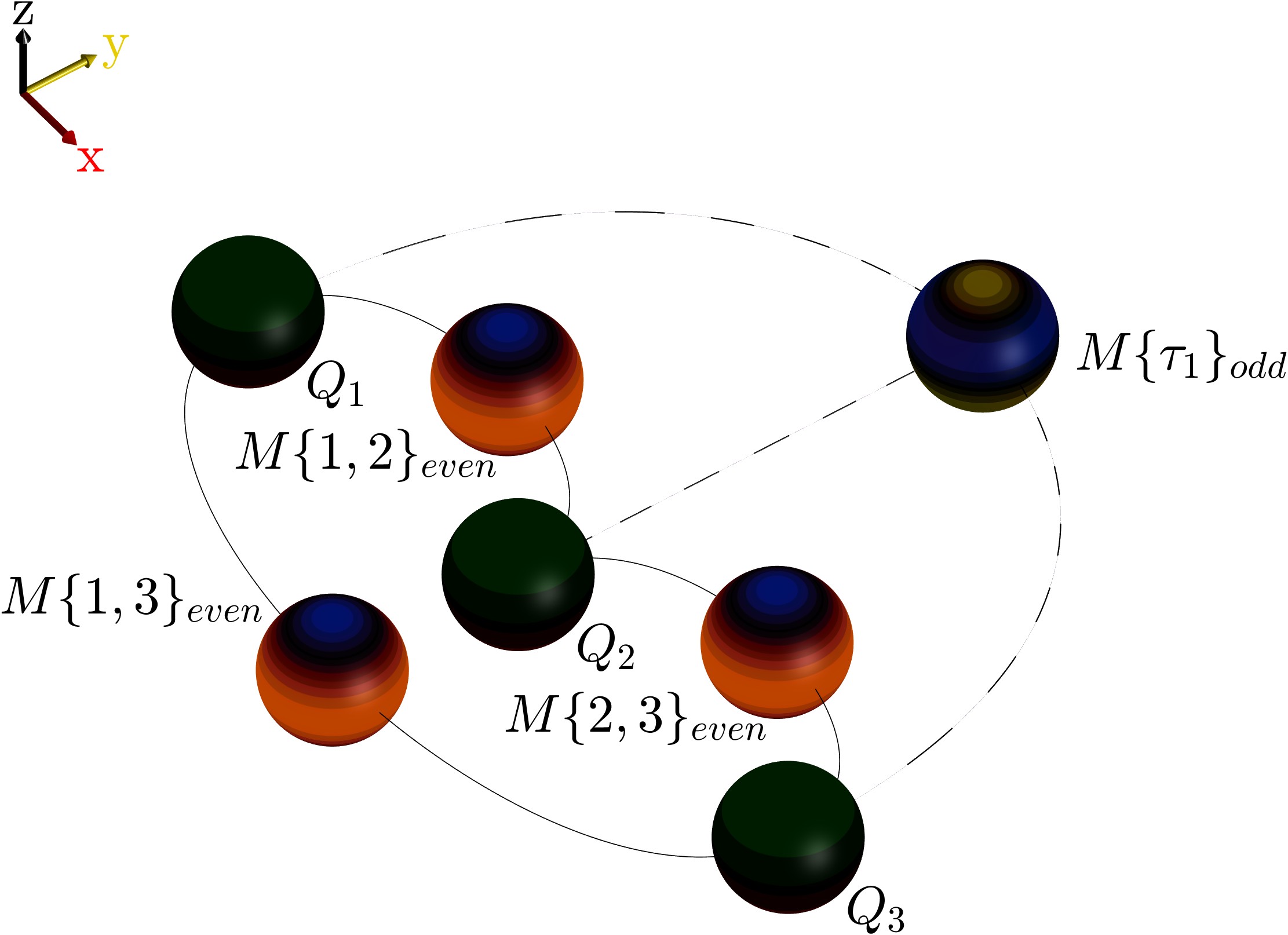}
				\subcaption{$p = 0.2$}
				\label{fig:quantum_without_class_0.2_M}
			\end{subfigure}
		\end{minipage}
	}
	
	\vspace{0.6cm}
	
	\makebox[\linewidth][c]{
		\begin{minipage}{1\textwidth}
			\begin{subfigure}[t]{0.5\textwidth}
				\centering
				\includegraphics[width=\textwidth]{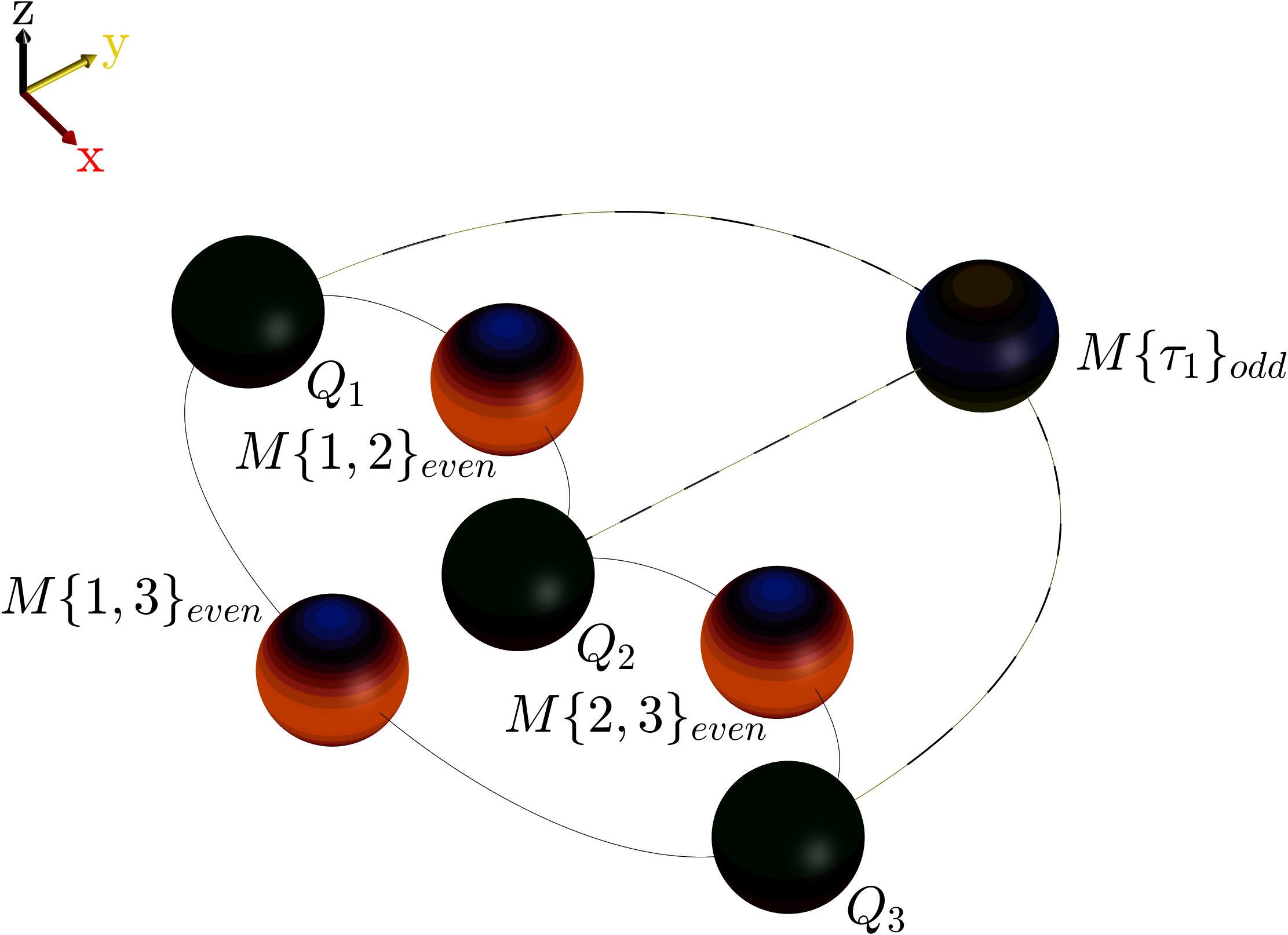}
				\subcaption{$p = 0.35$}
				\label{fig:quantum_without_class_0.35_M}
			\end{subfigure}
			\begin{subfigure}[t]{0.413\textwidth}
				\includegraphics[width=\textwidth]{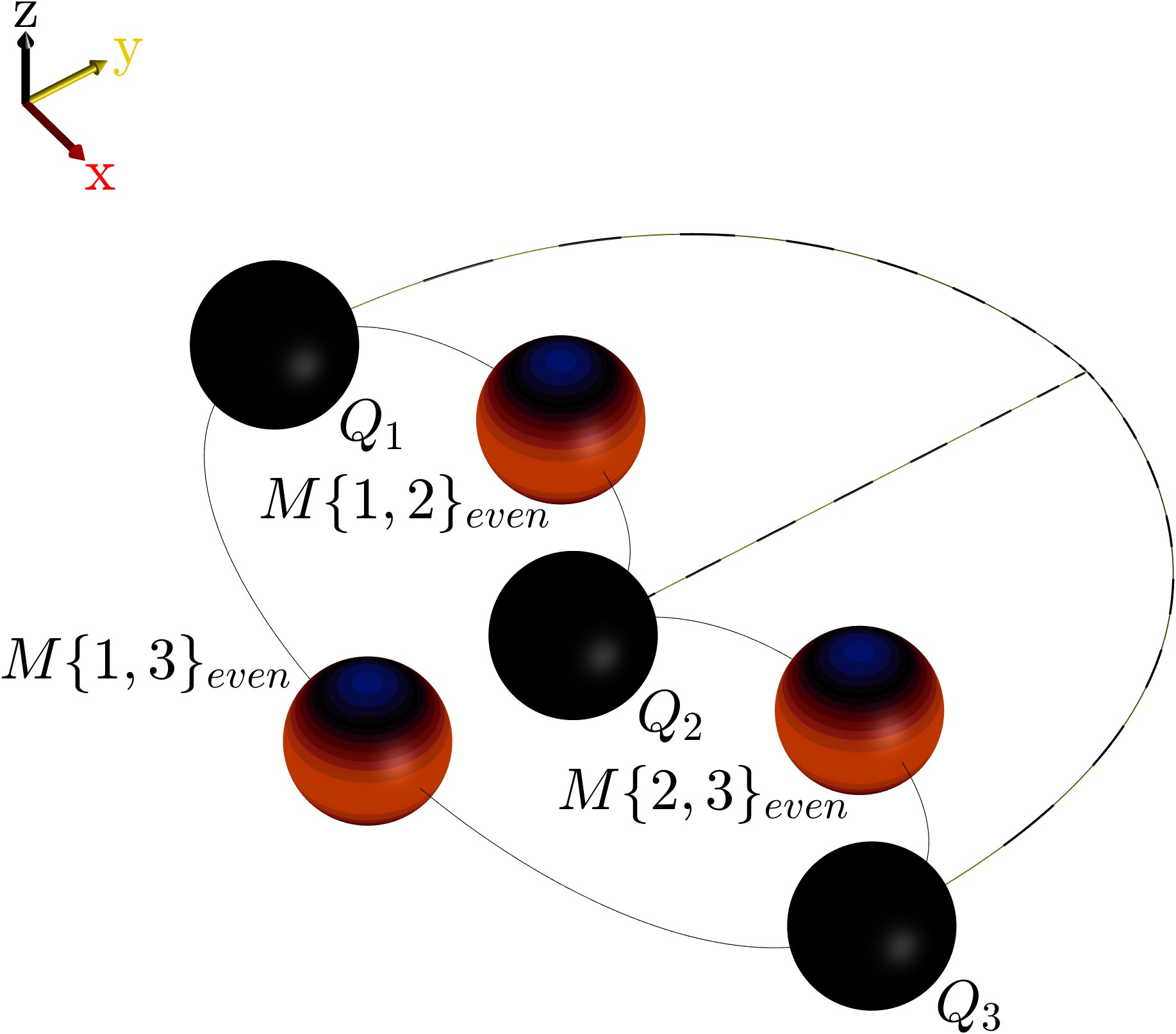}
				\subcaption{$p = 0.5$}
				\label{fig:quantum_without_class_0.5_M}
			\end{subfigure}
		\end{minipage}
	}
	
	\vspace{0.6cm}
	
	\makebox[\linewidth][c]{
		\begin{minipage}{1\textwidth}
			\begin{subfigure}[t]{0.5\textwidth}
				\centering
				\includegraphics[width=\textwidth]{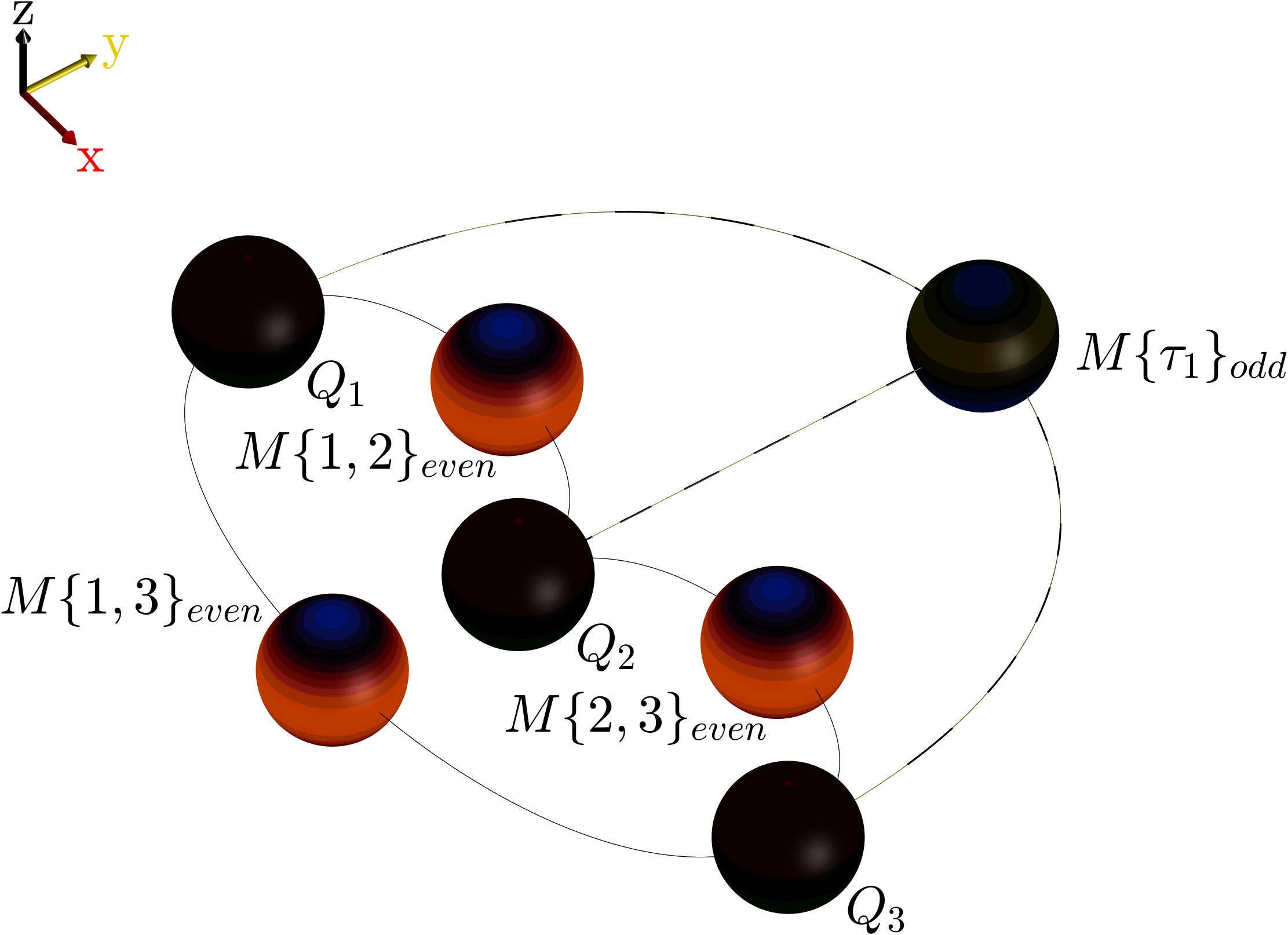}
				\subcaption{$p = 0.65$}
				\label{fig:quantum_without_class_0.65_M}
			\end{subfigure}
			\begin{subfigure}[t]{0.5\textwidth}
				\centering
				\includegraphics[width=\textwidth]{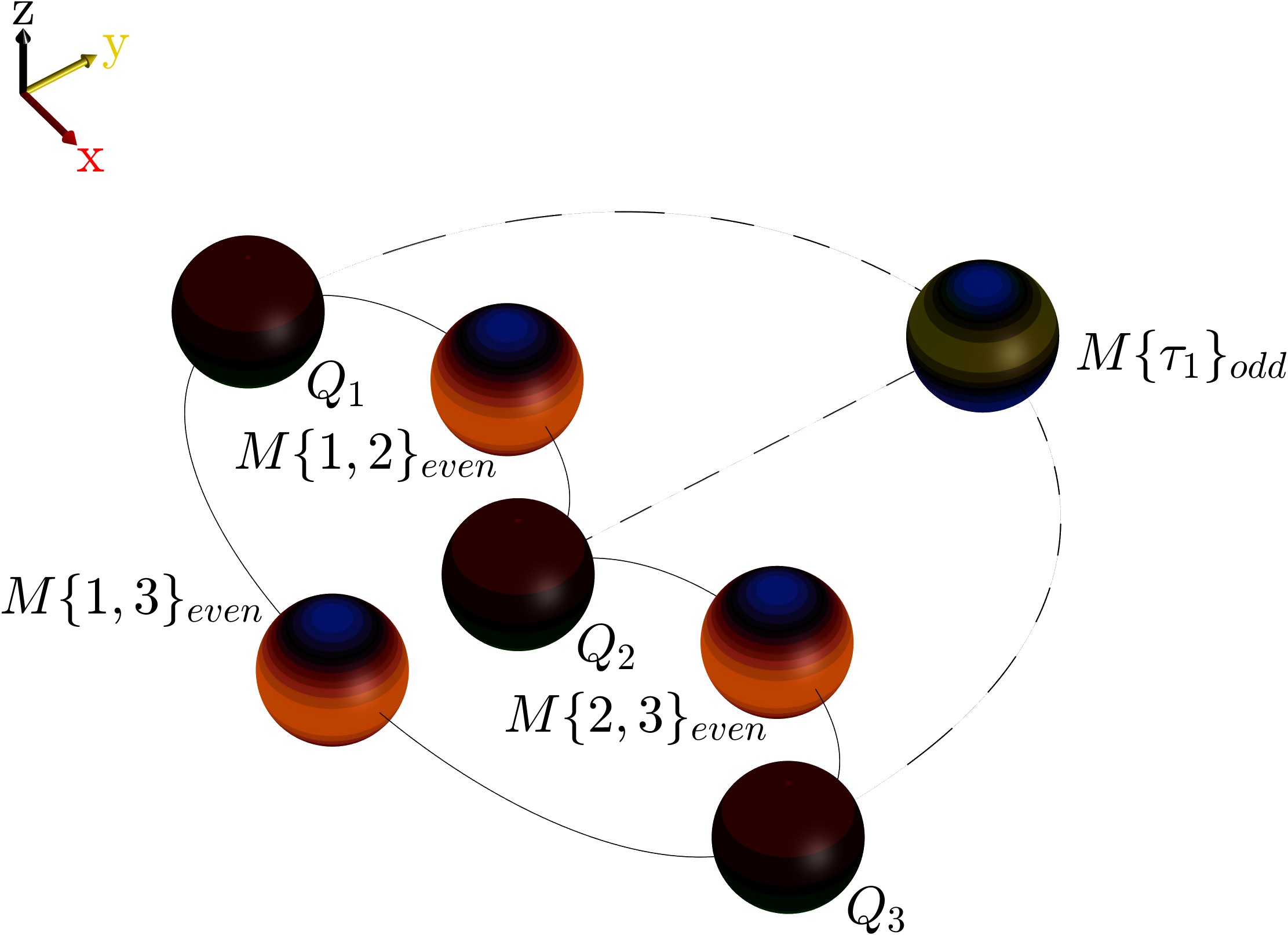}
				\subcaption{$p = 0.8$}
				\label{fig:quantum_without_class_0.8_M}
			\end{subfigure}
		\end{minipage}
	}
	\caption{$M$-Bead visualization of the parameterized mixed state $\rho_\text{W-mixture}\,(p)$ (see \cref{eq:w_wbar_mixed}). For $p<0.5$ $\overline{W}$ state contributions dominate, whereas for $p>0.5$ W state contributions prevail. In the case of $p=0.5$, the state is tripartite entangled without having tripartite classical correlation.}
	\label{fig:quantum_without_class_mixtures}
\end{figure}

Consider the parameterized mixture of the W and the anti-W state $\ket{\overline{\text{W}}} = 1/\sqrt{3}(\ket{110}+\ket{101}+\ket{011})$:
\begin{equation}
	\rho_\text{W-mixture}= p \left( \ket{\text{W}} \bra{\text{W}}  \right) + (1-p) \left( \ket{\overline{\text{W}}} \bra{\overline{\text{W}}} \right).
	\label{eq:w_wbar_mixed}
\end{equation}
\cref{fig:quantum_without_class_mixtures} displays a series of $M$-Bead representations of this mixture where the mixing parameter $p$ ranges from $p=0.05$ (a state close to $\ket{\overline{\text{W}}}$) to $p = 0.8$ (a state close to $\ket{\text{W}}$). Comparing \cref{fig:quantum_without_class_0.05_M} and \cref{fig:quantum_without_class_0.8_M} and recalling \cref{fig:ghz_w_pure}, it can be deduced that the tripartite $M\{\tau_1\}_{odd}$-Bead as well as the $Q$-Beads of the W and the $\overline{\text{W}}$ state have congruent but oppositely signed surface patterns. This is consistent with the anti-W state being the globally flipped relative of the W state. On the other hand, the bipartite beads must be identical for both pure states, as these have \textit{even} point symmetry for which a global flip (global $X$-gate operation) leaves them invariant. 

Therefore, when the mixing parameter approaches $p = 0.5$ the pure state contributions to the $\tau_1$- and the $Q$-Beads cancel each other progressively. The limiting case is reached at ${p=0.5}$ where the tripartite $M$-Bead vanishes and the $Q$-Beads become all-black, i.e., maximally mixed (compare \cref{fig:quantum_without_class_0.5_M}). On the other hand, the bipartite subsystems retain their corresponding $M$-Bead surface pattern independent of the choice of $p$, yet the colors change slightly to a more red-dominant hue, which, together with thinner entanglement arcs, suggests slightly less bipartite entanglement for mixtures closer to $p=0.5$. The state for $p=0.5$ is well-known and thoroughly discussed in literature \cite{tran_genuine_2017,kaszlikowski_quantum_2008} due to its special correlation properties. 

\begin{figure}[H]
	\centering
	\includegraphics[height=6.5cm]{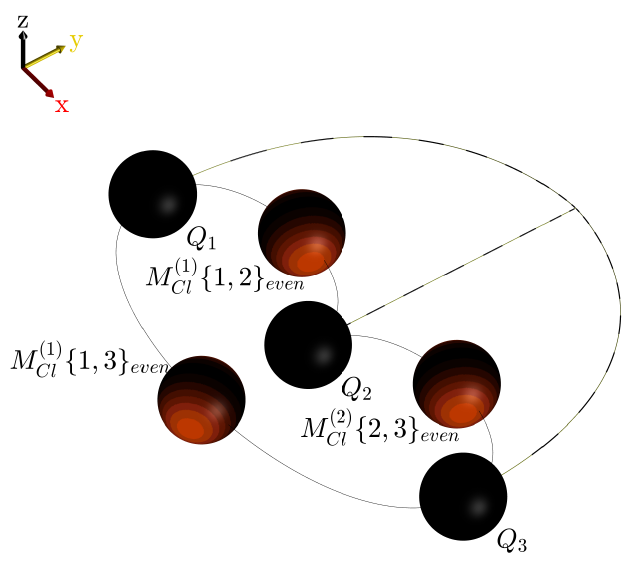}
	\caption{Classical correlation beads ($M_{Cl}$-Beads) of $\rho_\text{W-mixture}\,(p=0.5)$. It is clear from the representation that there are only bipartite contributions to the classically accessible correlation structure of the state.}
	\label{fig:quantum_without_class_CM}	
\end{figure}

Indeed, the state \cite{tran_genuine_2017, kaszlikowski_quantum_2008}
\begin{equation}
	\rho_\text{W-mixture}\,(p=0.5)= \frac{1}{2} \left( \ket{\text{W}} \bra{\text{W}} + \ket{\overline{\text{W}}} \bra{\overline{\text{W}}} \right)
	\label{eq:quantum_without_class}
\end{equation}
is of particular interest as it is genuinely tripartite entangled but does not have any genuine tripartite classical correlation. In the generalized BEADS representation of this state shown in \cref{fig:quantum_without_class_0.5_M}, both of these properties are directly evident. The non-existent tripartite classical correlation is intuitively reflected in the absence of tripartite $M$-Beads and $M_{Cl}$-Beads, which implies that no (joint) local measurement of arbitrary subsets of qubits will be able to access any genuinely tripartite correlation-based information. 
To support this finding, an $M_{Cl}$-Beads representation of $\rho_\text{W-mixture}\,(p=0.5)$ is depicted in \cref{fig:quantum_without_class_CM}. Clearly, only bipartite correlation-based information is classically accessible and is visualized by dedicated $M_{Cl}$-Beads. 
The tripartite entanglement of $\rho_\text{W-mixture}\,(p=0.5)$ becomes directly visible by corresponding tripartite entanglement arcs in the generalized BEADS representation of the mixed state even in the absence of tripartite $M$- and thus, also $E$-Beads.  Note that there are both, three-tangle and $\mathcal{W}_3$ arcs, in \cref{fig:quantum_without_class_0.5_M}, which reveals another important property of the state: Although it was seemingly constructed by incoherently mixing two W-class pure states, the resulting mixed state is a GHZ class mixed state.

\begin{figure}[H]          
	\centering
	\includegraphics[width=15cm]{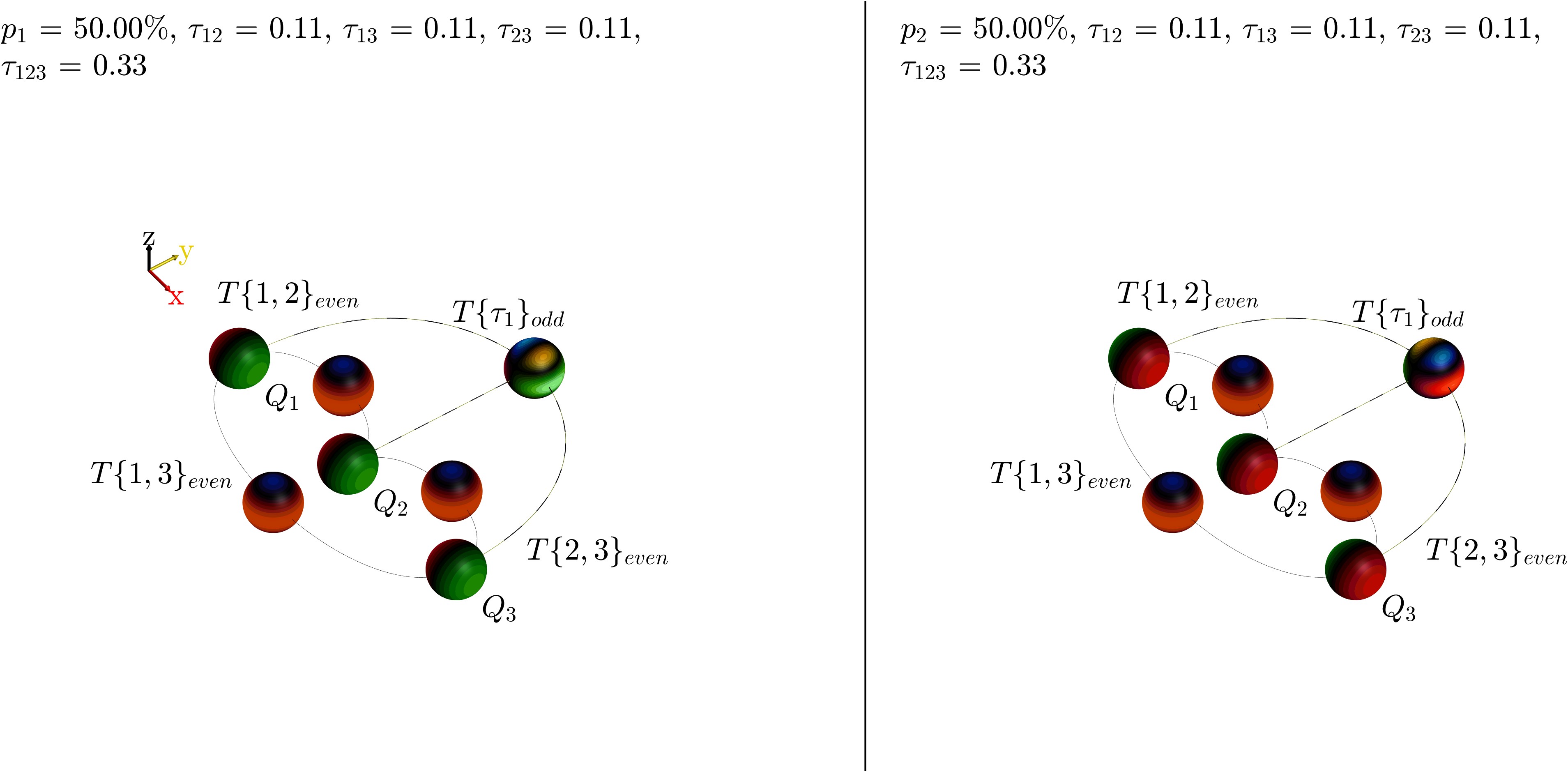}
	\caption{$T$-Beads representation of a possible least-entangled pure state ensemble realizing the state $\rho_\text{W-mixture}\,(p=0.5)$ (cf. \cref{eq:quantum_without_class}). The visualized pure states are  $\ket{\omega_3^{\pm}} = 1/\sqrt{2}(\ket{\text{W}} \pm \ket{\overline{\text{W}}})$, of which $\ket{\omega_3^{+}}$ is shown on the right. Both pure states are genuinely tripartite entangled and -- given the nonzero three-tangles $\tau_{123} = 1/3$ -- belong to the GHZ SLOCC class. Note that the states have oppositely signed $Q$- and tripartite $T$-Beads for which they cancel when both states are incoherently mixed using the denoted weights.}
	\label{fig:quantum_without_class_pure_decompT}        
\end{figure}

While this may seem counterintuitive at first glance, it can be well explained by considering the generalized BEADS representations of the contributing pure states in a possible least-entangled ensemble of $\rho_\text{W-mixture}\,(p=0.5)$ shown in \cref{fig:quantum_without_class_pure_decompT}. The displayed pure states are ${\ket{\omega_3^{\pm}} = 1/\sqrt{2}(\ket{\text{W}} \pm \ket{\overline{\text{W}}})}$ which are GHZ-class states with a three-tangle of $\tau_{3} = 1/3$. Clearly, both states are genuinely tripartite entangled, which can be directly inferred from the yellow and blue colors of the $T\{\tau_1\}_{odd}$-Beads. The corresponding BEADS representations point out another subtle aspect: A pure state belonging to a distinct SLOCC class does not contradict with the state simultaneously containing W- and GHZ-like hybrid entanglement structures. For $\ket{\omega_3^{\pm}}$, the bipartite marginals represented by corresponding $T\{j,k\}_{even}$-Beads are entirely W-like with uniform bipartite entanglement of all subsystems. However, the tripartite surface pattern visualized by $T\{\tau_1\}_{odd}$ much more (yet not perfectly) resembles the GHZ state surface pattern shown in \cref{{fig:ghz_w_pure}} up to a rotation than the corresponding tripartite structure of the W state. The states are still uniquely defined to belong to the GHZ-class based on their nonzero three-tangles and thus, can be transformed into the GHZ state with nonzero probability by invertible local operations.

Incoherently mixing the pure states $\ket{\omega_3^{\pm}}$ to form $\rho_\text{W-mixture}\,(p=0.5)$ then directly gives rise to the previously observed properties: The $Q$- and tripartite $T$-Beads (or equivalently, any other type of tripartite bead) cancel due to having opposite signs. Moreover, the mixed state two- and three-tangles result as the convex sums over the corresponding pure state tangles, and the nonzero three-tangle qualifies the state as being a GHZ-class mixed state, although it is structurally not ideally GHZ-type entangled. Note that $n$-partite entanglement in the absence of $n$-partite correlation functions is a phenomenon that was previously discovered \cite{tran_genuine_2017, kaszlikowski_quantum_2008} and shown to only occur for odd numbers of qubits $n$. In the generalized BEADS representation, the latter aspect becomes directly clear, as for odd $n$ the corresponding correlation function beads also have odd point symmetry, i.e., the sign changes under a flip operation of all $n$ qubits. Thus, as seen in \cref{fig:quantum_without_class_pure_decompT}, this allows to find combinations of pure states that induce the observed cancellation of correlation functions when mixed incoherently.

\clearpage
\begin{enumerate}[start = 14, label=\textbf{Result \arabic*:}, itemindent=1.15cm, labelindent=1cm]
	\item \label{res:BEADS12} A mixed state is $n$-partite entangled if at least one of the contributing pure states in its optimum decomposition is $n$-partite entangled (that is, shows an $n$-partite $E$-Bead). $n$-partite entanglement in the absence of $n$-partite mixed state $E$-Beads only occurs for odd $n$.
\end{enumerate}

\subsection{Applications}

In the following chapter, we discuss selected applications quantum information visualized by the generalized BEADS representation.

\subsubsection{Entanglement distillation}

The first use-case that shall be discussed is the BBPSSW entanglement distillation protocol.~\cite{bennett_purification_1995} The goal of entanglement distillation is to obtain few highly entangled states from many copies of a weakly entangled state. In the following, we choose the Bell state $\ket{\Phi^+}$ as target state for which the input density operator of the protocol must be a Werner state of the form
\begin{equation}
	W_F= F \ket{\Phi^+}\bra{\Phi^+} + \frac{1-F}{3} \left( \ket{\Phi^-}\bra{\Phi^-} + \ket{\Psi^+}\bra{\Psi^+}+  \ket{\Psi^-}\bra{\Psi^-} \right),
	\label{eq:distillation_werner_state}
\end{equation}
where $\ket{\Phi^\pm}$ and $\ket{\Psi^\pm}$ denote the four Bell basis states and $F=\bra{\Phi^+}\rho\ket{\Phi^+}$ is the fidelity.
Note that any bipartite density operator can be transformed into the form of \cref{eq:distillation_werner_state} by applying the twirling quantum channel $\varepsilon_\text{twirl} = \int dU (U\otimes U) \rho (U^\dagger \otimes U^\dagger)$  and subsequent application of a $Y$-gate on one of the involved qubits \cite{bennett_purification_1995}.

The BBPSSW protocol can be conducted starting with $2^n$ \textit{pairs} of $W_F$, where half of the pairs are measured and discarded in every round. At the beginning of every round, a bilateral CNOT gate (BXOR) is applied between the first and second qubits of each pair of Werner states $W_{F,j}$ and $W_{F,k}$, respectively. The target qubit pair $W_{F,k}$ is subsequently measured in the computational basis. If the parity of the measurement outcome is odd -- which is equivalent to measuring a $\ket{\Psi^{\pm}}$ state -- the control pair $W_{F,j}$ is discarded. If the outcome has even parity, $W_{F,j}$ survives and, as described before, the state for the next round is prepared by unitary twirling followed by a local $Y$-gate resulting in the Werner-state $W_{F,j}^{'}$. The procedure is repeated until the desired fidelity $F$ is reached.

\begin{figure}[H]
	\centering
	\begin{subfigure}[t]{0.48\textwidth}
		\centering
		\includegraphics[height=8cm]{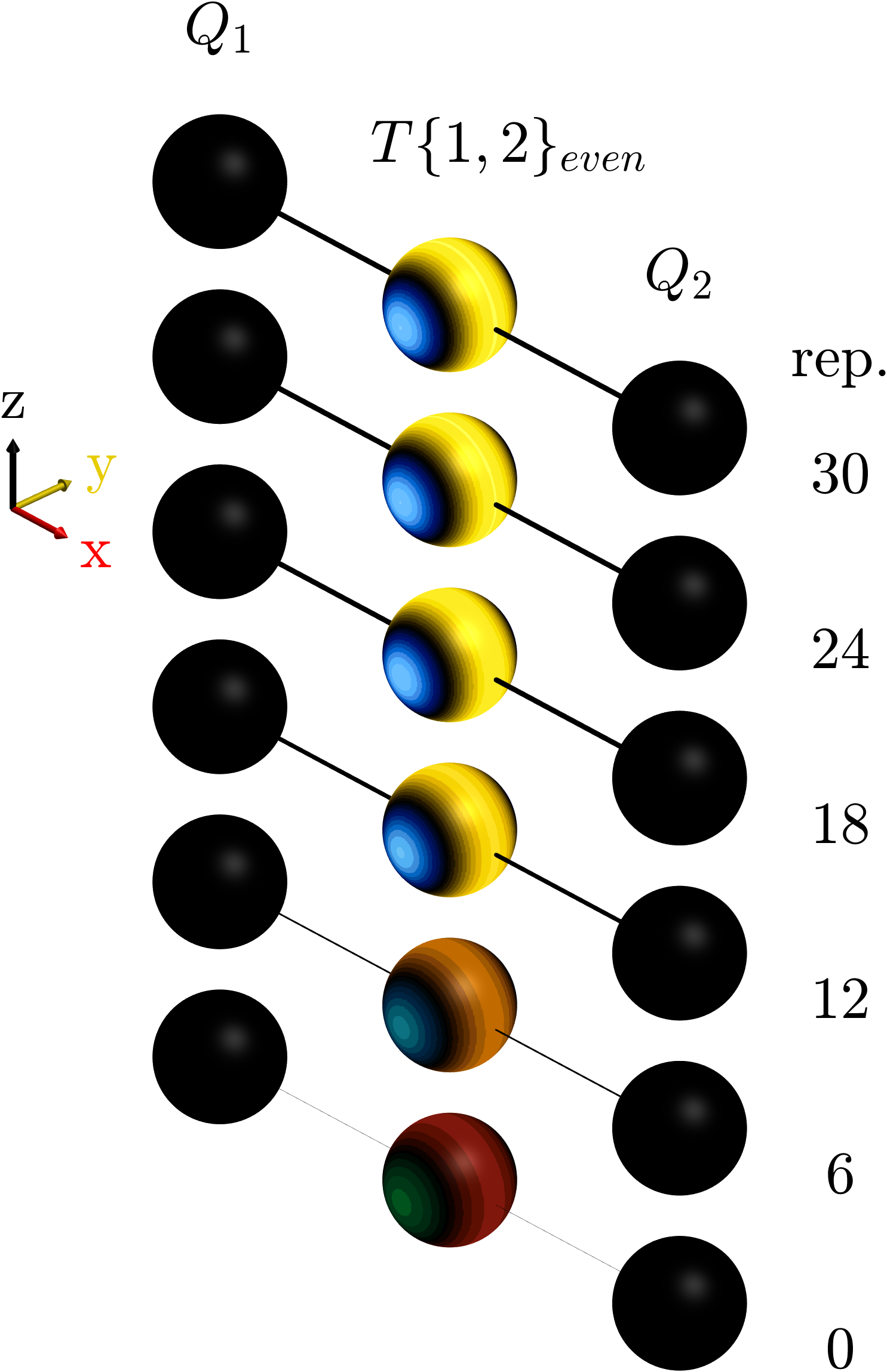}
		\subcaption{Total correlation function beads\newline(T-Beads).}
		\label{fig:distill_T}
	\end{subfigure}
	\hfill
	\begin{subfigure}[t]{0.48\textwidth}
		\centering
		\includegraphics[height=8cm]{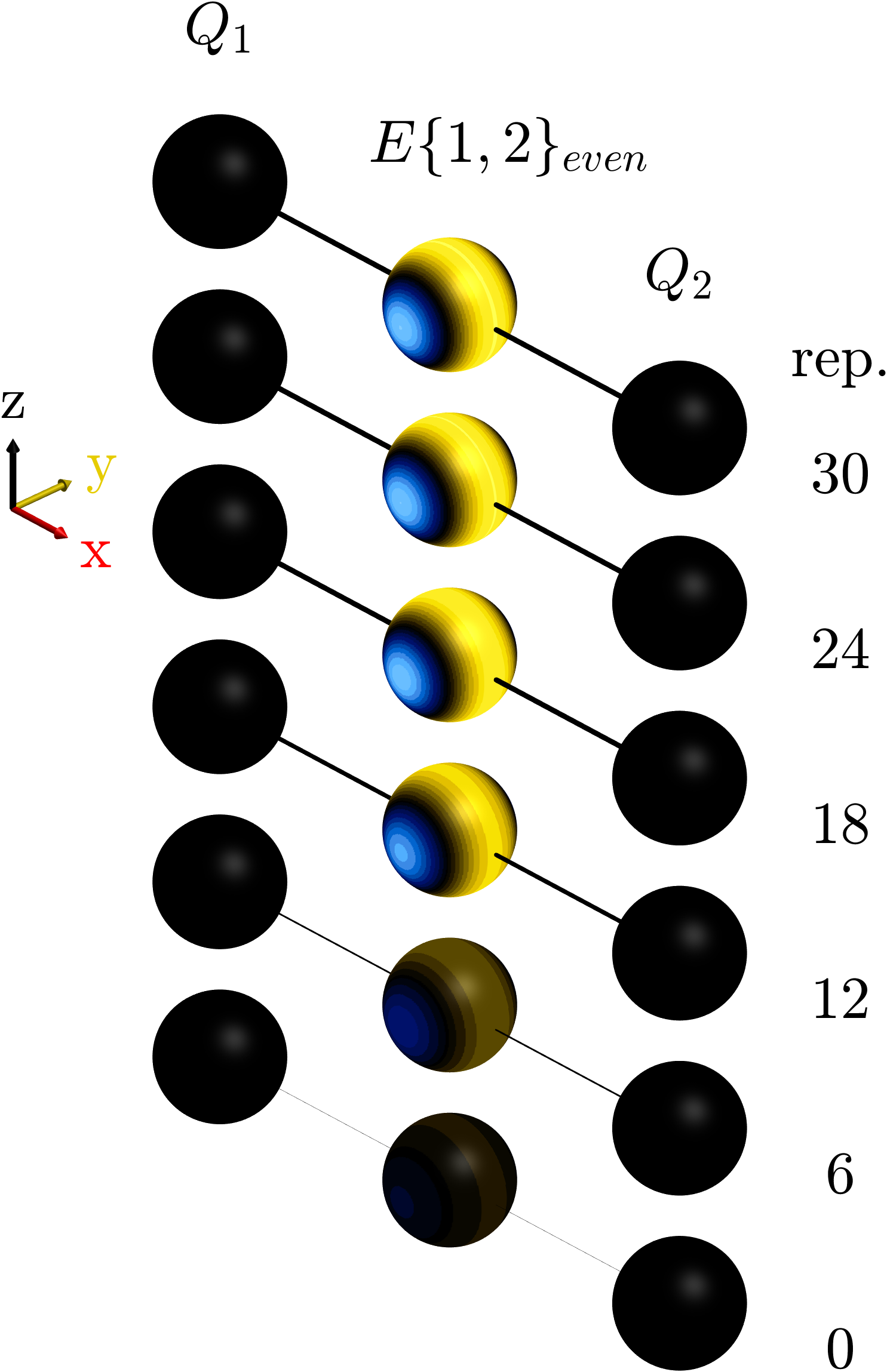}
		\subcaption{Entanglement-based correlation function beads (E-Beads).}
		\label{fig:distill_E}
	\end{subfigure}
	\caption{Generalized BEADS visualization of the quantum states obtained after different numbers of repetitions of the BBPSSW entanglement distillation protocol. The transition from a weakly entangled mixed state at 0 repetitions to the maximally entangled Bell state $\ket{\Phi^+}$ is directly clear in the visualization. The initial fidelity at repetition 0 is $F=0.58$.}
	\label{fig:ent_distillation_beads}
\end{figure}

\cref{fig:ent_distillation_beads} displays stacked-plot $T$-Bead and $E$-Beads visualizations of the Werner states $W_F$ obtained after the $n$-th repetition of the BBPSSW entanglement distillation protocol. The initial fidelity was chosen to be $F=0.58$. As can be directly seen from the visualization, the initial density operator at repetition 0 is a very weakly entangled mixed state. With increasing number of repetitions the color brightness of the $E$-Beads increases. At the same time, the color hue of the shown $T$-Beads changes from red and green (dominant separable structure of the underlying density operator) at repetition 0 to yellow and blue corresponding to entanglement at higher repetitions. The resulting state after 30 repetitions is approximately the maximally entangled Bell state $\ket{\Phi^+}$.

\subsubsection{Decoherence processes: Dephasing} \label{sec:DecoherenceMain}
In quantum systems, several different types of noise exist, which can cause a quantum state to decohere. In the following, we will discuss the effect of the phase-flip or \textit{dephasing} channel on three-qubit GHZ and W states. Further examples including depolarization and amplitude damping are discussed in supplementary \cref{sup:decoherence}. 

A phase-flip (Pauli Z-gate) occuring with probability $p$ can be described by corresponding single-qubit Kraus operators \cite{nielsen_quantum_2010}:
\begin{subequations}
	\begin{align}
		E_0 &= \sqrt{1-p}\,\,\mathbbm{1} = \sqrt{1-p} \begin{pmatrix} 1 & 0 \\ 0 & 1 \end{pmatrix}, \label{eq:kraus_dephasing_E0} \\
		E_1 &= \sqrt{p} \, \, Z = \sqrt{p} \begin{pmatrix} 1 & 0 \\ 0 & -1 \end{pmatrix},
		\label{eq:kraus_dephasing_E1}
	\end{align}
\end{subequations}
where $E_0$ describes the case for which no error occurs, whereas $E_1$ corresponds to the phase-flip taking place.

Geometrically, dephasing noise projects the single-qubit Bloch vectors onto the $z$-axis such that they lose their $x$- and $y$-components (coherences). Clearly, except for the case where the initial system already was in one of the computational basis states $\{\ket{0},\ket{1}\}$, the resulting state will be mixed as the projection will decrease the length (norm) of the Bloch vector to values smaller than 1 \cite{nielsen_quantum_2010}. For the following analysis, we assume that the single-qubit Kraus operators defined in \cref{eq:kraus_dephasing_E0,eq:kraus_dephasing_E1} are locally applied to all qubits of the observed three-qubit systems. That is, the three-qubit dephasing process is described by a set of eight three-qubit Kraus operators which correspond to tensor products of all possible combinations of the single-qubit Kraus operators.

\begin{figure}[H] 
	\vspace{-.4cm}
	\centering
	\makebox[\textwidth][c]{
		\begin{minipage}{1\textwidth}
			\begin{subfigure}[t]{0.5\textwidth}
				\centering
				\includegraphics[width=\textwidth,trim={6cm 4cm 5.8cm 5cm},clip]{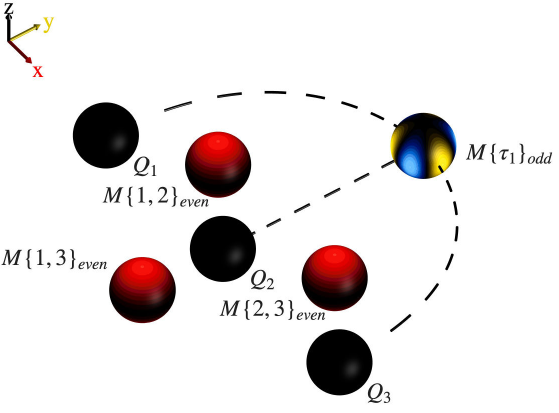}
				\subcaption{$p = 0$}
				\label{fig:GHZ_dph0_M}
			\end{subfigure}
			\begin{subfigure}[t]{0.5\textwidth}
				\centering
				\includegraphics[width=\textwidth, trim={6cm 4cm 5.8cm 5cm},clip]{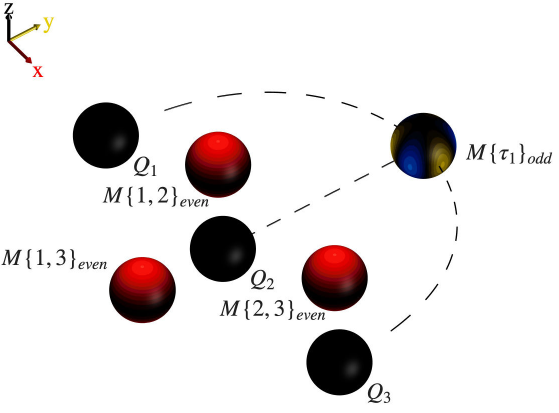}
				\subcaption{$p = 0.1$}
				\label{fig:GHZ_dph0.1_M}
			\end{subfigure}
		\end{minipage}
	}
	
	\vspace{0.3cm}
	
	\makebox[\textwidth][c]{
		\begin{minipage}{1\textwidth}
			\begin{subfigure}[t]{0.5\textwidth}
				\centering
				\includegraphics[width=\textwidth, trim={6cm 4cm 5.8cm 5cm},clip]{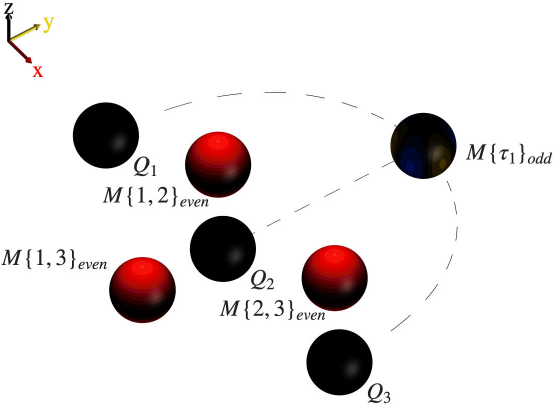}
				\subcaption{$p = 0.2$}
				\label{fig:GHZ_dph0.2_M}
			\end{subfigure}
			\begin{subfigure}[t]{0.5\textwidth}
				\centering
				\includegraphics[width=\textwidth, trim={6cm 4cm 5.8cm 5cm},clip]{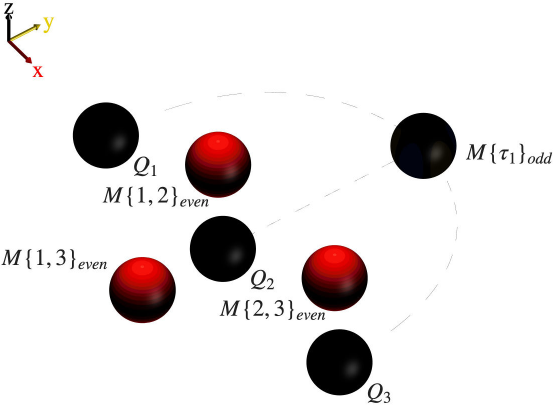}
				\subcaption{$p = 0.3$}
				\label{fig:GHZ_dph0.3_M}
			\end{subfigure}
		\end{minipage}
	}
	
	\vspace{0.3cm}
	
	\makebox[\textwidth][c]{
		\begin{minipage}{1\textwidth}
			\begin{subfigure}[t]{0.5\textwidth}
				\centering
				\includegraphics[width=\textwidth, trim={6cm 4cm 5.8cm 5cm},clip]{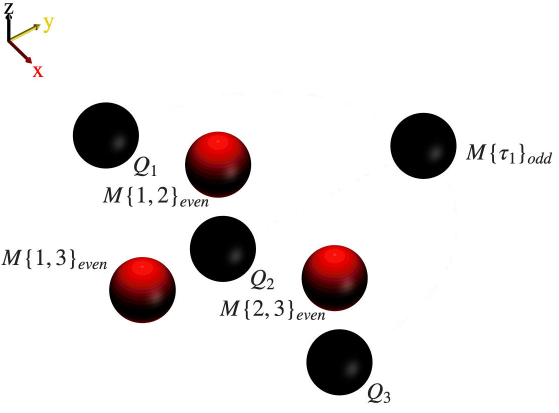}
				\subcaption{$p = 0.4$}
				\label{fig:GHZ_dph0.4_M}
			\end{subfigure}
			\begin{subfigure}[t]{0.5\textwidth}
				\centering
				\includegraphics[width=\textwidth, trim={6cm 4cm 5.8cm 5cm},clip]{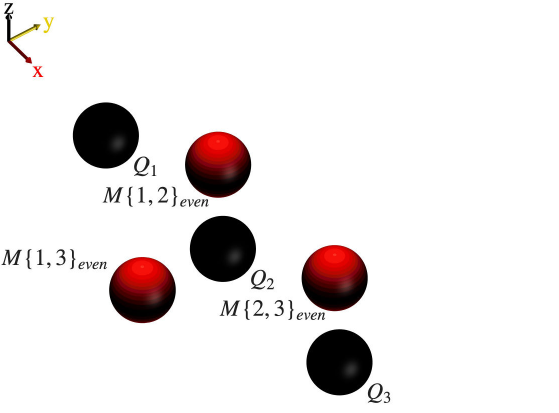}
				\subcaption{$p = 0.5$}
				\label{fig:GHZ_dph0.5_M}
			\end{subfigure}
		\end{minipage}
	}
	\caption{Total correlation ($M$-Beads) generalized BEADS representation visualizing the effect of the dephasing channel (compare \cref{eq:kraus_dephasing_E0,eq:kraus_dephasing_E1}) on the GHZ state. The genuine tripartite correlations (here, entanglement) vanish with increasing $p$. The final state at $p=0.5$ is $\rho_{p=0.5} = 1/2(\ket{000}\bra{000} + \ket{111}\bra{111})$, which is exclusively bipartite classically correlated.}
	\label{fig:ghz_dephasing}
	\vspace{-.3cm}
\end{figure}

\begin{figure}[H] 
	\centering
	\makebox[\textwidth][c]{
		\begin{minipage}{1\textwidth}
			\begin{subfigure}[t]{0.5\textwidth}
				\centering
				\includegraphics[width=\textwidth,trim={6cm 4cm 5.8cm 5cm},clip]{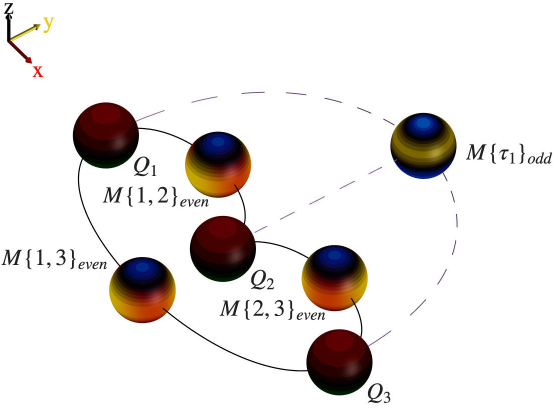}
				\subcaption{$p = 0$}
				\label{fig:W_dph0_M}
			\end{subfigure}
			\begin{subfigure}[t]{0.5\textwidth}
				\centering
				\includegraphics[width=\textwidth,trim={6cm 4cm 5.8cm 5cm},clip]{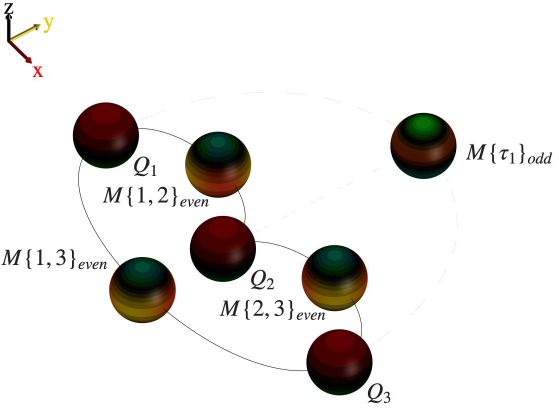}
				\subcaption{$p = 0.1$}
				\label{fig:W_dph0.1_M}
			\end{subfigure}
		\end{minipage}
	}
	
	\vspace{0.3cm}
	
	\makebox[\textwidth][c]{
		\begin{minipage}{1\textwidth}
			\begin{subfigure}[t]{0.5\textwidth}
				\centering
				\includegraphics[width=\textwidth,trim={6cm 4cm 5.8cm 5cm},clip]{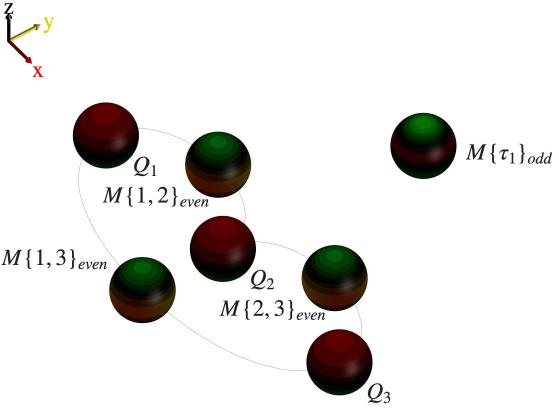}
				\subcaption{$p = 0.2$}
				\label{fig:W_dph0.2_M}
			\end{subfigure}
			\begin{subfigure}[t]{0.5\textwidth}
				\centering
				\includegraphics[width=\textwidth,trim={6cm 4cm 5.8cm 5cm},clip]{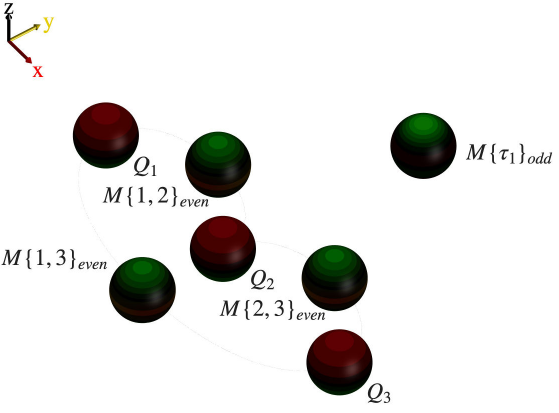}
				\subcaption{$p = 0.3$}
				\label{fig:W_dph0.3_M}
			\end{subfigure}
		\end{minipage}
	}
	
	\vspace{0.3cm}
	
	\makebox[\textwidth][c]{
		\begin{minipage}{1\textwidth}
			\begin{subfigure}[t]{0.5\textwidth}
				\centering
				\includegraphics[width=\textwidth,trim={6cm 4cm 5.8cm 5cm},clip]{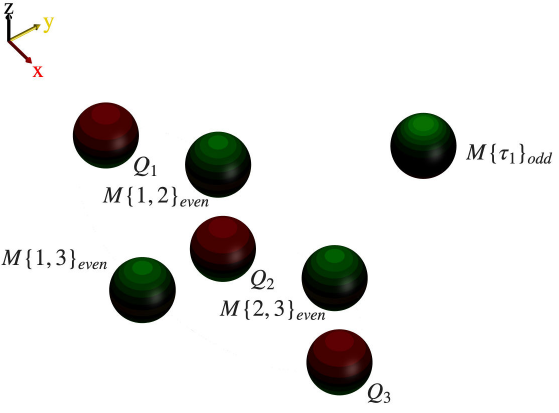}
				\subcaption{$p = 0.4$}
				\label{fig:W_dph0.4_M}
			\end{subfigure}
			\begin{subfigure}[t]{0.5\textwidth}
				\centering
				\includegraphics[width=\textwidth,trim={6cm 4cm 5.8cm 5cm},clip]{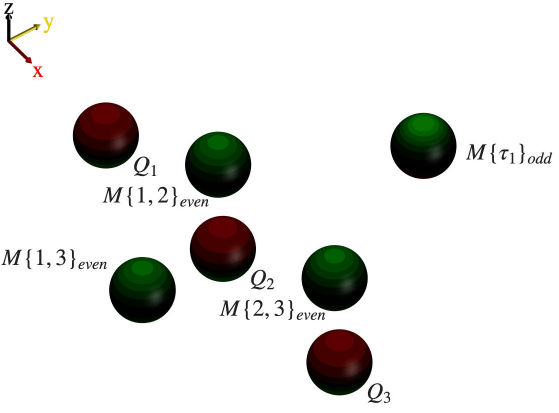}
				\subcaption{$p = 0.5$}
				\label{fig:W_dph0.5_M}
			\end{subfigure}
		\end{minipage}
	}
	\caption{Total correlation ($M$-Beads) representations of the states obtained after applying the local dephasing channel (compare \cref{eq:kraus_dephasing_E0,eq:kraus_dephasing_E1}) on all qubits of the tripartite W state with channel probability $p$. All spherical function components $x$- and $y$-direction vanish with increasing $p$. The final state at $p=0.5$ is exclusively classically correlated with all remaining genuinely correlation-based components being aligned axially symmetric with respect to the $z$-axis.}
	\vspace{-.3cm}
	\label{fig:w_dephasing}
\end{figure}

\cref{fig:ghz_dephasing} visualizes the effect of dephasing on the GHZ state in terms of $M$-Beads representations of the resulting states for different error probabilities $p$. Starting from the pure GHZ state at $p=0$ (see \cref{fig:GHZ_dph0_M}), only the tripartite correlations visualized by the $M\{\tau_1\}_{odd}$-Bead decrease with increasing $p$ until they completely vanish (cf. \cref{fig:GHZ_dph0.5_M}). The bipartite classical correlations, which correspond to $M$-Beads that are entirely axially symmetric with respect to the $z$-axis, remain unaffected by the dephasing channel. The final state at $p=0.5$ is the maximally bipartite classically correlated state $\rho_{p=0.5} = 1/2(\ket{000}\bra{000} + \ket{111}\bra{111})$. Notably, GHZ entanglement is quite robust against dephasing and the states remain genuinely tripartite entangled for all $p<0.5$. The entanglement arc at $p=0.4$ is marginally thin and thus, only visible on very close and careful inspection.
\cref{fig:W_dph0_M} shows an analogous series of $M$-Beads representations for dephasing of the three-qubit W~state. Unlike the GHZ state, the W state has both, two- and three-body $x$- and $y$-terms in its Pauli expansion which can be attributed to genuine correlations of the state. Thus, starting from the pure W~state at $p=0$ (see \cref{fig:W_dph0_M}), these bi-and tripartite diminish with increasing $p$ (cf. \cref{fig:W_dph0.1_M} to \cref{fig:W_dph0.5_M}). Note that at $p=0.2$ the genuine tripartite and only at $p = 0.5$ all entanglement has vanished. Thus, W-type tripartite entanglement is less robust against dephasing than GHZ-type entanglement. The resulting state for $p=0.5$ is $\rho = 1/3(\ket{001}\bra{001}+\ket{010}\bra{010}+\ket{100}\bra{100})$ and is classically correlated, which can be directly deduced from all beads being fully axially symmetric with respect to the $z$-axis.

Just by looking at these examples, the effect of the dephasing channel on arbitrary quantum states can be deduced phenomenologically using the generalized BEADS representation: The channel eliminates all but the $z$-components of each state. If the initial state was genuinely correlated, the residual state after full action of the channel will be classically correlated, as only the $z$-terms being invariant implies that all qubits in the final resulting state can be expressed in a single basis (the computational basis). In the BEADS representation this is directly reflected in only the $z$-components of all $Q$- and $M$-Beads being left after the full action of the channel.
\begin{enumerate}[start = 15, label=\textbf{Result \arabic*:}, itemindent=1.15cm, labelindent=1cm]
	\item \label{res:BEADS13} The generalized BEADS representation allows to deduce the effects and actions of quantum information algorithms and protocols as well as arbitrary quantum channels phenomenologically by considering the visualizations of input and (intermediate) output states.
\end{enumerate}

\section{Discussion}
As illustrated by the examples and results discussed in \cref{sec:Results}, the generalized BEADS representation of pure and mixed quantum states provides an intuitive and informationally complete view on quantum states. By visualizing the correlation structure of quantum states of interest, the generalized BEADS representation provides an illustrative access to otherwise abstract concepts of quantum information theory. In particular, we have shown how the visualization is capable of directly visualizing $n$-partite entanglement, total correlation, classical correlation, as well as nonclassical correlations. 

By recursively optimizing pure states ensembles of all possible marginals of a given state, the generalized BEADS representation enables to uniquely identify the entangling and separable structures of each marginal, respectively, and thus to characterize the entanglement and other types of correlations specific to this marginal. This is a crucial advancement compared to the original BEADS representation \cite{huber_beads_2025}, which uses the connected correlation function density operator components of a pure state without relying on any optimizations to visualize genuinely correlation-based information. While this is sufficient to identify $N$-partite entanglement of an $N$-partite pure state, it does not directly reveal $n$-partite entanglement ($n<N$) of reduced marginals which, in general, are mixed. Hence, the interpretation of such marginals in the original BEADS representation is confined to identifying genuine correlations -- without further characterization of the exact types -- which result as a direct consequence of the pure state being globally entangled. On the other hand, since the original representation variant does not require any optimizations, it is much less computationally expensive for which it is well-suited used for visualizing the dynamics of quantum systems in real-time simulations as, e.g., in the \textit{QuBeads} software \cite{huber_qubeads_2024}. 

Importantly, the generalized BEADS representation is always specific to a distinct pure state decomposition of a mixed quantum state. While this implies non-uniqueness of the representation, it must be stressed that this is of course a natural consequence of the properties of mixed quantum states which in general have non-unique decompositions. In fact, as we have shown in the results section, this intrinsic non-uniqueness does not always correspond to differing mixed state BEADS representations though, such that the latter only vary in the presence of special symmetry properties. The same applies to classically accessible and inaccessible information, which are also only non-unique if the state admits symmetry-induced degrees of freedom. Thus, the generalized BEADS representation is indeed able to directly reveal such degrees of freedom on a visual level, which are otherwise not directly clear.

The introduction of tangle-based and auxiliary W entanglement arcs in the generalized BEADS representation represents another improvement over the original BEADS representation. Not only do these arcs allow to quickly recognize any type of $n$-partite entanglement, but they also reveal further subtle features of the entanglement structure of multipartite states that goes beyond the well-known SLOCC classifications. For instance, as was shown for tripartite states, this allows to characterize how structurally close a state is to an ideal GHZ state, or conversely, to an ideal W state. In the original BEADS representation, the used entanglement or correlation arcs, are scaled proportionally to the $n$-partite connected correlation function norms, which are not sufficient for identifying and quantifying entanglement of reduced marginals let alone for in-depth characterizations of the entanglement on a structural level.

Moreover, as the generalized BEADS representation visualizes the structural components of a density operator that correspond, e.g., to specific correlation types, it exhibits additivity relations, which are usually not directly encountered in quantum information theory, especially when considering entropic correlation measures. For instance, the complete BEADS representation of a state intuitively corresponds to the sums of the representations of classical and nonclassical, or entangling and separable structures of the state.

\begin{figure}[H]          
	\centering
	\includegraphics[width=1.0\textwidth]{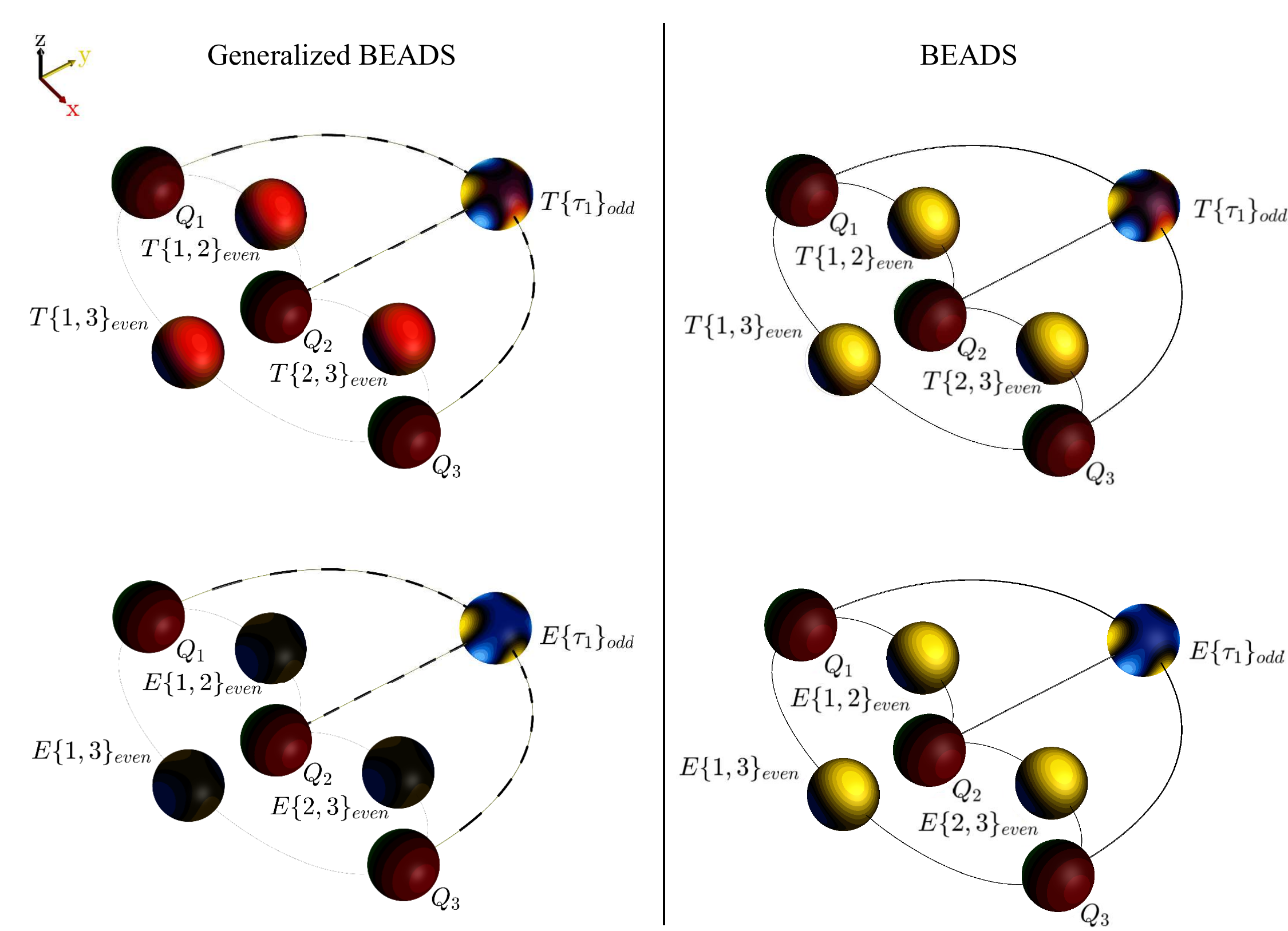}
	\caption{Generalized (right) and original (left) BEADS representations of the pure tripartite state $\ket{\psi} = \sqrt{0.8}\ket{\text{GHZ}} + \sqrt{0.2}\ket{\text{W}}$ using $T$- and $E$-Bead visualizations. While the tripartite beads coincide in both visualizations, the correlation structures of the mixed bipartite marginals are only correctly represented in the generalized BEADS representation. In the generalized BEADS representation, two- and three-tangle arcs further reveal the state to have weak bipartite and strong tripartite entanglement and to belong to the GHZ SLOCC class in the generalized BEADS representation. Moreover, the additional W-arc indicates a structural deviation from an ideal GHZ state, which coincides with the state being a coherent mixture of the GHZ and W states. Bipartite arcs in the original BEADS representation do not directly correspond to bipartite entanglement.}
	\label{fig:DisBEADSComp}        
\end{figure}

To further illustrate the discussed developments and improvements, consider \cref{fig:DisBEADSComp} which shows a side-by-side comparison of the pure state $\ket{\psi} = \sqrt{0.8}\ket{\text{GHZ}} + \sqrt{0.2}\ket{\text{W}}$ visualized by the generalized and the original BEADS representation in terms of $T$- and $E$-Beads. The tripartite correlation function beads coincide in both cases, as the state is globally pure. Hence, the tripartite $E\{\tau_1\}_{odd}$-Bead gives a direct representation of the tripartite entanglement of the state. Moreover, the tripartite entanglement arcs in the generalized BEADS representation allow to infer that the state is GHZ class due to the nonzero three-tangle arc but deviates structurally from an ideal GHZ state reflected in a green nonzero W-arc.
From the generalized BEADS representation, it is also clear that the mixed bipartite marginals of the state are weakly entangled which is directly reflected in the corresponding low-brightness bipartite $E$-Beads and thin two-tangle arcs. Here, the bipartite $E$-Beads directly visualize the entangling structures of the bipartite reduced density operators, i.e., they provide a direct visualization of bipartite entanglement. On the contrary, in the original BEADS representation, the $E$-Beads merely visualize the manifestations of genuine bipartite correlations which result from the global state being pure and entangled. The connected correlation function norm arc cannot be used to infer information about the genuine bipartite entanglement of the state.

Self-evidently, the generalized BEADS representation relies on nontrivial nonlinear optimizations that become more involved the larger the considered system size. It is still possible to obtain full BEADS representations of mixed states for systems of up to 
three qubits in a reasonable amount of time (seconds to hours depending on the state) on commercially available hardware. 
Note that for larger systems, the representation itself relies on a large number of beads, for which it is reasonable to examine and visualize smaller reduced subsystems instead. We currently investigate possibilities to incorporate machine learning approaches to speed up the optimization workflow.

In order to verify that the generalized BEADS representation and in particular, the underlying optimization framework yields correct results, we have validated the representation using parameterized states for which separability thresholds are well-known. These states include Werner states, and GHZ states with added white or classical noise. An overview of the obtained validation results is provided in \cref{app:benchmarks}.

Whereas the original BEADS representation \cite{huber_beads_2025} is specifically  well-suited for but not restricted to educational purposes, the generalized BEADS representation is considerably more powerful, and thus, appropriate for applications in research, where in-depth information theoretic analyses of correlation structures of states of interest are required. At the same time, the generalized BEADS representation can be used in a didactic setting to introduce the information theory of mixed quantum states on a visual and easily accessible level as demonstrated in supplementary \cref{sec:didactic_intro}.

The concept of analyzing and visualizing entire or special components of correlation functions to explore the correlation structure of a quantum state is closely related to analyzing $k$-body terms of the Pauli expansion of the corresponding density operator. A popular and easily accessible tool that is based on such $k$-body terms are sector lengths \cite{aschauer_2004, wyderka_2020}, which can be used for providing sufficient criteria, e.g., for biseperability or genuine multipartite entanglement of quantum states \cite{wyderka_2020}. Sector lengths are easy to compute and, unlike the generalized BEADS representation, do not require pure state ensemble optimizations to assess entanglement properties of mixed states. However, in the generalized BEADS representation, the occurrence of $n$-partite $E$-Beads uniquely indicates $n$-partite entanglement, that is, the existence of specific $E$-Beads is a necessary criterion for corresponding entanglement. Indeed, it is possible to use the generalized BEADS representation to construct cases of quantum states where the representation correctly represents all characteristic properties but sector-length-based sufficient entanglement (or separability) criteria fail to do so. An in-depth analysis is provided in supplementary \cref{sec:sector_length}.

In our article on the original BEADS representation \cite{huber_beads_2025} we have also briefly discussed the BEADS representation of pure quantum states as a generalization of the corresponding Husimi $\mathcal{Q}$ function \cite{husimi_1940, mintert_husimi_2004, sugita_husimi_2003} and how the representation is associated with the closely related Majorana star representation \cite{majorana_1932, devi_majorana_2012}. As the Husimi $\mathcal{Q}$ function of a pure state is the squared magnitude of the overlap of an $N$-qubit quantum state of interest with the fully symmetric spin coherent states $\ket{s(\theta, \phi)}$ in a spatial direction $(\theta, \phi)$, it generalizes in a straight-forward manner to mixed quantum states:
\begin{equation}
\mathcal{Q}(\theta,\phi)=\langle s(\theta, \phi) |\rho| s(\theta, \phi) \rangle.
\end{equation}
The mixed state Husimi $\mathcal{Q}$ function naturally has the same mathematical properties as its pure state counterpart. As a consequence, it is only a representation of the fully symmetric sector, i.e., the fully permutation-symmetric subspace, of a given density operator and thus, unlike the generalized BEADS representation, is not informationally complete. Self-evidently, the Husimi $\mathcal{Q}$ function of a mixed state cannot directly characterize its correlation structure, but it can still be directly obtained from the corresponding generalized BEADS representation as the average over all fully permutation-symmetric $T$-, $Q$-, and the identity Beads.

On the contrary, the Majorana representation does not generalize naturally to mixed states in a similarly simple way the $\mathcal{Q}$ function does. Nevertheless, a generalization to mixed state has been proposed in \cite{braun_majorana_2020}. However, this construction does not share the same analogies to the $\mathcal{Q}$ function as are present in the pure state case, and the representation is also not informationally complete let alone able to reveal the full correlation structure of a mixed state.

\section{Conclusion}
In this article, we extended the BEADS representation of pure quantum states \cite{huber_beads_2025} to the generalized BEADS representation of pure and mixed quantum states. Hereby, we particularly showed that it is indeed possible to intuitively visualize intricate correlation structures of mixed quantum states and mixed marginals of pure states uniquely with respect to a given constituting pure state ensemble. The introduced extensions comprise additional bead types and a comprehensive optimization framework in order to be able to represent different correlation types in a mathematically complete way. As the representation is based on visualizations of correlation function components, it provides a wealth of structural insights -- e.g., how specific correlation types manifest and influence in the outcomes of joint local measurements -- that go well beyond but are highly complementary to analyses using common entanglement and correlation measures.

We demonstrated how the generalized BEADS representation can be used to foster a deeper understanding of quantum algorithms, communication protocols, and even noise processes, which are based around mixed quantum states. Thus, the visualization sheds light on the otherwise abstract and subtle features of quantum states examined in quantum information theory.

\ack{The authors acknowledge helpful discussions with Barbara Kraus, Christian Schilling, and Thomas
Schulte-Herbr\"uggen.}

\funding{S G and D H acknowledge funding from the Digital Europe Programme for the project DigiQ: Digitally
Enhanced Quantum Technology Master (DigiQ) under Grant Agreement ID 10 108 4035, D H acknowledges funding by a Kekulé stipend of
the Verband der chemischen Industrie e.V. (VCI), TUM EDU grant 3170 573, S G and D H acknowledge funding from the Munich Center of Quantum
Science and Technology (MCQST), the Munich Quantum Valley (MQV), and the Bavarian NMR Center (BNMRZ).}

\roles{Examples: C B, S G, D H, Conceptualization: D H, Methodology: C B, D H Investigation: C B, D H,
Visualization: C B, S G, D H  Supervision: S G (secondary), D H (primary), Writing original draft: C B, D H, Writing review \& editing: C B, S G, D H.}

\data{All data are available in the main text or in the supplementary material.}

\suppdata{We provide supplementary information accompanying this article which has the following content:
\begin{itemize}
\item 
Mathematical and conceptual details and backgrounds for the construction of the generalized BEADS representation, \cref{sup:beads_overview}-\cref{sec:SuppMathDetailsEntQuant}.
\item
A didactic introduction to mixed states and the generalized BEADS representation, \cref{sec:didactic_intro}.
\item
An instruction on how to derive the entanglement threshold of parameterized Werner states phenomenologically using the generalized BEADS representation, \cref{sec:SuppEntThreshold}.
\item
Further bipartite examples of classically correlated, separable and entangled mixed states, \cref{supp:bipartite}.
\item 
Further tripartite examples: Parametrized incoherent and coherent mixtures of GHZ and W states, \cref{sup:tripartite}.
\item
An addendum to \cref{app:benchmarks}: $E$-Bead visualizations of parameterized tripartite noisy GHZ states \cref{sup:Benchmark}.
\item 
An addendum to \cref{sec:DecoherenceMain}: Decoherence processes of GHZ and W states, \cref{sec:DecoherenceMain}.
\item 
A comparison between the characterization of quantum states using the generalized BEADS representation and sector lengths \cite{wyderka_characterizing_2020}, \cref{{sec:sector_length}}.
\end{itemize}
}

\clearpage
\appendix
\titleformat{\section}
  {\normalfont\normalsize\bfseries}
  {Appendix \thesection}
  {1em}
  {}

\section{A brief discussion on the applicability of common \textit{n}-partite entanglement measures in ensemble optimizations of mixed quantum states}
\label{sec:app_tangles}

The average two-tangle \cite{coffman_distributed_2000,suter_entanglement_nodate} is given by:
\begin{equation}
	\overline{\tau_2}= \frac{C_{12}^2+C_{23}^2+C_{13}^2}{3},
	\label{eq:2_tangle}
\end{equation}
where the two-tangles $\tau_{2,{jk}} = C_{jk}^2$ are the squared concurrences of the bipartite reduced marginals formed by qubits $j$ and $k$. $\overline{\tau_2}$ measures the average pairwise entanglement as commonly encountered for Bell or W-type \cite{zeilinger_nasa_1997,dur_three_2000} entangled states and is easily generalizable to larger systems. However, the average two-tangle is zero for genuine three-way (or more general $n$-way) entanglement as present, e.g., in GHZ-states \cite{greenberger_going_2007,greenberger_bells_1990}, for which it is inapplicable as a cost function for least-$n$-partite-entangled ensemble optimizations.

A natural extension for the description of three-way entanglement is given by the bipartite concurrence
\begin{equation}
	C_{i(jk)}= \sqrt{2-2\,\text{Tr}(\rho_i^2)},
	\label{eq:bipartite_concurrence}
\end{equation}
which describes the entanglement between the bipartition defined by the cut between qubit $i$ and the bipartite marginal formed by qubits $j$ and $k$. This quantity, however, does not distinguish whether qubit $i$ is genuinely tripartite entangled with qubits $j$ and $k$, i.e., whether the entanglement is equally shared between all three qubits, or whether the entanglement is only bipartite, that is, qubit $i$ is only two-body entangled with either qubit $j$ or $k$. This also renders $C_{i(jk)}$ unfeasible to be used in pure state ensemble optimizations. 

The three-tangle $\tau_3$ is designed to quantify genuine three-way (GHZ-type) entanglement which is achieved by subtracting the contributions from pairwise entanglement measured by two-tangles from the bipartite concurrence (see Eq.~\ref{eq:bipartite_concurrence}). For a pure tripartite state, $\tau_3$ is defined as \cite{coffman_distributed_2000,suter_entanglement_nodate}
\begin{equation}
	\tau_3= C_{i(jk)}-(C_{ij}^2+C_{ik}^2).
	\label{eq:3_tangle}
\end{equation}
Opposite to the average two-tangle, the three-tangle is zero for pairwise and W-type entangled states. Hence, it is also unusable as a cost function for the desired optimizations.

While generalizations of the three-tangle to $n$-partite entanglement, called the $n$-tangle, have been proposed \cite{wong_ntangle_2001}, these are generally challenging to compute for mixed quantum states.

\section{A quantifier for capturing W-type entanglement}\label{sec:EntQuant}
As was shown in section~\ref{sec:Wmixtures}, it is possible that mixed quantum states in systems with odd numbers of qubits are genuinely multipartite entangled while showing no corresponding correlation function \cite{kaszlikowski_quantum_2008,tran_genuine_2017}. This can be partially compensated by augmenting the generalized BEADS representation by entanglement arcs which are scaled in diameter in dependence of an entanglement measure that captures the corresponding multipartite entanglement. In the tripartite case, the three tangle \cite{coffman_distributed_2000} represents a popular choice for quantifying GHZ-type entanglement. However, the three-tangle $\tau_3$ is a zero-measure with respect to W states, that is, the three-tangle is $\tau_3=0$ for all W states.

In order to also quantify the W-class entanglement it seems natural to make use of the entanglement-based connected correlation function norm 
\begin{equation}
\mathcal{N}_3 = || \rho_{E,123} ||_F = \sqrt{\text{Tr}(\rho_{E,123} \rho_{E,123}^\dagger)}
\end{equation} for pure states or the convex-roof extension
\begin{equation}
\mathcal{N}_3 = \inf_{\{\ket{\psi_{i,123}}\}}\left(\sum_i p_i || \rho_{E,i,123} ||_F\right)
\end{equation} in the case of mixed states, where $\rho_{E,i,123}$ is the density operator component corresponding to the tripartite connected correlation function of the $i$-th tripartite pure state $\ket{\psi_{i,123}}$ contributing with weight $p_i$ in a pure state ensemble realizing a density operator $\rho$. The infimum is again taken over all possible pure state ensembles $\{\ket{\psi_{i,123}}\}$. $\mathcal{N}_3$ is nonzero for both, GHZ- and W-type entanglement. Inconveniently, it does not distinguish between the two types of entanglement by itself. We can still isolate the $W$-type contribution to $\mathcal{N}_3$ as follows.

We first map $\mathcal{N}_3$ onto the three-tangle $\tau_3$ for all ideally GHZ-type entangled states. The required mapping can be derived by considering the parameterized pure state $\ket{\psi}_\theta = \cos(\theta) \ket{000} + \sin(\theta) \ket{111}$, which is a product state for $\theta = 0$ and structurally purely GHZ-type entangled for $0 < \theta \leq \frac{\pi}{4}$. As the degree of entanglement of $\ket{\psi}_\theta$ varies in dependence of $\theta$, we can determine the three tangle $\tau_3$ as a function of $\theta$. Recall that for a general pure state $\ket{\phi} = \sum_{i,j,k = 0}^1 a_{i,j,k} \ket{ijk}$, the three-tangle is given by \cite{coffman_distributed_2000}
\begin{equation}
\tau_3 = 4|\text{Det}(\ket{\phi})| = 4|d_1 - 2d_2 + 4d_3|,
\end{equation}
where $\text{Det}(\cdot)$ denotes the Cayley hyperdeterminant with terms
\begin{subequations}
\begin{align}
d_1 &= a_{000}^2 a_{111}^2 + a_{001}^2 a_{110}^2 + a_{010}^2 a_{101}^2 +a_{100}^2 a_{011}^2 \\
d_2 &= a_{000} a_{111}a_{001} a_{110} + a_{000} a_{111}a_{010} a_{101} +a_{000} a_{111}a_{100} a_{011} \nonumber\\
&+ a_{001} a_{110}a_{010} a_{101} + a_{001} a_{110}a_{100} a_{011} + a_{010} a_{101} a_{100} a_{011} \\
d_3 &= a_{000}a_{110}a_{101}a_{011} + a_{111}a_{001}a_{010}a_{100}.
\end{align}
\end{subequations}
Conveniently, for $\ket{\psi}_\theta$ the only nonzero terms are $a_{000} = \cos(\theta)$ and $a_{111} = \sin(\theta)$ and we get
\begin{equation}
\tau_3(\theta) = 4\cos^2(\theta)\sin^2(\theta) = \sin^2(2\theta) = \frac{1-\cos(4\theta)}{2}.
\label{eq:threetangle}
\end{equation}
Similarly, we can write out the parameterized density operator $\ket{\psi}_\theta \bra{\psi}_\theta$, compute its tripartite connected correlation function $\rho_{E,123}$ according to Eq.~\ref{eq:tripartite_E_correlationfunction}, and determine the entanglement-based connected correlation norm $\mathcal{N}_3$. As the derivation is quite long, we omit it here and refer the interested reader to supplementary section~\ref{sec:SuppMathDetailsEntQuant}, where auxiliary details are presented. For the subsequent steps, it is further advantageous to square and normalize  $\mathcal{N}_3$, i.e., $\tilde{\mathcal{N}}_3 = 2\mathcal{N}_3^2$ such that it adopts values in the interval $[0;1]$. For the state $\ket{\psi}_\theta$ we then arrive at
\begin{equation}
\tilde{\mathcal{N}}_3(\theta) = \frac{1}{8}(\cos^3(4\theta)-\cos^2(4\theta)-5\cos(4\theta)+5).
\label{eq:entnorm}
\end{equation}
We can now substitute $\epsilon = \cos(4\theta)$ in Eqs.~\ref{eq:threetangle} and~\ref{eq:entnorm}:
\begin{subequations}
\begin{align}
\tau_3(\epsilon) &= \frac{1-\epsilon}{2},\label{eq:threetangle2}\\
\tilde{\mathcal{N}}_3(\epsilon) &= \frac{1}{8}(\epsilon^3-\epsilon^2-5\epsilon + 5).
\end{align}
\end{subequations}
It is clear that $\tilde{\mathcal{N}}_3$ can then be mapped onto $\tau_3$ by finding the $\epsilon$ that is a real-valued solution to the equation
\begin{equation}
0 = \epsilon^3-\epsilon^2-5\epsilon + 5 - 8\tilde{\mathcal{N}}_3
\end{equation}
and which lies in $-1 \leq \epsilon \leq 1$.
There exists only one solution to the previous equation that always satisfies the required properties:
\begin{equation}
\hspace{-.5cm}
\epsilon_0 = \frac{1}{6} \left(-\sqrt[3]{4} (1 + i \sqrt3) \sqrt[3]{ \sqrt{81 \tilde{\mathcal{N}}_3^2 - 66 \tilde{\mathcal{N}}_3 - 15} + 27 \tilde{\mathcal{N}}_3 - 11} + \frac{8 i \sqrt[3]{2} (\sqrt3 + i)}{\sqrt[3]{3 \sqrt{81 \tilde{\mathcal{N}}_3^2 - 66 \tilde{\mathcal{N}}_3 - 15} + 27 \tilde{\mathcal{N}}_3 - 11}} + 2 \right).
\end{equation}
Note that $\epsilon_0$ is real-valued, i.e., $\mathfrak{I}(\epsilon_0) = 0$ for all physically allowed values of $\tilde{\mathcal{N}}_3 = [0;1]$. The modified entanglement-based connected correlation norm is obtained as
\begin{equation}
\overline{\mathcal{N}}_3 = \frac{1-\epsilon_0}{2},
\end{equation}
which is identical to the three-tangle $\tau_3$ for states that are structurally perfectly GHZ-type entangled, that is, states that have genuine tripartite entanglement but fully separable bipartite marginals, but is also nonzero in case of W-type entanglement. Moreover, $\overline{\mathcal{N}}_3$ is also bounded in the interval $[0;1]$. Crucially, it suffices to derive $\overline{\mathcal{N}}_3$ from the parameterized pure state $\ket{\psi}_{\theta}$ as $\tau_3$ and $\mathcal{N}_3$ are invariant under local unitary transformations, that is, any pure state with some degree of GHZ-type entanglement is equivalent to a state $\ket{\psi}_\theta$ up to some local unitary transformation.

For mixed states, the modified norm $\overline{\mathcal{N}}_3$ is again obtained as the convex roof construction over a constituting ensemble of pure states $\ket{\psi_i}$:
\begin{equation}
\overline{\mathcal{N}}_3 = \sum_i p_i \frac{1-\epsilon_{0,i}}{2}.
\end{equation}
In the generalized BEADS representation of genuinely multipartite entangled states, we can then quantify and visualize any structural deviation from ideal GHZ-type entanglement, that is, any structural W-like contribution to the entangling structure (even in case of GHZ-class pure states) by displaying secondary entanglement arcs which are scaled by the absolute difference
\begin{equation}
\mathcal{W}_3 = |\overline{\mathcal{N}}_3 - \tau_3|.
\label{eq:diff_ent_quantity}
\end{equation}

A graphical comparison of $\tau_3$, $\tilde{\mathcal{N}}_3$ and $\overline{\mathcal{N}}_3$ is provided in Fig.~\ref{fig:ent_quant_states}, where the three quantities are evaluated for $\ket{\psi}_\theta = \cos(\theta) \ket{000} + \sin(\theta) \ket{111}$ but also for the parameterized pure state {$\ket{\omega}_\theta = \cos{\left( \frac{4\ \text{acos}({1/\sqrt{3})}}{\pi} \theta \right)} \ket{001} + \frac{1}{\sqrt{2}} \sin{\left( \frac{4\ \text{acos}({1/\sqrt{3})}}{\pi} \theta \right)} \left(\ket{010}+\ket{100}\right)$}. It is clear that $\ket{\omega}_\theta$ is a product state ($\ket{001}$) only if $\theta = 0$ and -- being a coherent mixture of the states $\ket{001}$, $\ket{010}$, and $\ket{100}$ -- belongs to the W SLOCC class otherwise. That is, $\ket{\omega}_\theta$ has genuine tripartite W-type entanglement, for $0 < \theta \leq \pi/4$. Thus, this state serves as a suitable example for testing whether $\tilde{\mathcal{N}}_3$ and $\overline{\mathcal{N}}_3$ detect the genuine multipartite entanglement while $\tau_3$ fails to do so. Indeed, the plots shown in Fig.~\ref{fig:ent_quant_states} illustrate that the introduced quantities $\tilde{\mathcal{N}}_3$ and in particular $\overline{\mathcal{N}}_3$ exhibit the desired properties. In the case of GHZ-entangled states $\ket{\psi}_{\theta \neq 0}$, $\tilde{\mathcal{N}}_3$ and $\overline{\mathcal{N}}_3$ adopt values in the range $]0;\ 1]$ and $\overline{\mathcal{N}}_3$ is identical to the three-tangle $\tau_3$. On the other hand, $\tilde{\mathcal{N}}_3$ and $\overline{\mathcal{N}}_3$ are nonzero and adopt values in the ranges $]0;\ 0.3841]$ and $]0;\ 0.3158]$ (exact values rounded to four decimals), respectively, whereas $\tau_3$ is zero-valued for all possible values of $\theta$ for W-entangled states $\ket{\omega}_{\theta \neq 0}$.

\begin{figure}[h!]          
	\centering
	\includegraphics[width=\textwidth, trim= 0cm 0cm 0cm 0cm, clip]{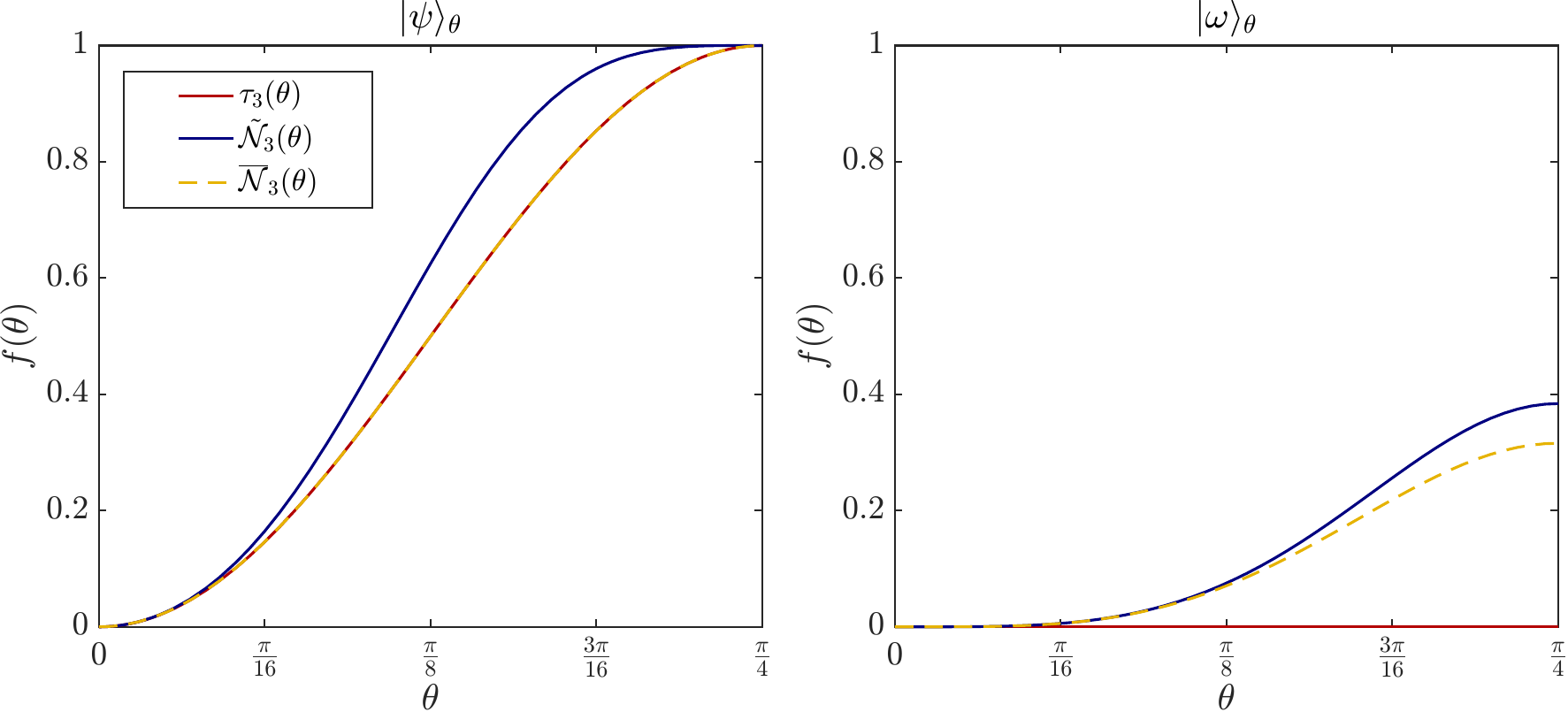}
	\caption{Comparison of the introduced entanglement quantifiers $\tilde{\mathcal{N}}_3$ and $\overline{\mathcal{N}}_3$, and the three-tangle $\tau_3$ for parameterized GHZ (left) and W (right) SLOCC class pure states. The quantities are evaluated for the states $\ket{\psi}_\theta = \cos(\theta) \ket{000} + \sin(\theta) \ket{111}$ and $\ket{\omega}_\theta = \cos{\left( \frac{4\ \text{acos}({1/\sqrt{3}})}{\pi} \theta \right)} \ket{001} + \frac{1}{\sqrt{2}} \sin{\left( \frac{4\ \text{acos}({1/\sqrt{3}})}{\pi} \theta \right)} \left(\ket{010}+\ket{100}\right)$. Note that for the case $\theta = 0$, the obtained states are product states whereas the states are genuinely tripartite entangled for  $0 < \theta \leq \pi/4$. It is clear that $\tilde{\mathcal{N}}_3$ and $\overline{\mathcal{N}}_3$ detect both types of entanglement while $\tau_3$ is zero-measure with respect to W-states, i.e., it is zero-valued for all $\ket{\omega}_\theta$. Moreover, $\overline{\mathcal{N}}_3$ and $\tau_3$ are identical for GHZ-entangled states.}
	\label{fig:ent_quant_states}
\end{figure}

To conclude this chapter, a selection of states $\ket{\psi}_\theta$ and $\ket{\omega}_\theta$ is visualized for $\theta = \{\pi/16,\ \pi/8,\ 3\pi/16\}$ using the generalized BEADS representation. Two different arc types are used to visually indicate genuine tripartite entanglement: black arcs visualize the three-tangle $\tau_3$, whereas the colored arcs, which are drawn such that they fill the gaps of the dashed three-tangle arcs, are scaled in diameter with respect to the quantity $\mathcal{W}_3$ (see Eq.~\ref{eq:diff_ent_quantity}), thus, visually quantifying W-type entanglement.

As a final remark, we do not claim that $\tilde{\mathcal{N}}_3$ and $\overline{\mathcal{N}}_3$ are entanglement (LOCC) monotones. Proofing these quantifiers to be tripartite entanglement monotones exceeds the scope of this article and may be addressed in a future work. 

\begin{figure}[ht!]          
	\centering
	\includegraphics[width=\textwidth, trim= 0cm 0cm 0cm 0cm, clip]{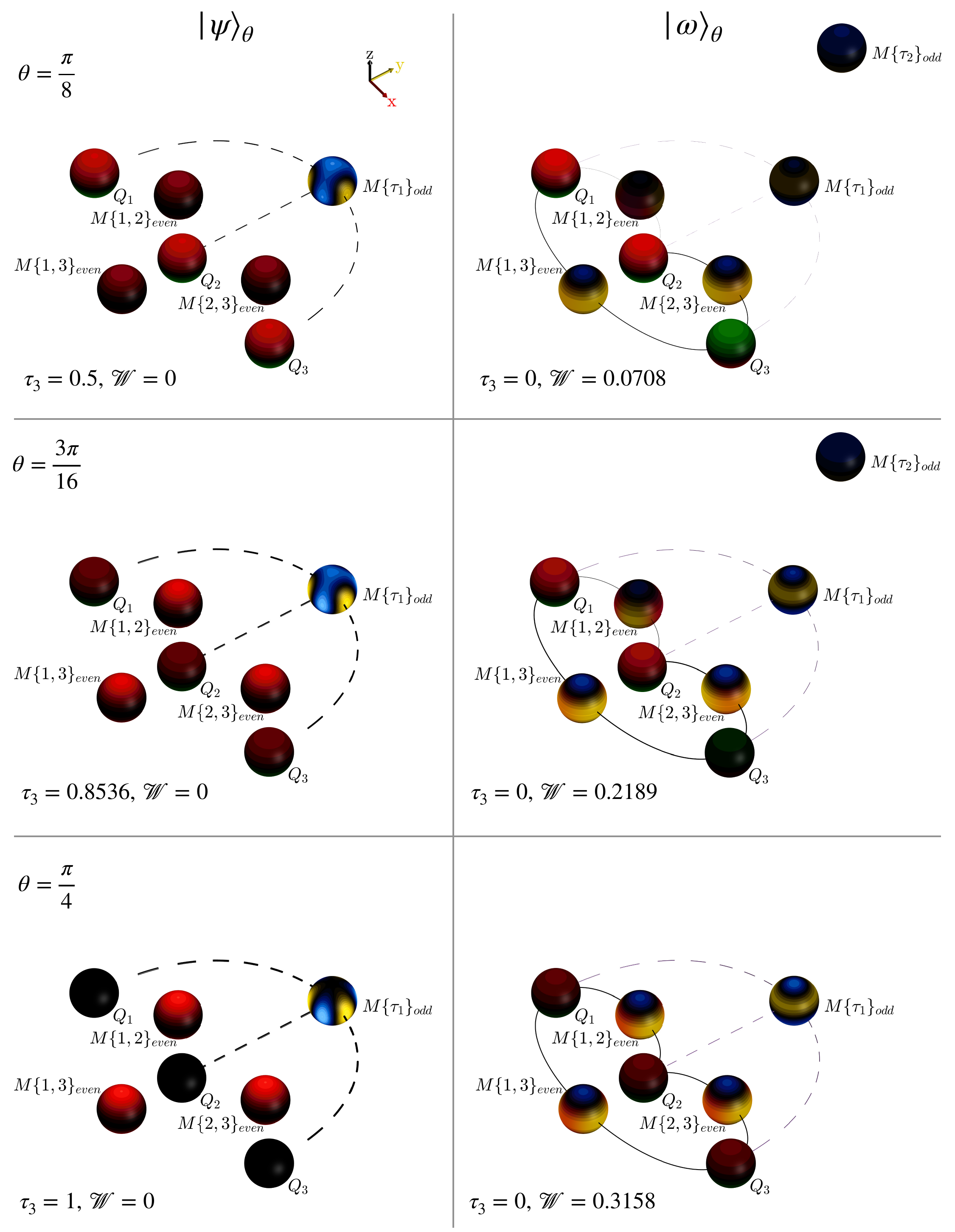}
	\caption{Generalized BEADS representation of parameterized states $\ket{\psi}_\theta$ and $\ket{\omega}_\theta$ for $\theta = \{ \pi/8,\ 3\pi/16,\ \pi /4\}$, which are strictly GHZ- or W-type multipartite entangled, respectively. The genuine tripartite entanglement is visually indicated by entanglement arcs connecting the $Q$-Beads and the fully permutation symmetric $\{1,2,3\ \tau_1\}$-Bead. The arcs are either scaled in diameter proportionally to the three-tangle $\tau_3$ (black arcs) to indicate GHZ-entanglement or the quantity $\mathcal{W}_3 = |\overline{\mathcal{N}}_3 - \tau_3|$ (violet arcs) representing W-type entanglement. Numerical values of $\tau_3$ and $\mathcal{W}_3$ are shown for every state.}
	\label{fig:ent_quant_states}
\end{figure}
\clearpage

\section{Additional details on the visualization of classical and nonclassical correlations}\label{app:class_corr}
\subsection{Fundamental concepts}
As discussed in \cref{sec:genuine_tripartite_$M_{Cl}$}, in the generalized BEADS representation we use the standard definition of classical correlation \cite{ollivier_quantum_2001} to optimize local projectors, i.e., a local measurement, that maximizes the information gain on the unmeasured system.  However, for visualizing the determined classical correlation, it is not intuitive to directly visualize the unmeasured reduced subsystem to reveal this information gain induced by the optimized measurement. In fact, the latter would require to separately visualize the unmeasured reduced marginals of all possible outcomes corresponding to the individual projectors associated with the measurement.

\begin{figure}[htbp]
\vspace{-.2cm}
	\centering
	\makebox[\textwidth][c]{
		\begin{minipage}{1\textwidth}
	\begin{subfigure}[t]{0.45\textwidth}
		\centering
		\includegraphics[width=\textwidth, trim={9cm 6cm 6cm 5cm},clip]{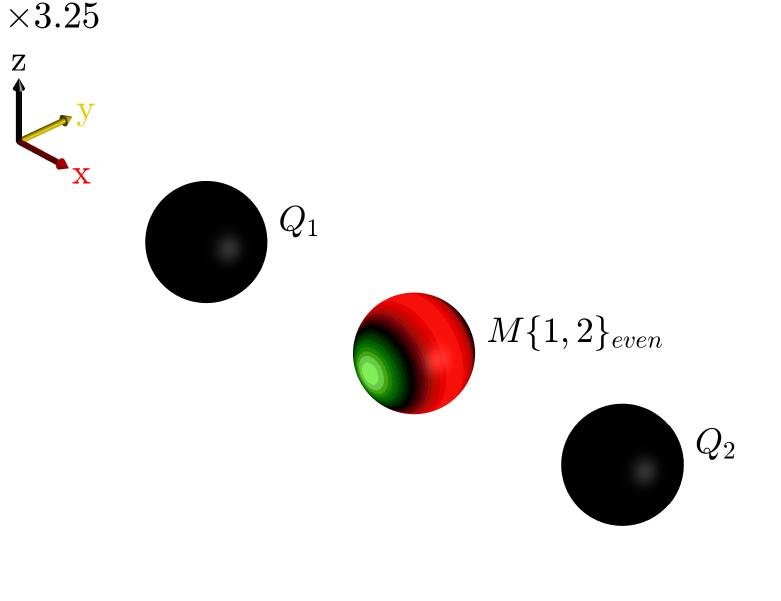}
		\caption{Pre-measurement state.}
		\label{fig:WernerPre}
	\end{subfigure}%
	\hspace{.1\textwidth}
	\begin{subfigure}[t]{0.45\textwidth}
		\centering
		\includegraphics[width=\textwidth, trim={9cm 6cm 6cm 5cm},clip]{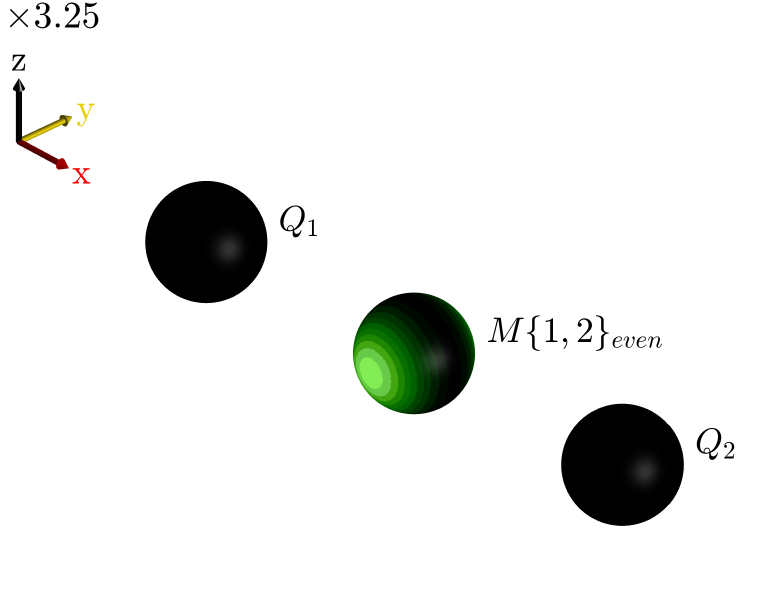}
		\caption{Post-measurement state.}
		\label{fig:WernerPost}
	\end{subfigure}
	\end{minipage}
	}
	\makebox[\textwidth][c]{
		\begin{minipage}{1\textwidth}
	\begin{subfigure}[t]{0.45\textwidth}
		\centering
		\includegraphics[width=\textwidth, trim={9cm 6cm 6cm 5cm},clip]{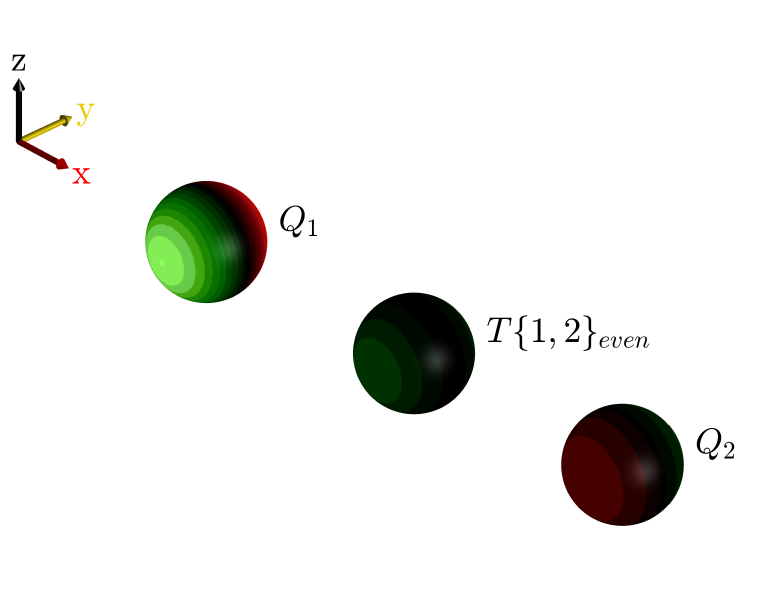}
		\caption{Outcome for projector $\Pi_0 = \frac{1}{2}{\setlength{\arraycolsep}{2pt}\begin{pmatrix}\begin{array}{cc}
		1 & -i\\
		i & 1
		\end{array}
		\end{pmatrix}} \otimes \mathbbm{1}$, ${p_0 = 50\%}$.}
		\label{fig:WernerOutcome0}
	\end{subfigure}%
	\hspace{.1\textwidth}
	\begin{subfigure}[t]{0.45\textwidth}
		\centering
		\includegraphics[width=\textwidth, trim={9cm 6cm 6cm 5cm},clip]{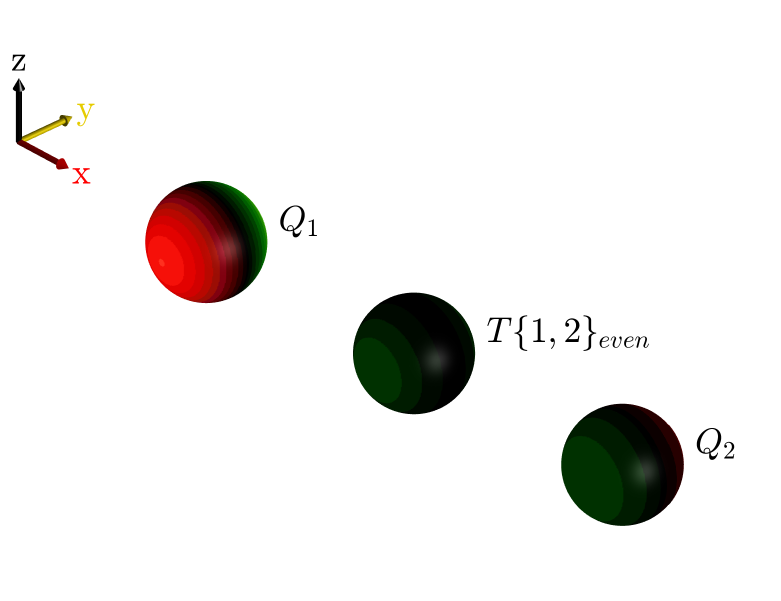}
		\caption{Outcome for projector $\Pi_1 = \frac{1}{2}{\setlength{\arraycolsep}{2pt}\begin{pmatrix}\begin{array}{cc}
		1 & i\\
		-i & 1
		\end{array}
		\end{pmatrix}} \otimes \mathbbm{1}$, ${p_1 = 50\%}$.}
		\label{fig:WernerOutcome1}
	\end{subfigure}
	\end{minipage}
	}
	\caption{Classical correlation analysis of the Werner state $\rho(0.3)= \frac{0.7}{4}\mathbbm{1} + 0.3\ket{\Phi^+}\bra{\Phi^+}$ visualized in \cref{fig:WernerPre}. Here, a local measurement of qubit 1 along the $y$-axis maximizes the classical correlation, which is directly reflected in information encoded on the $Q$-Beads $Q_2$ of the individual outcome BEADS representations (see \cref{fig:WernerOutcome0} and \cref{fig:WernerOutcome1}). However, the information gain on qubit 2 is not reflected in the combined post-measurement $Q_2$-Bead shown in \cref{fig:WernerPost}. While a visualization of classical correlation is thus possible in terms of the individual outcomes of optimum measurements, it is cumbersome and less intuitive. Note that spherical function values in the top visualizations are scaled by the denoted scaling factors to improve readability. The individual outcomes are visualized using $T$-Beads as the states are product states and do not show $M$-Beads. However, the convex sum over the outcomes gives the displayed post-measurement state, i.e., the weighted sum of $T$-Beads constitutes the $M$-Bead in \cref{fig:WernerPost}.}
	\vspace{-1cm}
	\label{fig:app_class_corr_bip_analysis}
\end{figure}

To give a concrete example, consider the Werner state $\rho(0.3)= \frac{0.7}{4}\mathbbm{1} + 0.3\ket{\Phi^+}\bra{\Phi^+}$, which was discussed in detail in \cref{sec:sep_states}. The individual outcomes for an optimum local measurement of qubit 1 along the $y$-axis, are shown in \cref{fig:app_class_corr_bip_analysis}. It is clear, that information on qubit 2 is gained by the measurement which is directly reflected in the local information encoded on the $Q$-Bead $Q_2$ of the individual outcome generalized BEADS representations. Crucially, the information gain is \underline{not} reflected in the overall mixed state describing the post measurement state, that is, the convex sum over all possible measurement outcomes. This always happens to be the case as local measurements of one qubit (here, qubit 1) are unable to manipulate the overall post-measurement reduced state of another unmeasured qubit.

In order to obtain a more compact visualization of classical correlation, which is in line with the main goal of the generalized BEADS representation of visualizing manifestations of correlation types in terms of correlation function components, we suggest a different approach that becomes directly evident when comparing \cref{fig:WernerPre} and \cref{fig:WernerPost}. Clearly, the post-measurement state in the provided example is not identical to the pre-measurement state and encodes less information on the bipartite $M\{1,2\}_{even}$. Notably, the shown Werner state is fully correlated in the sense that its reduced density operators are maximally mixed. On closer inspection, the same applies to the post-measurement state. In fact, the component of $M\{1,2\}_{even}$ which survives the local measurement is exactly the structure of the pre-measurement state, which is responsible for the observed information gain on the unmeasured qubit. In other words, the post-measurement $M$-Bead captures exactly the genuine correlation-based structure of $\rho(0.3)$ that underlies the observed classical correlation and thus, allows us to obtain a direct visualization of classical correlation as described in \cref{eq:rho_CM_projected}. Hence, the post-measurement $M$-Beads are exactly equivalent to the introduced $M_{Cl}$-Beads.

On the other hand, the genuinely correlation-based structure of a density operator, which does not survive the optimum classical-correlation-maximizing measurement, represents the classically inaccessible, or equivalently, the nonclassical structure with respect to this measurement. This serves as the foundation for a corresponding visualization in terms of $D$-Beads as described in \cref{eq:rho_D}.

As a last remark, note that in the generalized BEADS representation we confine the local measurements maximizing the classical correlation to local PVMs. While local POVMs in theory represent the more general approach, we deliberately choose PVMs to provide a direct connection to common experimental approaches and especially to align with the definition of classically correlated states as introduced in \cref{sec:classical_state}. This enables to directly identify classically correlated states based on their generalized BEADS representation as described in the Results section.

\subsection{Generalizations to multipartite systems}
In the multipartite case, there exists no unique definition of classical correlation. The most direct and natural extension to the bipartite classical correlation is given in terms of the maximum information gain obtained for a single unmeasured qubit while all other remaining qubits are measured. For tripartite systems $ABC$ this would for example correspond to classical correlation with respect to a bipartition between a single unmeasured and a set of measured qubits, that is, $C:AB$, $B:AC$, or $A:BC$. Thus, this generalization of the classical mutual information (cf. \cref{eq:classical_mutual}) corresponds to:
\begin{equation}
	M_{Cl}(k:\{\mathcal{M}\})_{\left\{ \Pi_l^{\{\mathcal{M}\}}\right\}} = \max_{\{\Pi_l^{\{\mathcal{M}\}}\}} \Big[ S(k) - S(k|\{\Pi_l^{\{\mathcal{M}\}}\}) \Big],
	\label{eq:classical_mutual_allbutone}
\end{equation}
where $k$ denotes the unmeasured qubit and $\{\mathcal{M}\} = \{1,\ 2,\hdots N\}\backslash \{k\}$ is the set of measured parties.
Again, the optimized measurement can in principle be a general POVM. However, we again advocate to restrict the considered measurements to PVMs to stay true to experimental implementations. Moreover, we deem it reasonable to further narrow the set of PVMs to joint local PVMs, motivated by the direct relationship to the definition of classically correlated multipartite states. 

A second possible notion of classical correlation in multipartite systems is given by the opposite approach of measuring a single qubit and quantifying the information gain on the unmeasured $N-1$-qubit system, i.e.:
\begin{equation}
	M_{Cl}(\{\mathcal{K}\}:m)_{\left\{ \Pi_l^{m}\right\}} = \max_{\{\Pi_l^{m}\}} \Big[ S(\mathcal{K}) - S(\mathcal{K}|\{\Pi_l^{m}\}) \Big].
	\label{eq:classical_mutual_justone}
\end{equation}
Here, $\{\mathcal{K}\}= \{1,\ 2,\hdots N\}\backslash \{m\}$ is the set of unmeasured qubits and $m$ denotes the single measured qubit.
To put this into context, in the tripartite case this would correspond to considering the bipartitions between a single measured and a set of unmeasured qubits  $A:BC$, $B:AC$, or $C:AB$. Here, the measurement automatically follows to be local and we again restrict it to PVMs in the generalized BEADS representation.

\begin{figure}[H]
	\centering
	\begin{subfigure}[t]{0.48\textwidth}
		\centering
		\includegraphics[width=\textwidth, trim={7.5cm 5.2cm 7cm 5.2cm},clip]{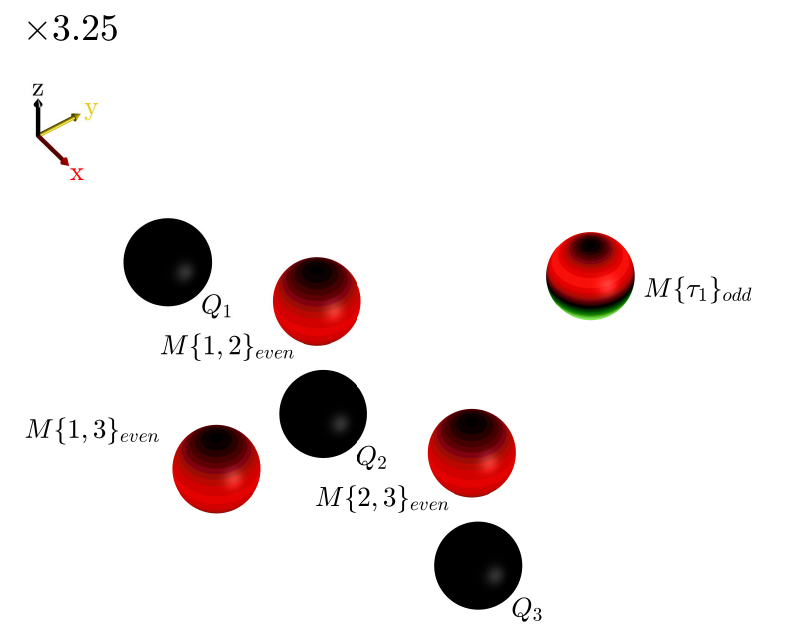}
		\caption{Generalized BEADS representation of $\rho_\text{tripartite}$.}
		\label{fig:trip_example_M}
	\end{subfigure}%
	\hspace{.04\textwidth}%
	\begin{subfigure}[t]{0.48\textwidth}
		\centering
		\includegraphics[width=\textwidth, trim={7.5cm 5.2cm 7cm 5.2cm},clip]{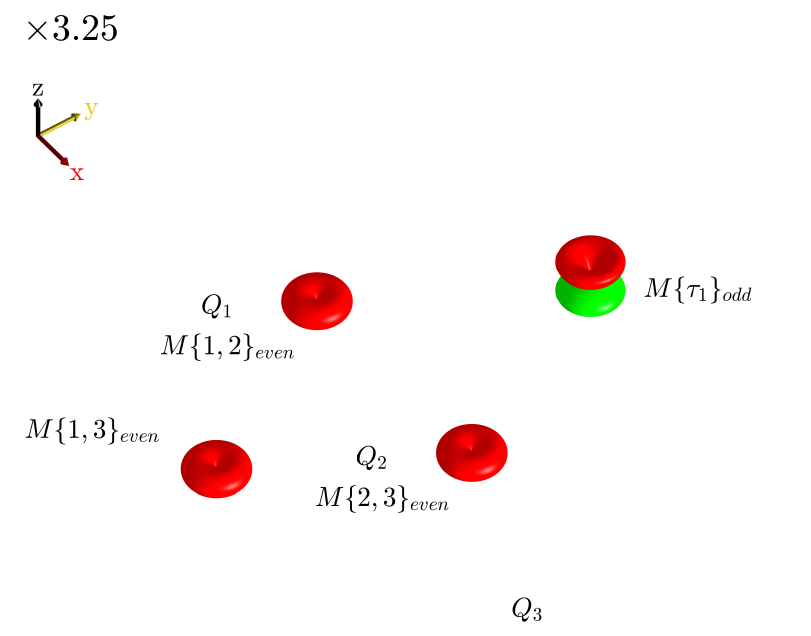}
		\caption{Generalized BEADS representation radial plot of $\rho_\text{tripartite}$.}
		\label{fig:trip_example_radial}
	\end{subfigure}
	
	\begin{subfigure}[t]{0.48\textwidth}
		\centering
		\includegraphics[width=\textwidth, trim={7.2cm 5.2cm 4.7cm 5cm},clip]{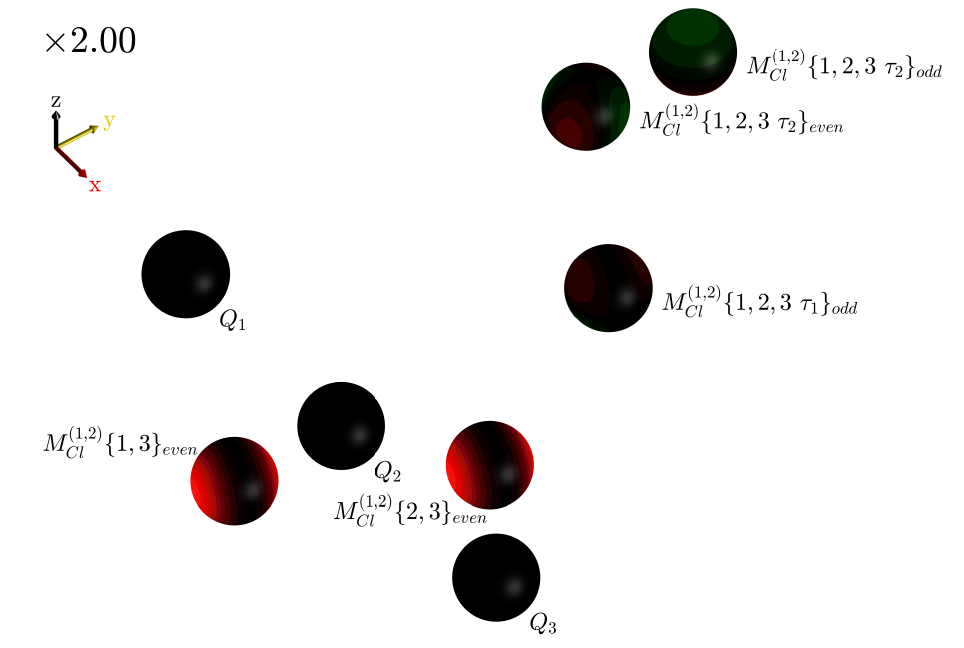}
		\caption{$M_{Cl}$-Beads representation of the classical correlation of $\rho_\text{tripartite}$ with respect to a joint local measurement of qubits 1 and 2.}
		\label{fig:trip_example_M_Cl_12}
	\end{subfigure}%
	\hspace{.04\textwidth}%
	\begin{subfigure}[t]{0.48\textwidth}
		\centering
		\includegraphics[width=\textwidth, trim={8cm 5.2cm 2.5cm 5cm},clip]{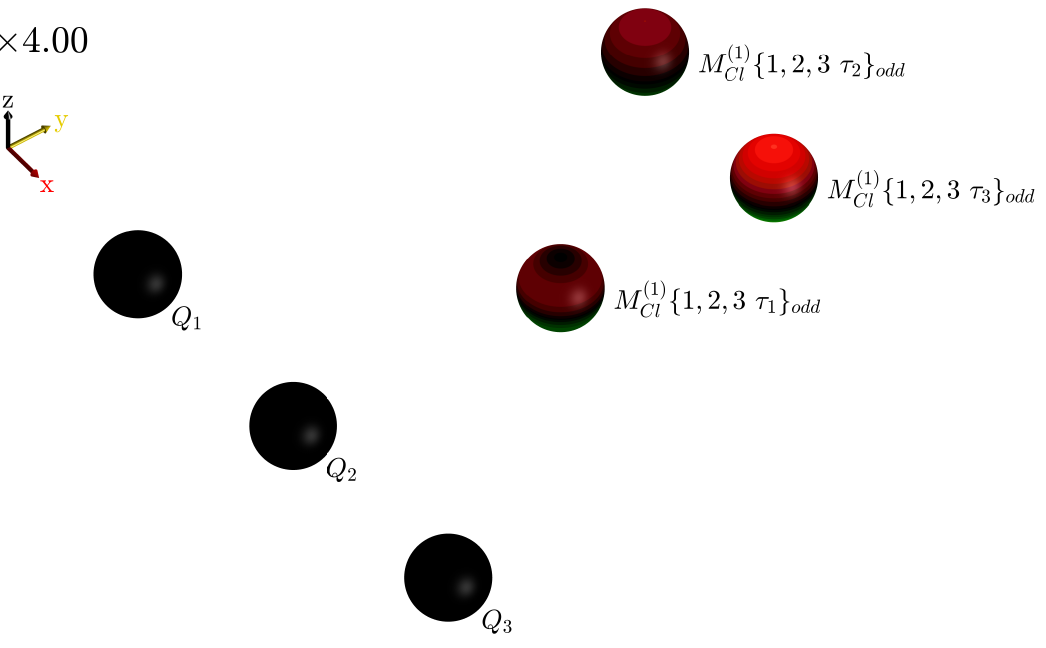}
		\caption{$M_{Cl}$-Beads representation of the classical correlation of $\rho_\text{tripartite}$ with respect to a single local measurement of qubit 1.}
		\label{fig:trip_example_M_Cl_1}
	\end{subfigure}
	
	\caption{Generalized BEADS representation of the tripartite state $\rho_\text{tripartite}$ (see \cref{eq:ClassCorExample}) visualizing its total correlation ($M$-Beads) and its classical correlation properties ($M_{Cl}$-Beads). Here, the classical correlation was determined using the discussed conceptually different approaches of measuring all but one, or alternatively, just one qubit such that the information gain (conditional entropy) on the unmeasured marginal is maximized, respectively. In the case of \cref{fig:trip_example_M_Cl_12}, the observed correlation is mostly bipartite, whereas it is genuinely tripartite in \cref{fig:trip_example_M_Cl_12}. Also note that in both cases, the measurements are symmetry-breaking, such that additional tripartite beads occur. Spherical function values are scaled by the denoted scaling factors to improve readability, respectively.}
	\vspace{-.5cm}
	\label{fig:ClassCorExampleStandard}
\end{figure}

The most general notion of classical correlation in multipartite systems is thus obtained as:
\begin{equation}
	M_{Cl}(\{\mathcal{K}\}:\{\mathcal{M}\})_{\left\{ \Pi_l^{\{\mathcal{M}\}}\right\}} = \max_{\{\Pi_l^{\{\mathcal{M}\}}\}} \Big[ S(\mathcal{K}) - S(\mathcal{K}|\{\Pi_l^{\{\mathcal{M}\}}\}) \Big],
	\label{eq:classical_mutual_justone}
\end{equation}
using the previously defined sets of measured and unmeasured qubits. For comparison of the different approaches, consider the mixed tripartite state
\begin{equation}
\rho_\text{tripartite}
=\frac{3}{8}\ket{\text{W}}\bra{\text{W}}
+
\frac{1}{8}
\left(
\ket{000}\bra{000}
+\ket{011}\bra{011}
+\ket{101}\bra{101}
+\ket{110}\bra{110}
+\ket{111}\bra{111}
\right).
\label{eq:ClassCorExample}
\end{equation}
$M_{Cl}$-Beads visualizations for both discussed notions of classical correlation are provided in \cref{fig:ClassCorExampleStandard}. Note the difference in the visualizations for the two- (\cref{fig:trip_example_M_Cl_12}) and single-qubit (\cref{fig:trip_example_M_Cl_1}) case. The former approach mostly detects bipartite correlations, whereas the latter case here results in capturing genuine tripartite correlations. This is however not a general rule for the two approaches, but a specific observation for the examined state.

Thus, while both of the discussed approaches give rise to valid classical correlation measures, correlations observed through the corresponding optimized measurements are not guaranteed to be genuinely multipartite. This effect, e.g., also occurs the tripartite GHZ state (see \cref{sec:ghz_w_pure} for a detailed overview). Although the state has genuine tripartite correlations, a valid measurement optimization for either of the discussed approaches will always converge to the computational basis. The detected correlations in this case are genuinely bipartite though and the genuine tripartite does not contribute to the observed classical correlation. 
The phenomenon of bipartite, or more general, lower-order correlations masking and suppressing higher-order multipartite correlations in the determination of classical correlations is a common and natural property of correspondingly affected quantum states. Isolating and quantifying genuine $n$-partite classical correlations is a more theoretical consideration. In the generalized BEADS representation, we can still visualize these isolated $n$-partite classical correlations as will be explained in the subsequent subsection.

\subsection{Visualizing genuine $n$-partite classical correlations}
As outlined in the preceeding section, in order to rigorously assess $n$-partite classical correlations, the (joint) local measurements underlying the correlation of interest cannot be optimized on the full examined density operator $\rho$ in order to avoid undesirable influences of lower-order correlations. Instead, the measurement can be optimized on a virtual operator $\tilde{\rho}_n$ that just encodes the genuine $n$-partite correlation structure of $\rho$. This operator is obtained by eliminating any undesired $k$-body ($k \neq n$) and adding an identity component to keep the operator positive-semidefinite and to obtain well-defined entropies in the following optimization. For the state $\rho_\text{tripartite}$ which was previously introduced in \cref{eq:ClassCorExample}, this virtual operator is
\begin{equation}
	\tilde{\rho}_{123} = \frac{1}{8} \, \mathbbm{1} + \frac{1}{32} \, \left( \mathrm{XXZ}+ \mathrm{XZX}+ \mathrm{YYZ}+ \mathrm{YZY}+ \mathrm{ZXX}+ \mathrm{ZYY}\right),
	\label{eq:trip_only_norm}
\end{equation}
where the capital letters denote corresponding Pauli product operators in short-hand notation.

Optimizing a joint local measurement on this operator then finds the genuine tripartite correlation. Note that the added identity component in the construction of $\tilde{\rho}_{123}$ is irrelevant, as the operator is only used to determine the measurement basis not the actual correlation.

\begin{figure}[htbp]
	\centering
	\begin{subfigure}[t]{0.48\textwidth}
		\centering
		\includegraphics[width=\textwidth, trim={7.5cm 5.2cm 7cm 5.2cm},clip]{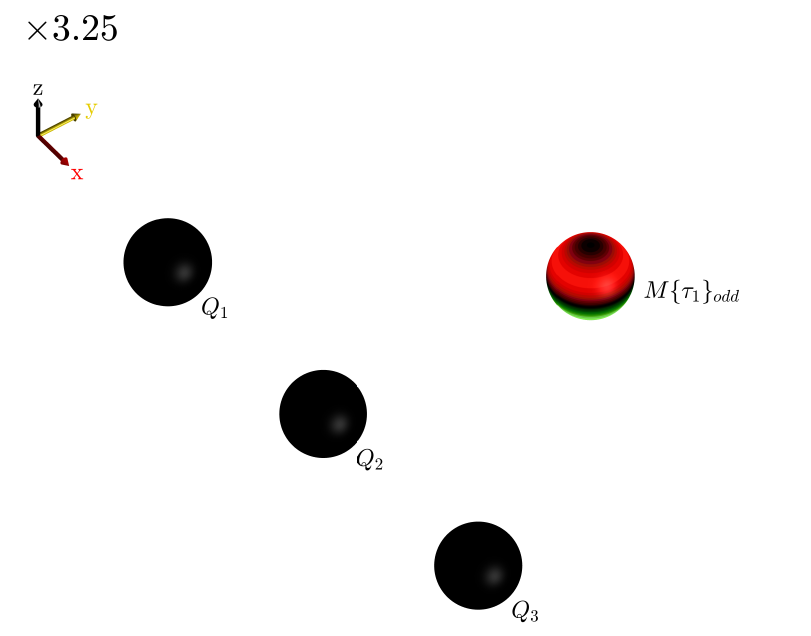}
		\label{fig:trip_only_M}
	\end{subfigure}%
	\hspace{.04\textwidth}%
	\begin{subfigure}[t]{0.48\textwidth}
		\centering
		\includegraphics[width=\textwidth, trim={7.5cm 5.2cm 7cm 5.2cm},clip]{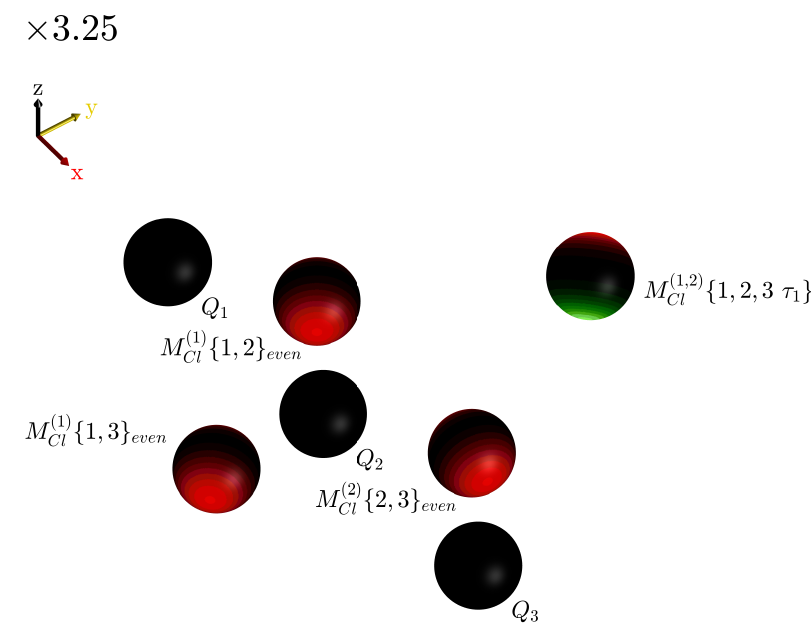}
		\label{fig:trip_only_norm_M_radial}
	\end{subfigure}
	\caption{Left: Total correlation generalized BEADS representation of the virtual density operator $\tilde{\rho}_{123}$ used to assess the genuine tripartite classical correlation of the state $\rho_\text{tripartite}$ (cf. \cref{eq:ClassCorExample}). Right: Classical correlation $M_{Cl}$-Beads representation of $\tilde{\rho}_{123}$. Here, the classical correlation was determined for each individual subsystem with respect to (joint) local measurements denoted by the superscript indexes of each bead.}
	\label{fig:trip_only_class_corr}
\end{figure}

Conveniently, deciding to visualize the genuine $n$-partite classical correlations as discussed in the preceding paragraphs allows to provide a compact representation where the classical correlation of each individual marginal can be visualized. Hence, we use this approach as a standard way of visualizing classical correlations in the generalized BEADS representation. A corresponding $M_{Cl}$-Beads representation for the state $\rho_\text{tripartite}$ is provided in \cref{fig:trip_only_class_corr}, which also provides an $M$-Bead representation of the utilized operator $\tilde{\rho}_{123}$.

\clearpage
\section{Validation of the pure state ensemble optimization framework}\label{app:benchmarks}
\subsection{Bipartite entanglement: Werner states}
\begin{figure}[htbp]          
	\begin{minipage}{0.5\linewidth}
		\centering
		\includegraphics[width=.98\textwidth, trim={11cm 3.5cm 8cm 2.7cm},clip]{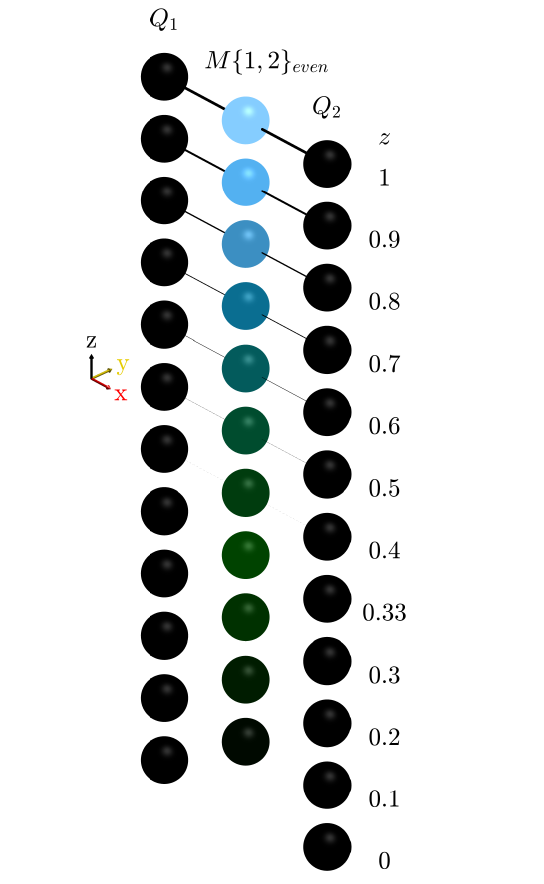}
	\end{minipage}%
	\begin{minipage}{0.5\linewidth}
		\centering
		\includegraphics[width=.98\textwidth, trim={11cm 3.5cm 8cm 2.7cm},clip]{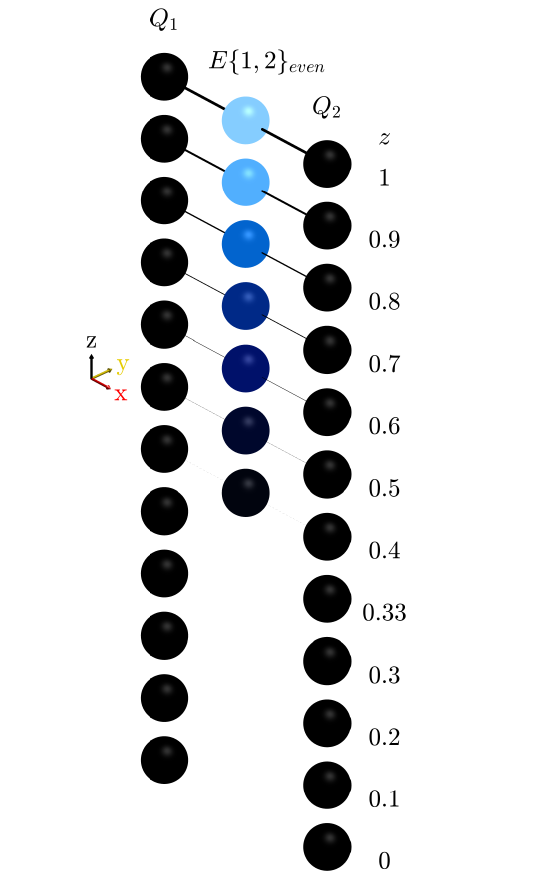}
	\end{minipage}
	\caption{Stacked $M$- (left) and $E$-Bead (right) BEADS visualizations of the parameterized singlet-centered Werner state $\rho_{\text{Werner},\ket{\Psi^-}}(z) = \frac{1-z}{4}\mathbbm{1}_4 + z\ket{\Psi^-}\bra{\Psi^-}$. The state is well-known for being separable for $z \leq 1/3$ \cite{werner_state_1989, ollivier_quantum_2001}. The generalized BEADS optimization framework reproduces this separability threshold consistently: States with $z< 1/3$ are genuinely correlated and thus, show an $M$-Bead except for the maximally mixed state at $z=0$. However, as is directly clear from the displayed entanglement arcs and especially the $E$-Bead visualization, the states are entangled only if $z > 1/3$.}
	\label{fig:benchmark_werner}
\end{figure}

To verify that the generalized BEADS representation correctly reproduces separability thresholds in the bipartite case, we have tested the visualization and the associated optimization framework on Werner states. In \cref{fig:benchmark_werner}, a stacked visualization of the parameterized singlet-centerd Werner state $\rho_{\text{Werner},\ket{\Psi^-}}(z) = \frac{1-z}{4}\mathbbm{1}_4 + z\ket{\Psi^-}\bra{\Psi^-}$ is shown. As is known from literature \cite{werner_state_1989, ollivier_quantum_2001}, the state is separable for $z \leq 1/3$. Clearly, the generalized BEADS representation reproduces this separability threshold and only the states for $z > 1/3$ show entanglement arcs and in particular, $E$-Beads.

\subsection{Tripartite entanglement: GHZ states with added white noise}
\begin{figure}[H] 
	\centering
	\makebox[\linewidth][c]{
		\begin{minipage}{1\textwidth}
			\begin{subfigure}[t]{0.5\textwidth}
				\centering
				\includegraphics[width=\textwidth, trim={6cm 4cm 5.8cm 5cm},clip]{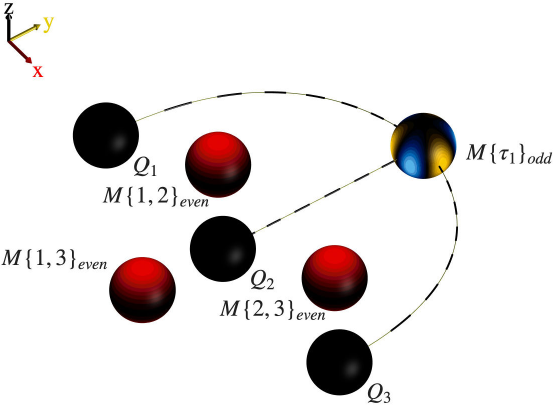}
				\subcaption{$p = 0.1$}
				\label{fig:GHZ_noise0.1_M}
			\end{subfigure}
			\begin{subfigure}[t]{0.5\textwidth}
				\centering
				\includegraphics[width=\textwidth, trim={6cm 4cm 5.8cm 5cm},clip]{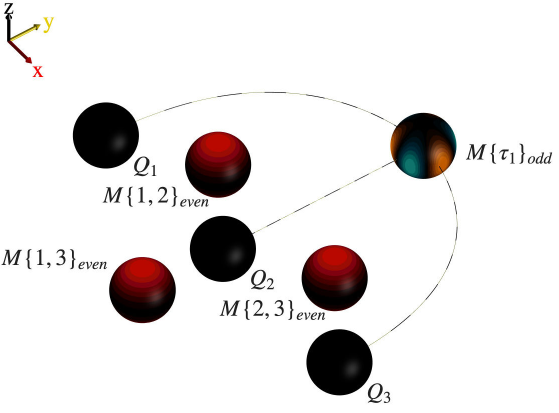}
				\subcaption{$p = 0.3$}
				\label{fig:GHZ_noise0.3_M}
			\end{subfigure}
		\end{minipage}
	}
	
	\vspace{0.6cm}
	
	\makebox[\linewidth][c]{
		\begin{minipage}{1\textwidth}
			\begin{subfigure}[t]{0.5\textwidth}
				\centering
				\includegraphics[width=\textwidth, trim={6cm 4cm 5.8cm 5cm},clip]{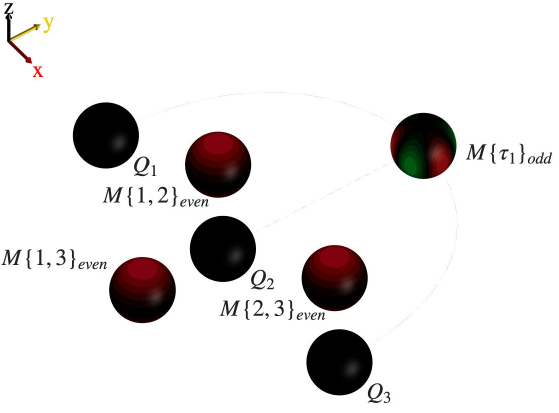}
				\subcaption{$p = 0.5$}
				\label{fig:GHZ_noise0.5_M}
			\end{subfigure}
			\begin{subfigure}[t]{0.5\textwidth}
				\centering
				\includegraphics[width=\textwidth, trim={6cm 4cm 5.8cm 5cm},clip]{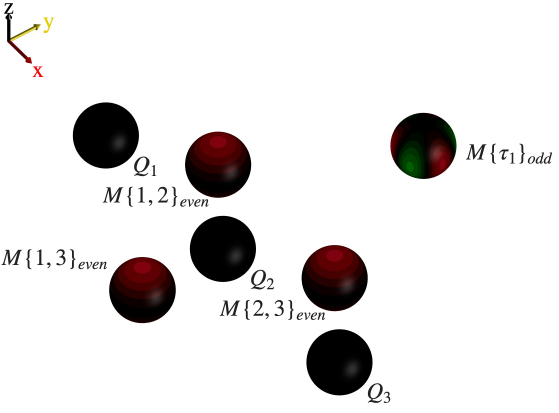}
				\subcaption{$p = 4/7 \approx 0.5714$}
				\label{fig:GHZ_noise0.57143_M}
			\end{subfigure}
		\end{minipage}
	}
	
	\vspace{0.6cm}
	
	\makebox[\linewidth][c]{
		\begin{minipage}{1\textwidth}
			\begin{subfigure}[t]{0.5\textwidth}
				\centering
				\includegraphics[width=\textwidth, trim={6cm 4cm 5.8cm 5cm},clip]{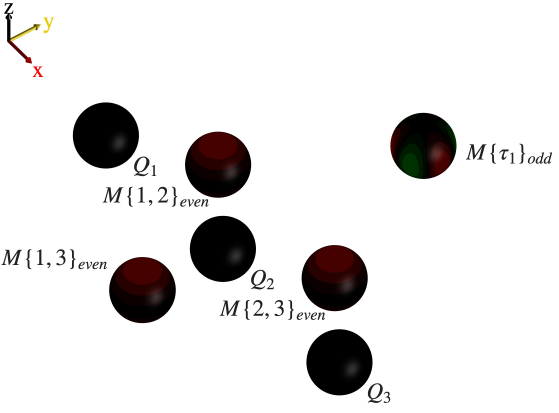}
				\subcaption{$r = 0.7$}
				\label{fig:GHZ_noise0.7_M}
			\end{subfigure}
			\hspace{0.4cm}
			\begin{subfigure}[t]{0.5\textwidth}
				\centering
				\includegraphics[width=\textwidth, trim={6cm 4cm 5.8cm 5cm},clip]{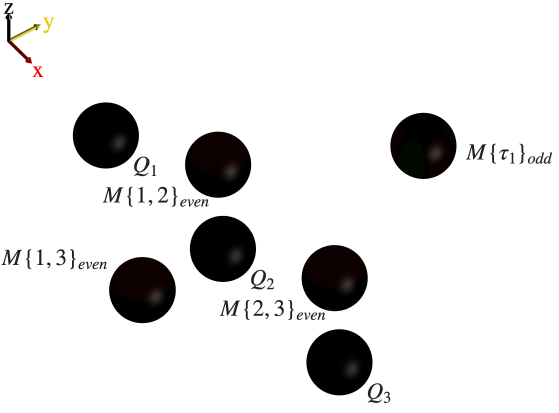}
				\subcaption{$r = 0.9$}
				\label{fig:GHZ_noise0.9_M}
			\end{subfigure}
		\end{minipage}
	}
	\caption{Total correlation ($M$-Beads) generalized BEADS representation of the noisy GHZ state $\rho(p) = (1-p)\ket{\text{GHZ}}\bra{\text{GHZ}} + p \mathbbm{1}_8$. As is clear from the colors of the $M\{\tau_1\}_{odd}$-Bead and the entanglement arcs, the state is genuinely tripartite entangled only for $p \leq 4/7$. The bipartite marginals are always classically correlated independent for all $p \neq 1$ and maximally correlated only for $p = 0$ (the pure GHZ state; not shown here). All properties are discussed in literature \cite{duer_1999, duer_2000, guhne_gme_2010} and correctly reproduced by the generalized BEADS representation and the underlying optimization framework.}
	\label{fig:benchmark_noisy_ghz}
\end{figure}

To validate the generalized BEADS representation in the case of multipartite states, we computed visualizations of the parameterized mixed tripartite state $\rho(p) = (1-p)\ket{\text{GHZ}}\bra{\text{GHZ}} + p \mathbbm{1}_8$, which is the GHZ state with added white noise that can be viewed as a generalization of Werner states to the tripartite case. Note that this state, despite its simple form, is structurally complicated due to the density operator having full rank for $p \neq 0$. The separability thresholds for this family of states are well-known: It is fully separable iff $0.8 \leq p \leq 1$, biseparable (yet not with respect to a single specific cut) iff $4/7 \leq p < 0.8$, and genuinely tripartite entangled iff $p < 4/7$ \cite{duer_1999, duer_2000, guhne_gme_2010}. Thus, we expect no tripartite $E$-Beads or entanglement arcs for $p < 4/7$. It is clear in \cref{fig:benchmark_noisy_ghz} that the generalized BEADS representation correctly reproduces the genuine multipartite entanglement threshold correctly and we only observe entanglement arcs and a shift from red and greed to yellow and blue colors of the tripartite $M$-Beads for $p < 4/7$. The complementary $E$-Beads visualizations are shown in supplementary \cref{fig:benchmark_noisy_ghz_E}.

At the same time, biseparability and full separability are properties that are not directly reflected in the mixed state BEADS representation but in the individual BEADS representations of the underlying pure state ensemble. Hence, for parameter values $p$ in the denoted biseparability or full separability ranges, we expect either only $E$-Beads for a single bipartite subsystem per state if the state is biseparable, although the subsystem need not be the same across all contributing pure states, or no $E$-Beads at all if it is fully separable. An exemplary selection of pure states for the biseperable state $\rho(p=0.4)$ is shown in supplementary \cref{fig:SLNoisy04_pure}.

\subsection{Tripartite entanglement: GHZ states with added classical noise}
\begin{figure}[H] 
	\centering
	\makebox[\linewidth][c]{
		\begin{minipage}{1\textwidth}
			\begin{subfigure}[t]{0.5\textwidth}
				\centering
				\includegraphics[width=\textwidth, trim={8cm 5cm 7.7cm 5cm},clip]{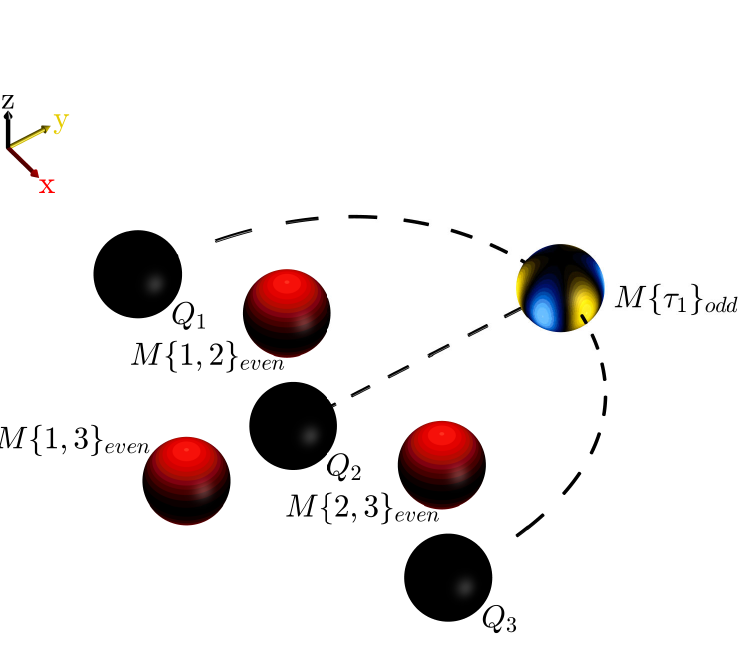}
				\subcaption{$r = 0.0$}
				\label{fig:GHZ_class0_M}
			\end{subfigure}
			\begin{subfigure}[t]{0.5\textwidth}
				\centering
				\includegraphics[width=\textwidth, trim={8cm 5cm 7.7cm 5cm},clip]{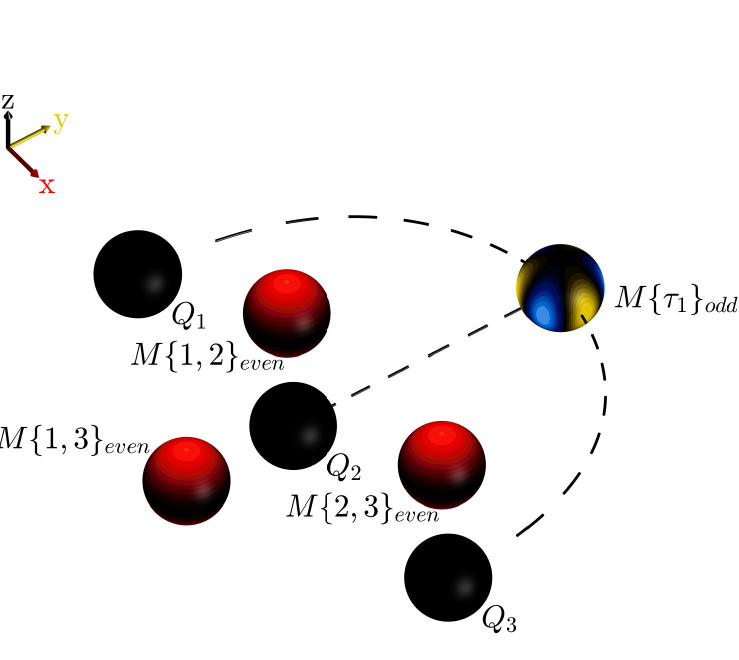}
				\subcaption{$r = 0.2$}
				\label{fig:GHZ_class0.2_M}
			\end{subfigure}
		\end{minipage}
	}
	
	\vspace{0.3cm}
	
	\makebox[\linewidth][c]{
		\begin{minipage}{1\textwidth}
			\begin{subfigure}[t]{0.5\textwidth}
				\centering
				\includegraphics[width=\textwidth, trim={8cm 5cm 7.7cm 5cm},clip]{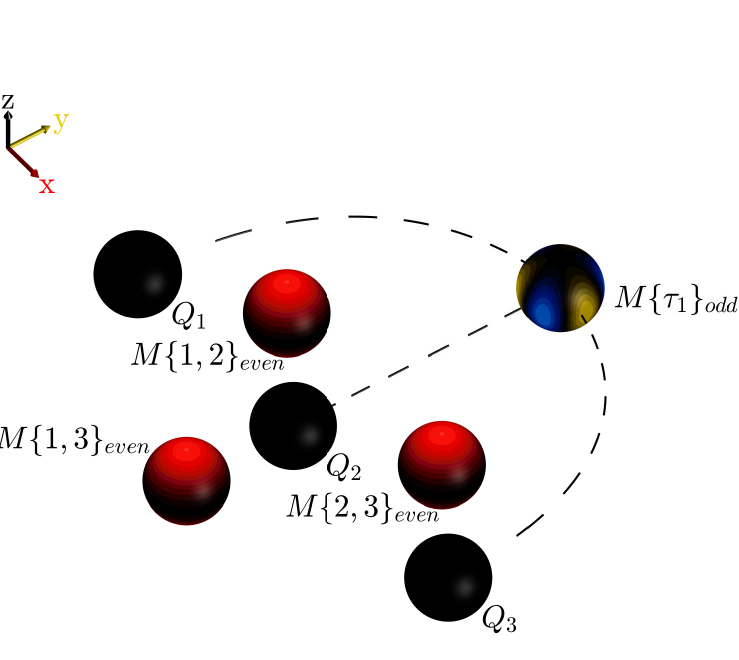}
				\subcaption{$r = 0.4$}
				\label{fig:GHZ_class0.4_M}
			\end{subfigure}
			\begin{subfigure}[t]{0.5\textwidth}
				\centering
				\includegraphics[width=\textwidth, trim={8cm 5cm 7.7cm 5cm},clip]{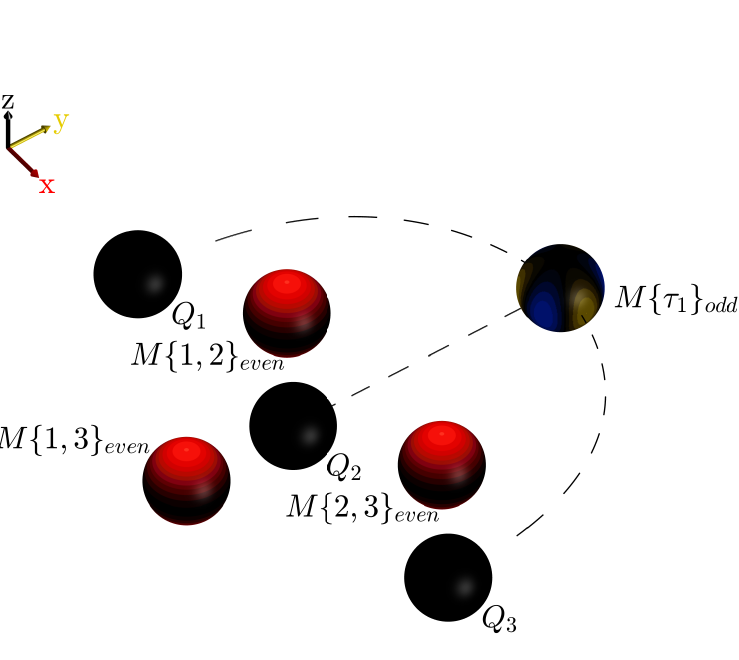}
				\subcaption{$r = 0.6$}
				\label{fig:GHZ_class0.6_M}
			\end{subfigure}
		\end{minipage}
	}
	
	\vspace{0.3cm}
	
	\makebox[\linewidth][c]{
		\begin{minipage}{1\textwidth}
			\begin{subfigure}[t]{0.5\textwidth}
				\centering
				\includegraphics[width=\textwidth, trim={8cm 5cm 7.7cm 5cm},clip]{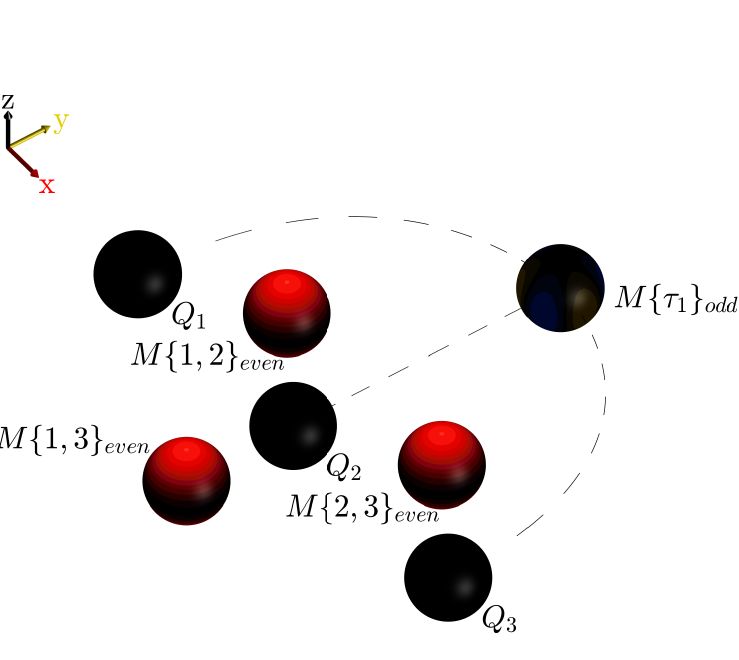}
				\subcaption{$r = 0.8$}
				\label{fig:GHZ_class0.8_M}
			\end{subfigure}
			\begin{subfigure}[t]{0.5\textwidth}
				\centering
				\includegraphics[width=\textwidth, trim={8cm 5cm 7.7cm 5cm},clip]{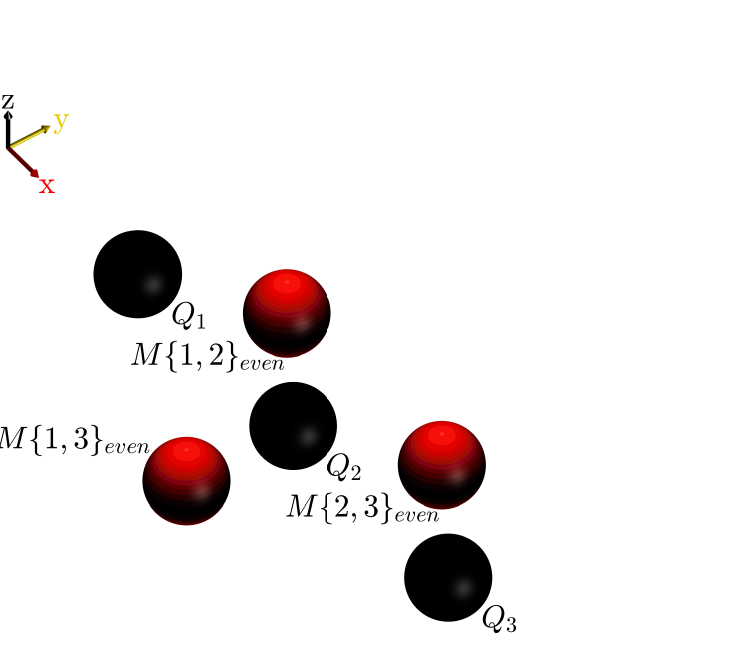}
				\subcaption{$r = 1.0$}
				\label{fig:GHZ_class1.0_M}
			\end{subfigure}
		\end{minipage}
	}
	\caption{Total correlation ($M$-Beads) generalized BEADS representation of the noisy GHZ state $\rho(r) = (1-r)\ket{\text{GHZ}}\bra{\text{GHZ}} + r/2 (\ket{000}\bra{000}+\ket{111}\bra{111})$. As is clear from the yellow and blue colored $M\{\tau_1\}_{odd}$-Bead and the entanglement arcs, the state is genuinely tripartite entangled for all $r \neq 1.0$, whereas the bipartite marginals are always maximally classically correlated independent of $r$. Both of these properties are correctly reproduced by the generalized BEADS representation and the underlying optimization framework.}
	\label{fig:benchmark_noisy_ghz_class}
\end{figure}

As a third test case for validation of the representation, we considered the mixed state $\rho(r) = (1-r)\ket{\text{GHZ}}\bra{\text{GHZ}} + r/2 (\ket{000}\bra{000}+\ket{111}\bra{111})= (1-r)\ket{\text{GHZ}}\bra{\text{GHZ}} + r/2 \ket{\text{GHZ}^-}\bra{\text{GHZ}^-}$, where $\ket{\text{GHZ}^-} = 1/\sqrt{2}(\ket{000}-\ket{111})$, and which can be interpreted as the GHZ state with added classical noise. This state has significantly different properties compared to the previously discussed GHZ state with added white noise. It is a structurally less involved rank-2 density operator, which is significantly easier to optimize with respect to its constituting pure state ensemble. This state is separable only in the fully classical case $r=1.0$, where $\rho(r) =1/2 (\ket{000}\bra{000}+\ket{111}\bra{111})$, and genuinely multipartite entangled otherwise, whereas the bipartite marginals are invariant under the given parametrization and are maximally classically correlated. As shown in \cref{fig:benchmark_noisy_ghz_class}, the generalized BEADS representation again exactly reproduces these properties. We also present a complementary $E$-Beads visualization in supplementary \cref{fig:benchmark_noisy_ghz_class_E}.

\clearpage
\bibliographystyle{unsrt} 
\bibliography{literature}  

\clearpage

\beginsupplement
\newgeometry{left=1cm, right=1cm, top=2cm, bottom=0cm}
\thispagestyle{empty}
	
\begin{center}
		\vspace*{3cm}
		
		{\LARGE \textbf{Supplementary Information}}\\[1.5cm]
		
		{\large The Generalized BEADS Representation: An Informationally Complete Visualization of Mixed Quantum States}\\[0.5cm]
		
		{\normalsize Celina Bräutigam, Steffen J. Glaser, Dennis Huber}\\[2cm]
		
		\textit{This document contains supplementary material supporting the main text.}
		
		\vfill
	\end{center}
	
\newpage
\newgeometry{left=2.5cm, right=2.5cm, top=5cm, bottom=0cm}
\pagestyle{plain}
\startcontents[supplementary]

\section*{Table of Contents}
\renewcommand{\theHsection}{supplementary.\arabic{section}}
\renewcommand{\theHsubsection}{supplementary.\arabic{section}.\arabic{subsection}}
\renewcommand{\theHsubsubsection}{supplementary.\arabic{section}.\arabic{subsection}.\arabic{subsubsection}}	
\setcounter{page}{1}

\titlecontents{lsection}
  [2.2em]
  {\bfseries}                          
  {\contentslabel{1.9em}}
  {}
  {\hfill\contentspage}

\titlecontents{lsubsection}
  [4.8em]                                
  {}
  {\contentslabel{2.5em}}
  {}
  {\titlerule*[.5pc]{.}\contentspage}

\titlecontents{lsubsubsection}
  [8.4em]
  {}
  {\contentslabel{3.5em}}
  {}
  {\titlerule*[.5pc]{.}\contentspage}
\printcontents[supplementary]{l}{1}{}

\clearpage
\section{Overview of bead sets and symmetry properties in systems of up to three qubits}\label{sup:beads_overview}
\begin{table}[h!]
\centering
\caption{\label{tab:TableSymmetries}Overview of BEADS sets $\ell^\prime$ and corresponding spherical function and permutation symmetries in systems consisting of up to three qubits. All symmetry properties of beads corresponding to linearity $g \geq 2$ apply to all types of correlation function beads. Fully permutation symmetric components are only given by beads $\{k,l\}_{\textit{even}}$ and $\{k,l,m\ \tau_1\}_{\textit{odd}}$. $\{\emptyset\}$ denotes the identity bead.}
\begin{tabular}{m{0.07\textwidth}<{\centering} m{0.15\textwidth}<{\centering} m{0.09\textwidth}<{\centering} m{0.15\textwidth}<{\centering} m{0.21\textwidth}<{\centering} m{0.12\textwidth}<{\centering}}
\hline\hline\\[-8pt]
Linearity $g$ & Bead $\ell^\prime$ & Ranks $j$ & Spherical function point symmetry & Permutation symmetry & Full permutation symmetry\\[14pt] \hline\hline \\[-8pt]
0 & $\{\emptyset\}$ & 0 & even & -- & -- \\[4pt] \cline{1-6} \\[-10pt]

1 & $\{k\}$ & 1 & odd & -- & -- \\[2pt]  \cline{1-6} \\[-12pt] 

\multirow{5}{*}{2} & \multirow{2}{*}{$\{k,l\}_{\textit{even}}$} & 0 & \multirow{2}{*}{even} & \multirow{2}{*}{{symmetric $\left(k,l\right)$}} & \multirow{2}{*}{\cmark} \\[2pt]
 & & 2 &  &  & \\[0pt]  \cline{2-6} \\[-8pt] 
 & $\{k,l\}_{\textit{odd}}$ & 1 & odd & {antisymmetric $\left(k,l\right)$} & \xmark \\[8pt]  \cline{1-6} \\[-12pt] 
\multirow{12}{*}{3} & \multirow{2}{*}{$\{k,l,m\ \tau_1\}_{\textit{odd}}$} & 1 & \multirow{2}{*}{odd} & \multirow{2}{*}{{symmetric $\left(k,l,m\right)$}} & \multirow{2}{*}{\cmark} \\[2pt]
 & & 3 &  &  & \\[0pt]  \cline{2-6} \\[-12pt]  
  & $\{k,l,m\ \tau_2\}_{\textit{odd}}$ & 1 & odd & \multirow{2}{*}{symmetric $\left(k,l\right)$} & \multirow{2}{*}{\xmark} \\[2pt]  \cline{2-4}\\[-10pt] 
  & $\{k,l,m\ \tau_2\}_{\textit{even}}$ & 2 & even &  &  \\[2pt] \cline{2-6}\\[-10pt] 
  & $\{k,l,m\ \tau_3\}_{\textit{odd}}$ & 1 & odd & \multirow{2}{*}{antisymmetric $\left(k,l\right)$} & \multirow{2}{*}{\xmark} \\[2pt]  \cline{2-4}\\[-10pt] 
  & $\{k,l,m\ \tau_3\}_{\textit{even}}$ & 2 & even &  &  \\[2pt]  \cline{2-6} \\[-10pt]
  & $\{k,l,m\ \tau_4\}_{\textit{even}}$ & 0 & even & {antisymmetric $\left(k,l,m\right)$}  & \xmark \\[4pt]  \hline\hline
\end{tabular}
\end{table}
\clearpage

\section{Overview of BEADS scaling factors}\label{sup:scaling_factors}
Recall from the publication of the original BEADS representation, that scaling factors $s_j^{\left(\ell^\prime\right)}\left(N,g\right)$ correspond to a product of three scaling factors \cite{huber_beads_2025}
\begin{equation}\label{eq:BEADSScalingFactor}
s_j^{\left(\ell^\prime\right)}\left(N,g\right)=\zeta\left(N\right)\cdot\xi_j^{\left(\ell^\prime\right)}\left(g\right)\cdot\eta_j,
\end{equation}
where $\zeta\left(N\right)=\sqrt{2^N}$ depends on the total number $N$ of qubits in a system, $\eta_j=\sqrt{4\pi/\left(2j+1\right)}$ is dependent on the rank $j$ of the corresponding LISA tensor operator, and $\xi_j^{\left(\ell^\prime\right)}\left(g\right)$ is an additional rank-$j$- and linearity-$g$-dependent factor which is specific for each bead. The latter type of scaling factors is chosen such that the resulting spherical function values directly correspond to experimentally measurable expectation values of Pauli product operators in case of fully permutation symmetric beads, or in dependence of the global unitary bound (GUB) of the underlying tensor operator as detailed in \cite{huber_beads_2025}. 
\begin{table}[h!]
\centering
\caption{\label{tab:TableSF}Overview of scaling factors $\xi_j^{\left(\ell^\prime\right)}\left(g\right)$ for up to trilinear LISA tensor operators. The scaling method specifies whether the displayed scaling factor was chosen to achieve a direct correspondence to experimentally measurable expectation values of Pauli product operators (canonical scaling method) or based on the global unitary bound (GUB) of the underlying LISA tensor operator.}
\begin{tabular}{p{0.15\textwidth}<{\centering} p{0.15\textwidth}<{\centering} m{0.15\textwidth}<{\centering} m{0.15\textwidth}<{\centering}}
\hline\hline\\[-5pt]
Bead $\ell^\prime$ & Rank $j$ & Scaling factor $\xi_j^{\left(\ell^\prime\right)}\left(g\right)$ & Scaling method\\[12pt] \hline\hline \\[-8pt]
$\{\emptyset\}$ & 0 & -- & -- \\[6pt]  \hline \\[-8pt]
$\{k\}$	& 1 & -- & -- \\[6pt] \hline \\[-8pt]
\multirow[c]{2}{*}{$\{k,l\}_{\textit{even}}$} & 0 & $\sqrt{\frac{1}{3}}$ & canonical\\[6pt]
	& 2 & $\sqrt{\frac{2}{3}}$ & canonical\\[6pt]  \hline \\[-8pt]
$\{k,l\}_{\textit{odd}}$	& 1	& $\frac{1}{\sqrt2}$ &	GUB\\[6pt]  \hline \\[-8pt]
\multirow[c]{2}{*}{$\{k,l,m\ \tau_1\}_{\textit{odd}}$} & 1 & $\sqrt{\frac{3}{5}}$ &	canonical\\[6pt]
	& 3 & $\sqrt{\frac{2}{5}}$ &	canonical\\[6pt]  \hline \\[-8pt]
$\{k,l,m\ \tau_2\}_{\textit{odd}}$ &	1 & $\frac{3}{3+\sqrt3}$ &	GUB\\[6pt]  \hline \\[-8pt]
$\{k,l,m\ \tau_2\}_{\textit{even}}$ &	2 & $\frac{1}{\sqrt2}$ &	GUB\\[6pt]  \hline \\[-8pt]
$\{k,l,m\ \tau_3\}_{\textit{odd}}$ &	1 & $\frac{1}{\sqrt2}$ &	GUB\\[6pt]  \hline \\[-8pt]
$\{k,l,m\ \tau_3\}_{\textit{even}}$ &	2 & $\frac{1}{\sqrt2}$ &	GUB\\[6pt]  \hline \\[-8pt]
$\{k,l,m\ \tau_4\}_{\textit{even}}$ &	0 & $\frac{1}{\sqrt2}$ &	GUB\\[6pt]  \hline\hline
\end{tabular}
\end{table}  
\clearpage

\section{Four-partite connected correlation functions}\label{sup:4partconn}
The four-partite connected correlation function of a four-partite quantum state $\rho$ is given by
\begin{align}
\Gamma_{E,1234}&= \Gamma_{T,1234} -\Gamma_{C,1234}\nonumber\\
&=\Gamma_{T,1234} - \sum_{\substack{\{j\}\sqcup\{k,l,m\}=\{1,2,3,4\}\\
\text{partitions}}}\Gamma_{C,(j)(klm)} - \sum_{\substack{\{j,k\}\sqcup\{l,m\}=\{1,2,3,4\}\\
\text{partitions}}}\Gamma_{C,(jk)(lm)} \nonumber\\
&\phantom{=}-\sum_{\substack{\{j\}\sqcup\{k\}\sqcup\{l,m\}=\{1,2,3,4\}\\
\text{partitions}}}\Gamma_{C,(j)(k)(lm)} \ \ - \Gamma_{C,(1)(2)(3)(4)} \nonumber\\
&=\Gamma_{T,1234} - \langle\sigma_{1\vec{r}_1}\rangle\Gamma_{E,234} - \langle\sigma_{2\vec{r}_2}\rangle\Gamma_{E,134} - \langle\sigma_{3\vec{r}_3}\rangle\Gamma_{E,124} - \langle\sigma_{4\vec{r}_4}\rangle\Gamma_{E,123} \nonumber\\
&\phantom{=}- \Gamma_{E,12}\Gamma_{E,34} - \Gamma_{E,13}\Gamma_{E,24} - \Gamma_{E,14}\Gamma_{E,23} \nonumber\\
&\phantom{=}- \langle\sigma_{1\vec{r}_1}\rangle\langle\sigma_{2\vec{r}_2}\rangle\Gamma_{E,34} - \langle\sigma_{1\vec{r}_1}\rangle\langle\sigma_{3\vec{r}_3}\rangle\Gamma_{E,24} - \langle\sigma_{1\vec{r}_1}\rangle\langle\sigma_{4\vec{r}_4}\rangle\Gamma_{E,23} \nonumber\\
&\phantom{=}- \langle\sigma_{2\vec{r}_2}\rangle\langle\sigma_{3\vec{r}_3}\rangle\Gamma_{E,14} - \langle\sigma_{2\vec{r}_2}\rangle\langle\sigma_{4\vec{r}_4}\rangle\Gamma_{E,13} - \langle\sigma_{3\vec{r}_3}\rangle\langle\sigma_{4\vec{r}_4}\rangle\Gamma_{E,12}\nonumber\\
&\phantom{=}- \left\langle\sigma_{1\vec{r}_1}\right\rangle\left\langle\sigma_{2\vec{r}_2}\right\rangle\left\langle\sigma_{3\vec{r}_3}\right\rangle\left\langle\sigma_{4\vec{r}_4}\right\rangle\nonumber\\
&=\left\langle\sigma_{1\vec{r}_1}\sigma_{2\vec{r}_2}\sigma_{3\vec{r}_3}\sigma_{4\vec{r}_4}\right\rangle \nonumber\\
&\phantom{=}- \langle\sigma_{1\vec{r}_1}\rangle \left\langle\sigma_{2\vec{r}_2}\sigma_{3\vec{r}_3}\sigma_{4\vec{r}_4}\right\rangle - \langle\sigma_{2\vec{r}_2}\rangle \left\langle\sigma_{1\vec{r}_1}\sigma_{3\vec{r}_3}\sigma_{4\vec{r}_4}\right\rangle - \langle\sigma_{3\vec{r}_3}\rangle \left\langle\sigma_{1\vec{r}_1}\sigma_{2\vec{r}_2}\sigma_{4\vec{r}_4}\right\rangle- \langle\sigma_{4\vec{r}_4}\rangle \left\langle\sigma_{1\vec{r}_1}\sigma_{2\vec{r}_2}\sigma_{3\vec{r}_3}\right\rangle\nonumber\\
&\phantom{=}-\langle\sigma_{1\vec{r}_1}\sigma_{2\vec{r}_2}\rangle \left\langle\sigma_{3\vec{r}_3}\sigma_{4\vec{r}_4}\right\rangle
-\langle\sigma_{1\vec{r}_1}\sigma_{3\vec{r}_3}\rangle \left\langle\sigma_{2\vec{r}_2}\sigma_{4\vec{r}_4}\right\rangle
-\langle\sigma_{1\vec{r}_1}\sigma_{4\vec{r}_4}\rangle \left\langle\sigma_{2\vec{r}_2}\sigma_{3\vec{r}_3}\right\rangle \nonumber\\
&\phantom{=}+2\left[\langle\sigma_{1\vec{r}_1}\rangle\langle\sigma_{2\vec{r}_2}\rangle \left\langle\sigma_{3\vec{r}_3}\sigma_{4\vec{r}_4}\right\rangle + \langle\sigma_{1\vec{r}_1}\rangle\langle\sigma_{3\vec{r}_3}\rangle \left\langle\sigma_{2\vec{r}_2}\sigma_{4\vec{r}_4}\right\rangle + \langle\sigma_{1\vec{r}_1}\rangle\langle\sigma_{4\vec{r}_4}\rangle \left\langle\sigma_{2\vec{r}_2}\sigma_{3\vec{r}_3}\right\rangle \right. \nonumber\\
&\phantom{=2[}\left.+ \langle\sigma_{2\vec{r}_2}\rangle\langle\sigma_{3\vec{r}_3}\rangle \left\langle\sigma_{1\vec{r}_1}\sigma_{4\vec{r}_4}\right\rangle + \langle\sigma_{2\vec{r}_2}\rangle\langle\sigma_{4\vec{r}_4}\rangle \left\langle\sigma_{1\vec{r}_1}\sigma_{3\vec{r}_3}\right\rangle + \langle\sigma_{3\vec{r}_3}\rangle\langle\sigma_{4\vec{r}_4}\rangle \left\langle\sigma_{1\vec{r}_1}\sigma_{2\vec{r}_2}\right\rangle \right] \nonumber\\
&\phantom{=}-6\left\langle\sigma_{1\vec{r}_1}\right\rangle\left\langle\sigma_{2\vec{r}_2}\right\rangle\left\langle\sigma_{3\vec{r}_3}\right\rangle\left\langle\sigma_{4\vec{r}_4}\right\rangle
\label{eq:fourpartite_E_correlationfunction}
\end{align}

\section{Tri- and four-partite connected correlation function density operator components}
\label{sec:tripartite_connected}

In the tripartite case, the connected correlation function density operator component $\rho_{E,123}$ corresponding to the connected correlation functions $\Gamma_{E,123}$ can be obtained as
\begin{align}
	\rho_{E,123}=\rho+\frac{1}{4}&\ \sum_{\alpha,\beta,\gamma=x,y,z}{\left\langle\sigma_{1\alpha}\right\rangle\left\langle\sigma_{2\beta}\right\rangle\left\langle\sigma_{3\gamma}\right\rangle\sigma_{1\alpha}\sigma_{2\beta}\sigma_{3\gamma}}\nonumber\\
	-\frac{1}{8}&\left[\sum_{\alpha,\beta=x,y,z}{\left\langle\sigma_{1\alpha}\sigma_{2\beta}\right\rangle\sigma_{1\alpha}\sigma_{2\beta}}\sum_{\gamma=x,y,z}{\left\langle\sigma_{3\gamma}\right\rangle\sigma_{3\gamma}}\right.\nonumber\\
	&+\sum_{\alpha,\gamma=x,y,z}{\left\langle\sigma_{1\alpha}\sigma_{3\gamma}\right\rangle\sigma_{1\alpha}\sigma_{3\gamma}}\sum_{\beta=x,y,z}{\left\langle\sigma_{2\beta}\right\rangle\sigma_{2\beta}}\nonumber\\
	&+\left.\sum_{\beta,\gamma=x,y,z}{\left\langle\sigma_{2\beta}\sigma_{3\gamma}\right\rangle\sigma_{2\beta}\sigma_{3\gamma}}\sum_{\alpha=x,y,z}{\left\langle\sigma_{1\alpha}\right\rangle\sigma_{1\alpha}}\right]\nonumber\\
	-\frac{1}{8}&\left[\sum_{\alpha,\beta=x,y,z}{\left\langle\sigma_{1\alpha}\sigma_{2\beta}\right\rangle\sigma_{1\alpha}\sigma_{2\beta}}\right.\nonumber\\
	&+\sum_{\alpha,\gamma=x,y,z}{\left\langle\sigma_{1\alpha}\sigma_{3\gamma}\right\rangle\sigma_{1\alpha}\sigma_{3\gamma}}\nonumber\\
	&+\left.\sum_{\beta,\gamma=x,y,z}{\left\langle\sigma_{2\beta}\sigma_{3\gamma}\right\rangle\sigma_{2\beta}\sigma_{3\gamma}}\right] \nonumber\\
	-\frac{1}{8}& \left[ \sum_{\alpha=x,y,z} \braket{\sigma_{1\alpha}} \sigma_{1\alpha} + \sum_{\beta=x,y,z} \braket{\sigma_{2\beta}} \sigma_{2\beta} + \sum_{\gamma=x,y,z} \braket{\sigma_{3\gamma}} \sigma_{3\gamma}+ \mathbbm{1}_8 \right].
	\label{eq:rho_E_pure_tripartite}
\end{align}

That is, individual components $\rho_{E,123}$ are obtained by isolating the genuinely tripartite connected correlation function density operator component according to \cref{eq:tripartite_E_correlationfunction} and by subtracting the full unwanted $k$-body terms, which involves all bipartite, single-qubit, and the identity terms, from $\rho$.

Analogously, following \cref{eq:fourpartite_E_correlationfunction}, the component $\rho_{E,1234}$ of a four-partite density operator corresponding to the four-partite connected correlation function, is obtained as

\begin{align}
	\rho_{E,1234}=\rho-\frac{3}{8}&\ \sum_{\alpha,\beta,\gamma,\delta=x,y,z}{\left\langle\sigma_{1\alpha}\right\rangle\left\langle\sigma_{2\beta}\right\rangle\left\langle\sigma_{3\gamma}\right\rangle\left\langle\sigma_{4\delta}\right\rangle\sigma_{1\alpha}\sigma_{2\beta}\sigma_{3\gamma}}\sigma_{4\delta}\nonumber\\
	-\frac{1}{16}&\left[\sum_{\alpha,\beta,\gamma=x,y,z}{\left\langle\sigma_{1\alpha}\sigma_{2\beta}\sigma_{3\gamma}\right\rangle\sigma_{1\alpha}\sigma_{2\beta}\sigma_{3\gamma}}\sum_{\delta=x,y,z}{\left\langle\sigma_{4\delta}\right\rangle\sigma_{4\delta}}\right.\nonumber\\
	&+\sum_{\alpha,\beta,\delta=x,y,z}{\left\langle\sigma_{1\alpha}\sigma_{2\beta}\sigma_{4\delta}\right\rangle\sigma_{1\alpha}\sigma_{2\beta}\sigma_{4\delta}}\sum_{\gamma=x,y,z}{\left\langle\sigma_{3\gamma}\right\rangle\sigma_{3\gamma}}\nonumber\\
&+\sum_{\alpha,\gamma,\delta=x,y,z}{\left\langle\sigma_{1\alpha}\sigma_{3\gamma}\sigma_{4\delta}\right\rangle\sigma_{1\alpha}\sigma_{3\gamma}\sigma_{4\delta}}\sum_{\beta=x,y,z}{\left\langle\sigma_{2\beta}\right\rangle\sigma_{2\beta}}\nonumber\\
	&+\left.\sum_{\beta,\gamma,\delta=x,y,z}{\left\langle\sigma_{2\beta}\sigma_{3\gamma}\sigma_{4\delta}\right\rangle\sigma_{2\beta}\sigma_{3\gamma}\sigma_{4\delta}}\sum_{\alpha=x,y,z}{\left\langle\sigma_{1\alpha}\right\rangle\sigma_{1\alpha}}\right]\nonumber\\	
-\frac{1}{16}&\left[\sum_{\alpha,\beta=x,y,z}{\left\langle\sigma_{1\alpha}\sigma_{2\beta}\right\rangle\sigma_{1\alpha}\sigma_{2\beta}}\sum_{\gamma,\delta=x,y,z}{\left\langle\sigma_{3\gamma}\sigma_{4\delta}\right\rangle\sigma_{3\gamma}\sigma_{4\delta}}\right.\nonumber\\
	&+\sum_{\alpha,\gamma=x,y,z}{\left\langle\sigma_{1\alpha}\sigma_{3\gamma}\right\rangle\sigma_{1\alpha}\sigma_{3\gamma}}\sum_{\beta,\delta=x,y,z}{\left\langle\sigma_{2\beta}\sigma_{4\delta}\right\rangle\sigma_{2\beta}\sigma_{4\delta}}\nonumber\\
&+\left.\sum_{\alpha,\delta=x,y,z}{\left\langle\sigma_{1\alpha}\sigma_{4\delta}\right\rangle\sigma_{1\alpha}\sigma_{4\delta}}\sum_{\beta,\gamma=x,y,z}{\left\langle\sigma_{2\beta}\sigma_{3\gamma}\right\rangle\sigma_{2\beta}\sigma_{3\gamma}}\right]\nonumber\\	
+\frac{1}{8}&\left[\sum_{\alpha=x,y,z}{\left\langle\sigma_{1\alpha}\right\rangle \sigma_{1\alpha}} \sum_{\beta=x,y,z}{\left\langle\sigma_{2\beta}\right\rangle\sigma_{2\beta}}\sum_{\gamma,\delta=x,y,z}{\left\langle\sigma_{3\gamma}\sigma_{4\delta}\right\rangle\sigma_{3\gamma}\sigma_{4\delta}}\right.\nonumber\\
&+\sum_{\alpha=x,y,z}{\left\langle\sigma_{1\alpha}\right\rangle \sigma_{1\alpha}} \sum_{\gamma=x,y,z}{\left\langle\sigma_{3\gamma}\right\rangle\sigma_{3\gamma}}\sum_{\beta,\delta=x,y,z}{\left\langle\sigma_{2\beta}\sigma_{4\delta}\right\rangle\sigma_{2\beta}\sigma_{4\delta}}\nonumber\\
&+\sum_{\alpha=x,y,z}{\left\langle\sigma_{1\alpha}\right\rangle \sigma_{1\alpha}} \sum_{\delta=x,y,z}{\left\langle\sigma_{4\delta}\right\rangle\sigma_{4\delta}}\sum_{\beta,\gamma=x,y,z}{\left\langle\sigma_{2\beta}\sigma_{3\gamma}\right\rangle\sigma_{2\beta}\sigma_{3\gamma}}\nonumber\\
&+\sum_{\beta=x,y,z}{\left\langle\sigma_{2\beta}\right\rangle \sigma_{2\beta}} \sum_{\gamma=x,y,z}{\left\langle\sigma_{3\gamma}\right\rangle\sigma_{3\gamma}}\sum_{\alpha,\delta=x,y,z}{\left\langle\sigma_{1\alpha}\sigma_{4\delta}\right\rangle\sigma_{1\alpha}\sigma_{4\delta}}\nonumber\\
&+\sum_{\beta=x,y,z}{\left\langle\sigma_{2\beta}\right\rangle \sigma_{2\beta}} \sum_{\delta=x,y,z}{\left\langle\sigma_{4\delta}\right\rangle\sigma_{4\delta}}\sum_{\alpha,\gamma=x,y,z}{\left\langle\sigma_{1\alpha}\sigma_{3\gamma}\right\rangle\sigma_{1\alpha}\sigma_{3\gamma}}\nonumber\\
&+\left.\sum_{\gamma=x,y,z}{\left\langle\sigma_{3\gamma}\right\rangle \sigma_{3\gamma}} \sum_{\delta=x,y,z}{\left\langle\sigma_{4\delta}\right\rangle\sigma_{4\delta}}\sum_{\alpha,\beta=x,y,z}{\left\langle\sigma_{1\alpha}\sigma_{2\beta}\right\rangle\sigma_{1\alpha}\sigma_{2\beta}}\right]\nonumber\\
-\frac{1}{16}&\left[\sum_{\alpha,\beta,\gamma=x,y,z}{\left\langle\sigma_{1\alpha}\sigma_{2\beta}\sigma_{3\gamma}\right\rangle\sigma_{1\alpha}\sigma_{2\beta}\sigma_{3\gamma}}+ \sum_{\alpha,\beta,\delta=x,y,z}{\left\langle\sigma_{1\alpha}\sigma_{2\beta}\sigma_{4\delta}\right\rangle\sigma_{1\alpha}\sigma_{2\beta}\sigma_{4\delta}}\right.\nonumber\\
&+\left.\sum_{\alpha,\gamma,\delta=x,y,z}{\left\langle\sigma_{1\alpha}\sigma_{3\gamma}\sigma_{4\delta}\right\rangle\sigma_{1\alpha}\sigma_{3\gamma}\sigma_{4\delta}}+ \sum_{\beta,\gamma,\delta=x,y,z}{\left\langle\sigma_{2\beta}\sigma_{3\gamma}\sigma_{4\delta}\right\rangle\sigma_{2\beta}\sigma_{3\gamma}\sigma_{4\delta}}\right]\nonumber\\
	-\frac{1}{16}&\left[\sum_{\alpha,\beta=x,y,z}{\left\langle\sigma_{1\alpha}\sigma_{2\beta}\right\rangle\sigma_{1\alpha}\sigma_{2\beta}}+ \sum_{\alpha,\gamma=x,y,z}{\left\langle\sigma_{1\alpha}\sigma_{3\gamma}\right\rangle\sigma_{1\alpha}\sigma_{3\gamma}}\right.\nonumber\\
	&+\sum_{\alpha,\delta=x,y,z}{\left\langle\sigma_{1\alpha}\sigma_{4\delta}\right\rangle\sigma_{1\alpha}\sigma_{4\delta}}+\sum_{\beta,\gamma=x,y,z}{\left\langle\sigma_{2\beta}\sigma_{3\gamma}\right\rangle\sigma_{2\beta}\sigma_{3\gamma}}\nonumber\\
	&+\left.\sum_{\beta,\delta=x,y,z}{\left\langle\sigma_{2\beta}\sigma_{4\delta}\right\rangle\sigma_{2\beta}\sigma_{4\delta}} + \sum_{\gamma,\delta=x,y,z}{\left\langle\sigma_{3\gamma}\sigma_{4\delta}\right\rangle\sigma_{3\gamma}\sigma_{4\delta}}\right] \nonumber\\
	-\frac{1}{16}& \left[ \sum_{\alpha=x,y,z} \braket{\sigma_{1\alpha}} \sigma_{1\alpha} + \sum_{\beta=x,y,z} \braket{\sigma_{2\beta}} \sigma_{2\beta} + \sum_{\gamma=x,y,z} \braket{\sigma_{3\gamma}} \sigma_{3\gamma}+ \sum_{\delta=x,y,z} \braket{\sigma_{4\delta}} \sigma_{4\delta} + \mathbbm{1}_{16} \right].
	\label{eq:rho_E_pure_fourpartite}
\end{align}

\newpage

\section{Mathematical details on the construction of the entanglement quantifiers $\tilde{\mathcal{N}}_3$ and $\overline{\mathcal{N}}_3$}\label{sec:SuppMathDetailsEntQuant}

In appendix~\ref{sec:EntQuant}, we introduced the entanglement quantifiers $\tilde{\mathcal{N}}_3$ and $\overline{\mathcal{N}}_3$, which can be used to reliably detect and distinguish tripartite GHZ- and W-type entanglement. In this section, we provide further details on the construction of  $\tilde{\mathcal{N}}_3$ and $\overline{\mathcal{N}}_3$. First, we outline how one arrives at the parametric expression $\tilde{\mathcal{N}}_3(\theta)$ for GHZ-entangled states $\ket{\psi}_\theta = \cos(\theta) \ket{000} + \sin(\theta) \ket{111}$. We start by writing out the full density operator 
\begin{equation}
\hspace{-3.5cm}
\rho_\theta = \ket{\psi}_\theta\bra{\psi}_\theta = \begin{pmatrix}
\cos^2(\theta) & 0 & 0 & 0 & 0 & 0 & 0 & \cos(\theta)\sin(\theta) \\
0 & 0 & 0 & 0 & 0 & 0 & 0 & 0 \\
0 & 0 & 0 & 0 & 0 & 0 & 0 & 0 \\
0 & 0 & 0 & 0 & 0 & 0 & 0 & 0 \\
0 & 0 & 0 & 0 & 0 & 0 & 0 & 0 \\
0 & 0 & 0 & 0 & 0 & 0 & 0 & 0 \\
0 & 0 & 0 & 0 & 0 & 0 & 0 & 0 \\
\cos(\theta)\sin(\theta) & 0 & 0 & 0 & 0 & 0 & 0 & \sin^2(\theta)
\end{pmatrix}
\end{equation}

\begin{figure}[h!]          
	\centering
	\includegraphics[width=\textwidth, trim= 0cm 0cm 0cm 0cm, clip]{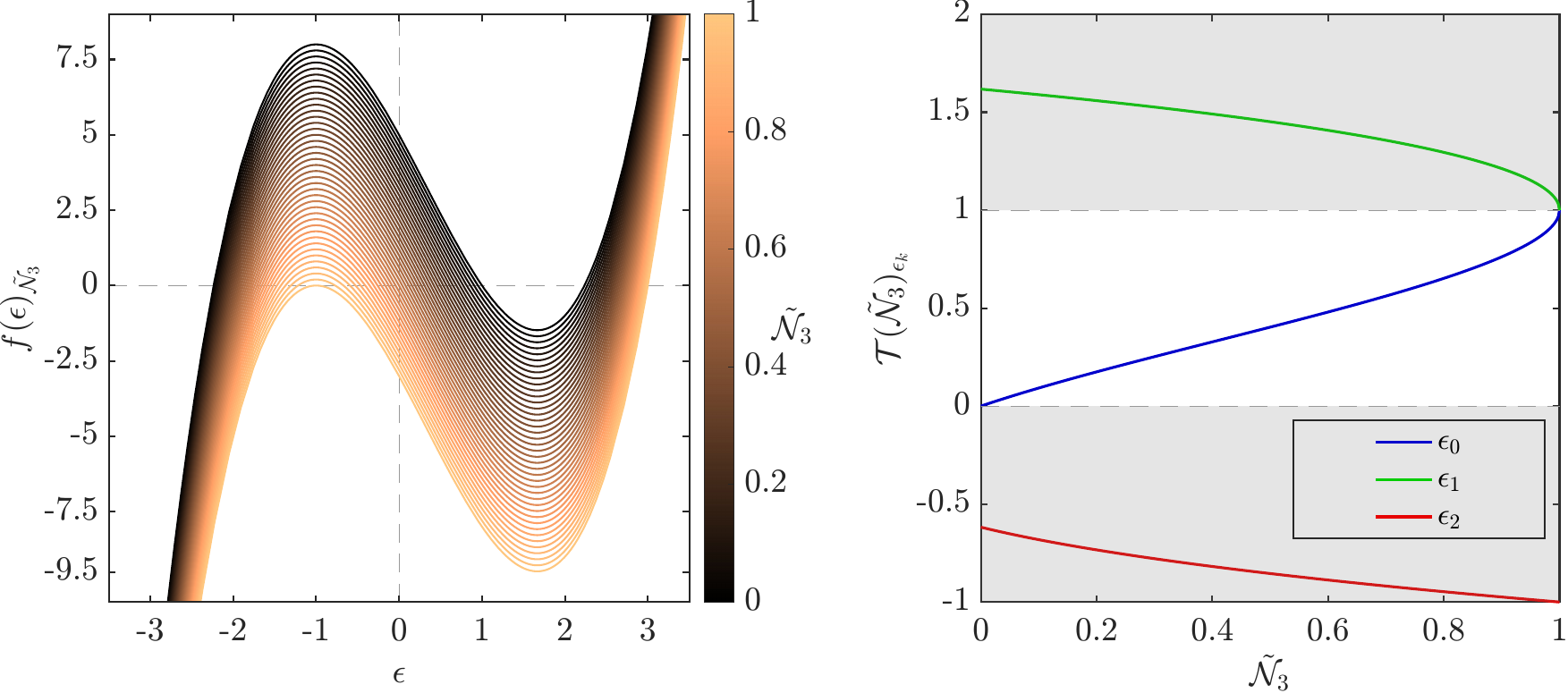}
	\caption{Left: Function plot of the polynomial $f(\epsilon)_{\tilde{\mathcal{N}}_3}$ for varying parameter values $\tilde{\mathcal{N}}_3$. The polynomial has three distinct roots unless $\tilde{\mathcal{N}}_3 = 1$, where there exists a repeated root at $\epsilon = \epsilon_0 = \epsilon_1 = -1$. Right: Graphical analysis of the resulting mapping $\mathcal{T}(\tilde{\mathcal{N}}_3)_{\epsilon_k} = (1-\epsilon_{k,\tilde{\mathcal{N}}_3})/2$ for all different roots of $f(\epsilon)_{\tilde{\mathcal{N}}_3}$. Undesired numerical ranges are greyed out. It is clear that only $\epsilon_0$ allows to achieve the desired mapping $\mathcal{T}: [0;\ 1] \rightarrow [0;\ 1]$.}
	\label{fig:PolynomialAnalysis}
	\vspace{-.5cm}
\end{figure}

Notably, this parameterized state, aside from the identity component, has only Pauli-Z lower-order terms, that is, only $\langle Z_k \rangle$ and $\langle Z_k Z_l \rangle$ are nonzero with respect to $\rho_\theta$, using the shorthand notation $Z_k  = \sigma_{kz}$ to express Pauli operators of qubit $k$, while all remaining single- and two-body Pauli terms have expectation value 0. 
One obtains the following nonzero expectation values of Pauli operators with respect to $\rho_\theta$:
\begin{subequations}
\begin{align}
&\langle \mathbbm{1} \rangle_\theta = \cos^2(\theta) + \sin^2(\theta),\\[8pt]
&\langle Z_1 \rangle_\theta = \langle Z_2 \rangle_\theta = \langle Z_3 \rangle_\theta = \cos^2(\theta) - \sin^2(\theta),\\[8pt]
&\langle Z_1 Z_2 \rangle_\theta = \langle Z_1 Z_3 \rangle_\theta = \langle Z_2 Z_3 \rangle_\theta = \cos^2(\theta) + \sin^2(\theta),\\[8pt]
&\langle X_1 X_2 X_3 \rangle_\theta = 2\ \cos(\theta)\sin(\theta), \\
&\langle X_1 Y_2 Y_3 \rangle_\theta = \langle Y_1 X_2 Y_3 \rangle_\theta = \langle Y_1 Y_2 X_3 \rangle_\theta = -2\ \cos(\theta)\sin(\theta), \\
&\langle Z_1 Z_2 Z_3 \rangle_\theta = \cos^2(\theta) - \sin^2(\theta).
\end{align}
\end{subequations}
The density operator component corresponding to the full genuine tripartite connected correlation function of $\rho_\theta$ is then
\begin{align}
\rho_{E,123,\theta} = \rho_\theta - \frac{1}{8}&\bigg( \langle \mathbbm{1} \rangle_\theta \mathbbm{1} + \langle Z_1 \rangle_\theta\ Z_1 + \langle Z_2 \rangle_\theta\ Z_2 + \langle Z_3 \rangle_\theta\ Z_3 \nonumber\\ &+ \langle Z_1 Z_2 \rangle_\theta\ Z_1 Z_2 + \langle Z_1 Z_3 \rangle_\theta\ Z_1 Z_3 + \langle Z_2 Z_3 \rangle_\theta\ Z_2 Z_3 \nonumber\\[6pt] &+
\langle Z_1 Z_2 \rangle_\theta \langle Z_3 \rangle_\theta\ Z_1 Z_2 Z_3 + \langle Z_1 Z_3 \rangle_\theta \langle Z_2 \rangle_\theta\ Z_1 Z_2 Z_3 + \langle Z_2 Z_3 \rangle_\theta \langle Z_1 \rangle_\theta\ Z_1 Z_2 Z_3 \nonumber\\
&- 2 \langle Z_1 \rangle_\theta \langle Z_2 \rangle_\theta \langle Z_3 \rangle_\theta\ Z_1 Z_2 Z_3 \bigg)\nonumber\\[0pt]
\phantom{\rho_{E,123,\theta}}=\phantom{\rho_\theta +} \frac{1}{8}&\bigg(\langle X_1 Y_2 Y_3 \rangle_\theta\ X_1 Y_2 Y_3 + \langle Y_1 X_2 Y_3 \rangle_\theta\ Y_1 X_2 Y_3 + \langle Y_1 Y_2 X_3 \rangle_\theta\ Y_1 Y_2 X_3 \nonumber\\
 &+\langle X_1 X_2 X_3 \rangle_\theta\ X_1 X_2 X_3 + \langle Z_1 Z_2 Z_3 \rangle_\theta\ Z_1 Z_2 Z_3 \nonumber\\[6pt]
&-\langle Z_1 Z_2 \rangle_\theta \langle Z_3 \rangle_\theta\ Z_1 Z_2 Z_3 - \langle Z_1 Z_3 \rangle_\theta \langle Z_2 \rangle_\theta\ Z_1 Z_2 Z_3 - \langle Z_2 Z_3 \rangle_\theta \langle Z_1 \rangle_\theta\ Z_1 Z_2 Z_3 \nonumber\\
&+ 2 \langle Z_1 \rangle_\theta \langle Z_2 \rangle_\theta \langle Z_1 \rangle_\theta\ Z_1 Z_2 Z_3 \bigg)\nonumber\\[0pt]
\phantom{\rho_{E,123,\theta}}=\phantom{\rho_\theta +} \frac{1}{4}&\begin{pmatrix}
a & 0 & 0 & 0 & 0 & 0 & 0 & b \\
0 & -a & 0 & 0 & 0 & 0 & 0 & 0 \\
0 & 0 & -a & 0 & 0 & 0 & 0 & 0 \\
0 & 0 & 0 & a & 0 & 0 & 0 & 0 \\
0 & 0 & 0 & 0 & -a & 0 & 0 & 0 \\
0 & 0 & 0 & 0 & 0 & a & 0 & 0 \\
0 & 0 & 0 & 0 & 0 & 0 & a & 0 \\
b & 0 & 0 & 0 & 0 & 0 & 0 & -a
\end{pmatrix}\\[6pt]
\text{with } &a = \cos^3(2\theta)-\cos(2\theta) \text{ and } b=2\sin(2\theta)\nonumber
\end{align}
Finally, we arrive at $\tilde{\mathcal{N}}_3(\theta)$ by calculating the doubled square of the Frobenius norm of $\rho_{E,123,\theta}$ which after simplification by using well-known trigonometric identities is obtained as
\begin{equation}
\tilde{\mathcal{N}}_3(\theta) = 2\ \text{Tr}(\rho_{E,123}\ \rho_{E,123}^\dagger) = \frac{[\cos^2(4\theta) - 5][\cos(4\theta) - 1]}{8} = \frac{1}{8}(\cos^3(4\theta)-\cos^2(4\theta)-5\cos(4\theta)+5).
\label{eq:N3detailed}
\end{equation}
As outlined in appendix~\ref{sec:EntQuant}, we then substitute $\epsilon = \cos(4\theta)$ to find a mapping $\mathcal{T}$ of $\tilde{\mathcal{N}}_3$ onto the three-tangle $\tau_3$ for GHZ-entangled states. Thus, we first query which $\epsilon$ corresponds to a given value of $\tilde{\mathcal{N}}_3$ and then reinsert this $\epsilon$ into Eq.~\ref{eq:threetangle2} to achieve the intended mapping. It is clear that this requires finding the roots of the polynomial
\begin{equation}
f(\epsilon)_{\tilde{\mathcal{N}}_3} = \epsilon^3-\epsilon^2-5\epsilon + 5 - 8\tilde{\mathcal{N}}_3,
\end{equation}
which is Eq.~\ref{eq:N3detailed} rewritten as a function of $\epsilon$ in dependence of $\tilde{\mathcal{N}}_3$.
As can be seen in Fig.~\ref{fig:PolynomialAnalysis}, this polynomial has three distinct real-valued roots for the allowed parameter range of $\tilde{\mathcal{N}}_3$ $[0;\ 1[$ and a root with multiplicity two only when $\tilde{\mathcal{N}}_3 = 1$. The roots have the following analytic expressions:
\begin{subequations}
\begin{align}
\hspace{-.5cm}
\epsilon_0 &= \frac{1}{6} \left(-\sqrt[3]{4} (1 + i \sqrt3) \sqrt[3]{ \sqrt{81 \tilde{\mathcal{N}}_3^2 - 66 \tilde{\mathcal{N}}_3 - 15} + 27 \tilde{\mathcal{N}}_3 - 11} + \frac{8 i \sqrt[3]{2} (\sqrt3 + i)}{\sqrt[3]{3 \sqrt{81 \tilde{\mathcal{N}}_3^2 - 66 \tilde{\mathcal{N}}_3 - 15} + 27 \tilde{\mathcal{N}}_3 - 11}} + 2 \right),\\
\hspace{-.5cm}
\epsilon_1 &= \frac{1}{6} \left(
i\sqrt[3]{4}(\sqrt{3}+i)
\sqrt[3]{ \sqrt{81 \tilde{\mathcal{N}}_3^2 - 66 \tilde{\mathcal{N}}_3 - 15} + 27 \tilde{\mathcal{N}}_3 - 11}
-
\frac{8i\sqrt[3]{2}(\sqrt{3}-i)}
{\sqrt[3]{3\sqrt{81 \tilde{\mathcal{N}}_3^2 - 66 \tilde{\mathcal{N}}_3 - 15} + 27 \tilde{\mathcal{N}}_3 - 11}}
+ 2
\right),\\
\hspace{-.5cm}
\epsilon_2 &= \frac{1}{3} \left(
\sqrt[3]{4}
\sqrt[3]{\sqrt{81 \tilde{\mathcal{N}}_3^2 - 66 \tilde{\mathcal{N}}_3 - 15} + 27 \tilde{\mathcal{N}}_3 - 11}
+
\frac{8\sqrt[3]{2}}
{\sqrt[3]{3\sqrt{81 \tilde{\mathcal{N}}_3^2 - 66 \tilde{\mathcal{N}}_3 - 15} + 27 \tilde{\mathcal{N}}_3 - 11}}
+ 1
\right),
\end{align}
\end{subequations}
and have imaginary parts $\mathfrak{I}(\epsilon_k) = 0$ as the imaginary units in the above expressions always cancel.
The desired mapping $\mathcal{T}: \tilde{\mathcal{N}}_3 \rightarrow \overline{\mathcal{N}}_3$, or equivalently $\mathcal{T}: \tilde{\mathcal{N}}_3 \rightarrow \tau_3$ for GHZ-entangled states, maps intervals $[0;\ 1] \rightarrow [0;\ 1]$. It is clear in Fig.~\ref{fig:PolynomialAnalysis} that only $\epsilon_0$ allows to achieve the desired properties while the remaining roots provide mappings outside of the allowed numerical range and thus, can be neglected.
\clearpage

\section{A didactic introduction to mixed states and the generalized BEADS representation}
\label{sec:didactic_intro}
In the following, we provide a phenomenological introduction to mixed states for non-experts. For simplicity, whenever possible, we confine ourselves to two-qubit states using computational basis states only (i.e., $\ket{0}$ and $\ket{1}$) for constructing the introduced types of mixed states.

\subsection*{\textbf{Pure product states}}
Before we actually get to mixed states, we shall briefly rediscover pure product states, which serve as the simplest type of quantum states overall. First, recall that pure product states are those that are separable and that can be written as a tensor product of pure single-qubit states, i.e., $\ket{\psi} = \ket{\phi_1} \otimes \ket{\phi_2} \hdots \ket{\phi_N}$ \cite{nielsen_quantum_2010}.

In the BEADS picture, for such states the full information is given by the $Q$-Beads, which correspond to the reduced single-qubit density operators. Pure product states are not entangled and thus, do not show $E$-Beads (entanglement-based correlation function beads). Instead they only show $C$-Beads (compound correlation function beads) which represent redundant information given by the product of $Q$-Beads and/or lower-order $E$-Beads (only for systems consisting of more than two qubits). $T$-Beads (\textit{t}otal correlation function beads), which correspond to the sum of $E$- and $C$-Beads, are consequently identical to $C$-Beads for product states \cite{huber_beads_2025}. As genuine correlations of a pure state always emerge from entanglement, pure product states are strictly uncorrelated.

\begin{figure}[htbp]
\centering
\includegraphics[width=\linewidth]{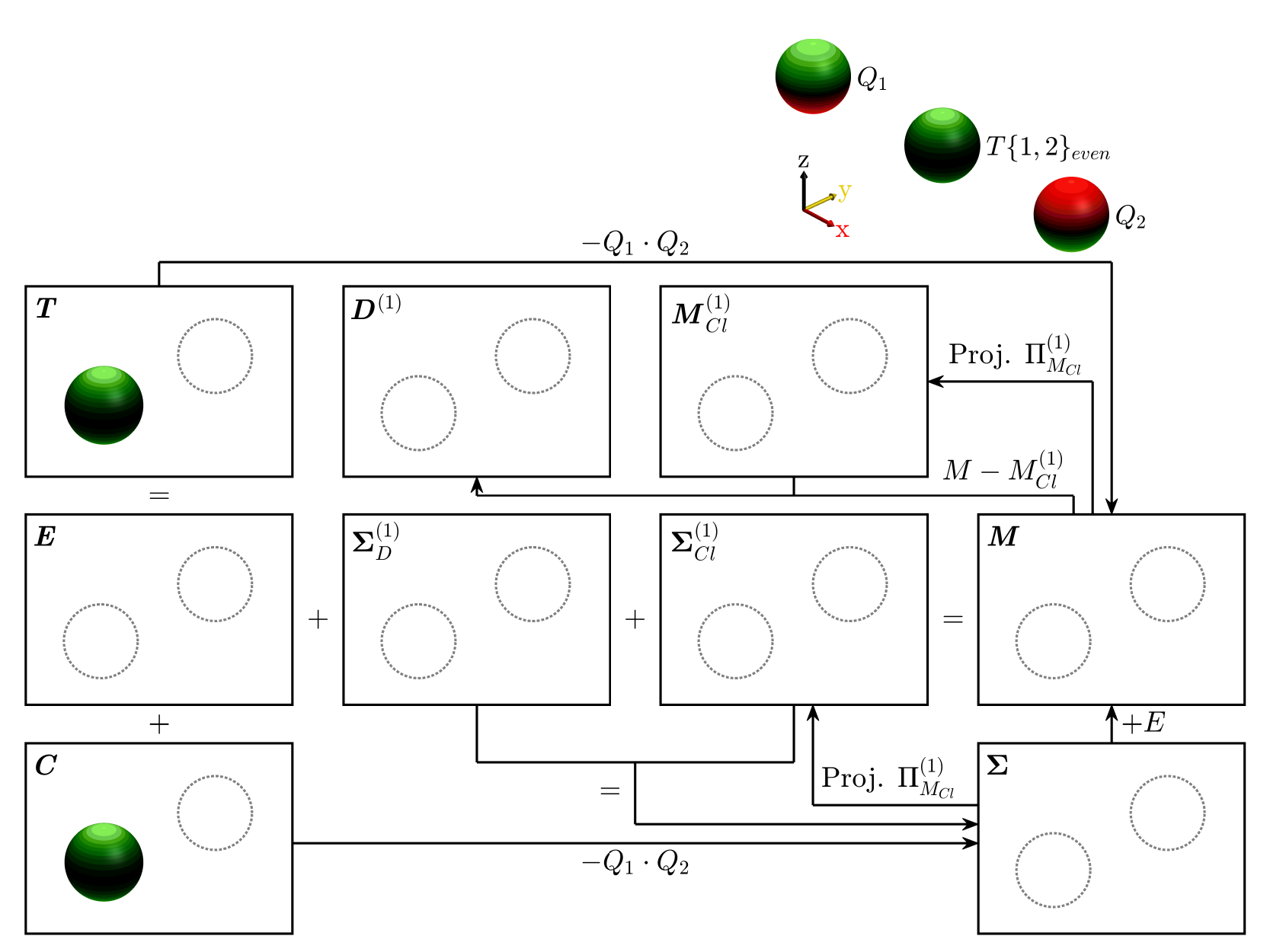}
\caption{Generalized BEADS representation of the pure product state $\ket{\psi} = \ket{01}$. It is clear that the $T$- and $C$-Beads only represent the products of both $Q$-Beads, i.e., redundant information. All other beads vanish due to the state being fully uncorrelated.}
\label{fig:DidProductPure}
\end{figure}

As an example, consider the state $\ket{10} = \ket{1} \otimes \ket{0}$ visualized in Fig.~\ref{fig:DidProductPure}. It is clear, that in this case, as for all pure product states, the \textit{color brightness} of all $Q$-Beads is \textit{maximum} and it can be easily seen, that here the $T$- and $C$-Beads are \textit{identical} and correspond to the product of $Q$-Bead values along all possible directions. Moreover, the state does not show any other types of beads. Note that these observations apply to all pure product states, and are thus sufficient to identify this class of states unambiguously using the BEADS representation. 

To prevent confusion for the remainder of this chapter, note that \textit{total} in the term total correlation function bead (abbrev. $T$-Bead) refers to the fact that this bead represents the complete correlation function of a state, but must not be understood as the bead visualizing the genuine total correlation (total mutual information). The correlation function, on the other hand, quantifies experimentally measurable expectation values of Pauli operators corresponding to joint local measurements performed on a given set of qubits but does only indirectly relate to correlations of a state as an information theoretic resource. While it is possible that $T$-Beads also capture genuine correlation properties of an arbitrary quantum state, they strictly represent non-correlation-based redundant information for product states.

\subsection*{\textbf{Mixed product states}}
We can now generalize these observations to mixed product states. As specified in Eq.~\ref{eq:product_state}, a mixed product state is defined as a product of single-qubit density operators $\rho_{\text{prod}} = \rho_1 \otimes \rho_2 \hdots \rho_N$ \cite{ding_quantum_2023}. While this looks very similar to the product of state vectors in the previous subsection, in the mixed state case, the individual density operators $\rho_k$ are generally mixed, that is, $\rho_k = \sum_i p_i \ket{\psi_i}\bra{\psi_i}$ with weights $p_i$. In other words, while the overall state is given by a simple tensor product of single-qubit density operators, the latter are obtained as weighted sums (called convex sums, i.e., the weights sum to 1) of pure single-qubit states. This implies that a mixed product state can actually be written as a convex sum of pure $N$-qubit product states $\rho_{\text{prod}} = \sum_i p_i \, \ket{\psi_i}\bra{\psi_i} \otimes \sum_j q_j \, \ket{\phi_j}\bra{\phi_j} \otimes \hdots \otimes \sum_l w_l \, \ket{\omega_l}\bra{\omega_l} = \sum_{i,j\hdots l}p_iq_j \hdots w_l \, \ket{\psi_i}\bra{\psi_i} \otimes \ket{\phi_j}\bra{\phi_j} \otimes \hdots \otimes \ket{\omega_l}\bra{\omega_l}$ 
with weights $p$, $q$, and $w$ corresponding to single-qubit pure states $\ket{\psi}$, $\ket{\phi}$, and $\ket{\omega}$, respectively. As mixed states are defined to be separable if they can be expressed as convex sums of pure product states, mixed product states are indeed fully separable.

\begin{figure}[htbp]
\centering
\vspace{-.3cm}
\includegraphics[width=.9\textwidth]{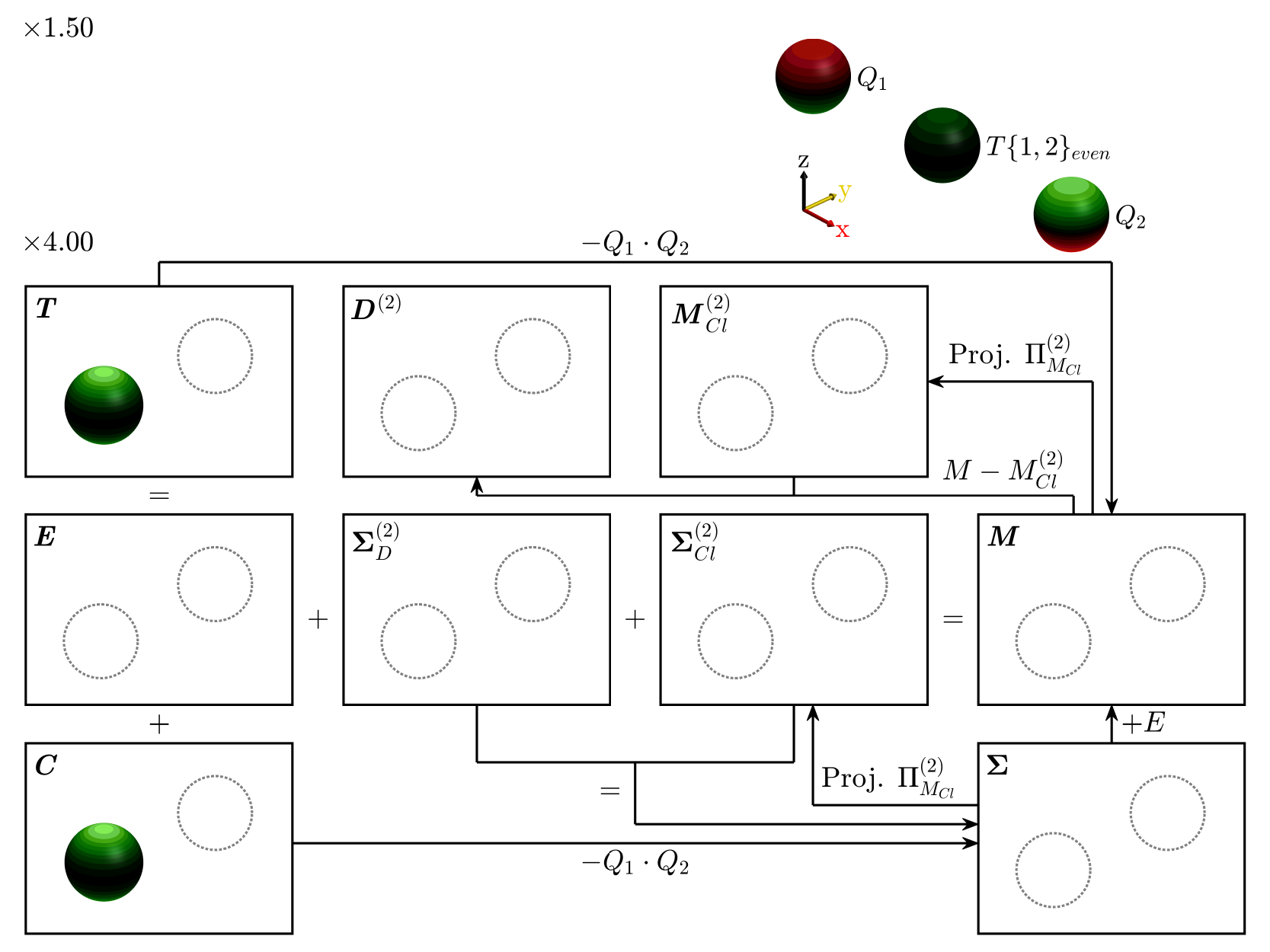}
\caption{Generalized BEADS representation of the mixed product state  $\rho = (0.7\ket{0}\bra{0}+0.3\ket{1}\bra{1}) \otimes (0.2\ket{0}\bra{0}+0.8\ket{1}\bra{1})$. For better visibility, bead values are scaled by the specified scaling factors. As in the pure state case, the $T$- and $C$-Beads only represent redundant information. All other beads vanish due to the state being fully uncorrelated. The $Q$-Beads do not have full color brightness due to the corresponding reduced density operators being mixed.}
\vspace{-2cm}
\label{fig:DidProductMixed}
\end{figure}

\begin{figure}[h!]
\centering
\includegraphics[width=1.\textwidth]{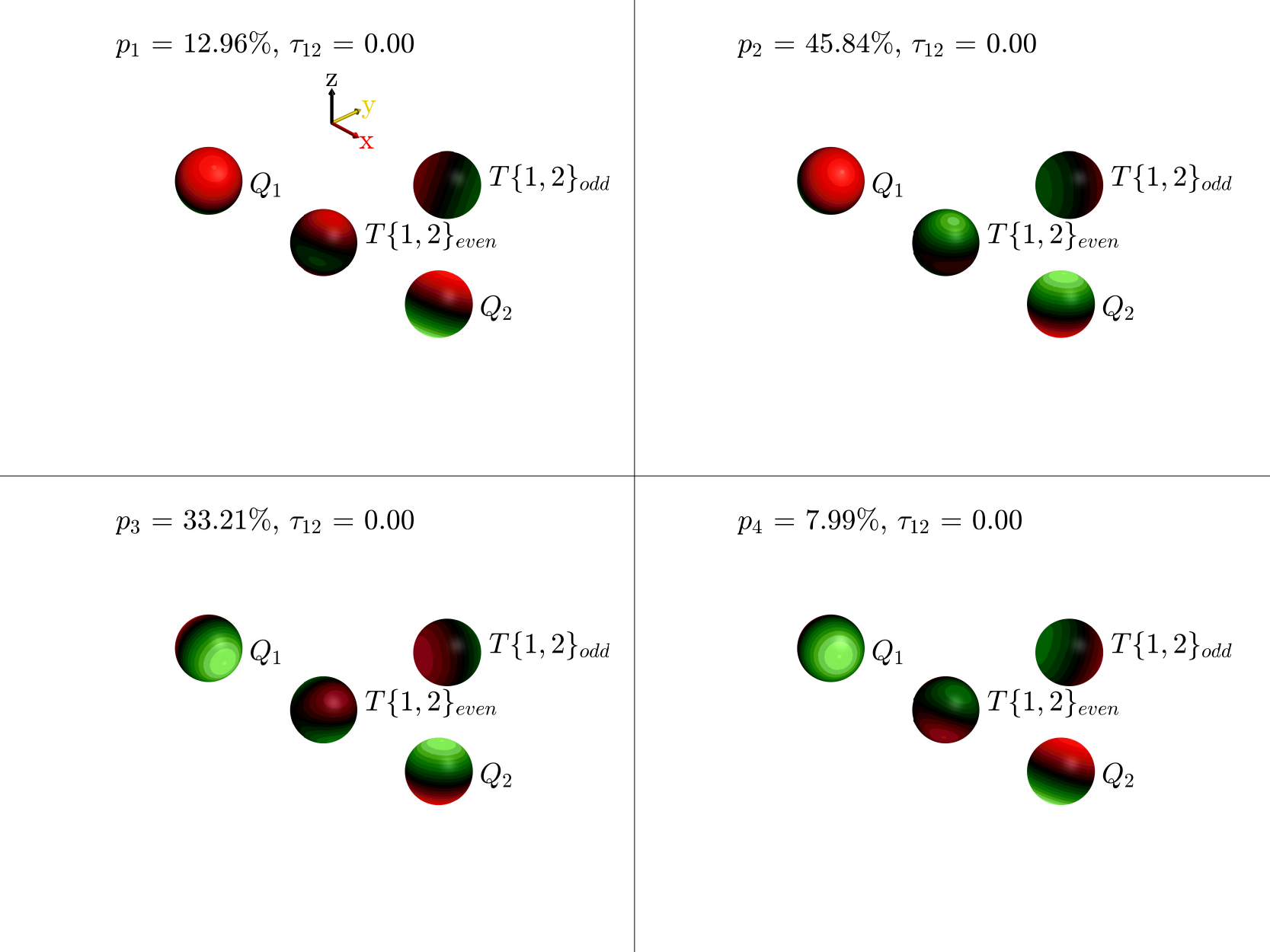}
\caption{BEADS representation of a possible pure state decomposition of the state $\rho = (0.7\ket{0}\bra{0}+0.3\ket{1}\bra{1}) \otimes (0.2\ket{0}\bra{0}+0.8\ket{1}\bra{1})$. All contributing pure states are product states which can be deduced from the $Q$-Beads maximum color brightness. Note that forming the weighted sum of these states gives $T$- and $C$-Beads that equal the product of the summed $Q$-Beads.}
\label{fig:DidProductMixedPure}
\end{figure}

As can be seen in Fig.~\ref{fig:DidProductMixed}, mixed product states (here, $\rho = (0.7\ket{0}\bra{0}+0.3\ket{1}\bra{1}) \otimes (0.2\ket{0}\bra{0}+0.8\ket{1}\bra{1})$) result in similar BEADS representations as their pure state relatives. The only types of correlation function beads occurring are $T$- and $C$-Beads, both of which are identical and only represent redundant information. An easy distinction from pure product states can be made by looking at the $Q$-Beads which generally do \textit{not have full color brightness} in the mixed state case. For completeness, a possible pure state decomposition of the state displayed in Fig.~\ref{fig:DidProductMixed} is shown in Fig.~\ref{fig:DidProductMixedPure}. It is clear from the BEADS representation that all contributing pure states are product states.


\subsection*{\textbf{Classically correlated mixed states}}
We will now become acquainted with the first type of genuinely correlated mixed states, classically correlated states. The general form of this type of states is given by $\rho_{\text{class}} = \sum_{r,s,...w} p_{rs...w} \, \ket{r}\bra{r} \otimes \ket{s}\bra{s} \otimes ... \otimes \ket{w}\bra{w}$ \cite{ding_quantum_2023} as previously specified in Eq.~\ref{eq:classically_correlated}. This again looks very similar to the definition of mixed product states in the previous section. However, the individual pure states $\ket{r}$, $\ket{s}$, and $\ket{w}$ are restricted to single orthonormal bases, respectively, and unlike in the previous case, the weights $p_{rs...w}$ are not formed as products of single-qubit state weights. This means, all vectors $\ket{r}$ must be elements of the same orthonormal basis although the bases may be different for different qubits, here e.g., $\ket{r}$ compared to $\ket{s}$. These constraints may seem insignificant at first glance, yet they greatly influence the correlation properties of such states. In general, classically correlated states $\rho_{\text{class}}$ can not be written as a simple tensor product of mixed single-qubit density operators. 

\begin{figure}[h!]
\centering
\includegraphics[width=.9\textwidth]{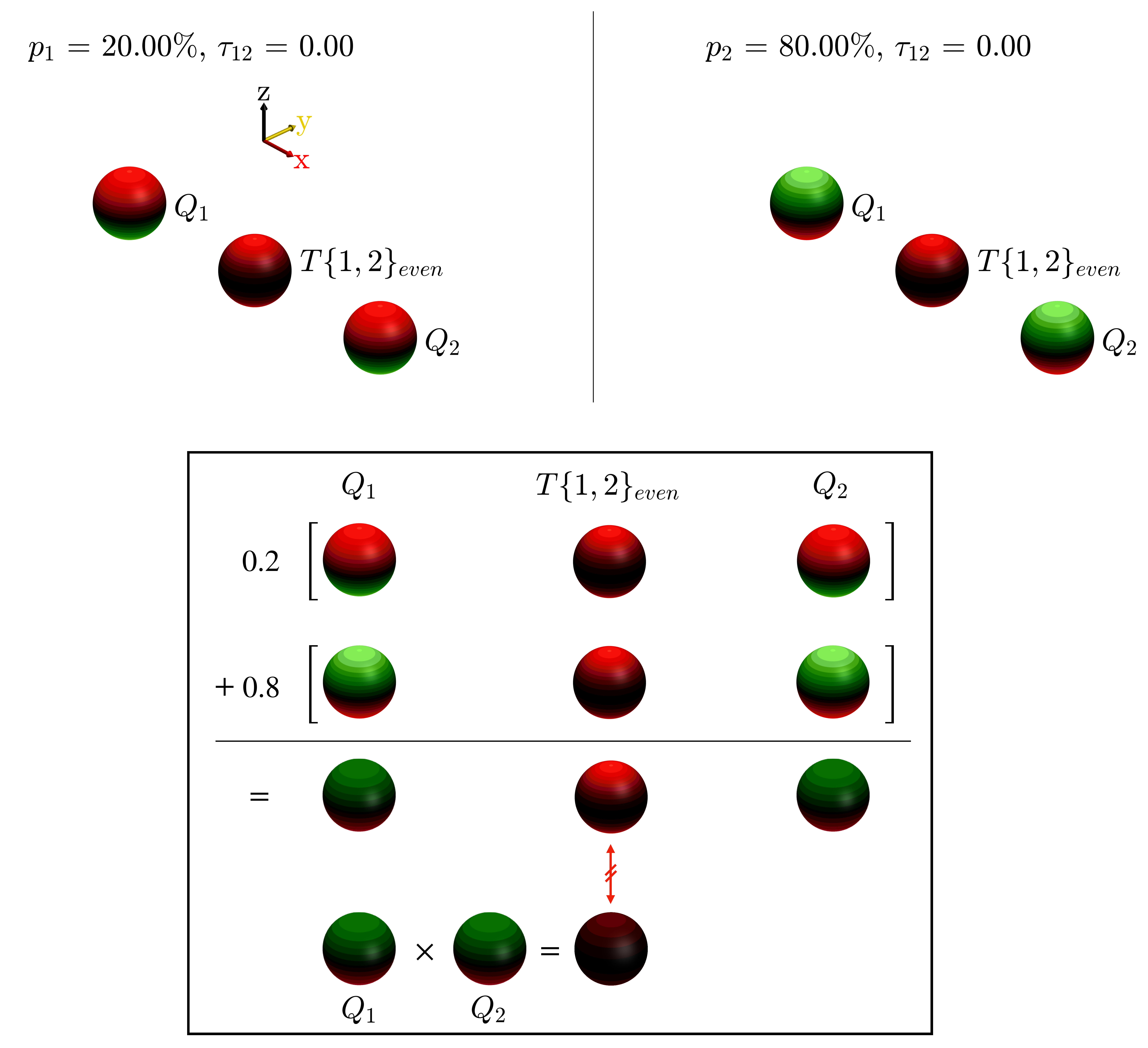}
\caption{BEADS representation of pure states constituting the state $\rho_{\text{class}} = 0.2\ket{00}\bra{00} + 0.8\ket{11}\bra{11}$. Forming the weighted sum of both the pure states partially cancels the $Q$-Beads but leaves the $T$-Bead invariant for which the information encoded on the $T$-Bead exceeds the information given by the product of $Q$-Beads as indicated by the product at the bottom.}
\label{fig:DidClassPureDecomp}
\end{figure}

Consider the state $\rho_{\text{class}} = 0.2\ket{00}\bra{00} + 0.8\ket{11}\bra{11}$. A BEADS representation of both contributing pure states is given in Fig.~\ref{fig:DidClassPureDecomp}. It is obvious that the $Q$-Beads are coaligned for the individual states, but aligned oppositely when comparing the pure states with each other. On the contrary, the $T$-Bead is identical in both cases. As shown, when combining both states including their respective weights, this leads to a partial cancellation of the information encoded on the $Q$-Beads but preserves the $T$-Bead. As a consequence, the product of the \textit{mixed state} $Q$-Beads no longer equals the $T$- or $C$-Bead which is illustrated at the bottom of Fig.~\ref{fig:DidClassPureDecomp}. Thus, the $T$- and $C$-Beads now carry information, which cannot be recovered from the $Q$-Beads and the state must be genuinely correlated. To make this more concrete, the mixed product state that shows the same $Q$-Beads as $\rho_{\text{class}}$ but $T$- and $C$-Beads identical to the product bead shown in Fig.~\ref{fig:DidClassPureDecomp} is given by $\tilde{\rho} = (0.2\ket{0}\bra{0} + 0.8\ket{1}\bra{1}) \otimes (0.2\ket{0}\bra{0} + 0.8\ket{1}\bra{1}) = 0.04 \ket{00}\bra{00} + 0.16 \ket{01}\bra{01} + 0.16 \ket{10}\bra{10} + 0.64\ket{11}\bra{11}$, which is clearly not identical to $\rho_{\text{class}}$.
\pagebreak

\begin{figure}[h!]
\centering
\includegraphics[width=\linewidth]{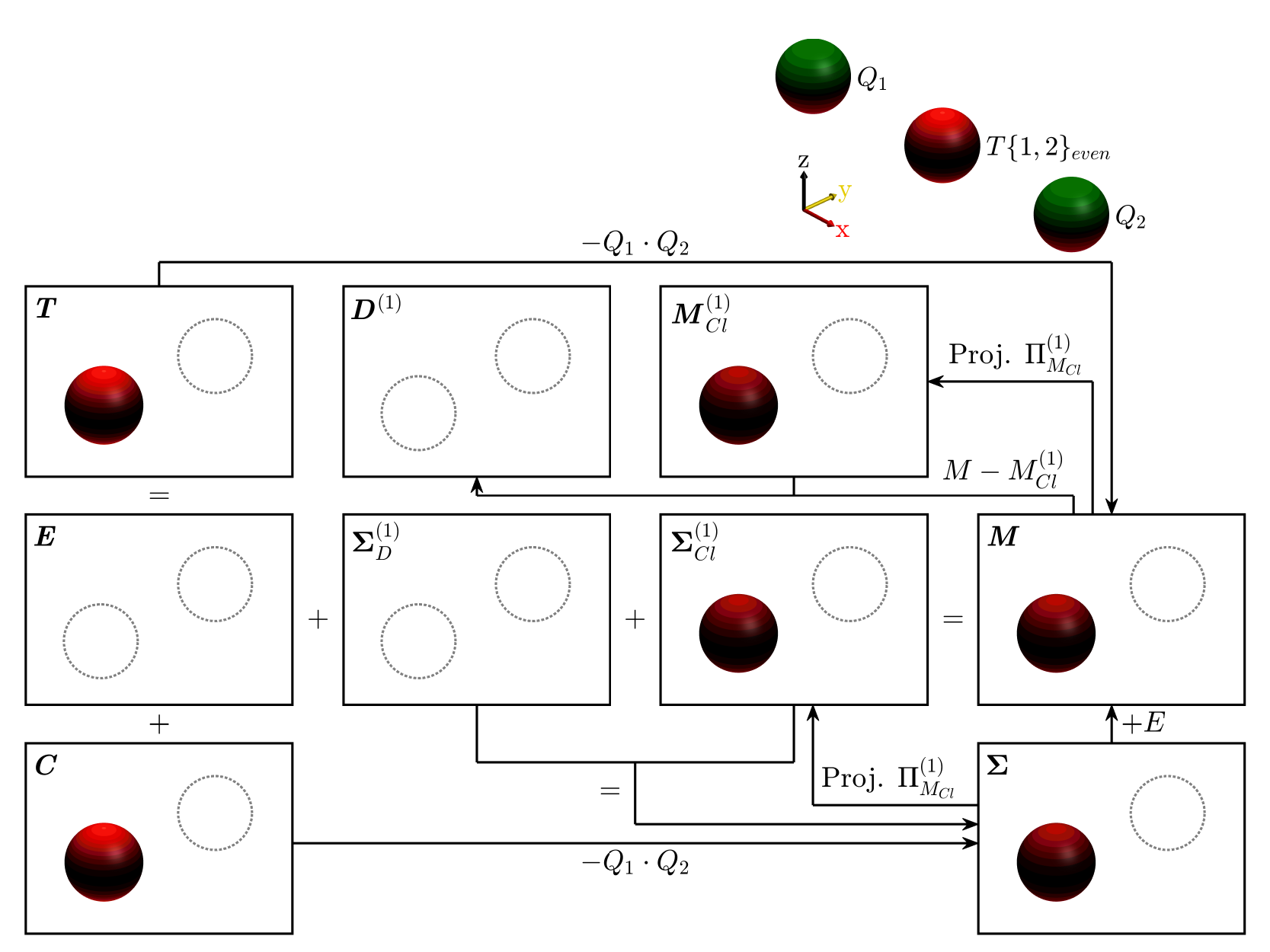}
\caption{Generalized BEADS representation of the state $\rho_{\text{class}} = 0.2\ket{00}\bra{00} + 0.8\ket{11}\bra{11}$. The $T$- and $C$-Beads are identical and carry more information than given by the product $Q_1 \times Q_2$ which can be directly inferred from the bead color brightness. The excess information is a direct consequence of genuine correlations and is visualized by the $M$- and $\Sigma$-Beads, which are identical due to the state being fully separable. Directly assessing the classical correlation of the state by an optimum measurement (here, of qubit 1) gives the displayed $M_{Cl}^{(1)}$- and $\Sigma_{Cl}^{(1)}$-Beads which are again identical to the $M$- and $\Sigma$-Beads. All other bead types are zero. Thus, the state is classically correlated when accessed from qubit 1.}
\label{fig:DidClassClass1}
\end{figure}

\begin{figure}[h!]
\centering
\includegraphics[width=\linewidth]{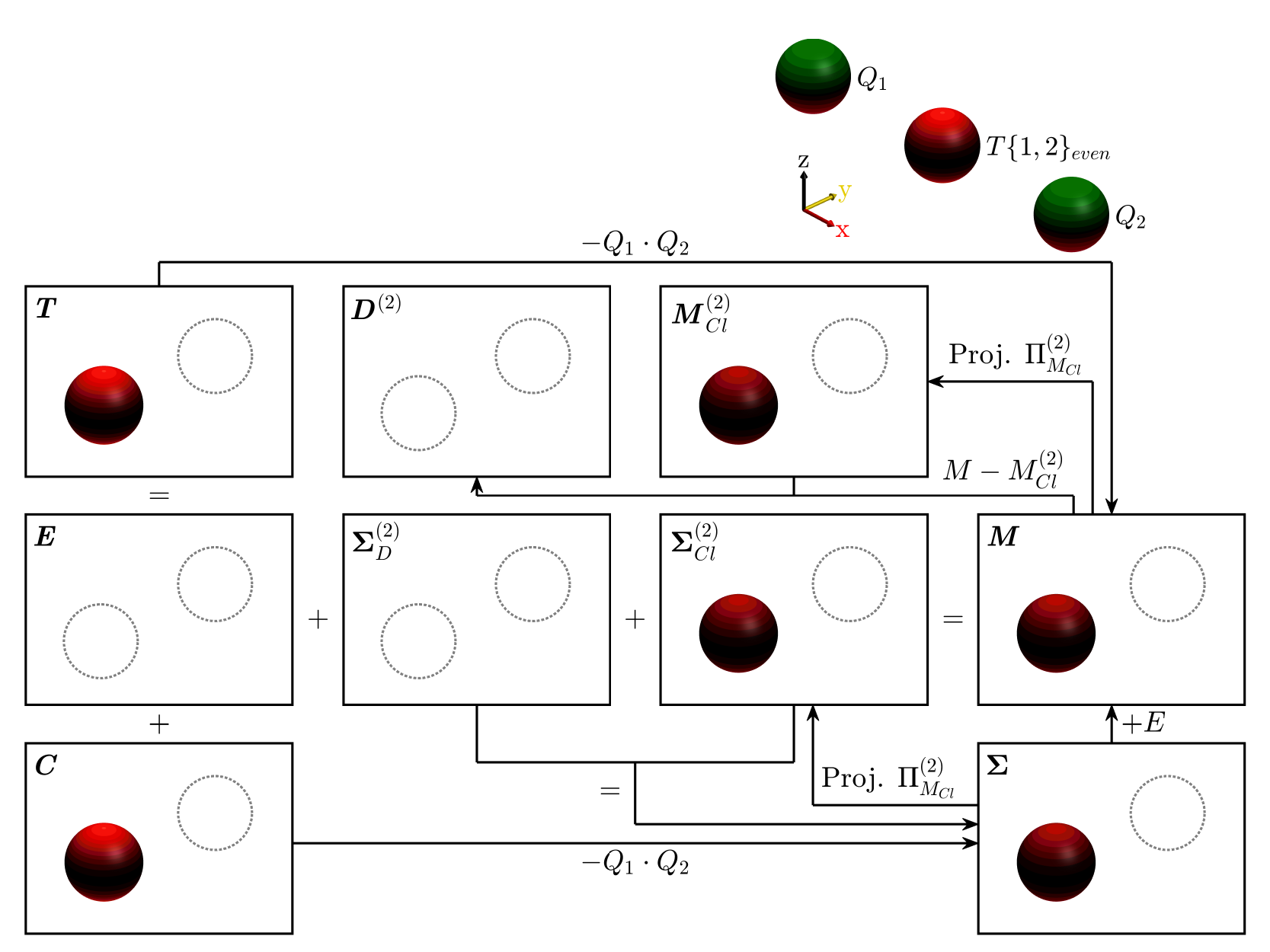}
\caption{Generalized BEADS representation of the state $\rho_{\text{class}} = 0.2\ket{00}\bra{00} + 0.8\ket{11}\bra{11}$. In this case, the classical correlation is assessed by measuring qubit 2. The resulting BEADS representation is identical to Fig.~\ref{fig:DidClassClass1}. Hence, $\rho$ is a classical-classical state.}
\label{fig:DidClassClass2}
\end{figure}

In the generalized BEADS representation of classically correlated states, $T$- and $C$-Beads are again \textit{identical} and $E$-Beads vanish due to these states also being fully separable. However, as illustrated in Figs.~\ref{fig:DidClassClass1} and~\ref{fig:DidClassClass2}, the $T$- and $C$-Beads now encode additional information that cannot be recovered from the $Q$-Beads. In the two-qubit case, it is possible to extract this excess information by simply subtracting the product of $Q$-Beads (of the mixed state) as indicated in the following figures. This gives so-called $M$- or $\Sigma$-Beads when subtracting from $T$- or $C$-Beads, respectively. $M$-Beads visualize the total \textit{m}utual information, i.e., the total correlation of a mixed state. On the other hand, $\Sigma$-Beads only visualize correlations of separable parts of a quantum state. Classically correlated states being fully separable, consequently have identical $M$- and $\Sigma$-Beads due to the $T$- and $C$-Beads being identical as well.

In Figs.~\ref{fig:DidClassClass1} and~\ref{fig:DidClassClass2}, two additional types of beads occur: $M_{Cl}$- and $\Sigma_{Cl}$-Beads. These beads directly assess the \underline{cl}assical correlation, also called \textit{classical mutual information}, of a state and are obtained as projections of the $M$- and $\Sigma$-, respectively. For the shown two-qubit case this projection follows directly from the definition of classical correlation (cf. Eq.~\ref{eq:classical_mutual}) and involves an optimized single-qubit measurement of \textit{one} of the two qubits. In particular, this measurement is chosen such, that we gain the maximum amount of information on the \textit{entire} state. In simple terms, this measurement allows to get a mixed state after the measurement, i.e., the sum of all outcomes weighted by their probabilities, that is as similar to the state before the measurement as possible. As becomes clear in the subsequent figures, classically correlated states represent the limiting case in which this optimum measurement fully recovers the entire state, for which the $M_{Cl}$- and $\Sigma_{Cl}$-Beads result in being identical to the $M$- and $\Sigma$-Beads. This gives a set of criteria to unambiguously identify classically correlated states based on their generalized BEADS representation: A state is classically correlated (with respect to measurement of a specific qubit) if its $T$- and $C$-Beads are identical, if the $M$-, $\Sigma$-, $M_{Cl}$- and $\Sigma_{Cl}$-Beads are nonzero and identical, and if all other bead types vanish.

Note that in Figs.~\ref{fig:DidClassClass1} and~\ref{fig:DidClassClass2} $M_{Cl}$- and $\Sigma_{Cl}$-Beads (and the projectors $\Pi_{M_{Cl}}$) have superscript indices, which denote the measured qubit. In fact, as only one of the qubits is measured (in the two-qubit case), classical correlation is an asymmetric property and may vary in dependence of which qubit is measured. For the displayed examples it is clear from comparing the $M_{Cl}$- and $\Sigma_{Cl}$-Beads in both figures, that the classical correlation must be the same in both cases as the resulting beads are identical. It is conventional to use the more refined nomenclature \textit{classical-classical} state to indicate that this state is strictly classically correlated when accessed from any of the two qubits. As we will see in the following subsection, this is not always the case, and a state may well behave classically when measuring one qubit but nonclassically when measuring another.

\pagebreak

\subsection*{\textbf{Nonclassically correlated separable mixed states}}
Whereas entanglement is the only source of nonclassical correlation in pure quantum states, mixed states can be fully separable (i.e., there exists a decomposition of the state into pure product states) but still nonclassically correlated. Indeed, any separable mixed state for which one of the qubits cannot be expressed in a single orthonormal basis, as would be required for classically correlated states (see previous subsection), is nonclassically correlated and has nonzero quantum discord \cite{ollivier_quantum_2001} when measured with respect to this particular qubit. 
Mathematically such states can be expressed as a convex sum of \textit{mixed} product states (cf. Eq.~\ref{eq:separable_state}) or equivalently into pure product states (see Eq.~\ref{eq:classically_correlated}) with the relaxed constraint that each single-qubit pure state can be chosen arbitrarily from any basis.

As we have seen for classically correlated states, classical correlation (classical mutual information) is an asymmetric property and so is nonclassical (quantum) correlation, which is commonly measured in terms of quantum discord \cite{ollivier_quantum_2001}. As quantum discord quantifies the difference between the total mutual information of a state and its classical mutual information, it equally depends on finding an optimum measurement to determine the latter. This implies that mixed quantum states can be classically correlated when measuring one qubit, i.e., the measurement does not cause any disturbance and fully restores the pre-measurement state, but behave nonclassically when measuring another qubit. Such states are called classical-quantum or vice versa in the two-qubit case.

\subsubsection*{Example A: Classical-quantum states with fully classically accessible correlation function components}
\begin{figure}[h!]
\centering
\includegraphics[width=\linewidth]{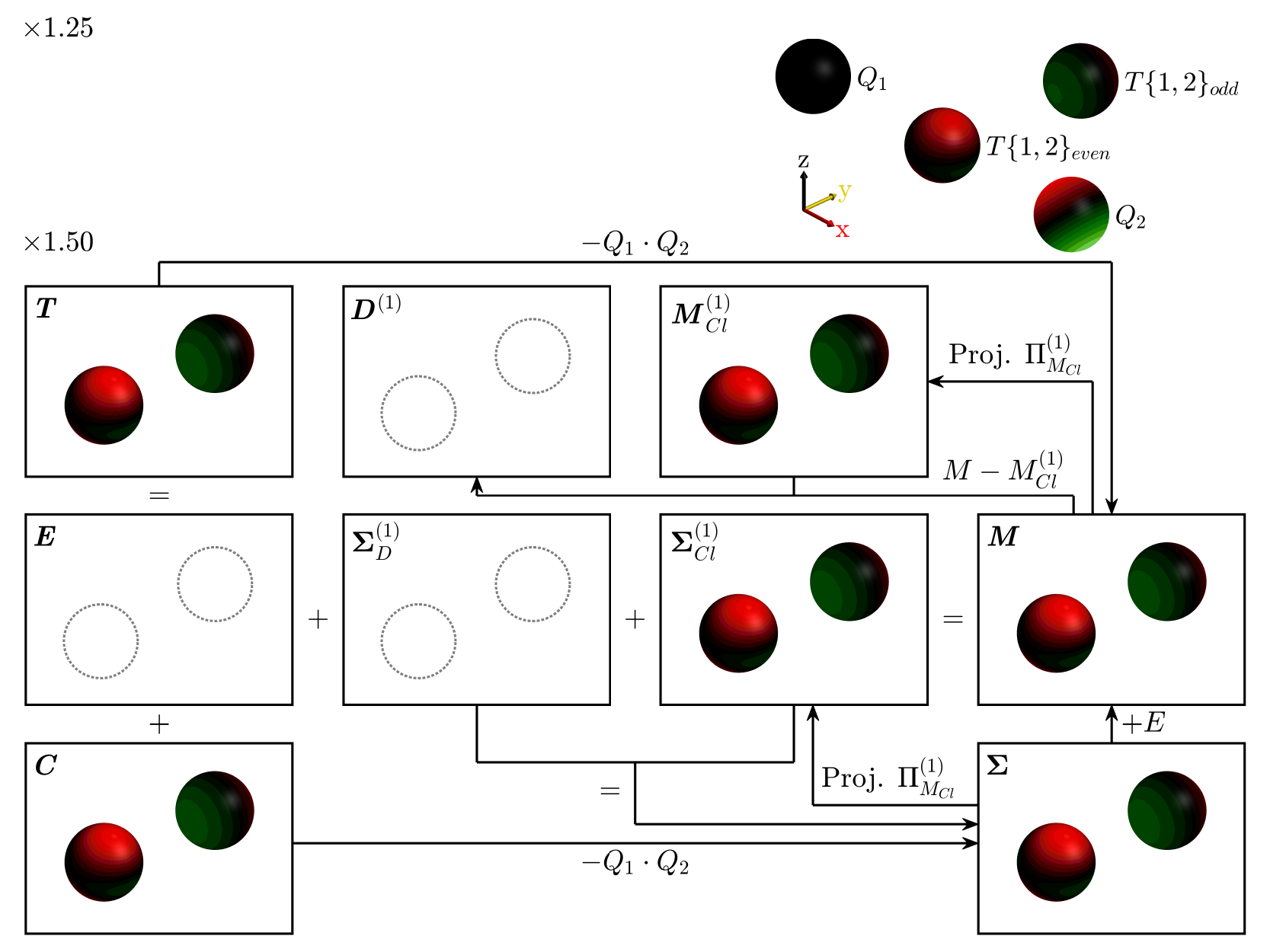}
\caption{Generalized BEADS representation of the state $\rho_{\text{class-quant}} = 0.5(\ket{00}\bra{00} + \ket{1-}\bra{1-})$. The $T$- and $C$-Beads as well as the $M$-, $\Sigma$-, $M_{Cl}^{(1)}$-, and $\Sigma_{Cl}^{(1)}$-Beads are identical, respectively. No other correlation function bead types are visible. Thus, the state is classically correlated when accessed from qubit 1.}
\label{fig:DidClassQuant1_1}
\end{figure}

To understand how this nonclassicality becomes visible in the generalized BEADS representation, we start from a classical-classical state similar to the one shown in the previous subsection and modify it to be quantum-correlated with respect to qubit 2. This is achieved by choosing different weights and exchanging one of the computational basis states, here $\ket{11}$, with a Hadamard-basis (x-basis) state: ${\rho_{\text{class-quant}} = 0.5(\ket{00}\bra{00} + \ket{1-}\bra{1-})}$. It is clear that qubit 1 is expressed only in terms of computational basis states, for which the overall mixed state is expected to be classically correlated with respect to being accessed from qubit 1. In fact, this can be directly seen in the corresponding generalized BEADS representation shown in Fig.~\ref{fig:DidClassQuant1_1}: The $T$- and $C$-Beads are identical for which the state must be fully separable. In addition, the $M$-, $M_{Cl}$-, $\Sigma$-, and $\Sigma_{Cl}$-Beads are also identical. No other types of beads are visible. Hence, we can identify this state to be classically correlated when accessed from qubit~1. 

\begin{figure}[h!]
\centering
\includegraphics[width=\linewidth]{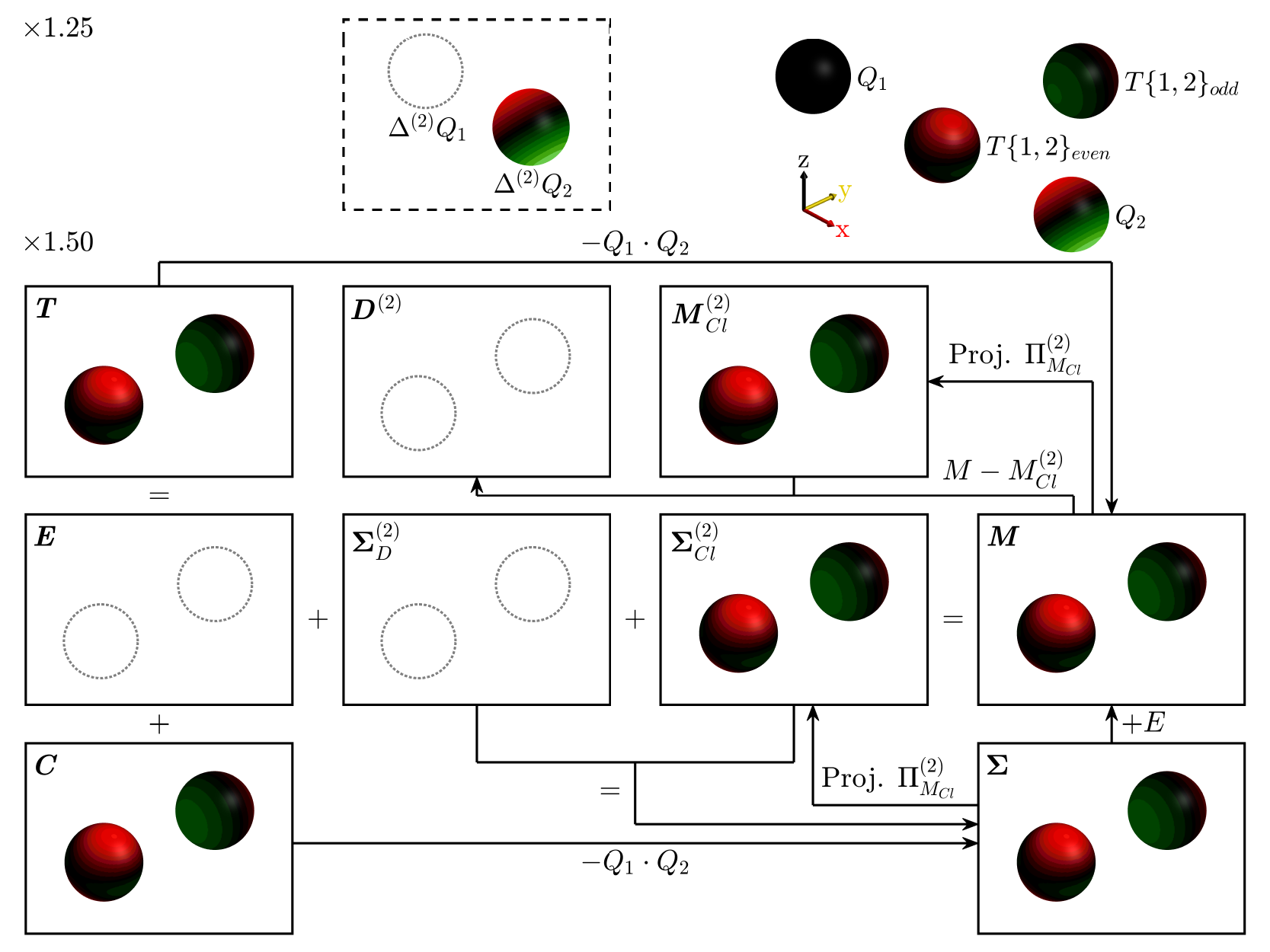}
\caption{Generalized BEADS representation of the state $\rho_{\text{class-quant}} = 0.5(\ket{00}\bra{00} + \ket{1-}\bra{1-})$. The $T$- and $C$-Beads as well as the $M$-, $\Sigma$-, $M_{Cl}^{(2)}$-, and $\Sigma_{Cl}^{(2)}$-Beads are again identical, respectively. No other correlation function bead types are visible. Unlike in Fig.~\ref{fig:DidClassQuant1_1}, the optimum measurement maximizing the classical mutual information induces a loss of the entire information encoded on Q-Bead $Q_2$, which is visualized by $\Delta^{(2)} Q_2$. Hence, the state is nonclassically correlated when accessed from qubit 2.}
\label{fig:DidClassQuant1_2}
\end{figure}

The picture subtly changes if we access the state from qubit 2. As can be seen in Fig.~\ref{fig:DidClassQuant1_2}, while all types of correlation function beads remain unchanged when compared to Fig.~\ref{fig:DidClassQuant1_1}, which would suggest that the state was classically correlated also in this scenario, we find additional so-called difference-$Q$-Beads $\Delta^{(k)}Q$ occuring at the top of Fig.~\ref{fig:DidClassQuant1_2}. These beads directly visualize the $Q$-Bead information that is lost when the optimum measurement that maximizes the classical mutual information is performed. Recall that if the state was truly classically correlated with respect to qubit 2, this measurement would restore the entire state. This example may appear counterintuitive at first but reveals an important property of nonclassically correlated separable states: They are not necessarily associated with higher degrees of correlation that manifest in additional correlation function components. Difference-$Q$-Beads thus serve as a first direct visual indicator in the generalized BEADS representation for a quantum state being nonclassically correlated.

\pagebreak

\subsubsection*{Example B: Classical-quantum states with classically inaccessible correlation function components}
\begin{figure}[h!]
\centering
\includegraphics[width=\linewidth]{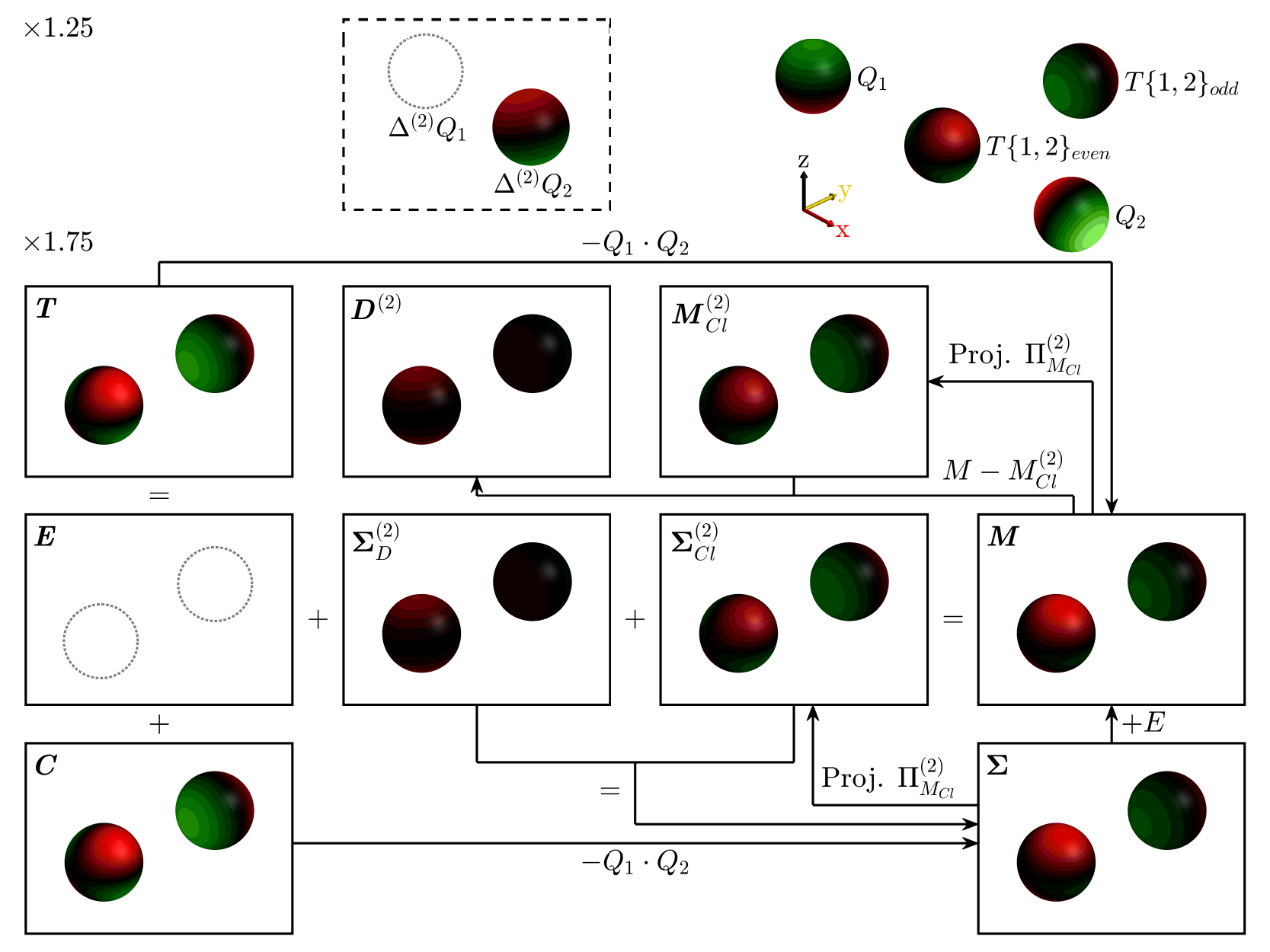}
\caption{Generalized BEADS representation of the state $\rho_{\text{class-quant}}' = 0.25\ket{00}\bra{00} + 0.75\ket{1-}\bra{1-}$. The $T$- and $C$-Beads are identical, whereas here the $M$-, $\Sigma$-, $M_{Cl}^{(2)}$-, and $\Sigma_{Cl}^{(2)}$ do not coincide. Additional $D^{(2)}$ and $\Sigma_D^{(2)}$-Beads are visible, which provide direct visualizations of the classically inaccessible correlation function components of the state. Similar to Fig.~\ref{fig:DidClassQuant1_2}, the optimum measurement maximizing the classical mutual information induces a loss of information encoded on Q-Bead $Q_2$ for this state, which is visualized by $\Delta^{(2)} Q_2$. Hence, this state is also nonclassically correlated when accessed from qubit 2.}
\label{fig:DidClassQuant2_2}
\end{figure}

It is still possible, and indeed much more common, to find a separable mixed state being quantum correlated and showing correlation function contributions that cannot be accessed classically. An example is given in Fig.~\ref{fig:DidClassQuant2_2}, which shows the generalized BEADS representation of the state $\rho_{\text{class-quant}}' = 0.25\ket{00}\bra{00} + 0.75\ket{1-}\bra{1-}$. This is thus constructed using the same contributing pure states but different weight as in the previous example. Again, we obtain a nonzero difference-bead $\Delta^{(2)}Q_2$, which is however not identical to $Q_2$ indicating that only a part of the information is lost when optimally measuring qubit 2. In addition, this state also shows additional correlation function beads $D^{(2)}$ and $\Sigma_D^{(2)}$, which visualize the correlation function component that is classically (by means of an optimized single-qubit measurement) inaccessible. In other words, these beads visualize the parts of the correlation function of a state which exist in addition to what would be achievable only by classical correlation. These beads can generally occur if a state has nonzero quantum discord. In analogy to the $M_{Cl}$- and $\Sigma_{Cl}$-Beads, $D$-Beads visualize any genuine correlation-based components of the state that cannot be related to its classical mutual information, i.e., the part of the $M$-Beads which are not covered by the $M_{Cl}$-Beads, whereas $\Sigma_D$-Beads only represent the separable parts thereof. As the examined state is again fully separable, the $D^{(2)}$ and $\Sigma_D^{(2)}$ coincide in this case.

\pagebreak

\subsubsection*{Example C: Quantum-quantum state}
We can now test the discussed criteria for identifying quantum correlated separable mixed states on the quantum-quantum state $\rho_{\text{quant-quant}} = 0.15 \ket{00}\bra{00} + 0.05 \ket{11}\bra{11} + 0.15\ket{-0}\bra{-0} + 0.25\ket{+1}\bra{+1} + 0.1\ket{0+}\bra{0+} + 0.3\ket{1-}\bra{1-}$. Clearly, this state is constructed using computational and x-basis states for both qubits, for which we can directly infer that this must be a separable quantum-quantum state. 

\begin{figure}[h!]
\centering
\vspace{-1.8cm}
\includegraphics[width=.95\linewidth]{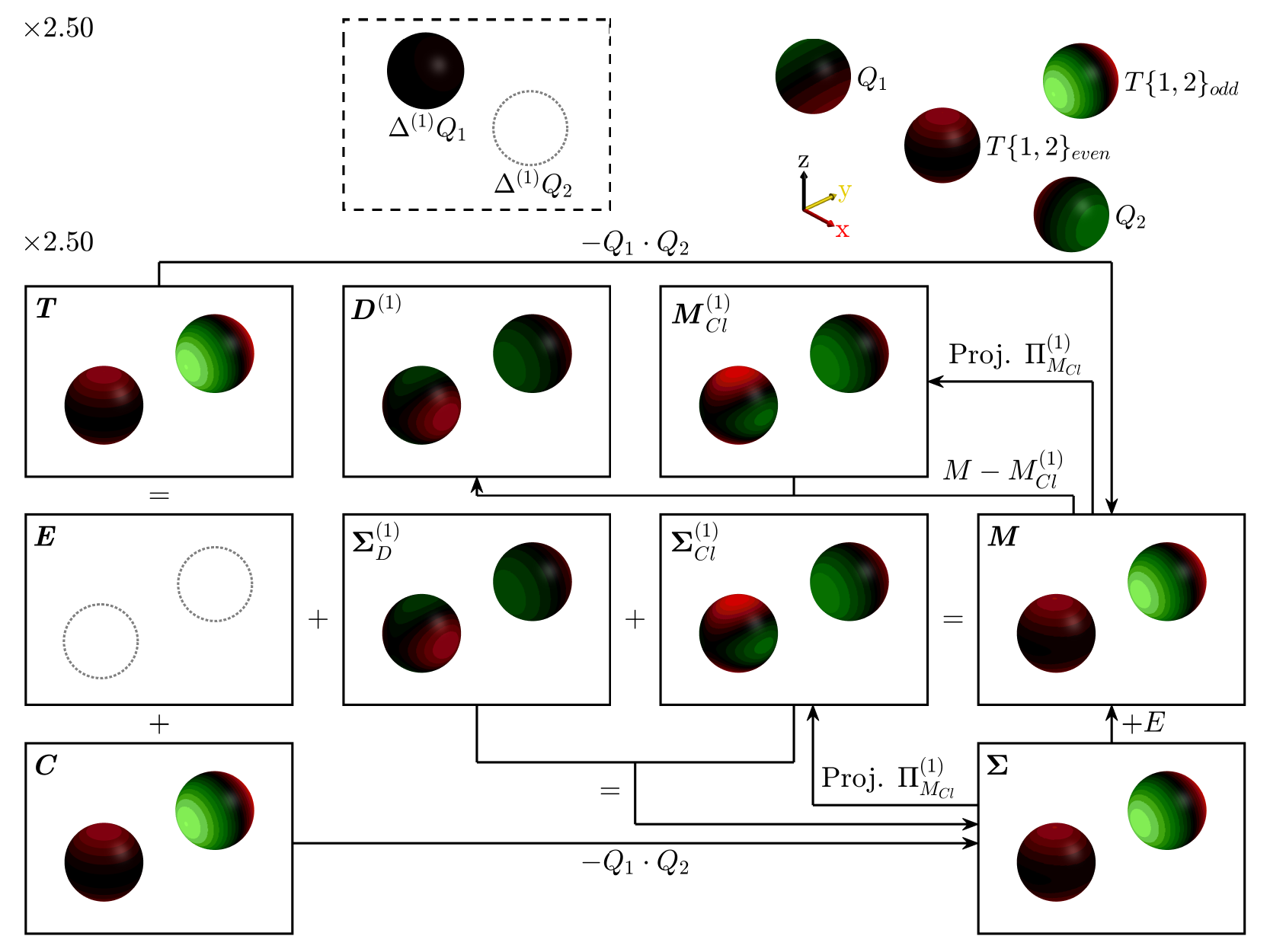}
\caption{Generalized BEADS representation of the state $\rho_{\text{quant-quant}} = 0.15 \ket{00}\bra{00} + 0.05 \ket{11}\bra{11} + 0.15\ket{-0}\bra{-0} + 0.25\ket{+1}\bra{+1} + 0.1\ket{0+}\bra{0+} + 0.3\ket{1-}\bra{1-}$. Here, the classical mutual information is evaluated with respect to a measurement of qubit 1. Following the discussed criteria, the state is nonclassically correlated under the chosen conditions.}
\label{fig:DidQuantQuant1}
\vspace{-.4cm}
\end{figure}

\begin{figure}[h!]
\centering
\includegraphics[width=.95\linewidth]{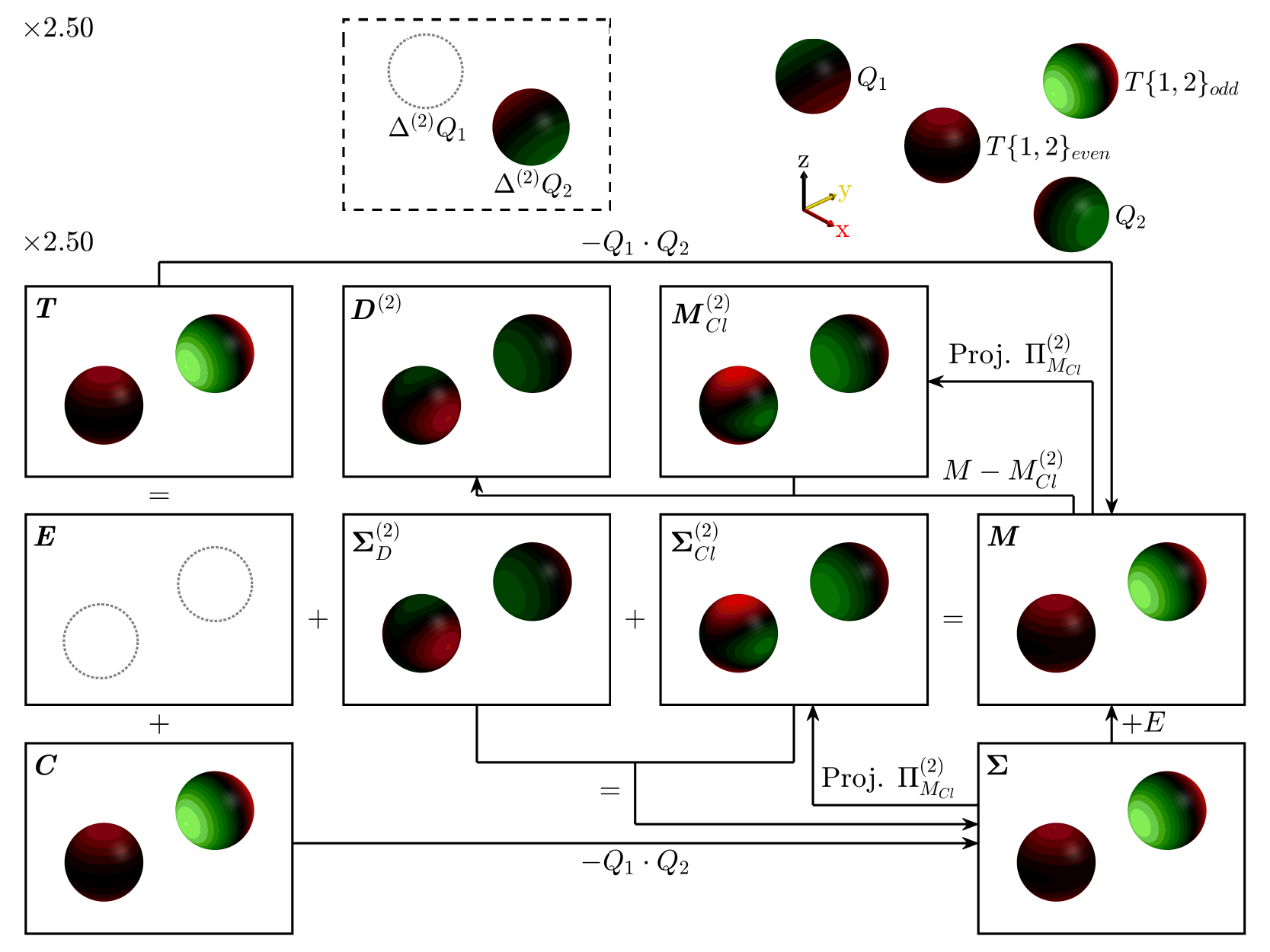}
\caption{Generalized BEADS representation of the state $\rho_{\text{quant-quant}} = 0.15 \ket{00}\bra{00} + 0.05 \ket{11}\bra{11} + 0.15\ket{-0}\bra{-0} + 0.25\ket{+1}\bra{+1} + 0.1\ket{0+}\bra{0+} + 0.3\ket{1-}\bra{1-}$. The classical correlation is evaluated with respect to a measurement of qubit 2. The state can be identified to be nonclassically correlated based on the occurring $D$-, $\Sigma_D$, and $\Delta Q$-beads. However, the correlation properties behave asymmetrically which becomes evident as the $\Delta^{(2)} Q$-, $M^{(2)}_{Cl}$-, $\Sigma^{(2)}_{Cl}$-, $D^{(2)}$-, and $\Sigma^{(2)}_{D}$-Beads differ from the corresponding beads in Fig.~\ref{fig:DidQuantQuant1}.}
\vspace{-2cm}
\label{fig:DidQuantQuant2}
\end{figure}

The generalized BEADS representations of the state accessed from qubits 1 and 2 are shown in Figs.~\ref{fig:DidQuantQuant1} and~\ref{fig:DidQuantQuant2}, respectively. In both cases, the $T$- and $C$-Beads coincide and there is no $E$-Bead for which the state must be fully separable. Identical $M$- and $\Sigma$-Beads can be observed, which allows to directly identify this state to be genuinely correlated. The $M^{(k)}_{Cl}$- and $\Sigma^{(k)}_{Cl}$-Beads coincide (for each specific choice of the measured qubit $k$) but differ from the $M$- and $\Sigma$-Beads. Hence, the state is not classically correlated. This can be further confirmed by the occurrence of $\Delta^{(k)} Q$-, $D^{(k)}$-, and $\Sigma^{(k)}_{D}$-Beads, where the latter two coincide in each case, respectively. Note that the  $\Delta^{(k)} Q$-, $M^{(k)}_{Cl}$-, $\Sigma^{(k)}_{Cl}$-, $D^{(k)}$-, and $\Sigma^{(k)}_{D}$-Beads differ slightly in Figs.~\ref{fig:DidQuantQuant1} and~\ref{fig:DidQuantQuant2}. This suggests that although the visualized state is nonclassically correlated with respect to both qubits, the correlation is asymmetric and again depends on the choice of the measured qubit.

\newpage
\subsection*{\textbf{Entangled mixed states}}
We finally arrive at the last general class of mixed quantum states: entangled mixed states. These are the states which cannot be written as a convex combination of pure product states, that is, their pure state decomposition does contain at least one entangled pure state. It is important to understand that -- while any state that can be expressed only by pure product states is fully separable -- being able to write a mixed state as a convex combination of entangled pure states does not guarantee the state to be entangled. As a canonical example, consider the state $\rho = 1/2( \ket{\Phi^+}\bra{\Phi^+} + \ket{\Phi^-}\bra{\Phi^-})$ with $\ket{\Phi^\pm} = 1/\sqrt{2}(\ket{00}\pm\ket{11})$ which can be rewritten as $\rho = 1/2(\ket{00}\bra{00}+\ket{11}\bra{11})$ and is thus a fully separable classical-classical state. As a consequence, the possible pure state decompositions of a given mixed state thus need to be explored, commonly by optimization techniques that find the least-entangled decomposition into pure states, to characterize its entanglement (and separability) properties \cite{hughston_complete_1993}.

In the generalized BEADS representation, entangled states are the only class of states that show E-Beads (entanglement-based correlation function) alongside all other previously discussed bead types and hence, can be easily identified. This is a natural consequence of the pure states in the decomposition of entangled mixed states being entangled and thus having E-Beads as well. The mixed state E-Bead simply results as the weighted (convex) sum of all pure state E-Beads.

\begin{figure}[h!]
\centering
\vspace{-1.5cm}
\includegraphics[width=.95\linewidth]{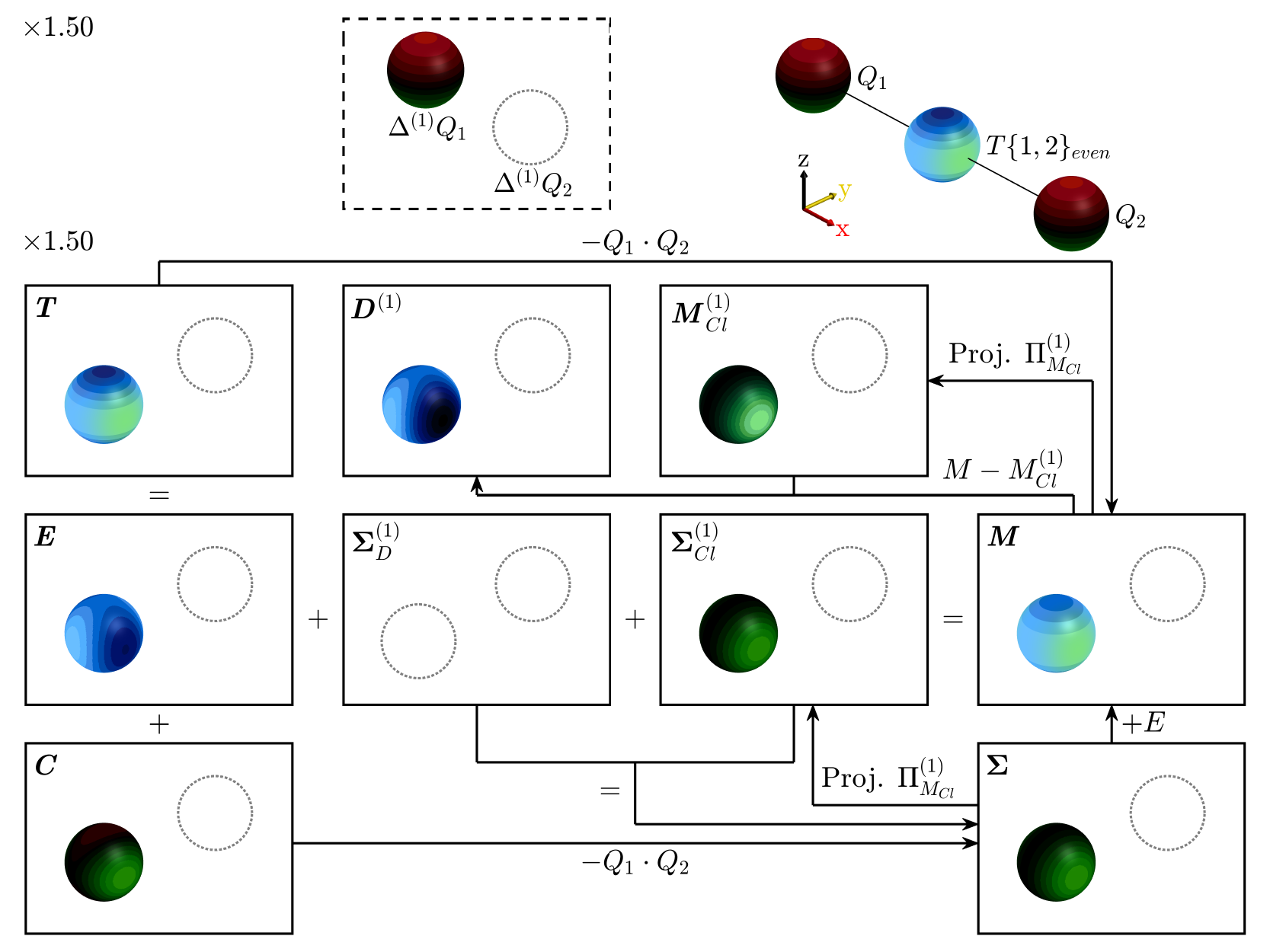}
\caption{Generalized BEADS representation of the state $\rho_{\text{ent}} = 0.65 \ket{\Psi^-}\bra{\Psi^-} + 0.35 \ket{00}\bra{00}$. The state has a prominent $E$-Bead and is thus clearly entangled. Any single-qubit measurement in the $xy$-plane maximizes the classical mutual information of this state. Here, the $M_{Cl}$-Beads with the largest separable component is visualized for which the $\Sigma^{(1)}_{D}$-Bead vanishes.}
\label{fig:DidEnt1}
\end{figure}

\begin{figure}[h!]
\centering
\includegraphics[width=.95\linewidth]{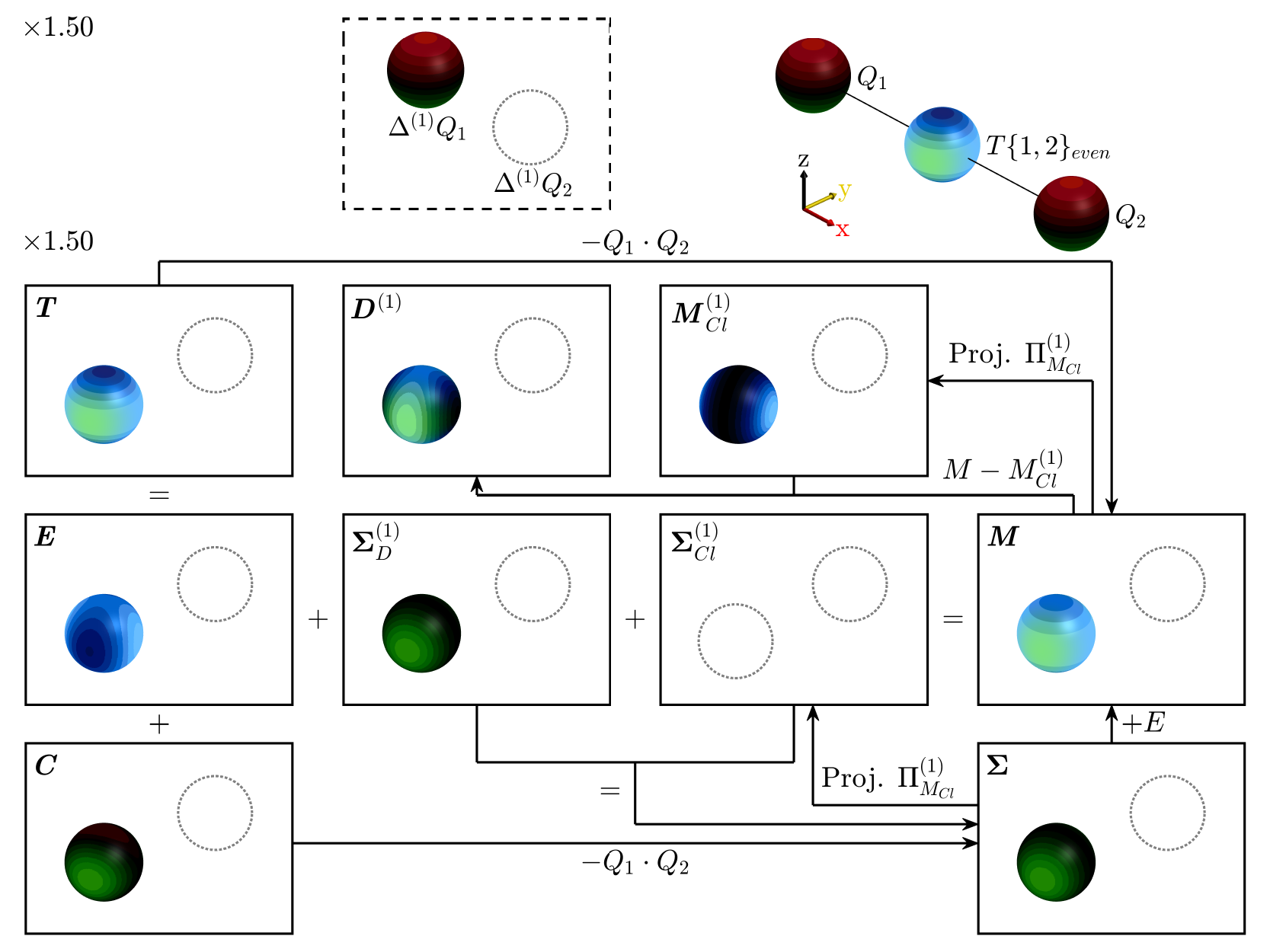}
\caption{Generalized BEADS representation of the state $\rho_{\text{ent}} = 0.65 \ket{\Psi^-}\bra{\Psi^-} + 0.35 \ket{00}\bra{00}$. Here, the classical mutual information is again evaluated with respect to qubit 1. The projection corresponds to the $M_{Cl}$-Beads with the smallest separable component for which the $\Sigma^{(1)}_{D}$-Bead becomes maximum.}
\vspace{-2cm}
\label{fig:DidEnt2}
\end{figure}

As a simple example, we consider $\rho_{\text{ent}} = 0.65 \ket{\Psi^-}\bra{\Psi^-} + 0.35 \ket{00}\bra{00}$, a mixture of the singlet state $\ket{\Psi^-}=1/\sqrt{2}(\ket{01}-\ket{10})$ and the product state $\ket{00}$. In its generalized BEADS representation (see Figs.~\ref{fig:DidEnt1} and~\ref{fig:DidEnt2}), we can directly see that the predominant color of the $T$-Bead is blue originating from the E-Bead contribution. Recall that the maximally entangled singlet state $ \ket{\Psi^-}$ has an easily recognizable BEADS representation consisting of black $Q$-Beads (maximally mixed reduced density operators) and $T$- and $E\{1,2\}_{even}$-Beads that corresponds to an isotropic sphere with spherical function value -1 (bright blue color) in all directions \cite{huber_beads_2025}. The BEADS representation of $\ket{00}$ can be seen in Fig.~\ref{fig:DidClassPureDecomp}. 

We can now analyze the generalized BEADS visualization of $\rho_{\text{ent}}$ in close detail. Considering the spherical function values (color brightness), the BEADS representation shows an axially symmetric surface pattern with respect to the $z$-axis. However, the hue is non-uniform in the $xy$-plane: A bright green region occurs which indicates some separable contribution to the correlation function along this specific direction. It also turns out that this green region adopts different orientations in Figs.~\ref{fig:DidEnt1} and~\ref{fig:DidEnt2}. Indeed, symmetries as in this case allow for an additional degree of freedom where the optimum entanglement-minimizing pure state decomposition is fixed up to a global rotation around the symmetry axis (here, the $z$-axis).

\begin{figure}[h!]
\centering
\vspace{-1.5cm}
\includegraphics[width=1.\linewidth]{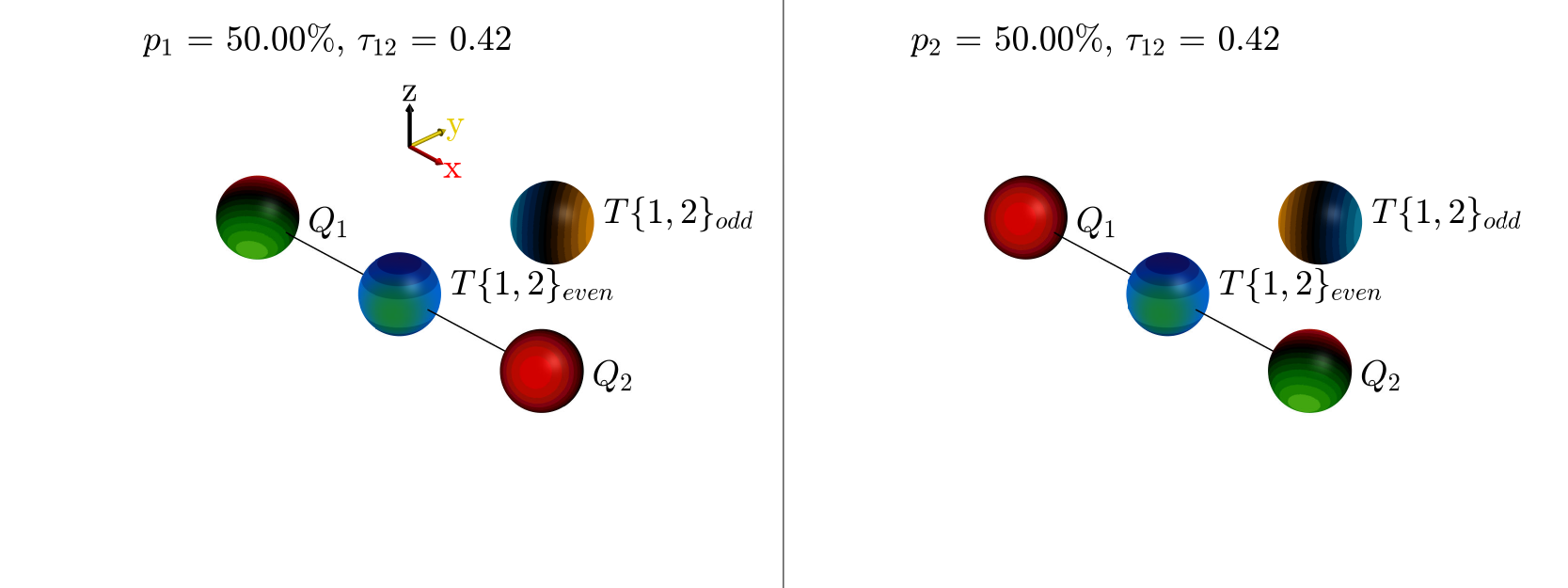}
\caption{A possible minimally entangled pure state decomposition of the mixed state $\rho_{\text{ent}} = 0.65 \ket{\Psi^-}\bra{\Psi^-} + 0.35 \ket{00}\bra{00}$. Both pure states have the same two-tangle $\tau_{12}$. The antisymmetric $T\{1,2\}_{odd}$ beads are oriented oppositely for which they cancel each other when both pure states are combined to give the mixed state representation. On the other hand, the $T\{1,2\}_{even}$ beads are identical in both cases and add up. Note that performing the same global rotation of both pure states by an arbitrary angle around the $z$-axis, gives another valid minimum-entanglement decomposition of $\rho_{\text{ent}}$.}
\label{fig:DidEntPureDecomp}
\end{figure}

The pure state decomposition underlying the mixed state visualization of $\rho_{\text{ent}}$ shown in Fig.~\ref{fig:DidEnt1} is revealed in Fig.~\ref{fig:DidEntPureDecomp}. Note that these pure states strongly deviate from the states used to formulate $\rho_{\text{ent}}$ originally, which emphasizes that finding least-entangled decompositions of a mixed state is a highly non-trivial and non-intuitive task. One aspect that becomes clear though, is that both of the displayed pure states have the same two-tangle $\tau_{12}$, which is always the case for optimum decompositions of \textit{two-qubit} mixed states \cite{wootters_entanglement_1998}.

In the generalized BEADS representation of the overall entangled mixed states we find all possible types of beads. As for nonclassically correlated separable states we can also determine the classical mutual information of the entangled state. A common misconception is that entanglement and classical correlation are distinct incompatible types of correlation. However, as becomes clear in the following figures this is by no means the case. Finding the optimum measurement to maximize the classical mutual information does not query which property of the state, be it entanglement-based or separable characteristics, underlie the classically recovered information. As a consequence, the classical mutual information may be associated with a projection that accesses entanglement-based correlation function components. 

In the generalized BEADS representation we conventionally visualize the $M_{Cl}^{(k)}$-Bead that covers the largest separable (and least entangled) component when symmetry-based degrees of freedoms occur such as for $\rho_{\text{ent}}$. In Fig.~\ref{fig:DidEnt1} the $M_{Cl}^{(k)}$-Beads still contain small parts of the $E$-Bead. This becomes best evident by comparing the $D^{(k)}$-Beads with the $E$-Beads, respectively, which turn out to be not identical. Note that, as shown in Fig.~\ref{fig:DidEnt2}, any arbitrary projection in the $xy$-plane would also maximize the classical-mutual-information for the given state but shift the composition of the $M_{Cl}^{(k)}$-Bead to contain a larger part of the $E$-Bead but a smaller separable contribution, thus reducing the $\Sigma^{(k)}_{Cl}$-Bead component. The opposite effect is observed for $D^{(k)}$- and $\Sigma^{(k)}_{D}$-Beads.


\pagebreak

All discussed types of quantum states are summarized in Table~\ref{tab:state_types_characteristics} of the main text which provides recipes for quickly and unambiguously identifying the individual classes of states discussed in this section.

\clearpage

\section{Deriving entanglement thresholds based on the generalized BEADS representation}\label{sec:SuppEntThreshold}

The generalized BEADS representation is not only useful for visualizing and understanding the correlation structure of a given arbitrary quantum state. Indeed we can also apply it for deriving entanglement thresholds of parameterized mixed quantum states. In the following, we show how the minimum entanglement-admitting weight of a singlet-centered bipartite Werner state can be derived solely based on visual constructions using the BEADS representation. The state of interest has the parametric expression
\begin{equation}
\rho_W = p\ket{\Psi^-}\bra{\Psi^-} + \frac{1-p}{4} \mathbbm{1}_4 = p\ket{\Psi^-}\bra{\Psi^-} + \frac{(1-p)}{4}(\ket{\Phi^+}\bra{\Phi^+}+\ket{\Phi^-}\bra{\Phi^-}+\ket{\Psi^+}\bra{\Psi^+}+\ket{\Psi^-}\bra{\Psi^-}),
\label{eq:werner_param}
\end{equation}

where $\ket{\Phi^+}=1/\sqrt{2}(\ket{00}+\ket{11})$, $\ket{\Phi^-}=1/\sqrt{2}(\ket{00}-\ket{11})$, $\ket{\Psi^+}=1/\sqrt{2}(\ket{01}+\ket{10})$, and $\ket{\Psi^-}=1/\sqrt{2}(\ket{01}-\ket{10})$ are the standard Bell basis states, $\mathbbm{1}_4$ is the two-qubit identity matrix, and  $p \in [0;\ 1]$ is the admixture weight. Thus, such a Werner state can be interpreted as the singlet state with added white noise. We are now interested in finding the minimum value of $p$ which admits entanglement, or equivalently, the maximum value of $p$ which admits a separable decomposition of $\rho_W$. Note that these threshold values are well-known in literature \cite{ollivier_quantum_2001}. We can still derive the threshold $p$ empirically using simple  BEADS-based constructions only. 

\begin{figure}[h!]
\centering
\includegraphics[trim={9cm 7cm 6cm 5.5cm},
  clip, width=.45\textwidth]{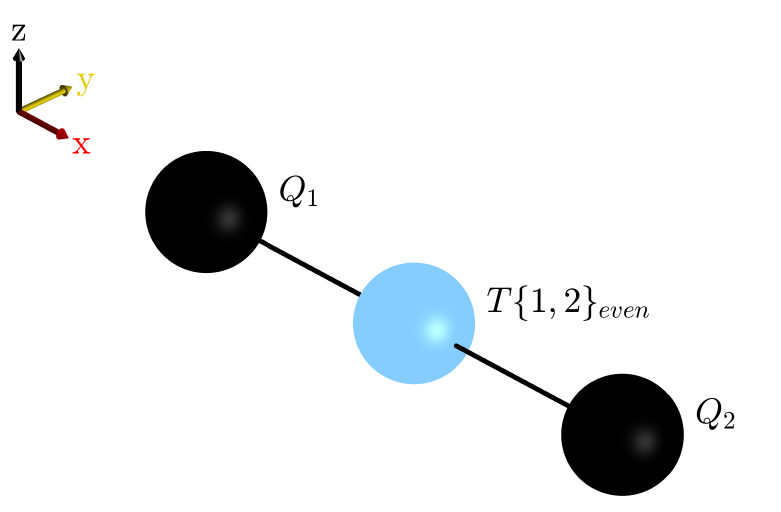}
\caption{Generalized BEADS representation of the pure maximally entangled singlet state $\ket{\Psi^-}=1/\sqrt{2}(\ket{01}-\ket{10})$. The reduced density operators are maximally mixed, which is represented by all-black $Q$-Beads. The $T\{1,2\}_{even}$-, or in this case equivalently, the $E\{1,2\}_{even}$-Bead has an isotropic surface pattern with spherical function value $-1$ in all spatial directions.}
\label{fig:supp_thresh_singlet}
\end{figure}

We start by viewing the generalized BEADS representation of the pure singlet state $\ket{\Psi^-}$ in Fig.~\ref{fig:supp_thresh_singlet}. It is immediately clear that this state is maximally entangled, as it is pure and the $Q$-Beads (reduced density operators) are all-black, that is, maximally mixed. Moreover, based on its colors the $T$-Bead which is equivalent to the $E$-Bead, in this case, has an isotropic distribution of spherical function values, i.e., the spherical function value is $-1$ in all spatial directions. Since incoherently adding white noise (the identity matrix, or equivalently the maximally mixed state) to another state does not change the surface pattern morphology and only reduces the spherical function values in magnitude (cf. Fig.~\ref{fig:SLNoisy04} for the same effect observed for GHZ states), the Werner states $\rho_W$ are expected to have maximally mixed $Q$-Beads and a $T\{1,2\}_{even}$-Bead with an isotropic surface pattern for all possible values of $p$.

\begin{figure}[h!]
\centering
\vspace{-1.5cm}
\includegraphics[width=1\textwidth]{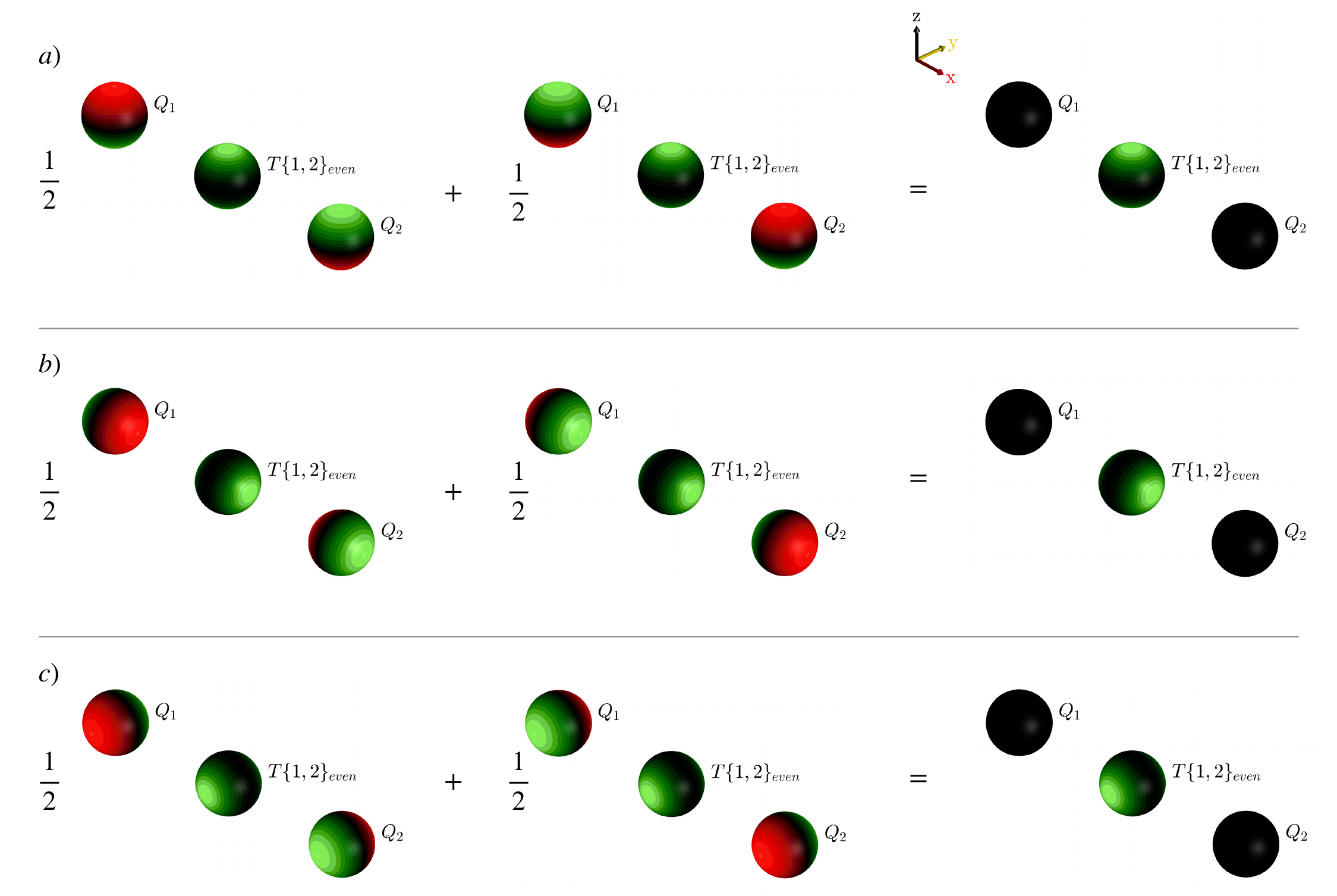}
\caption{Construction protocol of maximally classically correlated states using the generalized BEADS representation. Here, (a) computational-, (b) x-, and (c) y-basis pure product states are combined at equal weights to give mixed states, respectively. These states are maximally classically correlated, that is, the reduced density operators are maximally mixed (black $Q$-Beads) and the $T$-Bead has a maximum spherical function value of $-1$ along a basis-dependent axis. The resulting states are (a) $\rho_{cl,1} = 1/2(\ket{01}\bra{01}+\ket{10}\bra{10})$, (b) $\rho_{cl,2} = 1/2(\ket{+-}\bra{+-}+\ket{-+}\bra{-+})$, and (c) $\rho_{cl,3} = 1/2(\ket{RL}\bra{RL}+\ket{LR}\bra{LR})$, where $\ket{+} = 1/\sqrt{2}(\ket{0}+\ket{1})$, $\ket{-} = 1/\sqrt{2}(\ket{0}-\ket{1})$, $\ket{R} = 1/\sqrt{2}(\ket{0}+i\ket{1})$, $\ket{L} = 1/\sqrt{2}(\ket{0}-i\ket{1})$.} 
\label{fig:supp_thresh_1}
\end{figure}

Thus, it is convenient to approach the entanglement threshold problem from the perspective of separable pure state decompositions of $\rho_W$. Knowing that the final Werner state has zero-valued $Q$-Beads, an isotropic $T$-Bead pattern, and only a fully permutation symmetric $T$-Bead component, we can first construct a mixed state for which the $Q$-Beads are maximally mixed and the $T\{1,2\}_{even}$-Bead has a maximum spherical function value $-1$ value along a single axis. For instance, we choose the computational basis. As the $T$-Bead spherical function value needs to be negative, the only suitable basis states are $\ket{01}$ and $\ket{10}$. As can be seen in Fig.~\ref{fig:supp_thresh_1}, mixing both of these pure states at equal weights allows to cancel the $Q$-Beads while leaving a $T\{1,2\}_{even}$-Bead with a maximum spherical function value of $-1$ along the $z$-axis. We repeat this approach using basis states corresponding to orthogonal spatial directions, here, x- and y-basis states. As a side remark, note that each of the resulting mixed states in Fig.~\ref{fig:supp_thresh_1} is a separable, maximally classically correlated state constructed from pure product states. Crucially, we have maximized the $T$-Bead spherical function value to the physically allowed boundary value $-1$ along the basis-dependent principle axis. 

\begin{figure}[h!]
\centering
\includegraphics[width=1\textwidth]{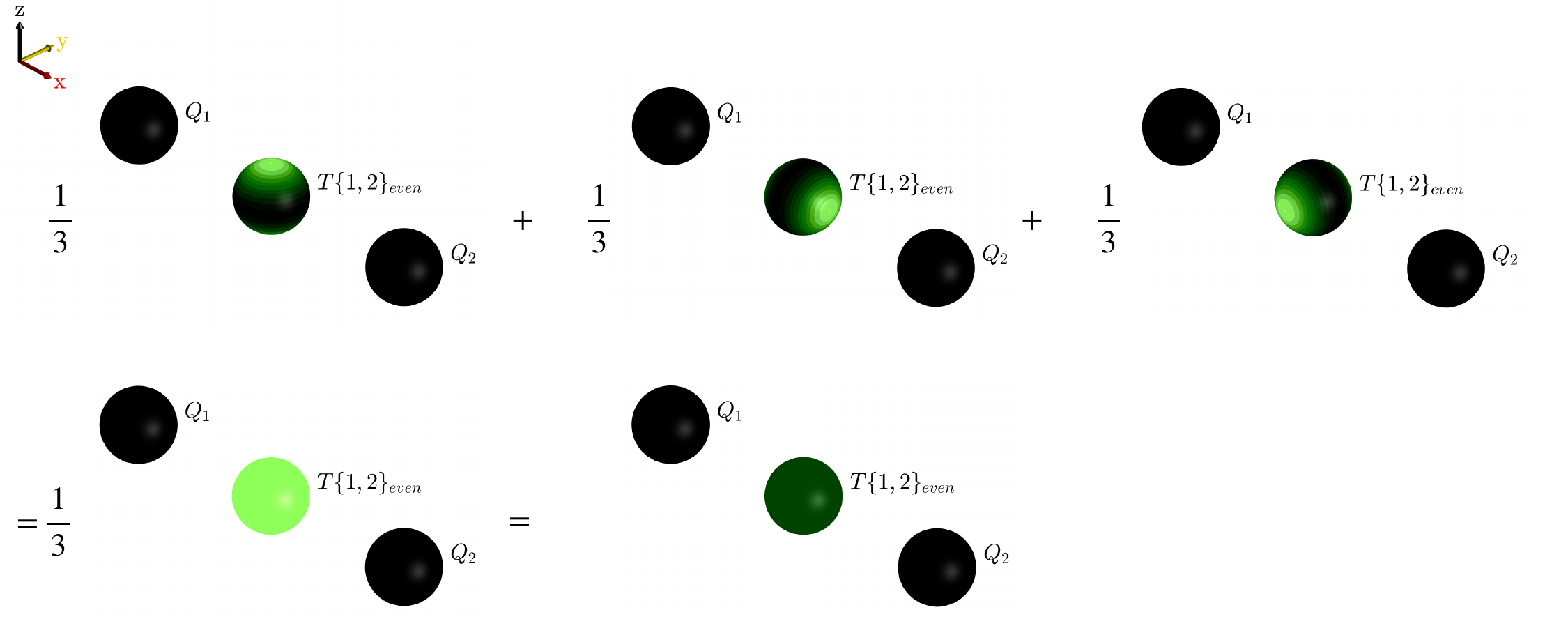}
\caption{Combining the mixed states obtained in Fig.~\ref{fig:supp_thresh_1} at equal weights $w = 1/3$ gives the separable threshold Werner state $\rho_{W, \text{thresh}} = 1/3(\rho_{cl, 1}+\rho_{cl, 2}+\rho_{cl, 3})$ which has a $T$-Bead spherical function value of $-1/3$ in all directions. As the $T$-Beads of $\rho_{cl, k}$ have maximum magnitude (spherical function value $-1$) along distinct directions before forming $\rho_{W, \text{thresh}}$, there exists no separable decomposition that allows to achieve larger absolute $T$-Bead spherical function values.} 
\label{fig:supp_thresh_2}
\vspace{-.9cm}
\end{figure}

We now can generate the desired isotropic surface pattern and thus, arrive at the separable threshold state $\rho_{W, \text{thresh}}$ by combining the mixed states $\rho_{cl, k}$ obtained in Fig.~\ref{fig:supp_thresh_1} at equal weights $w = 1/3$. It is directly clear in Fig.~\ref{fig:supp_thresh_2} that, since we start from states that have a maximum $T$-Bead value of $-1$, the resulting state has a maximum spherical function value of  $(-1)w = -1/3$ in all spatial directions. As we only use mixed states in the construction that adopt the physically allowed maximum spherical function value, there must not be a separable decomposition that admits larger spherical function values in magnitude and the resulting state must be indeed the separable threshold state $\rho_{W, \text{thresh}}$.

Conveniently, we can now directly read off the threshold admixture weight $p$ as the absolute spherical function value of the $T\{1,2\}_{even}$-Bead in an arbitrary directions, which results in $p = |-1/3| = 1/3$. As a last step, we can determine the overlap of $\rho_{W, \text{thresh}}$ with the singlet state, which is not identical to $p$, as the singlet state has nonzero overlap with the white noise term in Eq.~\ref{eq:werner_param}. This results in $\langle\Psi^-\rangle = \text{Tr}(\ket{\Psi^-}\bra{\Psi^-}\ \rho_{W, \text{thresh}}) = p + \frac{1-p}{4} = 1/3 + 1/6 = 1/2$. Thus, singlet-centered Werner states are entangled for all $p>1/3$ (or $\langle\Psi^-\rangle>1/2$) and separable for all $p\leq 1/3$ ($\langle\Psi^-\rangle\leq1/2$).
\clearpage

\section{Additional bipartite examples}\label{supp:bipartite}
In this supplementary section, we discuss additional generalized BEADS representations of mixed bipartite quantum states, which serve to further reinforce the understanding of the visualization and for showcasing alternative plotting variants.

\subsection{Separable states}\label{sup:bipartite}
\begin{figure}[H]
	\centering
	\includegraphics[width=\linewidth]{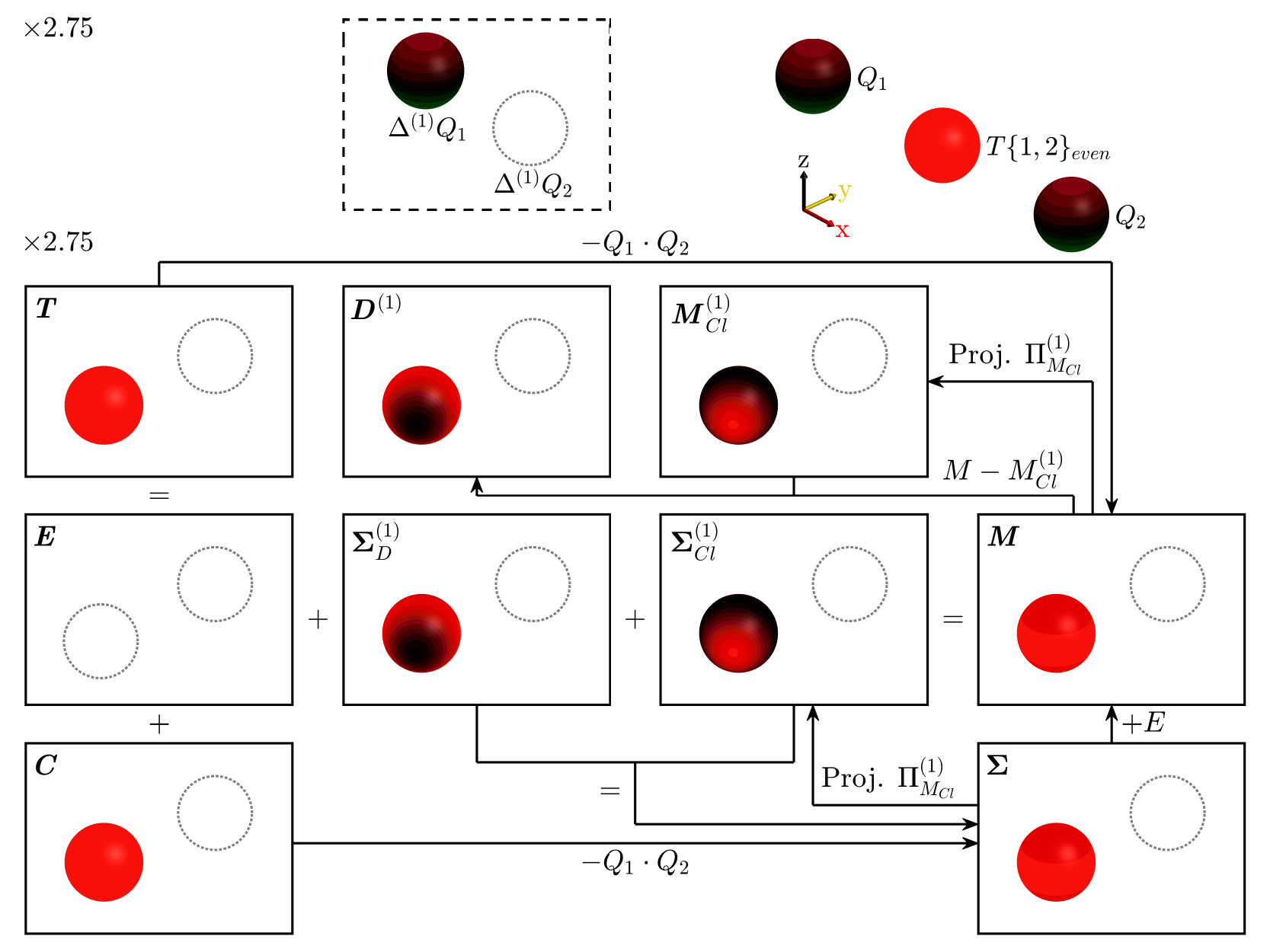}	
	\caption{Full generalized BEADS representation of the state $\rho_\text{sep}$ (cf. \cref{eq:ghz_w_mixed_traced}). The $T$- and $C$-Beads are identical as the state is separable. Whereas the surface patterns of the $T$- and $C$-Beads are fully isotropic, the $M$- and $\Sigma$-Beads adopt an ellipsoid pattern that is axially symmetric with respect to the $z$-axis. As a consequence, the maximum classically accessible information is obtained by projections in the $xy$-plane. Here, $M_{Cl}$-, $\Sigma_{Cl}$-, $D$-, and $\Sigma_{Cl}$-Beads are shown for one exemplary projection of qubit~1.}
	\label{fig:red_ghzw}
\end{figure}

The following example of a nonclassically correlated separable state illustrates the difference between $T$- and $M$-Beads for non-maximally correlated separable states. Consider the state
\begin{equation}
	\rho_{\text{sep}}=\text{Tr}_3\left[\frac{1}{2}\left(\ket{\text{W}}\bra{\text{W}}+ \ket{\text{GHZ}}\bra{\text{GHZ}} \right)\right],
	\label{eq:ghz_w_mixed_traced}
\end{equation}
with the $\ket{\text{W}}$- and $\ket{\text{GHZ}}$-state as defined in \cref{sec:ghz_w_pure}. $\rho_{\text{sep}}$ is visualized in \cref{fig:red_ghzw}.

\begin{figure}[H]
	\centering
	\includegraphics[width=\linewidth]{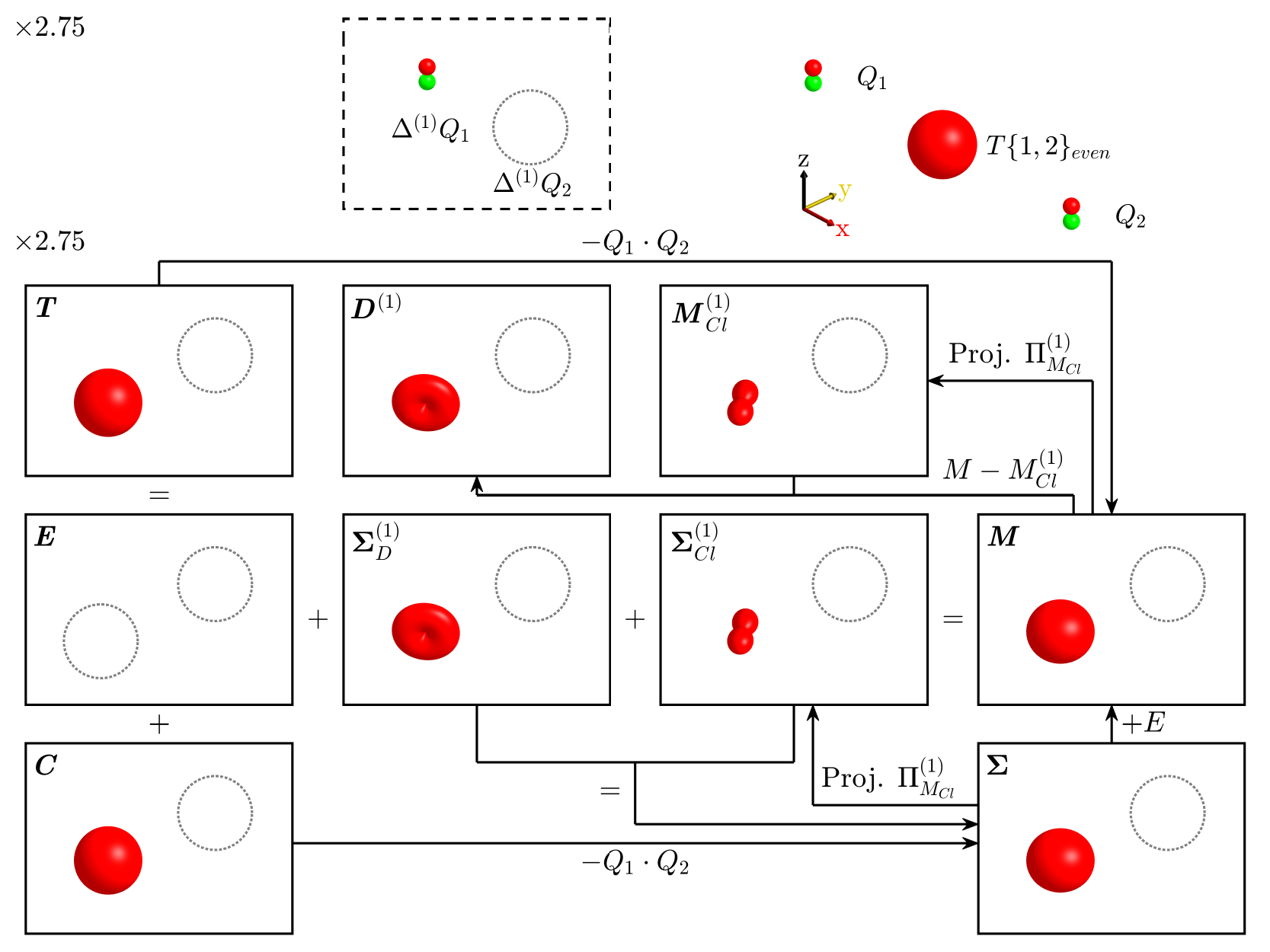}
	\caption{Full generalized BEADS representation of the state $\rho_\text{sep}$ (cf. \cref{eq:ghz_w_mixed_traced}) using radial plots. The $T$- and $C$-Beads are identical as the state is separable. In the shown radial plots, marginal differences in shape and small spherical function values are easier to recognize and distinguish than in the standard plotting variant shown in \cref{fig:red_ghzw}.}
	\label{fig:red_ghzw_radial}
\end{figure}

Here, the $T$- and $C$-Beads, which are identical as the state is separable, appear to be completely isotropic, that is, the spherical function value of these beads is the same in all spatial directions which results in a homogeneously colored bead. As the reduced density operators ($Q$-Beads) are not maximally mixed though, the $T$- and $C$-Beads encode redundant information, such that the $M$- and $\Sigma$-Beads surface patterns turn out slightly compressed and thus, ellipsoid with smaller-valued poles along the $z$-axis. 
This can be seen more clearly in the radial plotting variant shown in \cref{fig:red_ghzw_radial}, where the spherical function value in a specific direction corresponds to the distance from the origin. Indeed, the $T$- and $C$-Beads are genuinely isotropic resulting in perfectly spherical beads, whereas the $M$- and $\Sigma$-Beads are ellipsoid. Note that we have previously introduced the radial plotting method in \cite{huber_beads_2025}. In particular, this variant of the generalized BEADS representation allows to better recognize small spherical function values due to avoiding dark colors.

\begin{figure}[H]          
	\centering
	\includegraphics[width=10cm, trim= 2cm 2cm 2cm 2cm, clip]{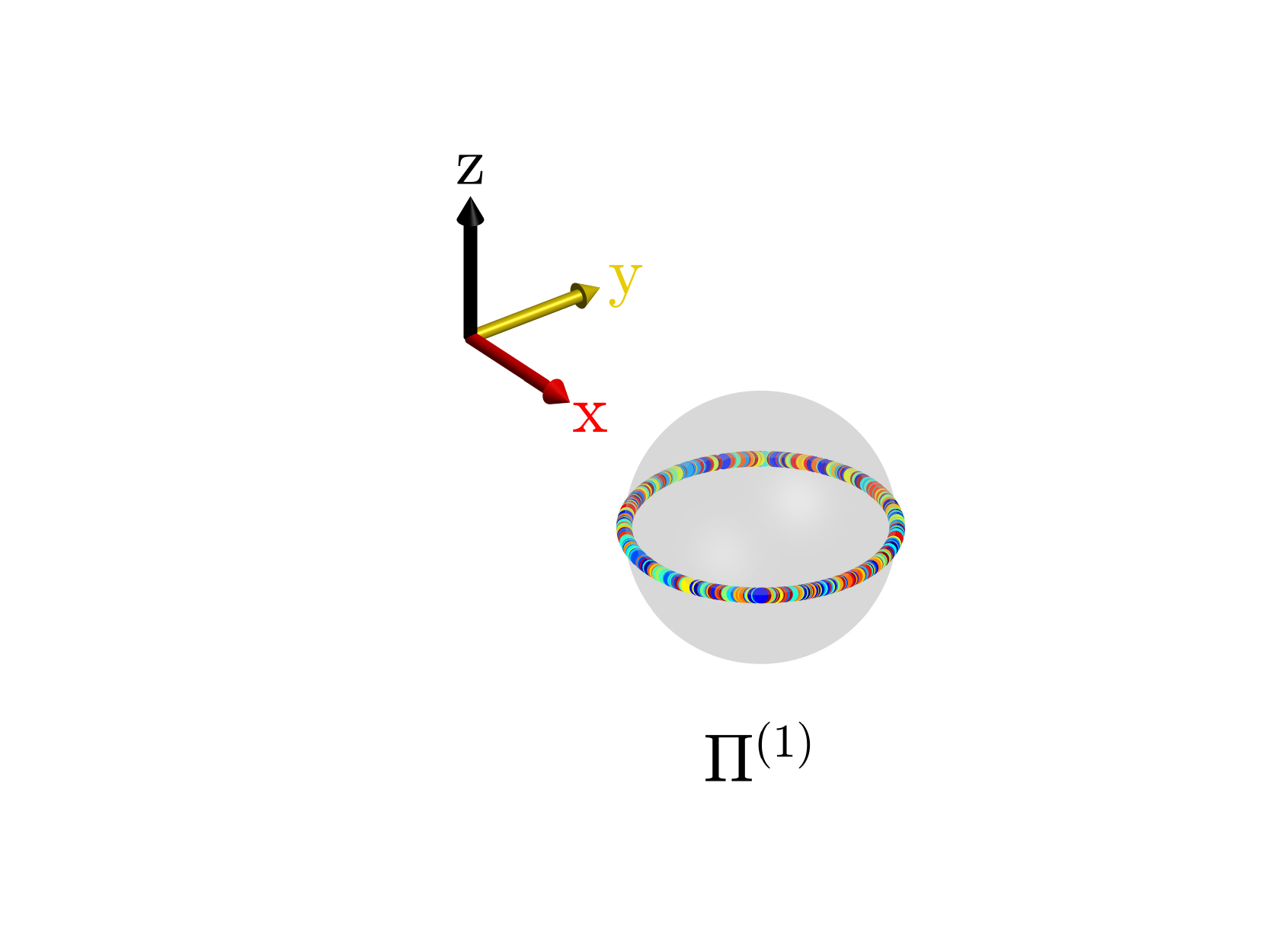}
	\caption{Bloch sphere analysis of optimum measurement directions that maximize the classically accessible mutual information of the state $\rho_\text{sep}$ (cf. \cref{eq:ghz_w_mixed_traced}). The figure shows the optimization results for 500 Haar-distributed initial unitary seeds. The optimum projection directions are visualized as points on the Bloch sphere. The preferred orientation of projectors exhibits a rotational degree of freedom around the $z$-axis.}
	\label{fig:blochvectors_ghz_w_mixed_traced_500}
\end{figure}

The axial symmetry induced by the ellipsoid surface pattern of the $M$- and $\Sigma$-Beads causes the optimum local measurement that maximizes the classically accessible information to correspond to an arbitrary direction in the transverse $xy$-plane similar to what was discussed for the entangled mixed state in \cref{sec:entbipartite}. To illustrate this planar degree of freedom in the choice of local measurements, \cref{fig:blochvectors_ghz_w_mixed_traced_500} visualizes the obtained projection directions after optimization of 500 Haar-measure-drawn seeds. The found $M_{Cl}^{(1)}$- and $\Sigma_{Cl}^{(1)}$-Beads in \cref{fig:red_ghzw,fig:red_ghzw_radial} therefore correspond to a possible exemplary projection in the $xy$-plane. Since the state is fully separable, $M_{Cl}^{(1)}$- and $\Sigma_{Cl}^{(1)}$-, as well as $D^{(1)}$- and $\Sigma_D^{(1)}$-Beads coincide, respectively. As the density operator components corresponding to the classically accessible information are fully permutation symmetric (even $M_{Cl}$- and $\Sigma_{Cl}$-Bead point symmetry), the surface patterns of the $D$- and $\Sigma_D$-Beads exhibit a distinctive distorted ring-like structure. These characteristic dumbbell ($M_{Cl}$- and $\Sigma_{Cl}$-Beads) and squashed torus shapes ($D$- and $\Sigma_D$-Beads) become even more apparent in the radial plot. 

\subsection{Entangled states}
\subsubsection{Example A}
\begin{figure}[htbp]
	\centering
	\includegraphics[width=\linewidth]{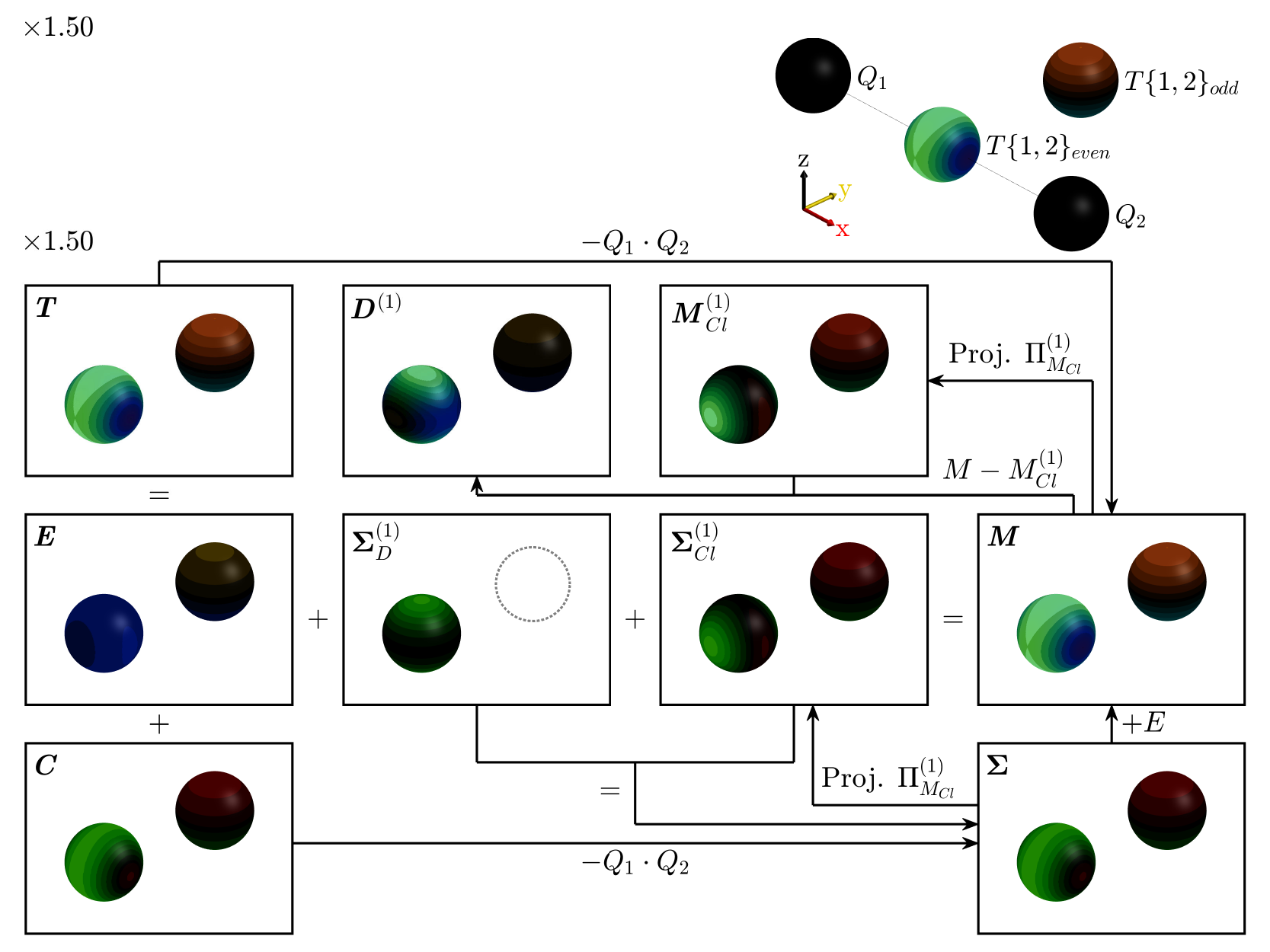}
	\caption{Full generalized BEADS representation of the entangled mixed state $\rho_\text{ent}$ (cf. \cref{eq:gisela}). Clearly the state has a residual separable structure ($C$- and $\Sigma$-Beads) alongside its entangling structure visualized by $E$-Beads. Hence, it is not maximally entangled.}
	\label{fig:gisela_all}
\end{figure}

The following example is constructed as a mixture of four maximally entangled states: 
\begin{equation}
	\rho_\text{ent}=\frac{1}{3}\ket{\psi}\bra{\psi}+\frac{1}{5}\ket{\Phi^+}\bra{\Phi^+}+\frac{6}{15}\ket{\Psi^-}\bra{\Psi^-}+\frac{1}{15}\ket{\Psi^+}\bra{\Psi^+}
	\label{eq:gisela}
\end{equation}
consisting of the three Bell states $\ket{\Phi^+}$, $\ket{\Psi^-}$, and $\ket{\Psi^+}$ as well as
\begin{equation} \ket{\psi}=\frac{1}{\sqrt{2}}
	\begin{pmatrix}
		0 \\
		1 \\
		-1i \\
		0
	\end{pmatrix}.
\end{equation}
As can be seen immediately in \cref{fig:gisela_all}, the $T$-Bead of $\rho_\text{ent}$ is mostly blue and green colored, which indicate entanglement and an underlying separable structure, respectively. Since the $Q$-Beads are fully black (maximally mixed), the reduced marginals do not encode local information and the $T$- and $M$-, as well as the $C$- and $\Sigma$-Beads coincide, respectively. It is clear that the state, despite being constructed from maximally entangled pure states, is not maximally entangled. Moreover, the separable structure of the underlying density operator contributes significantly to the genuine bipartite correlation of the state, as is directly evident from the $\Sigma$-Beads. Note that the state is asymmetric and thus, has correlation function beads of $even$ and $odd$ symmetry.

\begin{figure}[htbp]          
	\centering
	\includegraphics[width=10cm, trim= 2cm 2cm 2cm 2cm, clip]{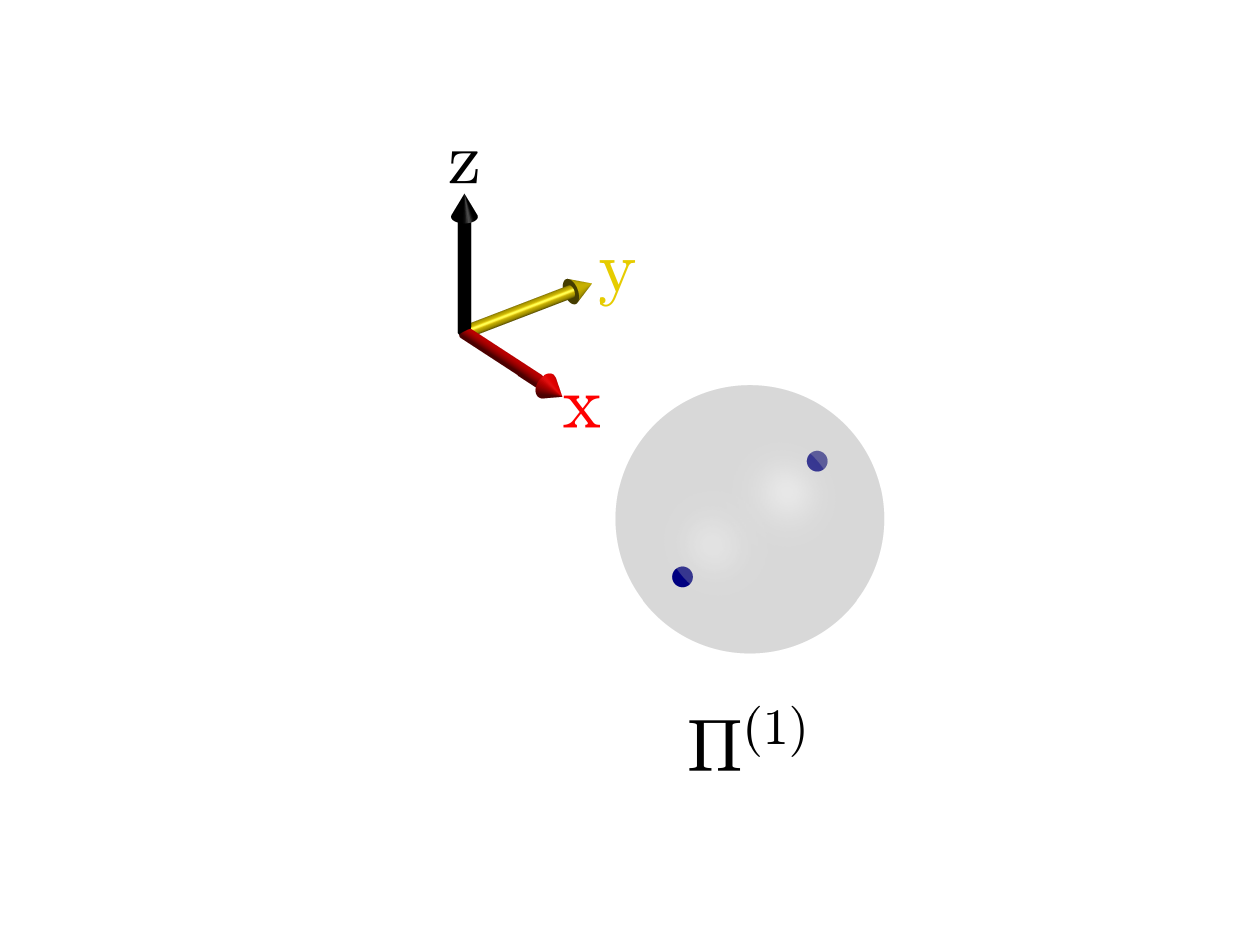}
	\caption{Bloch sphere analysis of optimum local measurement directions that maximize the classically accessible mutual information of the state $\rho_\text{ent}$ (cf. \cref{eq:ghz_w_mixed_traced}). The figure shows the optimization results for 500 Haar-distributed initial unitary seeds. The optimum projector orientations are visualized as points on the Bloch sphere. Only a single preferred measurement direction exists. Note that this direction is not parallel to the $y$-axis, as the optimum gain in information is obtained for a projection along a skew axis which allows to capture the maximum components of the symmetric bead $M\{1,2\}_{even}$ and the antisymmetric bead $M\{1,2\}_{odd}$.}
	\label{fig:blochvectors_gisela_10}
\end{figure}

Unlike for most other cases discussed in this article, there is a unique local measurement direction that maximizes the classically accessible mutual information of the state. This is further illustrated by the corresponding Bloch sphere analysis is depicted in \cref{fig:blochvectors_gisela_10}. Notably, this preferred measurement direction is not aligned parallel to the $y$-axis as one might conclude from the surface patterns of the $M\{1,2\}_{even}$- and $\Sigma\{1,2\}_{even}$-Beads. In fact, the optimum measurement captures the largest possible component of the symmetric bead $M\{1,2\}_{even}$ and the entire antisymmetric component arising from the separable structure ($\Sigma\{1,2\}_{odd}$). The latter can be directly deduced from the missing antisymmetric $\Sigma_D^{(1)}\{1,2\}_{odd}$-Bead.  Note that the nonclassical genuinely correlation-based structure, i.e., the classically inaccessible information of the state represented by $D^{(1)}$-Beads, contains parts of the entangling and separable structures of $\rho_\text{ent}$.

\subsubsection{Example B}
The parameterized state
\begin{equation}
	\rho(z)= \frac{1}{2} (\ket{00}\bra{00} + \ket{11} \bra{11}) + \frac{z}{2} (\ket{00} \bra{11}+ \ket{11} \bra{00}),
	\label{eq:zurek_example_stacked}
\end{equation}
was analyzed in detail in the work of Ollivier and Zurek \cite{ollivier_quantum_2001}. The state
describes the transition from a classically correlated mixed state for $z=0$ to a maximally entangled pure state for $z=1$ by introducing off-diagonal terms in the density operator (coherences) with increasing values of $z$. A stacked plot serves as the ideal visualization tool to capture the correlation structure of the parameterized state in dependence of the parameter $z$. 

\begin{figure}[htbp]
	\centering
	\makebox[\linewidth][c]{
		\begin{minipage}{1\textwidth}
			\begin{subfigure}[t]{0.48\textwidth}
				\centering
				\includegraphics[height=9cm, trim={10cm 4.5cm 4.5cm 3.65cm}, clip]{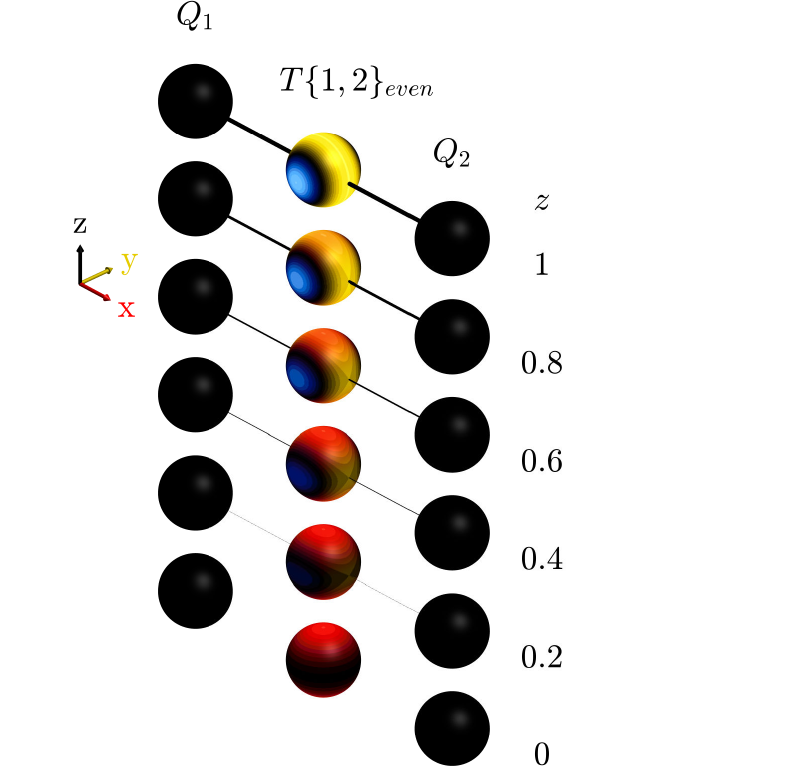}
				\subcaption{Total correlation function beads ($T$-Beads).}
				\label{fig:zurek_stacked_T5}
			\end{subfigure}
			\hspace{0.04\textwidth}
			\begin{subfigure}[t]{0.48\textwidth}
				\centering
				\includegraphics[height=9cm, trim={10cm 4.5cm 4.5cm 3.65cm}, clip]{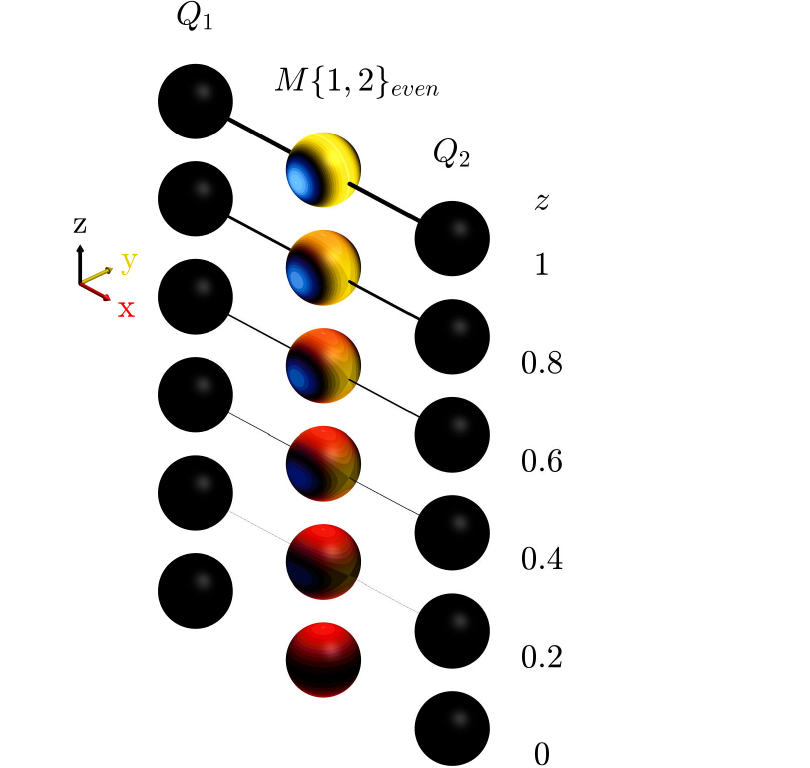}
				\subcaption{Total correlation beads ($M$-Beads).}
				\label{fig:zurek_stacked_M5}
			\end{subfigure}
		\end{minipage}
	}
	
	\vspace{0.5cm}
	
	\makebox[\linewidth][c]{
		\begin{minipage}{1\textwidth}
			\begin{subfigure}[t]{0.48\textwidth}
				\centering
				\includegraphics[height=9cm, trim={10cm 4.5cm 4.5cm 3.65cm}, clip]{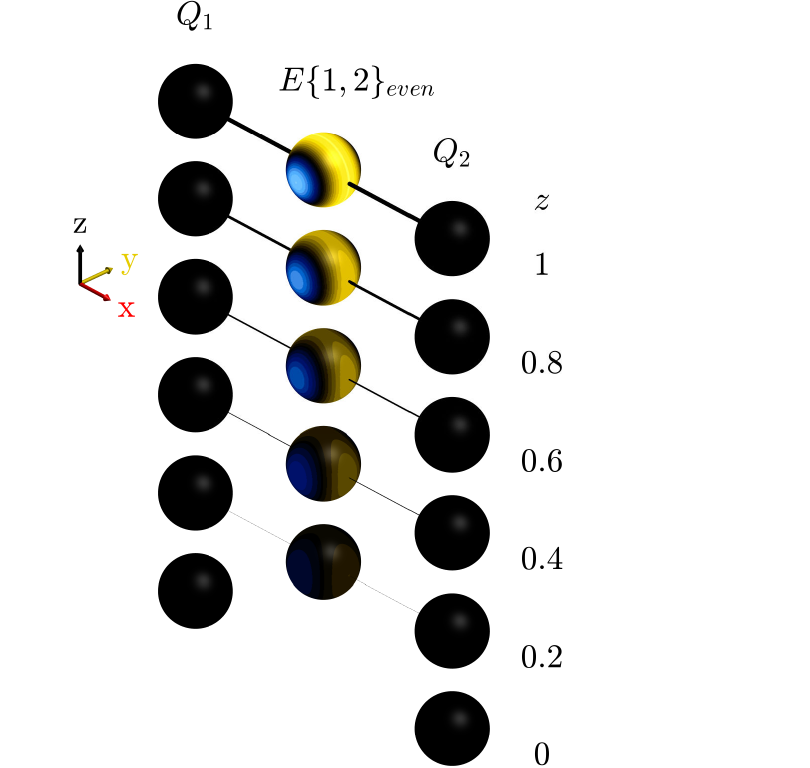}
				\subcaption{Entanglement-based correlation function beads ($E$-Beads).}
				\label{fig:zurek_stacked_E5}
			\end{subfigure}
			\hspace{0.04\textwidth}
			\begin{subfigure}[t]{0.48\textwidth}
				\centering
				\includegraphics[height=9cm,trim={10cm 4.5cm 4.5cm 3.65cm}, clip]{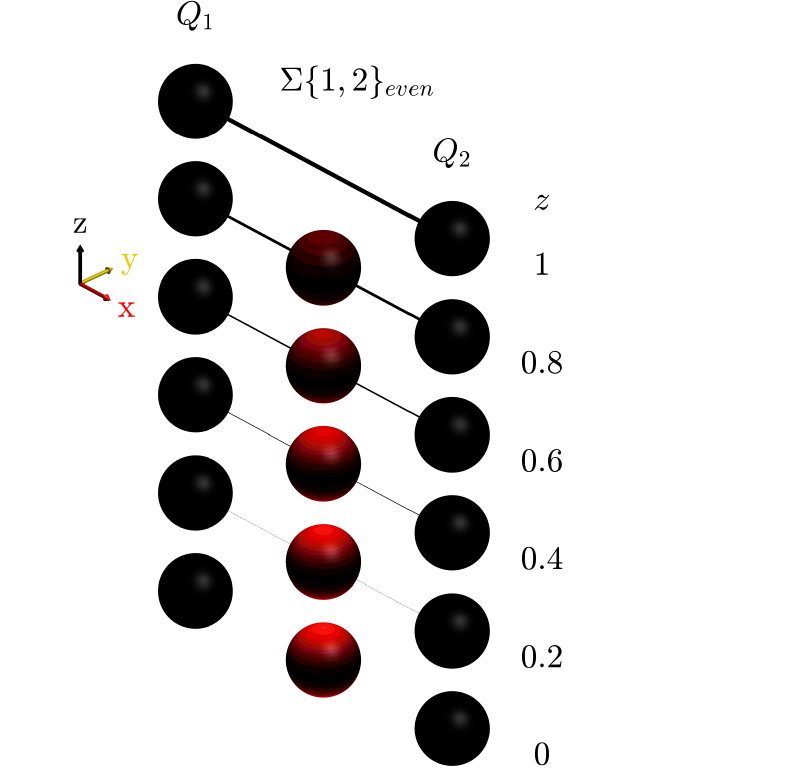}
				\subcaption{Genuine correlations arising from the separable structure of the state ($\Sigma$-Beads).}
				\label{fig:zurek_stacked_CD5}
			\end{subfigure}
		\end{minipage}
	}
	
	\caption{Stacked plot generalized BEADS representations of the parameterized state $\rho(z)$ (\cref{eq:zurek_example_stacked}). The $T$- and $M$-Bead visualizations of the states are identical as the reduced density operators are maximally mixed, for which the entire information can be attributed to genuine correlations. Due to the same reason, $C$- and $\Sigma$-Beads coincide in this case, for which the $C$-Bead visualization is omitted. As can be seen from the $E$-Bead representations, all states for $z\neq 0 $ are entangled. The state for $z=1$ is the maximally entangled pure Bell state $\ket{\Phi^+}$.}
	\label{fig:zurek_stacked_all}
\end{figure}

\cref{fig:zurek_stacked_T5} displays the $T$-Bead representations as a function of the parameter $z$ at an increment of $\Delta z = 0.2$. Due to its characteristic surface pattern, the maximally mixed Bell state can be readily identified at $z=1$. As the blue-yellow color scheme is exclusively used to indicate entanglement in the generalized BEADS representation, it is clear that states with larger $z$ are more strongly entangled. This is even more obvious when viewing the isolated $E$- and $\Sigma$-Bead representations in \cref{fig:zurek_stacked_E5} and \cref{fig:zurek_stacked_CD5}. Evidently, the $E$-Beads become brighter whereas the $\Sigma$-Beads darken for growing $z$ corresponding to an increasing entangling and decreasing separable structure, respectively. As the reduced density operators of $\rho(z)$ do not encode local information, the $C$- and $\Sigma$-Beads representations are identical for which the $C$-Bead visualization is omitted to avoid redundancy. Due to the same reason, $T$- and $M$-Beads (cf. \cref{fig:zurek_stacked_T5} and \cref{fig:zurek_stacked_M5}) also coincide, respectively. 

\begin{figure}[htbp]
	\centering
	\makebox[\linewidth][c]{
		\begin{minipage}{1\textwidth}
			\begin{subfigure}[t]{0.48\textwidth}
				\centering
				\includegraphics[height=9cm,trim={10cm 4.5cm 4.5cm 3.65cm}, clip]{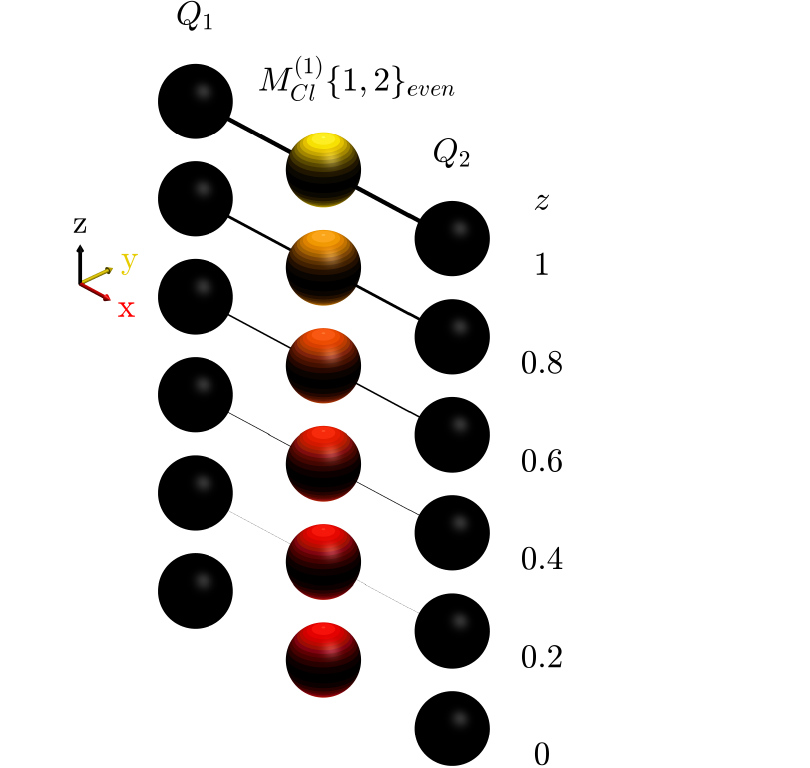}
				\subcaption{Classically accessible genuinely correlation-based information ($M_{Cl}^{(1)}$-Beads).}
				\label{fig:zurek_stacked_M_Cl5}
			\end{subfigure}
			\hspace{0.04\textwidth}
			\begin{subfigure}[t]{0.48\textwidth}
				\centering
				\includegraphics[height=9cm,trim={10cm 4.5cm 4.5cm 3.65cm}, clip]{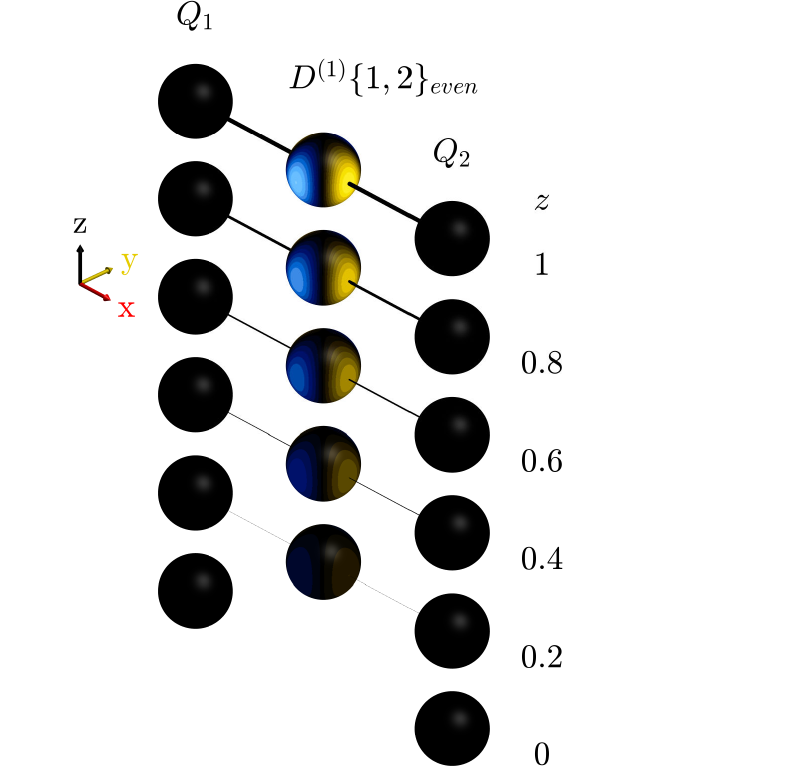}
				\subcaption{Nonclassical (i.e., classically inaccessible) genuinely correlation-based information ($D^{(1)}$-Beads).}
				\label{fig:zurek_stacked_D5}
			\end{subfigure}
		\end{minipage}
	}
	
	\vspace{0.5cm}
	
	\makebox[\linewidth][c]{
		\begin{minipage}{1\textwidth}
			\begin{subfigure}[t]{0.48\textwidth}
				\centering
				\includegraphics[height=9cm,trim={10cm 4.5cm 4.5cm 3.65cm}, clip]{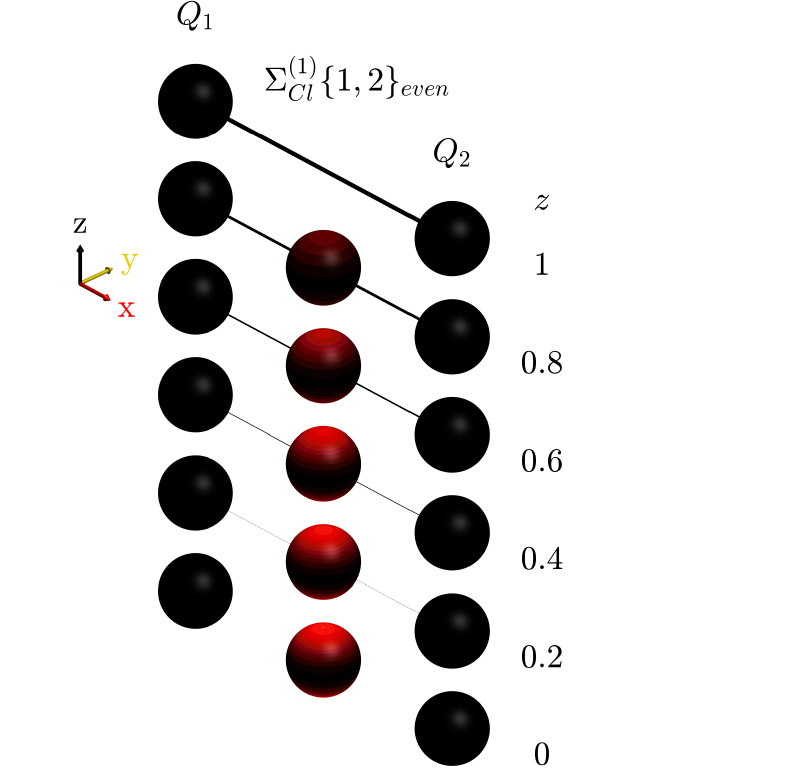}
				\subcaption{Contribution to classically accessible genuinely correlation-based information arising from the separable structure of the state ($\Sigma_{Cl}^{(1)}$-Beads).}
				\label{fig:zurek_stacked_CL5}
			\end{subfigure}
			\hspace{0.04\textwidth}
			\begin{subfigure}[t]{0.48\textwidth}
				\centering
				\includegraphics[height=9cm,trim={10cm 4.5cm 4.5cm 3.65cm}, clip]{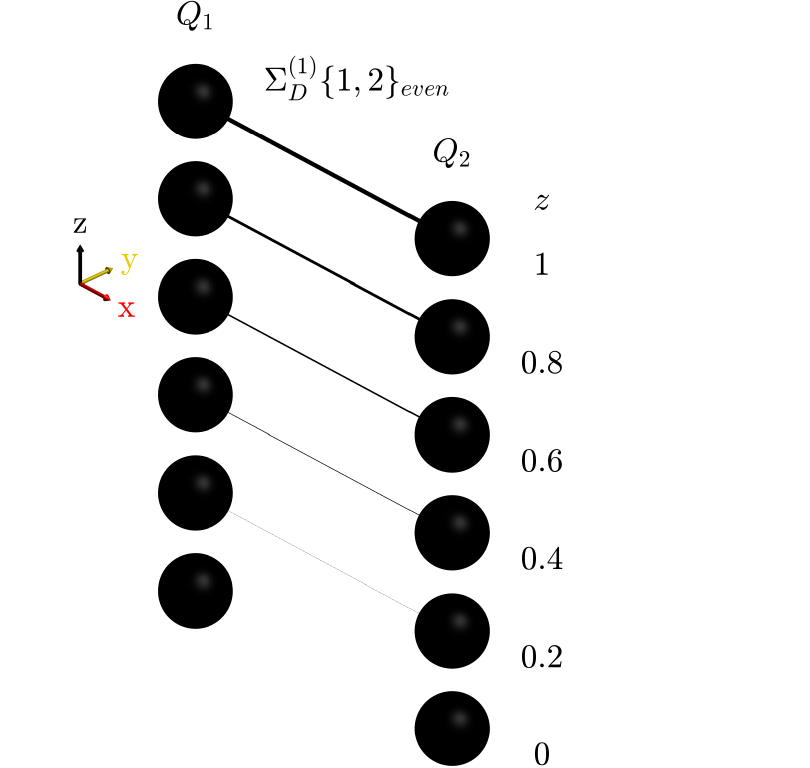}
				\subcaption{Contribution to nonclassical genuinely correlation-based information arising from the separable structure of the state ($\Sigma_{D}^{(1)}$-Beads).}
				\label{fig:zurek_stacked_SigmaD5}
			\end{subfigure}
		\end{minipage}
	}
	\caption{Stacked plot generalized BEADS representations of the parameterized state $\rho(z)$ (\cref{eq:zurek_example_stacked}). Whereas the classically accessible information ($M_{Cl}^{(1)}$-Beads) captures increasingly more components of the entangling structure of $\rho(z)$ for larger $z$, the nonclassical information in this case strictly arises from the entangling structure. Hence, $\Sigma_{Cl}^{(1)}$-Beads exist for any except the maximally entangled pure state case, whereas none of the states shows $\Sigma_{D}^{(1)}$-Beads.}
	\label{fig:zurek_stacked_all_2}
\end{figure}

Similar to the preceding example, there exists a unique optimum local measurement along the $z$-axis that maximizes the classical mutual information for all $z\neq 1$. For the pure Bell state ($z=1$), the local measurement can be conducted along an arbitrary direction in the $xz$-plane or along the $y$-axis. As is clear from the corresponding $M_{Cl}^{(1)}$-Beads (see \cref{fig:zurek_stacked_M_Cl5}), the classically accessible information increasingly captures larger components of the entangling structure of $\rho(z)$ for larger $z$. In the case $z=0$, the $M_{Cl}$-Bead is identical to the corresponding $\Sigma$- and $M$-Beads. Thus, $\rho(0)$ must be classically correlated, whereas $\rho(z)$ is nonclassically correlated for any other $z$. Note that this property is directly reflected in nonzero quantum discord for all $z\neq0$ \cite{ollivier_quantum_2001}.

Whereas the $M_{Cl}^{(1)}$-Beads of the examined states vary in their composition with respect to involved contributions from entangling and separable structures of the state, the $D^{(1)}$- and $\Sigma_D^{(1)}$-Beads (see \cref{fig:zurek_stacked_D5} and \cref{fig:zurek_stacked_SigmaD5}) directly reveal, that the nonclassical structure of states $\rho(z)$ exclusively arises from the states' entanglement.

\clearpage

\vspace{-3cm}
\section{Additional tripartite examples}\label{sup:tripartite}

\subsection{Parameterized incoherent mixtures of GHZ and W states} \label{sup:GHZWIncoherent}
\begin{figure}[H] 
	\vspace{-0.2cm}
	\centering
	\makebox[\textwidth][c]{
		\begin{minipage}{1\textwidth}
			\begin{subfigure}[t]{0.5\textwidth}
				\centering
				\includegraphics[width=1\textwidth, trim={7.7cm 5.2cm 7.3cm 6.6cm},clip]{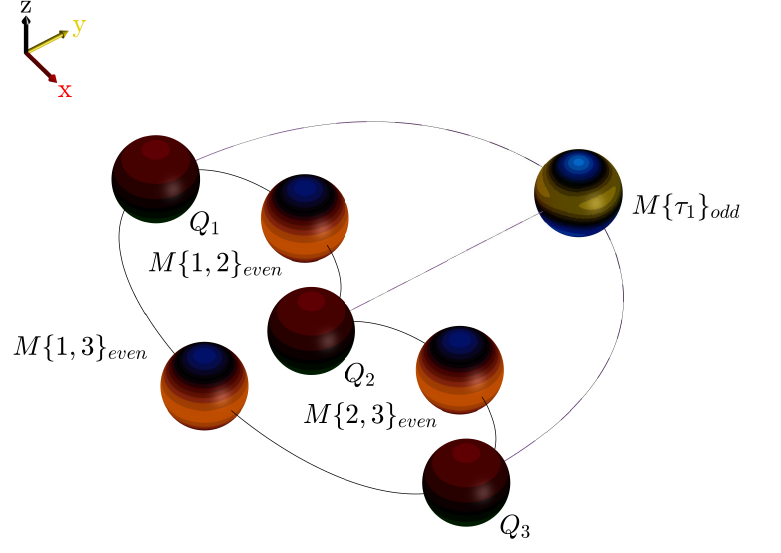}
				\subcaption{$p = 0.05$}
				\label{fig:ghz_w_mixed_0.05_M}
			\end{subfigure}
			\begin{subfigure}[t]{0.5\textwidth}
				\centering
				\includegraphics[width=1\textwidth, trim={7.7cm 5.2cm 7.3cm 6.6cm},clip]{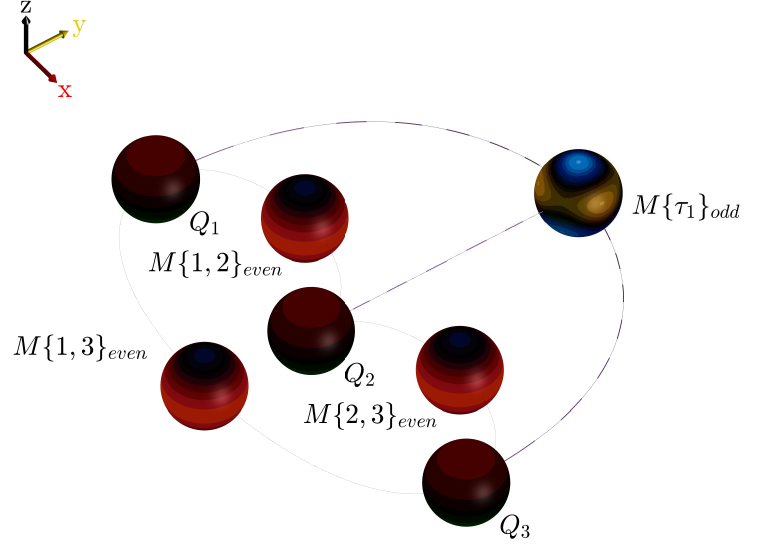}
				\subcaption{$p = 0.2$}
				\label{fig:ghz_w_mixed_0.2_M}
			\end{subfigure}
		\end{minipage}
	}
	
	\vspace{0.3cm}
	
	\makebox[\textwidth][c]{
		\begin{minipage}{1\textwidth}
			\begin{subfigure}[t]{0.5\textwidth}
				\centering
				\includegraphics[width=1\textwidth, trim={7.7cm 5.2cm 7.3cm 6.6cm},clip]{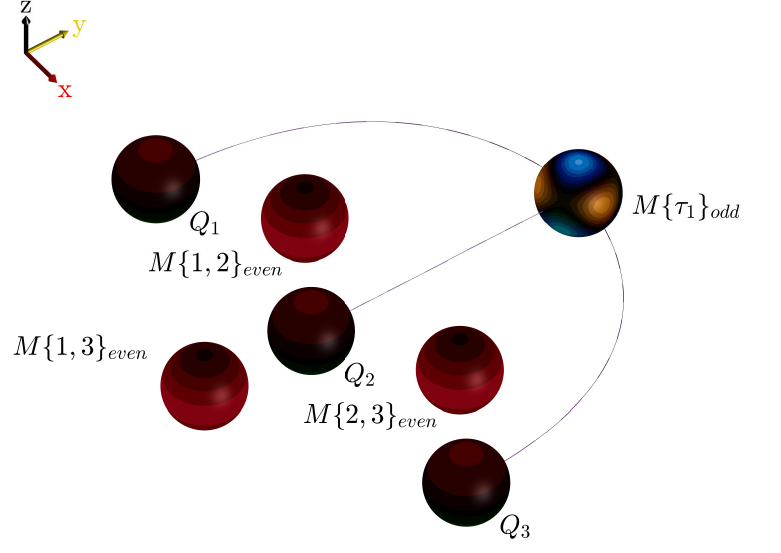}
				\subcaption{$p = 0.35$}
				\label{fig:ghz_w_mixed_0.35_M}
			\end{subfigure}
			\begin{subfigure}[t]{0.5\textwidth}
				\centering
				\includegraphics[width=1\textwidth, trim={7.7cm 5.2cm 7.3cm 6.6cm},clip]{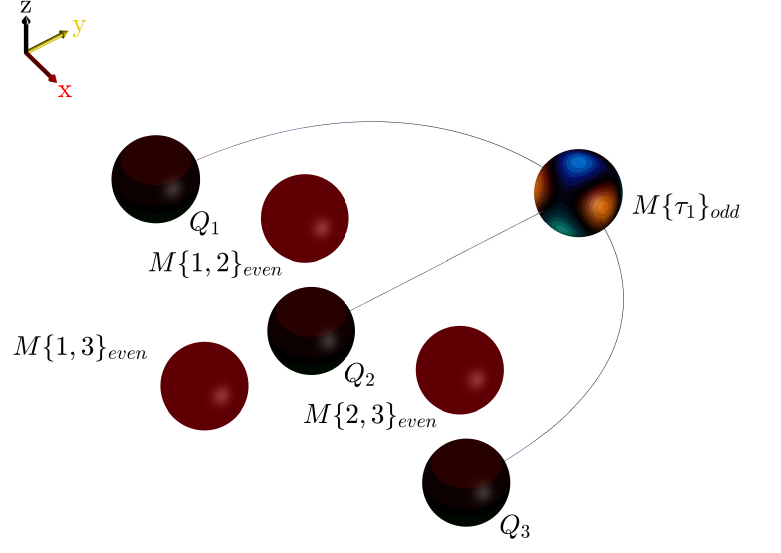}
				\subcaption{$p = 0.5$}
				\label{fig:ghz_w_mixed_0.5_M}
			\end{subfigure}
		\end{minipage}
	}
	
	\vspace{0.3cm}

	\makebox[\textwidth][c]{
		\begin{minipage}{1\textwidth}
			\begin{subfigure}[t]{0.5\textwidth}
				\centering
				\includegraphics[width=1\textwidth, trim={7.7cm 5.2cm 7.3cm 6.6cm},clip]{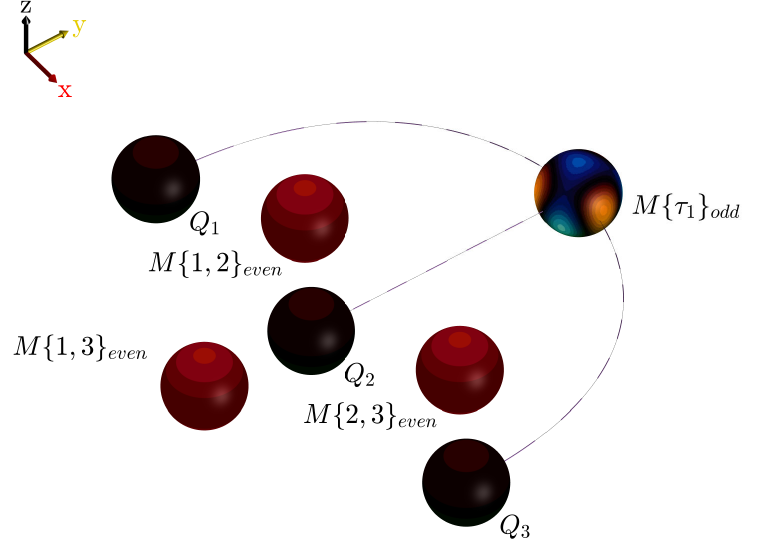}
				\subcaption{$p = 0.65$}
				\label{fig:ghz_w_mixed_0.65_M}
			\end{subfigure}
			\begin{subfigure}[t]{0.5\textwidth}
				\centering
				\includegraphics[width=1\textwidth, trim={7.7cm 5.2cm 7.3cm 6.6cm},clip]{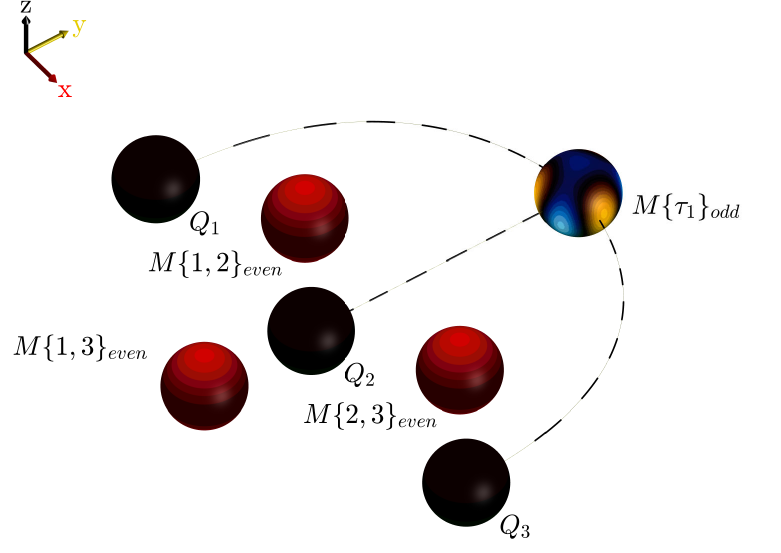}
				\subcaption{$p = 0.8$}
				\label{fig:ghz_w_mixed_0.8_M}
			\end{subfigure}
		\end{minipage}
	}
	\caption{Generalized BEADS representation of the parameterized incoherent mixture of tripartite GHZ and W states $\rho_\text{GHZ-W}(p)$ (see \cref{eq:ghz_w_mixed}). Here, the visualization displays the total correlation properties in terms of $M$-Beads. The bipartite marginals are already fully separable for $p>0.3$, but retain a W-like axially symmetric surface pattern for $p < 0.5$. The three-tangle of the shown states does not increase monotonically and is smaller but nonzero at $p=0.65$ than at $p=0.5$. All of the shown states belong to the GHZ mixed state class.}
	\label{fig:ghz_w_mixed_mixtures}
\end{figure}

In \cref{sec:ghz_w_pure}, we discussed characteristic features of GHZ and W states that become visible in their generalized BEADS representations. This includes surface pattern morphologies of bi- and tripartite correlation function beads and entanglement properties which directly follow from the BEADS representation. In the following, we examine the parameterized incoherent mixture of the GHZ and the W state
\begin{equation}
	\rho_\text{GHZ-W}(p) = p \left( \ket{\text{GHZ}} \bra{\text{GHZ}} \right) + (1-p) \left( \ket{\text{W}} \bra{\text{W}} \right).
	\label{eq:ghz_w_mixed}
\end{equation}
Generalized BEADS representations of $\rho_\text{GHZ-W}$ using $M$-Beads for visualizing the total correlation properties are shown in \cref{fig:ghz_w_mixed_mixtures}. Here, we omit visualizations of the pure W ($p=0$) and GHZ ($p=1$) states, which have already been provided in \cref{fig:ghz_w_pure}.

We start by considering the W-dominant state at $p=0.05$ in \cref{fig:ghz_w_mixed_0.05_M}. Clearly, this state displays distinctive features of the W state, e.g., axially symmetric surface patterns of all bipartite $M\{j,k\}_{even}$ and the tripartite $M\{\tau_1\}_{odd}$-bead with respect to the $z$-axis. Moreover, the bipartite marginals are equally entangled and the tripartite entanglement arcs indicate genuine tripartite entanglement. On closer inspection, the three-tangle arc has nonzero diameter and thus, the state belongs to the GHZ mixed state class but is structurally very close to an ideal W state.

As the GHZ contribution slowly increases in \cref{fig:ghz_w_mixed_0.2_M} and \cref{fig:ghz_w_mixed_0.35_M}, the generalized BEADS representation reveals that the bipartite entanglement decreases rapidly and vanishes completely at $p=0.3$. At the same time the surface pattern of the $M\{\tau_1\}_{odd}$-Bead shifts away from the axially symmetric morphology characteristic for W states and adopts a spotted appearance. This is accompanied by a rapid decrease of the $\mathcal{W}_3$ entanglement arc diameter. It can be concluded that W-type entanglement -- especially the characteristic entanglement of the bipartite marginals -- must be rather unstable under mixing with a GHZ state. 

However, as can be seen, for instance, in \cref{fig:ghz_w_mixed_0.35_M}, even though the bipartite marginals are no longer entangled, the axial symmetry and a torus-like surface pattern of the bipartite $M$-Beads is conserved for $p<0.5$. As a consequence, the bipartite subsystems preserve a nonclassical correlation structure similar to that of the W state while becoming separable. A full analysis of the bipartite marginal for the case $p=0.5$ shown in \cref{fig:ghz_w_mixed_0.5_M} can be found in supplementary \cref{sup:bipartite}.  

\begin{figure}[H]
\begin{minipage}[c]{.5\textwidth}
	\includegraphics[width=\textwidth]{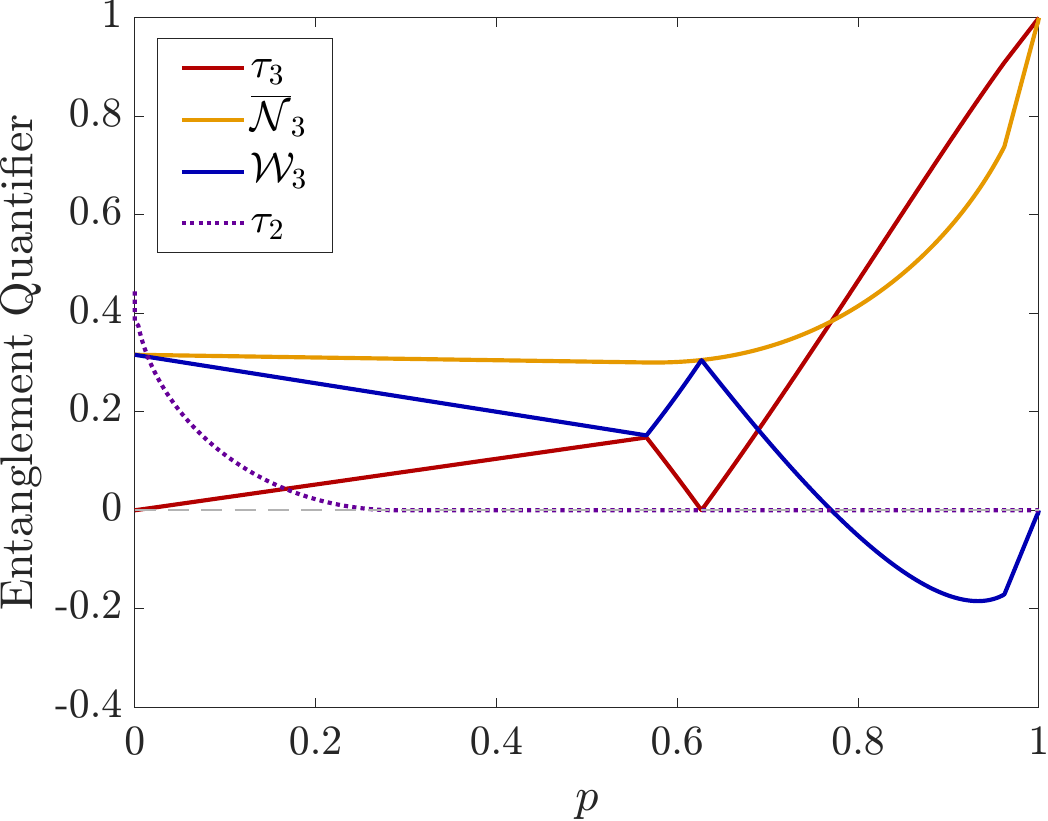}
	\phantomsubcaption
	\label{fig:GHZWMeasures}
\end{minipage}%
\begin{minipage}[c]{.5\textwidth}
	\includegraphics[width=\textwidth, trim={7.5cm 5.2cm 7.3cm 6.6cm},clip]{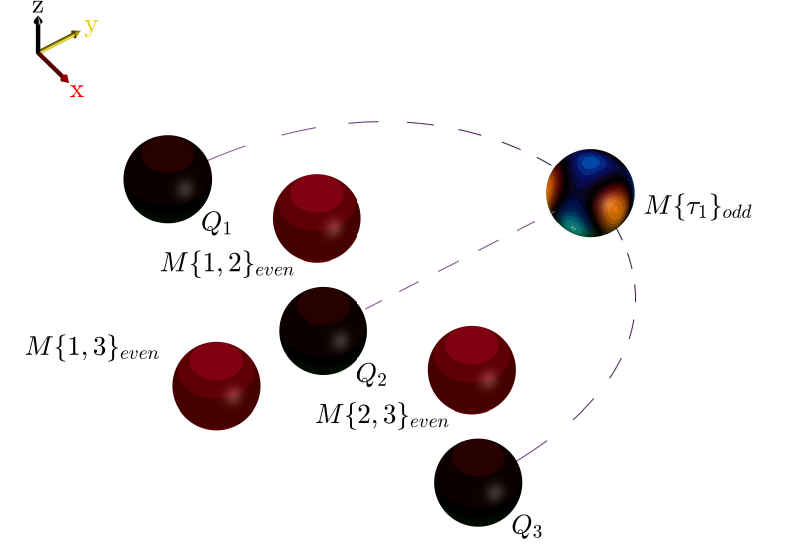}
	\phantomsubcaption
	\label{fig:ghz_w_mixed_0.627_M}
\end{minipage}
\vspace{-.3cm}
\caption{Left: Comparison of different entanglement measures and quantifiers for states $\rho_\text{GHZ-W}(p)$. Right: Generalized BEADS representation of the state $\rho_\text{GHZ-W}(p\approx0.627)$. The state has a zero-valued three-tangle while being genuinely tripartite entangled (nonzero $\mathcal{W}_3$-arc) and thus, belongs to the W mixed state class.}
\label{fig:GHZWMixOverview}
\vspace{-2cm}
\end{figure}

The case $p=0.5$ marks the point, where the structural GHZ state features become evidently dominant. For any $p>0.5$, the bipartite $M$-Beads adopt surface patterns that have maximum spherical function values along the $z$-axis, while at the same time, the subsystems stay fully separable. It is noteworthy though, that the three-tangle of $\rho_\text{GHZ-W}$ is not monotonically increasing with $p$. It shows a non-smooth behavior and increases or decreases for increasing $p$ in certain intervals as can be seen in \cref{fig:GHZWMixOverview}. Crucially, the state obtained for $p \approx 0.627$ has a vanishing three-tangle but nonzero $\mathcal{W}_3$. Hence, this state belongs to the W mixed state class and highlights the important property that mixed states do not necessarily belong to the GHZ mixed state class just because one possible pure state decomposition contains a GHZ state.
\newpage

At $p=0.8$ the $M\{\tau_1\}_{odd}$-Bead already closely resembles the ideal tripartite GHZ state $M$-Bead (see \cref{sec:ghz_w_pure}) and the three-tangle strongly increases. An alternative representation that gives a different perspective on the BEADS representation of the discussed states $\rho_\text{GHZ-W}$ is given in the supplementary radial plot depicted in \cref{fig:ghz_w_mixed_mixtures_radial}.

\begin{figure}[htbp] 
	\centering
	\makebox[\textwidth][c]{
		\begin{minipage}{1\textwidth}
			\begin{subfigure}[t]{0.5\textwidth}
				\centering
				\includegraphics[width=1\textwidth, trim={7.7cm 5.2cm 7.3cm 6.6cm},clip]{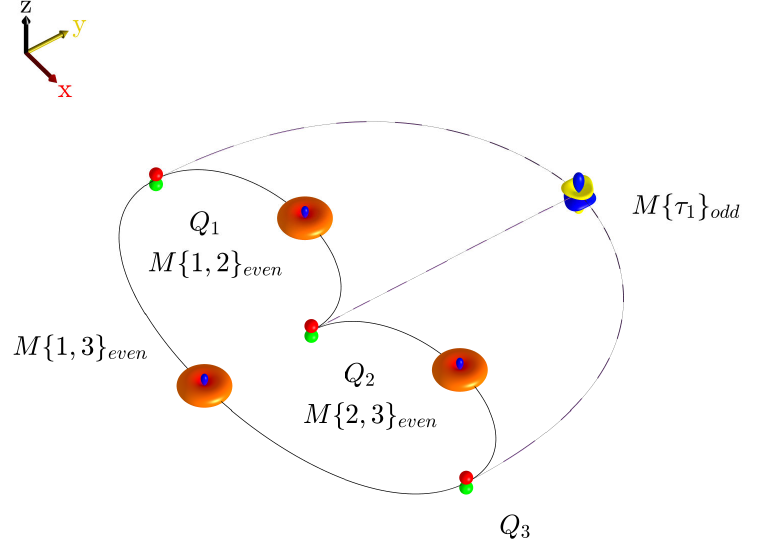}
				\subcaption{$p = 0.05$}
				\label{fig:WGHZ0.05_M_radial}
			\end{subfigure}
			\begin{subfigure}[t]{0.5\textwidth}
				\centering
				\includegraphics[width=1\textwidth, trim={7.7cm 5.2cm 7.3cm 6.6cm},clip]{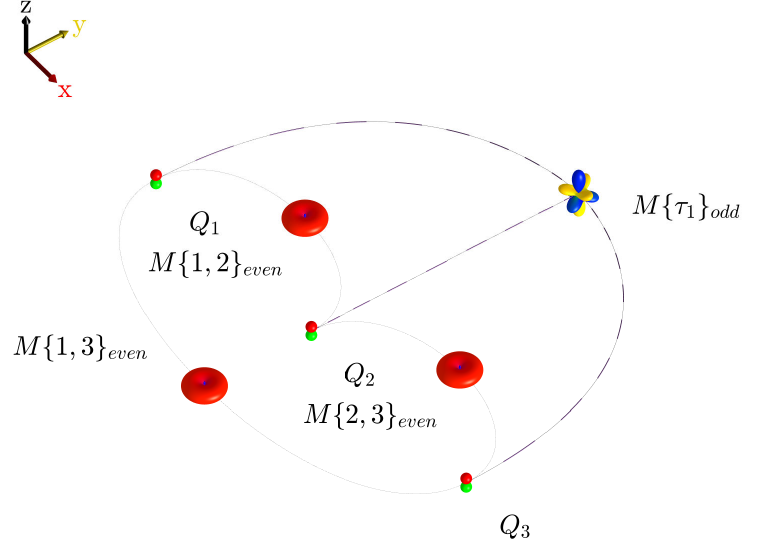}
				\subcaption{$p = 0.2$}
				\label{fig:WGHZ0.2_M_radial}
			\end{subfigure}
		\end{minipage}
	}
	
	\vspace{0.3cm}
	
	\makebox[\textwidth][c]{
		\begin{minipage}{1\textwidth}
			\begin{subfigure}[t]{0.5\textwidth}
				\centering
				\includegraphics[width=1\textwidth, trim={7.7cm 5.2cm 7.3cm 6.6cm},clip]{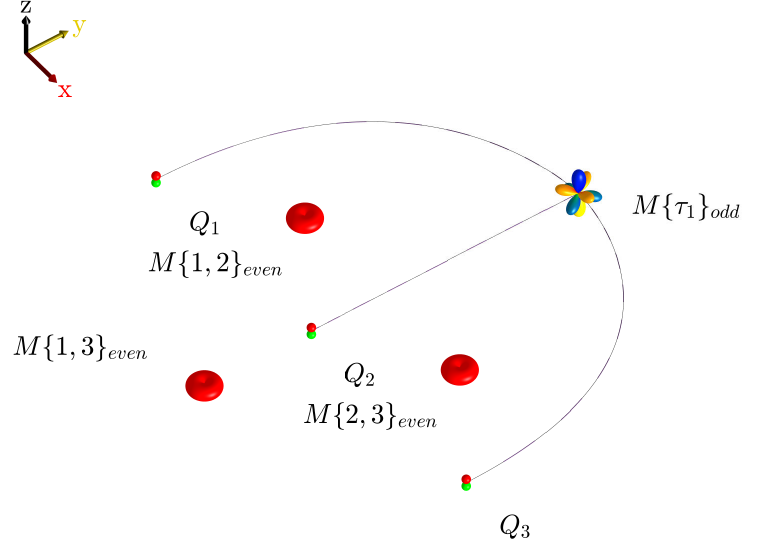}
				\subcaption{$p = 0.35$}
				\label{fig:WGHZ0.35_M_radial}
			\end{subfigure}
			\begin{subfigure}[t]{.5\textwidth}
				\centering
				\includegraphics[width=1\textwidth, trim={7.7cm 5.2cm 7.3cm 6.6cm},clip]{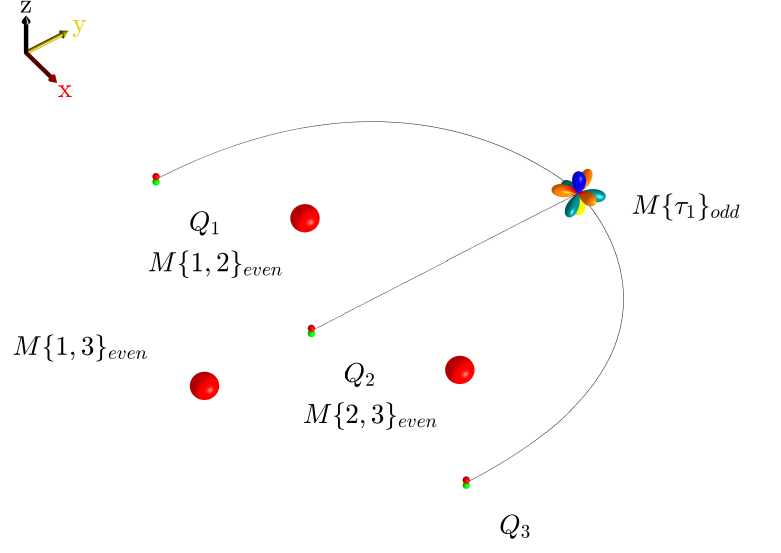}
				\subcaption{$p = 0.5$}
				\label{fig:WGHZ0.5_M_radial}
			\end{subfigure}
		\end{minipage}
	}
	
	\vspace{0.3cm}
	
	\makebox[\textwidth][c]{
		\begin{minipage}{1\textwidth}
			\begin{subfigure}[t]{0.5\textwidth}
				\centering
				\includegraphics[width=1\textwidth, trim={7.7cm 5.2cm 7.3cm 6.6cm},clip]{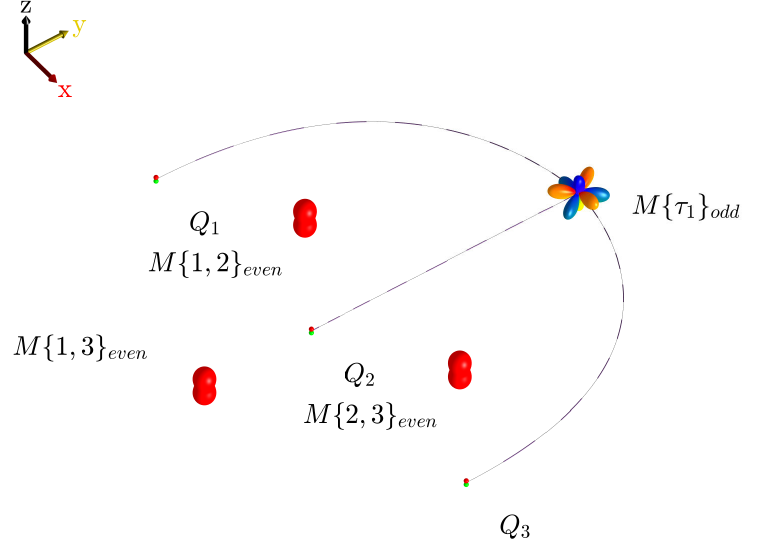}
				\subcaption{$p = 0.65$}
				\label{fig:WGHZ0.65_M_radial}
			\end{subfigure}
			\begin{subfigure}[t]{0.5\textwidth}
				\centering
				\includegraphics[width=1\textwidth, trim={7.7cm 5.2cm 7.3cm 6.6cm},clip]{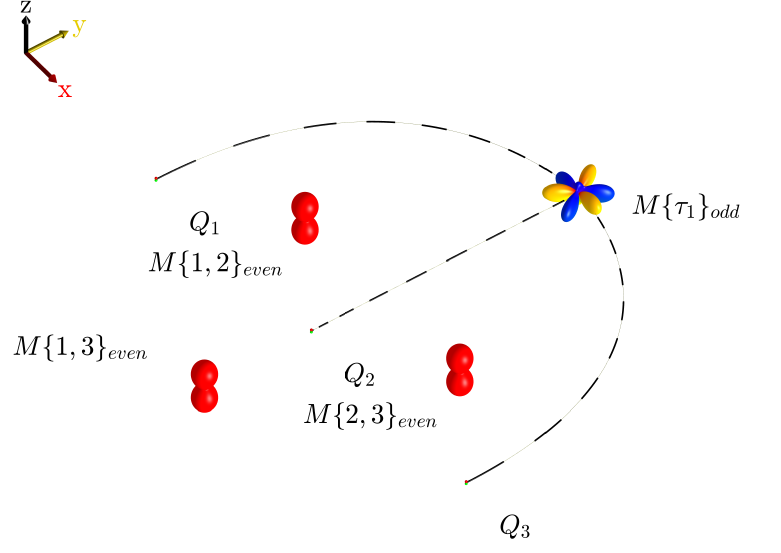}
				\subcaption{$p = 0.8$}
				\label{fig:WGHZ0.8_M_radial}
			\end{subfigure}
		\end{minipage}
	}
	\caption{Complementary radial plotting variant of \cref{fig:ghz_w_mixed_mixtures} that provides enhanced visibility of characteristic bead shapes (corresponding to surface patterns in the standard BEADS plots).}
	\label{fig:ghz_w_mixed_mixtures_radial}
\end{figure}

\clearpage
\subsection{Coherent mixtures of GHZ and W state}\label{sup:GHZWCoherent}
\begin{figure}[H]
\centering
	\includegraphics[width=0.55\textwidth]{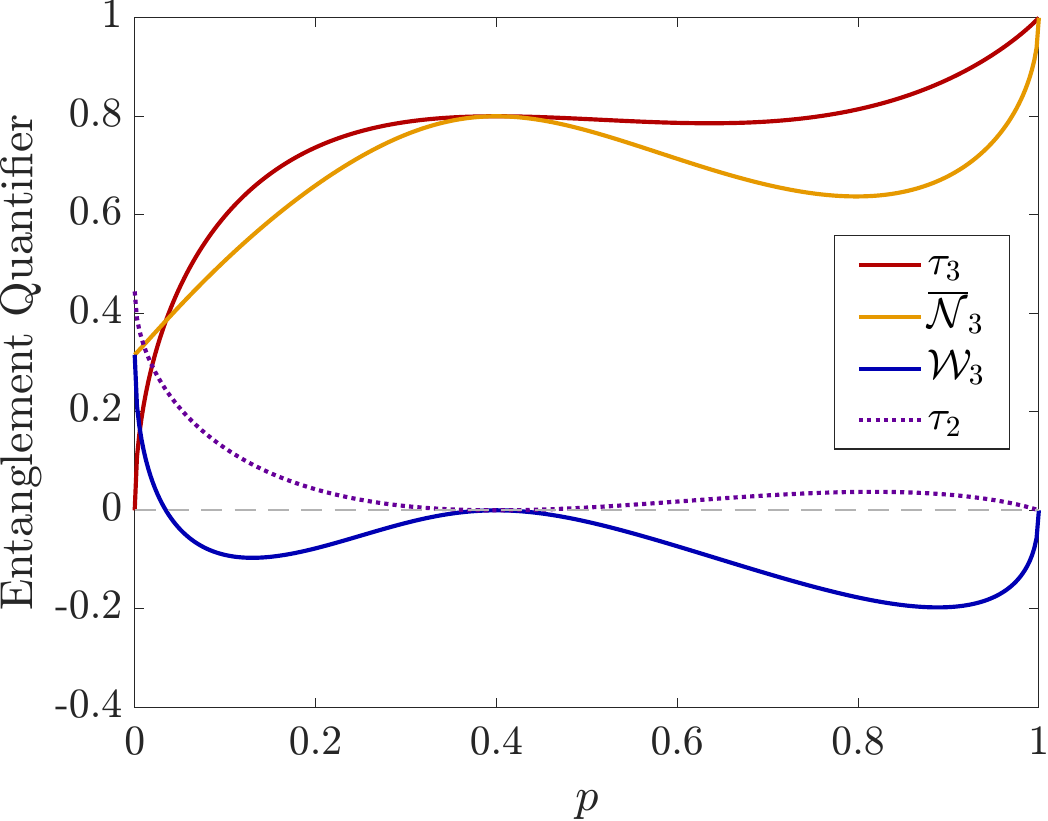}
	\caption{Comparison of different entanglement measures and quantifiers for coherently mixed pure states $\ket{\psi_\text{GHZ-W,coh.}}(p)$ (see \cref{eq:ghz_w_coherent}). The three-tangle increases monotonically with increasing admixture weight $p$. At $p=0.4$, the bipartite marginals of the corresponding state become fully separable ($\tau_2 = 0$).}
	\label{fig:GHZWCohOverview}
\end{figure}

\begin{figure}[htbp] 
	\centering
	\makebox[\textwidth][c]{
		\begin{minipage}{1\textwidth}
			\begin{subfigure}[t]{0.5\textwidth}
				\centering
				\includegraphics[width=1\textwidth, trim={7.7cm 5.2cm 7.3cm 6.6cm},clip]{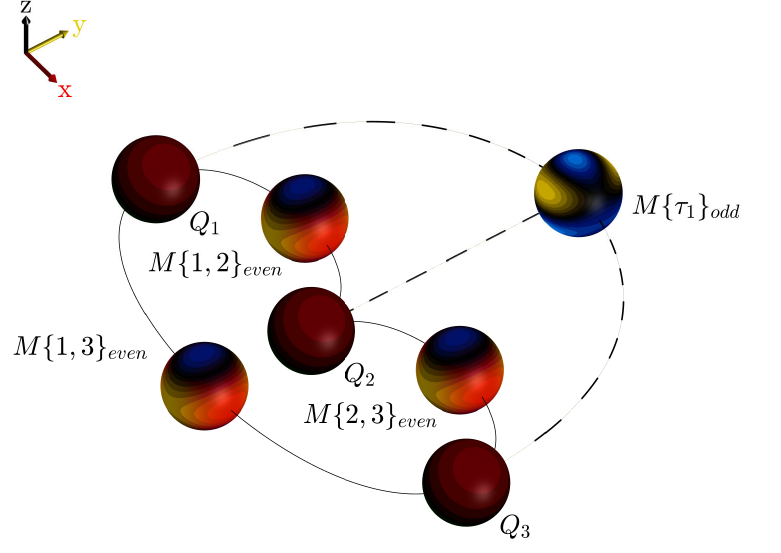}
				\subcaption{$p = 0.05$}
				\label{fig:ghz_w_coherent_0.05_M}
			\end{subfigure}
			\begin{subfigure}[t]{0.5\textwidth}
				\centering
				\includegraphics[width=1\textwidth, trim={7.7cm 5.2cm 7.3cm 6.6cm},clip]{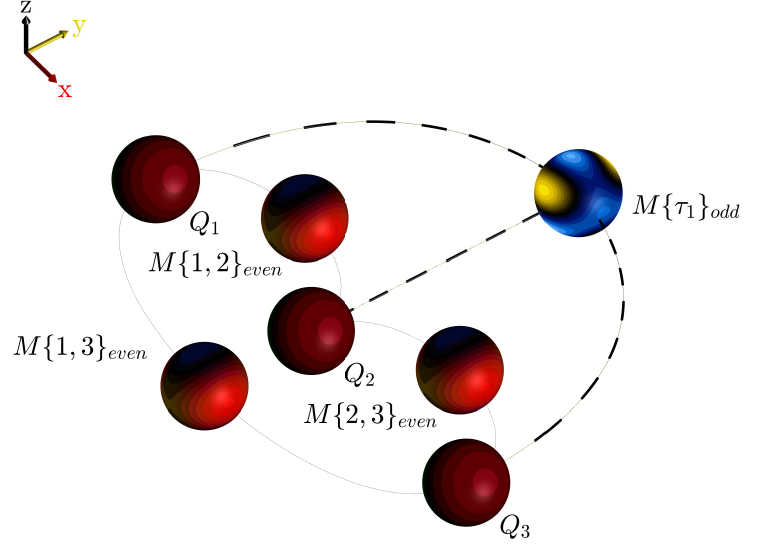}
				\subcaption{$p = 0.2$}
				\label{fig:ghz_w_coherent_0.2_M}
			\end{subfigure}
		\end{minipage}
	}
	
	\vspace{0.3cm}
	
	\makebox[\textwidth][c]{
		\begin{minipage}{1\textwidth}
			\begin{subfigure}[t]{0.5\textwidth}
				\centering
				\includegraphics[width=1\textwidth, trim={7.7cm 5.2cm 7.3cm 6.6cm},clip]{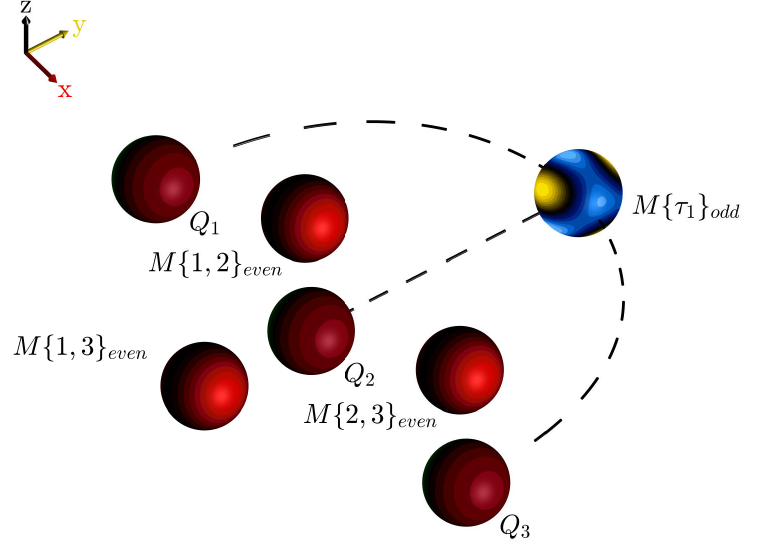}
				\subcaption{$p = 0.4$}
				\label{fig:ghz_w_coherent_0.4_M}
			\end{subfigure}
			\begin{subfigure}[t]{0.5\textwidth}
				\centering
				\includegraphics[width=1\textwidth, trim={7.7cm 5.2cm 7.3cm 6.6cm},clip]{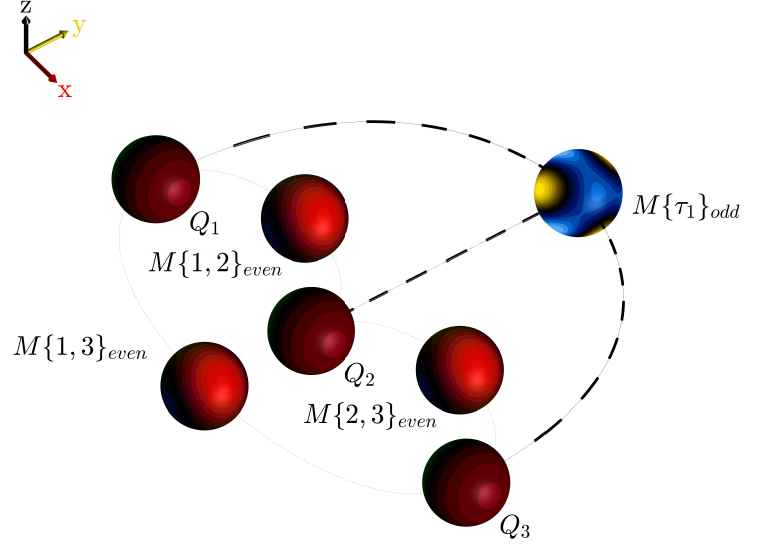}
				\subcaption{$p = 0.6$}
				\label{fig:ghz_w_coherent_0.6_M}
			\end{subfigure}
		\end{minipage}
	}
	
	\vspace{0.3cm}
	
	\makebox[\textwidth][c]{
		\begin{minipage}{1\textwidth}
			\begin{subfigure}[t]{0.5\textwidth}
				\centering
				\includegraphics[width=1\textwidth, trim={7.7cm 5.2cm 7.3cm 6.6cm},clip]{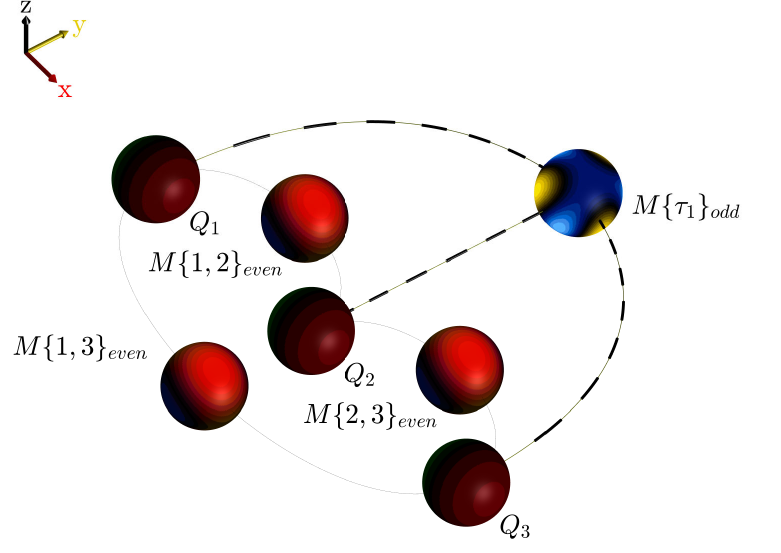}
				\subcaption{$p = 0.8$}
				\label{fig:ghz_w_coherent_0.8_M}
			\end{subfigure}
			\begin{subfigure}[t]{0.5\textwidth}
				\centering
				\includegraphics[width=1\textwidth, trim={7.7cm 5.2cm 7.3cm 6.6cm},clip]{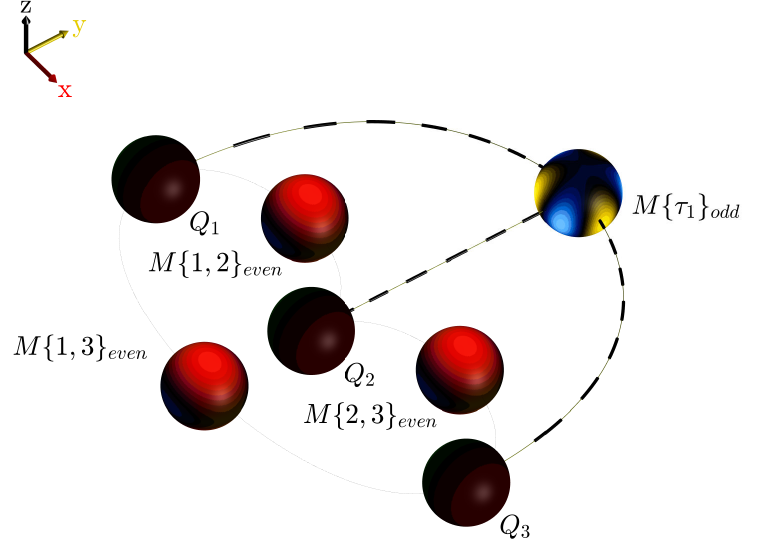}
				\subcaption{$p = 0.95$}
				\label{fig:ghz_w_coherent_0.95_M}
			\end{subfigure}
		\end{minipage}
	}
	\caption{Generalized BEADS representation of the parameterized coherent mixture of tripartite GHZ and W states $\ket{\psi_\text{GHZ-W,coh.}}(p)$ (see \cref{eq:ghz_w_coherent}). Here, the visualization displays the total correlation properties in terms of $M$-Beads. All shown states belong to the GHZ class. The W-type structural features, especially with respect to the tripartite $M\{\tau_1\}_{odd}$, disappear already at small admixture weights such as $p=0.05$. The bipartite entanglement vanishes specifically at $p=0.4$, where the state is structurally ideally GHZ-entangled with fully separable, classically correlated bipartite marginals and a fully permutation symmetric tripartite $M$-Bead with three-fold rotational symmetry. States with $0.4 < p < 1$ show weak bipartite entanglement but the bead surface patterns are close to ideally GHZ-entangled states, respectively.}
	\label{fig:ghz_w_coherent}
\end{figure}

\begin{figure}[htbp] 
	\centering
	\makebox[\textwidth][c]{
		\begin{minipage}{1\textwidth}
			\begin{subfigure}[t]{0.5\textwidth}
				\centering
				\includegraphics[width=1\textwidth, trim={7.7cm 5.2cm 7.3cm 6.6cm},clip]{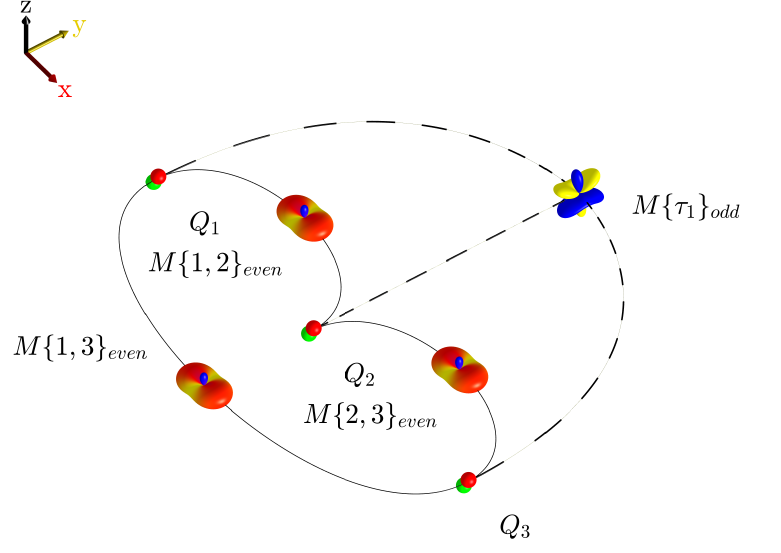}
				\subcaption{$p = 0.05$}
				\label{fig:WGHZ0.05coh_M_radial}
			\end{subfigure}
			\begin{subfigure}[t]{0.5\textwidth}
				\centering
				\includegraphics[width=1\textwidth, trim={7.7cm 5.2cm 7.3cm 6.6cm},clip]{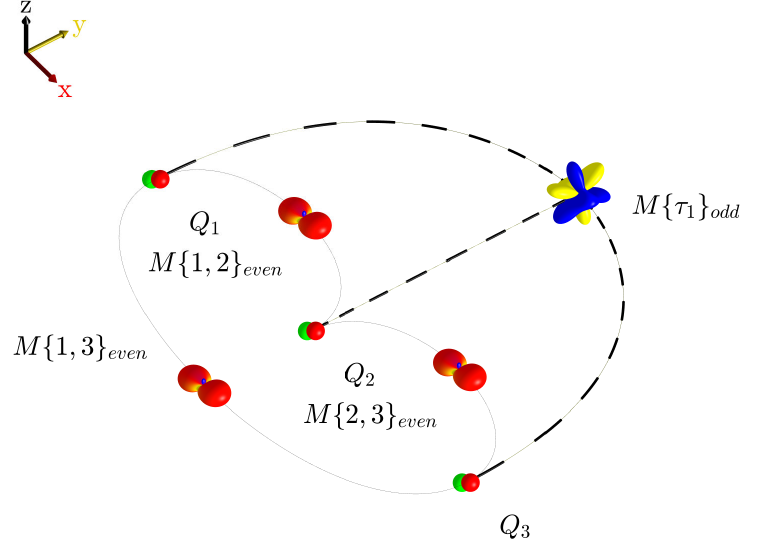}
				\subcaption{$p = 0.2$}
				\label{fig:WGHZ0.2coh_M_radial}
			\end{subfigure}
		\end{minipage}
	}
	
	\vspace{0.3cm}
	
	\makebox[\textwidth][c]{
		\begin{minipage}{1\textwidth}
			\begin{subfigure}[t]{0.5\textwidth}
				\centering
				\includegraphics[width=1\textwidth, trim={7.7cm 5.2cm 7.3cm 6.6cm},clip]{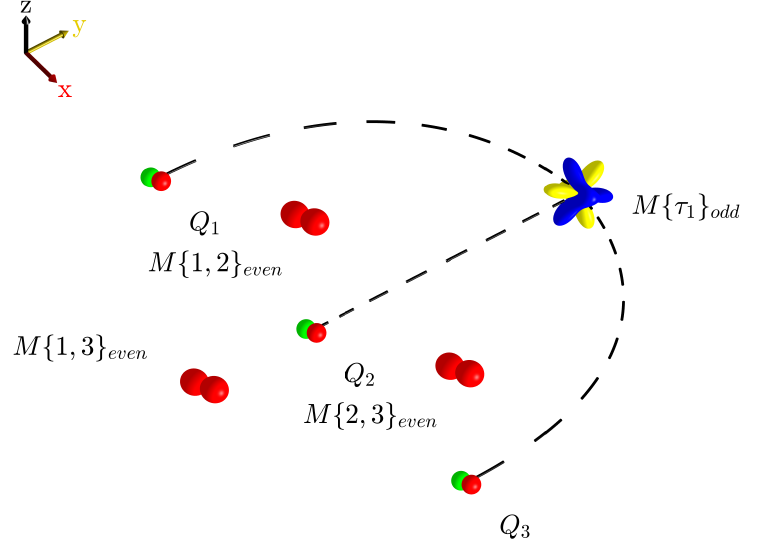}
				\subcaption{$p = 0.4$}
				\label{fig:WGHZcoh0.4_M_radial}
			\end{subfigure}
			\begin{subfigure}[t]{.5\textwidth}
				\centering
				\includegraphics[width=1\textwidth, trim={7.7cm 5.2cm 7.3cm 6.6cm},clip]{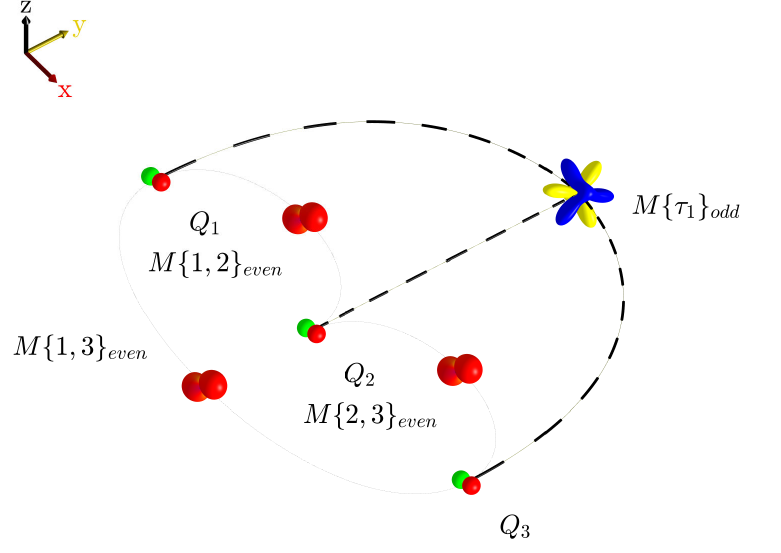}
				\subcaption{$p = 0.6$}
				\label{fig:WGHZcoh0.6_M_radial}
			\end{subfigure}
		\end{minipage}
	}
	
	\vspace{0.3cm}
	
	\makebox[\textwidth][c]{
		\begin{minipage}{1\textwidth}
			\begin{subfigure}[t]{0.5\textwidth}
				\centering
				\includegraphics[width=1\textwidth, trim={7.7cm 5.2cm 7.3cm 6.6cm},clip]{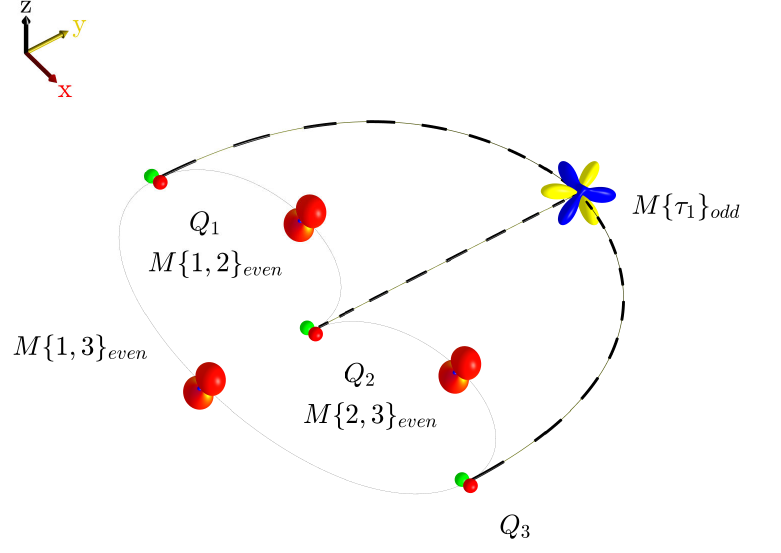}
				\subcaption{$p = 0.8$}
				\label{fig:WGHZcoh0.8_M_radial}
			\end{subfigure}
			\begin{subfigure}[t]{0.5\textwidth}
				\centering
				\includegraphics[width=1\textwidth, trim={7.7cm 5.2cm 7.3cm 6.6cm},clip]{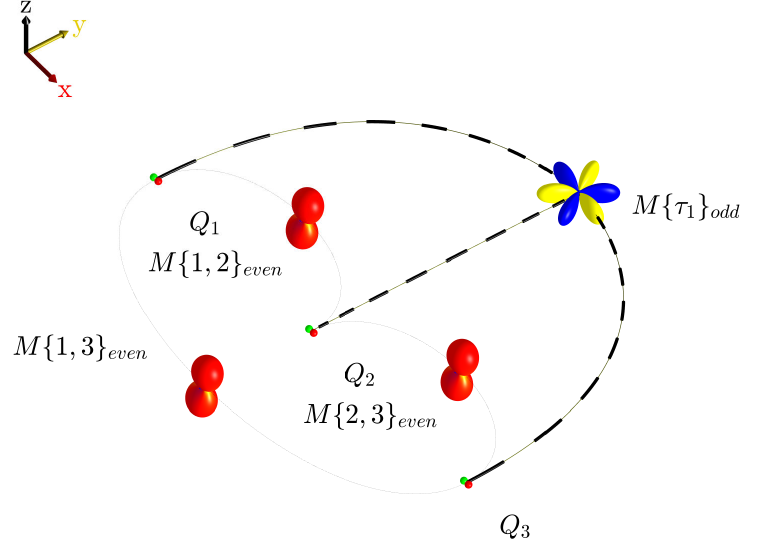}
				\subcaption{$p = 0.95$}
				\label{fig:WGHZcoh0.95_M_radial}
			\end{subfigure}
		\end{minipage}
	}
	\caption{Complementary radial plotting variant of \cref{fig:ghz_w_coherent} that provides enhanced visibility of characteristic bead shapes (corresponding to surface patterns in the standard BEADS plots).}
	\label{fig:ghz_w_coherent_mixtures_radial}
\end{figure}

The pure tripartite state
\begin{equation}
	\ket{\psi_\text{GHZ-W,coh.}}(p) = \sqrt{p}  \ket{\text{GHZ}^{+}} + \sqrt{1-p} \ket{\text{W}}
	\label{eq:ghz_w_coherent}
\end{equation}
describes a parameterized coherent mixture of the GHZ and the W state. Note the difference to the state discussed in \cref{sup:GHZWIncoherent}, which represented an incoherent mixture of the GHZ and W states, that is, a mixed state. Generalized BEADS representations of states $\ket{\psi_\text{GHZ-W,coh.}}(p)$ for different admixture weights $p$ are shown in \cref{fig:ghz_w_coherent} and using the radial plotting variant in \cref{fig:ghz_w_coherent_mixtures_radial}. Clearly, all pure states for $p\neq 0$ show a three-tangle arc and thus, belong to the GHZ SLOCC class. This is further highlighted in \cref{fig:GHZWCohOverview} which provides an overview of the measures and quantifiers underlying the entanglement arcs in the corresponding generalized BEADS representations.

For $p=0.05$, the state already exhibits a pronounced influence from the GHZ state. The characteristic axially symmetric surface patterns of the W state are not recognizable anymore. Thus, unlike in the previous case of an incoherent GHZ-W mixture, adding a marginally small GHZ contribution strongly suppresses the W state features. At this mixing ratio, both types of tripartite entanglement arcs are observable. Notably, the three-tangle is already large given the small GHZ weight, whereas the $\mathcal{W}_3$ arc is small indicating structural proximity to an ideal GHZ state. The bipartite subsystems remain entangled at $p=0.05$. 

With increasing $p$, the entanglement of the bipartite subsystems gradually decreases and vanishes selectively at $p=0.4$ (see \cref{fig:ghz_w_coherent_0.2_M} and \cref{fig:ghz_w_coherent_0.4_M}). At this point, the tripartite entanglement is strictly GHZ-type as evidenced by the disappearance of the $\mathcal{W}_3$ and a pronounced three-tangle arc. Thus, with the bipartite marginals being fully separable (and in fact only classically correlated), the state for $p=0.4$ is structurally ideally GHZ-entangled. 

As $p$ further increases, subtle W-type structural features reemerge (compare \cref{fig:ghz_w_coherent_0.6_M} and \cref{fig:ghz_w_coherent_0.8_M}). In particular, weak bipartite entanglement and nonclassical bipartite correlation structures indicated by slightly distorted surface patterns of the bipartite $M$-Beads become apparent. Both types of tripartite entanglement arcs are visible. 

For $p=0.95$, the state closely resembles the $\text{GHZ}$-state. The tripartite $M\{\tau_1\}_{odd}$-bead displays a characteristic three-fold rotational symmetry and the bipartite marginals show correlations predominantly along the $z$-axis (see \cref{fig:ghz_w_coherent_0.95_M}). Compared to the incoherent mixture discussed in the preceeding section, coherent mixtures of the GHZ and W states are structurally strongly dominated by the GHZ state for all mixing ratios $p \neq 0$. 

\clearpage
\section{Addendum to \cref{app:benchmarks}: $E$-Bead visualizations of parameterized tripartite noisy GHZ states} \label{sup:Benchmark}
\begin{figure}[H] 
	\centering
	\makebox[\linewidth][c]{
		\begin{minipage}{1\textwidth}
			\begin{subfigure}[t]{0.48\textwidth}
				\centering
				\includegraphics[height=6cm,trim={6cm 3.5cm 5cm 5cm},clip]{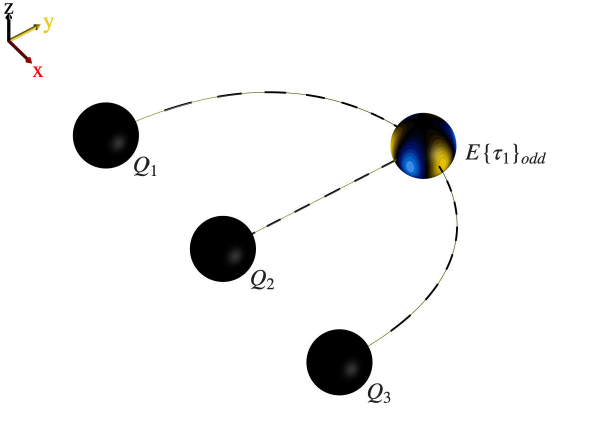}
				\subcaption{$p = 0.1$}
				\label{fig:GHZ_noise0.1_E}
			\end{subfigure}
			\hspace{0.4cm}
			\begin{subfigure}[t]{0.48\textwidth}
				\centering
				\includegraphics[height=6cm,trim={6cm 3.5cm 5cm 5cm},clip]{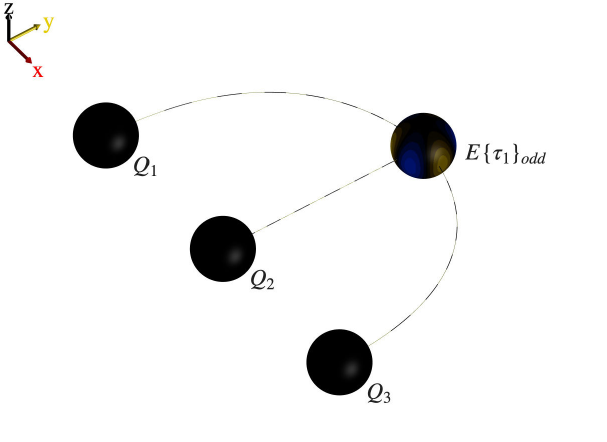}
				\subcaption{$p = 0.3$}
				\label{fig:GHZ_noise0.3_E}
			\end{subfigure}
		\end{minipage}
	}
	
	\vspace{0.3cm}
	
	\makebox[\linewidth][c]{
		\begin{minipage}{1\textwidth}
			\begin{subfigure}[t]{0.48\textwidth}
				\centering
				\includegraphics[height=6cm,trim={6cm 3.5cm 5cm 5cm},clip]{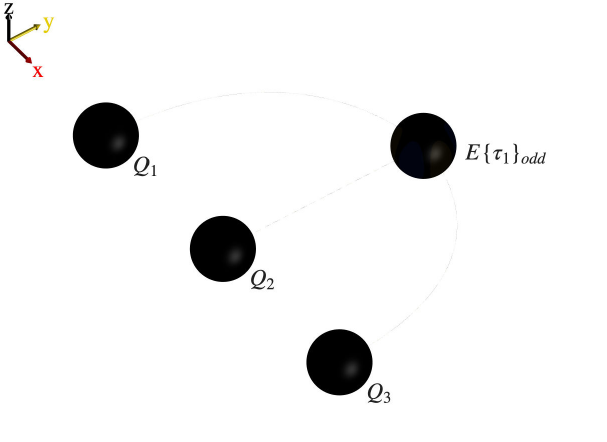}
				\subcaption{$p = 0.5$}
				\label{fig:GHZ_noise0.5_E}
			\end{subfigure}
			\hspace{0.4cm}
			\begin{subfigure}[t]{0.48\textwidth}
				\centering
				\includegraphics[height=6cm,trim={6cm 3.5cm 5cm 5cm},clip]{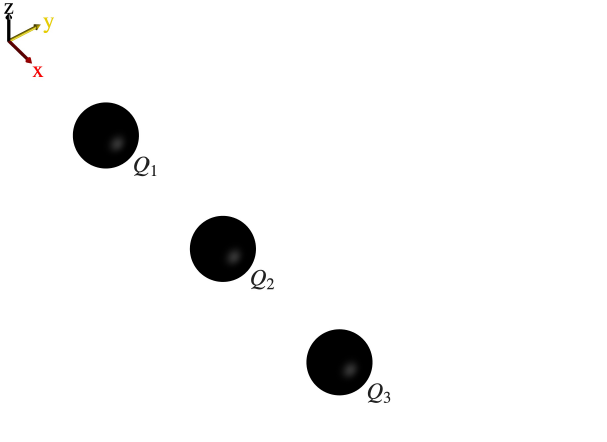}
				\subcaption{$p = 4/7 \approx 0.5714$}
				\label{fig:GHZ_noise0.57143_E}
			\end{subfigure}
		\end{minipage}
	}
	
	\vspace{0.3cm}
	
	\makebox[\linewidth][c]{
		\begin{minipage}{1\textwidth}
			\begin{subfigure}[t]{0.48\textwidth}
				\centering
				\includegraphics[height=6cm,trim={6cm 3.5cm 5cm 5cm},clip]{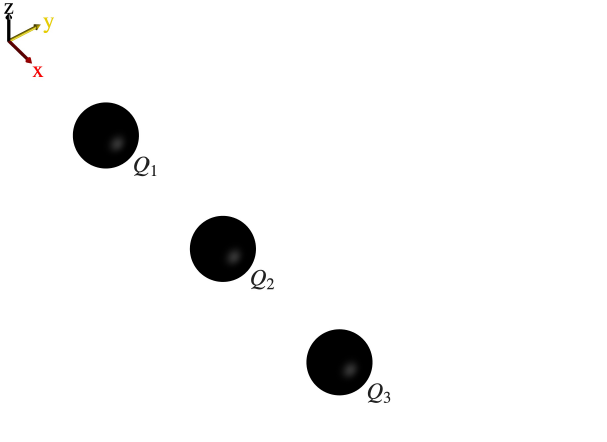}
				\subcaption{$r = 0.7$}
				\label{fig:GHZ_noise0.7_E}
			\end{subfigure}
			\hspace{0.4cm}
			\begin{subfigure}[t]{0.48\textwidth}
				\centering
				\includegraphics[height=6cm,trim={6cm 3.5cm 5cm 5cm},clip]{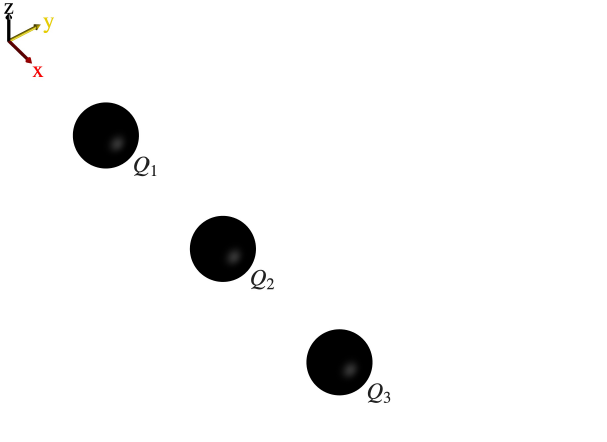}
				\subcaption{$r = 0.9$}
				\label{fig:GHZ_noise0.9_E}
			\end{subfigure}
		\end{minipage}
	}
	\caption{$E$-Bead representation of the noisy GHZ state $\rho(p) = (1-p)\ket{\text{GHZ}}\bra{\text{GHZ}} + p \mathbbm{1}_8$. As is directly clear from the $E\{\tau_1\}_{odd}$-Beads and the entanglement arcs, the state is genuinely tripartite entangled only for $p \leq 4/7$. The bipartite marginals are not entangled for any $p$. All properties are discussed in literature \cite{duer_1999, duer_2000, guhne_gme_2010} and correctly reproduced by the generalized BEADS representation and the underlying optimization framework.}
	\label{fig:benchmark_noisy_ghz_E}
	\vspace{-2cm}
\end{figure}

\begin{figure}[H] 
	\centering
	\vspace{-0.7cm}
	\makebox[\linewidth][c]{
		\begin{minipage}{1\textwidth}
			\hspace*{-1cm}
			\begin{subfigure}[t]{0.5\textwidth}
				\centering
				\includegraphics[height=6cm,trim={6cm 5.3cm 5.3cm 6cm},clip]{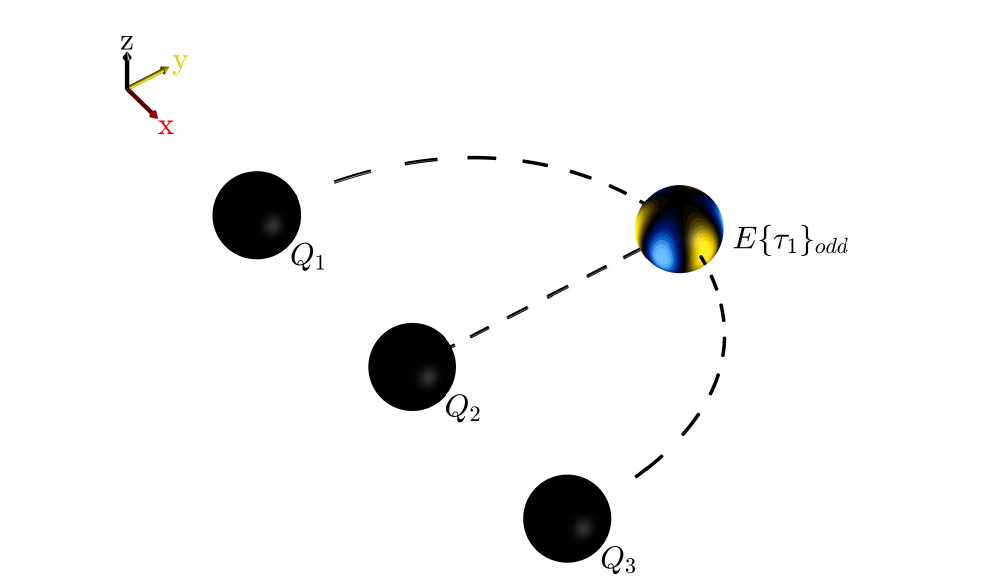}
				\subcaption{$r = 0.0$}
				\label{fig:GHZ_class0_E}
			\end{subfigure}
			\hspace{0.8cm}
			\begin{subfigure}[t]{0.5\textwidth}
				\centering
				\includegraphics[height=6cm,trim={6cm 5.3cm 5.3cm 6cm},clip]{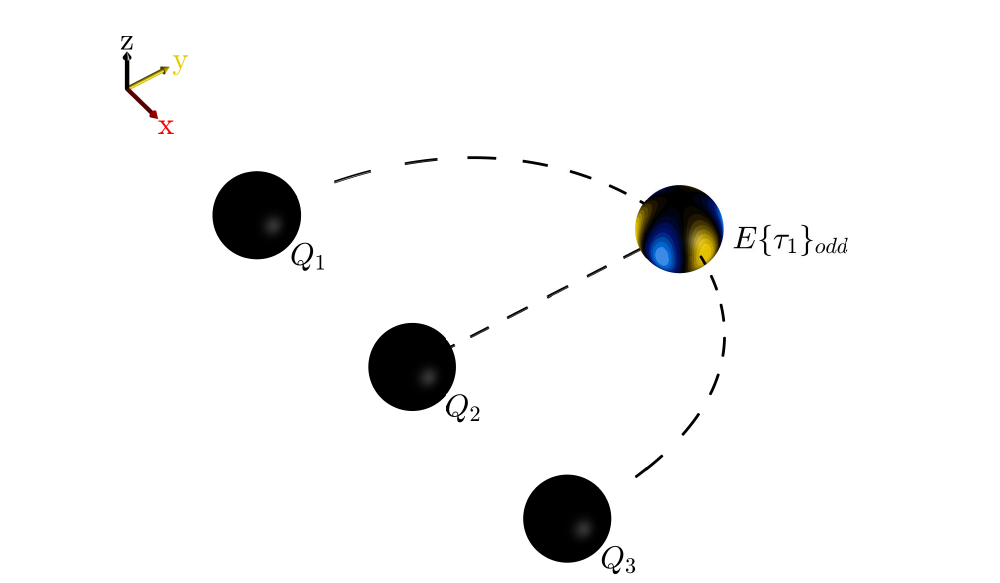}
				\subcaption{$r = 0.2$}
				\label{fig:GHZ_class0.2_E}
			\end{subfigure}
		\end{minipage}
	}
	
	\vspace{0.3cm}
	
	\makebox[\linewidth][c]{
		\begin{minipage}{1\textwidth}
			\hspace*{-1cm}
			\begin{subfigure}[t]{0.5\textwidth}
				\centering
				\includegraphics[height=6cm,trim={6cm 5.3cm 5.3cm 6cm},clip]{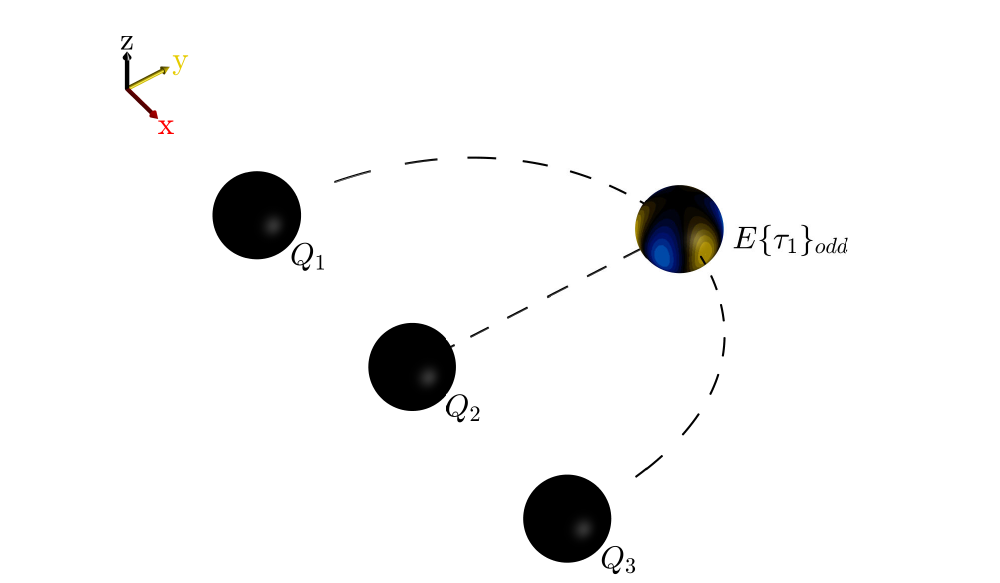}
				\subcaption{$r = 0.4$}
				\label{fig:GHZ_class0.4_E}
			\end{subfigure}
			\hspace{0.8cm}
			\begin{subfigure}[t]{0.5\textwidth}
				\centering
				\includegraphics[height=6cm,trim={6cm 5.3cm 5.3cm 6cm},clip]{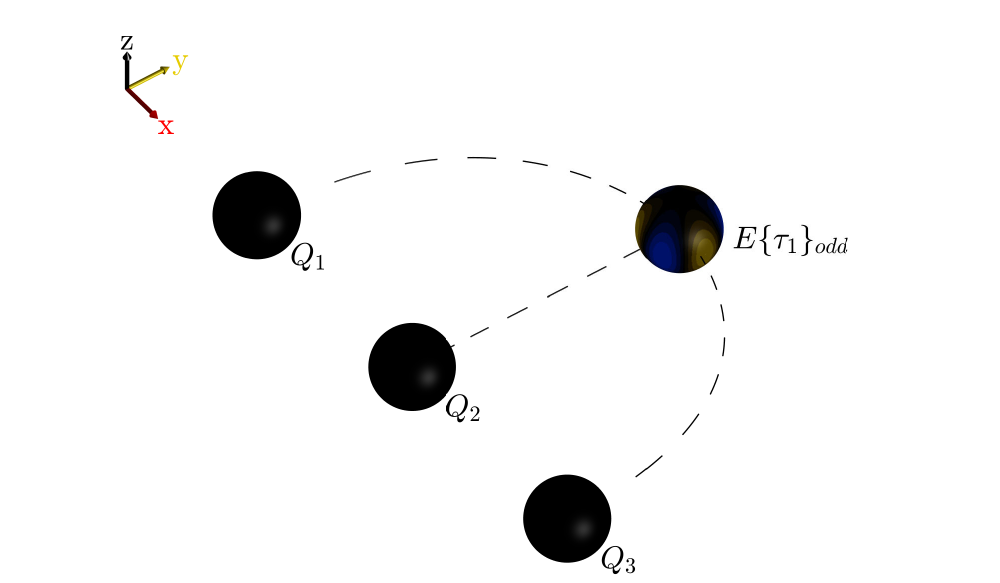}
				\subcaption{$r = 0.6$}
				\label{fig:GHZ_class0.6_E}
			\end{subfigure}
		\end{minipage}
	}
	
	
	\makebox[\linewidth][c]{
		\begin{minipage}{1\textwidth}
			\hspace{-1cm}
			\begin{subfigure}[t]{0.5\textwidth}
				\centering
				\includegraphics[height=6cm,trim={6cm 5.3cm 5.3cm 6cm},clip]{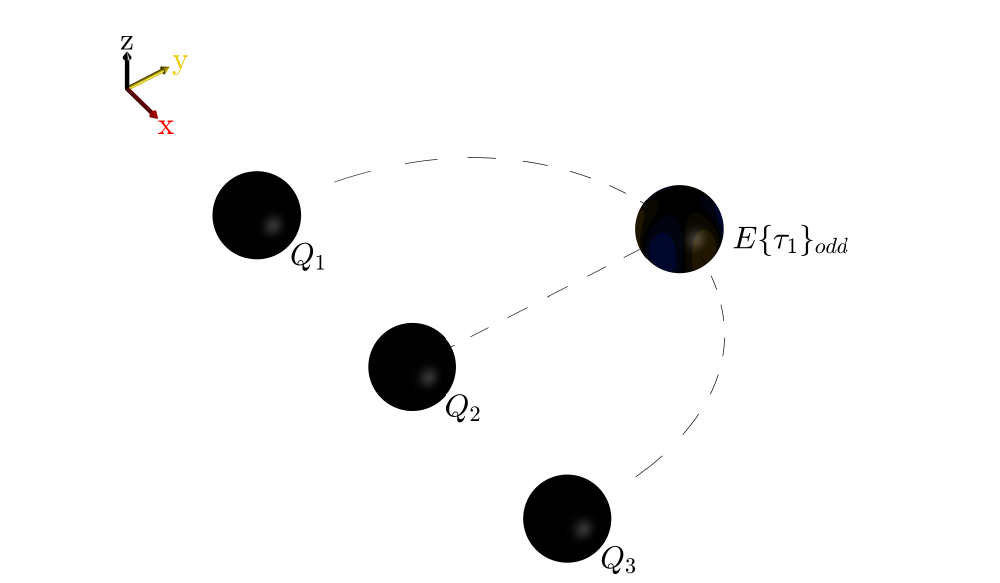}
				\subcaption{$r = 0.8$}
				\label{fig:GHZ_class0.8_E}
			\end{subfigure}
			\hspace{0.8cm}
			\begin{subfigure}[t]{0.5\textwidth}
				\centering
				\includegraphics[height=6cm,trim={6cm 5.3cm 5.3cm 6cm},clip]{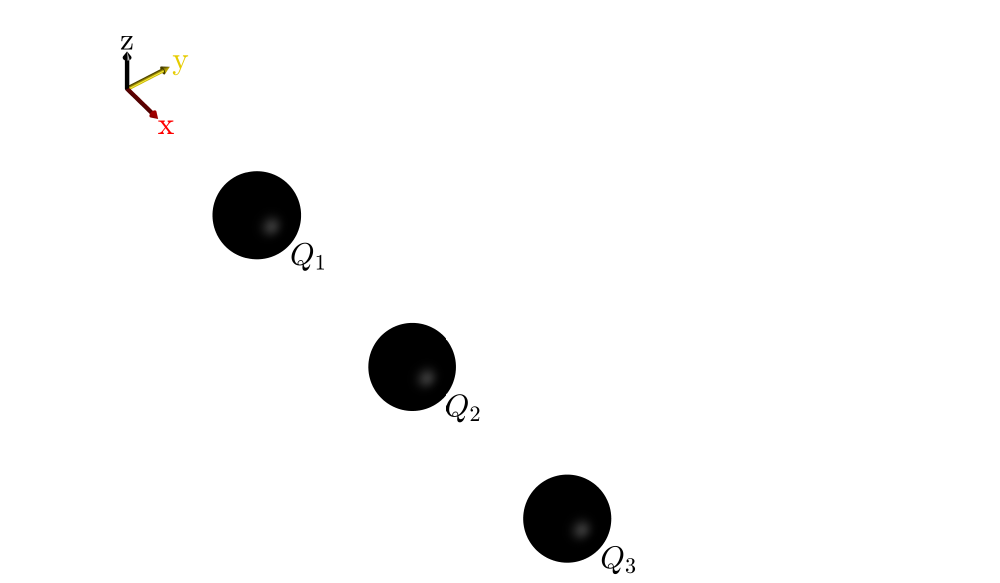}
				\subcaption{$r = 1.0$}
				\label{fig:GHZ_class1.0_E}
			\end{subfigure}
		\end{minipage}
	}
	\caption{$E$-Bead representation of the noisy GHZ state $\rho(r) = (1-r)\ket{\text{GHZ}}\bra{\text{GHZ}} + r/2 (\ket{000}\bra{000}+\ket{111}\bra{111})$. As is directly clear from the $E$-Bead and the entanglement arcs, the state is genuinely tripartite entangled for all $r \neq 1.0$, whereas the bipartite marginals are not entanglement independent of $r$. Both of these properties are correctly reproduced by the generalized BEADS representation and the underlying optimization framework.}
	\label{fig:benchmark_noisy_ghz_class_E}
\end{figure}

Here, we show $E$-Bead visualizations complementary to the $M$-Bead representations of the noisy GHZ states discussed in \cref{app:benchmarks}. The following figures include $E$-Bead representations of the GHZ state with added white noise $\rho(p) = (1-p)\ket{\text{GHZ}}\bra{\text{GHZ}} + p \mathbbm{1}_8$ (cf. \cref{fig:benchmark_noisy_ghz_E}) and classical noise $\rho(r) = (1-r)\ket{\text{GHZ}}\bra{\text{GHZ}} + r/2 (\ket{000}\bra{000}+\ket{111}\bra{111})$ (cf. \cref{fig:benchmark_noisy_ghz_class_E}).

\clearpage
\section{Addendum to \cref{sec:DecoherenceMain}: Decoherence processes of GHZ and W states}\label{sup:decoherence}

\subsection{Depolarization}
Besides dephasing, which was discussed in \cref{sec:DecoherenceMain} of the main text, \textit{depolarization} represents another common type of noise. Upon depolarization a state $\rho$ is transformed into the maximally mixed state $\frac{1}{2}\mathbbm{1}_{2^N}$ with probability $p$. The underlying quantum channel is defined as \cite{nielsen_quantum_2010}
\begin{equation}
	\mathcal{E}(\rho)= (1-p) \, \rho+ p \, \frac{\mathbbm{1}}{2} = (1-p) \, \rho + \frac{p}{3} \left(X\rho X+ Y\rho Y+ Z\rho Z\right),
	\label{eq:depolarizing_channel}
\end{equation}
using the single-qubit Kraus operators
\begin{subequations}
	\begin{align}
		E_0 &= \sqrt{1-p}\,\,\mathbbm{1}_2,  \label{eq:kraus_depolarization_E0} \\
		E_1 &= \sqrt{\frac{p}{3}} \, \, X, 
		\label{eq:kraus_depolarization_E1} \\
		E_2 &=\sqrt{\frac{p}{3}} \, \, Y, 
		\label{eq:kraus_depolarization_E2} \\
		E_3 &= \sqrt{\frac{p}{3}}\, \, Z,
		\label{eq:kraus_depolarization_E3}
	\end{align}
\end{subequations}
where $\mathbbm{1}_2$, $X$, $Y$, $Z$, are the single-quit Pauli matrices.

\begin{figure}[htbp] 
	\centering
	\makebox[\textwidth][c]{
		\begin{minipage}{1\textwidth}
			\begin{subfigure}[t]{0.5\textwidth}
				\centering
				\includegraphics[width=\textwidth, trim={6cm 4cm 5.8cm 5cm},clip]{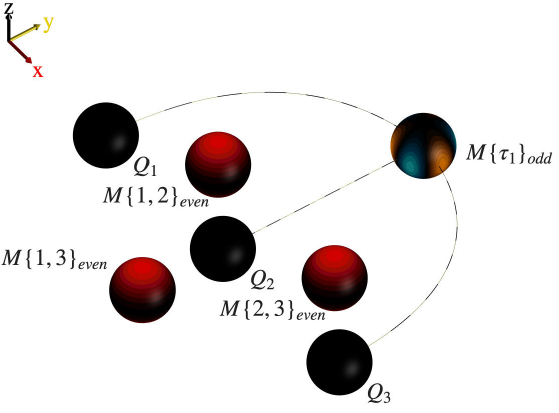}
				\subcaption{$p = 0.1$}
				\label{fig:GHZ_dep0.1_M}
			\end{subfigure}
			\begin{subfigure}[t]{0.5\textwidth}
				\centering
				\includegraphics[width=\textwidth, trim={6cm 4cm 5.8cm 5cm},clip]{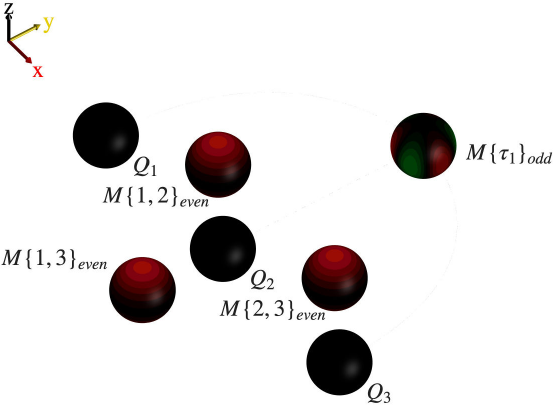}
				\subcaption{$p = 0.2$}
				\label{fig:GHZ_dep0.2_M}
			\end{subfigure}
		\end{minipage}
	}
	
	\vspace{0.3cm}
	
	\makebox[\textwidth][c]{
		\begin{minipage}{1\textwidth}
			\begin{subfigure}[t]{0.5\textwidth}
				\centering
				\includegraphics[width=\textwidth, trim={6cm 4cm 5.8cm 5cm},clip]{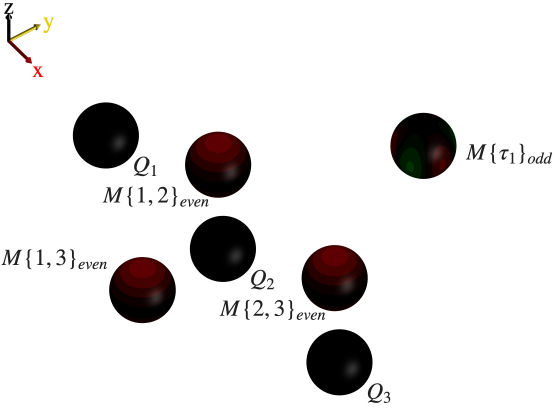}
				\subcaption{$p = 0.3$}
				\label{fig:GHZ_dep0.3_M}
			\end{subfigure}
			\begin{subfigure}[t]{0.48\textwidth}
				\centering
				\includegraphics[width=\textwidth, trim={6cm 4cm 5.8cm 5cm},clip]{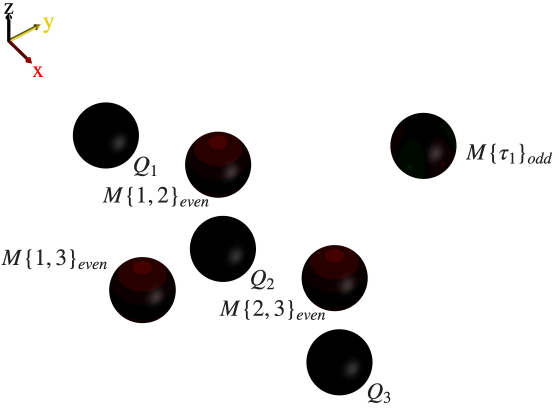}
				\subcaption{$p = 0.4$}
				\label{fig:GHZ_dep0.4_M}
			\end{subfigure}
		\end{minipage}
	}
	
	\vspace{0.3cm}
	
	\makebox[\textwidth][c]{
		\begin{minipage}{1\textwidth}
			\begin{subfigure}[t]{0.5\textwidth}
				\centering
				\includegraphics[width=\textwidth, trim={6cm 4cm 5.8cm 5cm},clip]{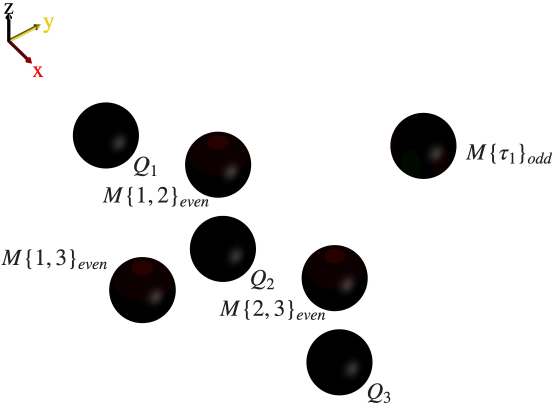}
				\subcaption{$p = 0.5$}
				\label{fig:GHZ_dep0.5_M}
			\end{subfigure}
			\begin{subfigure}[t]{0.48\textwidth}
				\centering
				\includegraphics[width=\textwidth, trim={6cm 4cm 5.8cm 5cm},clip]{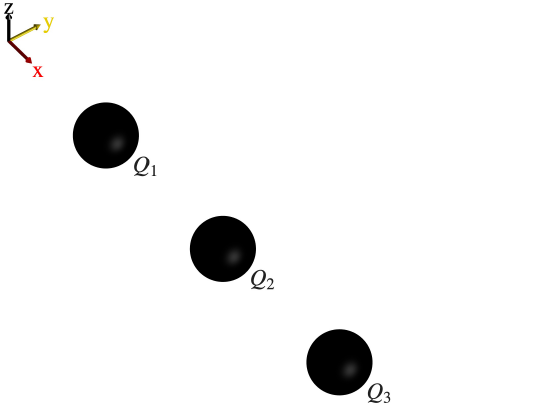}
				\subcaption{$p = 0.75$}
				\label{fig:GHZ_dep0.75_M}
			\end{subfigure}
		\end{minipage}
	}
	\caption{Generalized BEADS representation visualizing the effect of the depolarization channel on the GHZ state (compare \cref{eq:kraus_depolarization_E0}-\cref{eq:kraus_depolarization_E3}). States for different channel parameters $p$ are represented in terms of their total correlation properties using $M$-Beads. Depolarization causes any polarization as well as genuine correlations to vanish with increasing $p$ which directly corresponds to darkening $Q$- and $M$-Beads. In the shown visualizations, the genuine tripartite entanglement is destroyed for $p\geq 0.3$. The final state at $p=0.75$ is the maximally mixed state.}
	\label{fig:ghz_depolarization}
\end{figure}

\cref{fig:ghz_depolarization} illustrates the effect of the depolarization channel on the GHZ state by corresponding generalized BEADS representations. In the following, we assume that the channels are always applied to all qubits simultaneously. It can be seen that for increasing $p$, all $Q$- and $M$-Beads become gradually darker. Thus, any polarization as well as all bipartite and tripartite correlations vanish, resulting in the maximally mixed state for $p=0.75$, which only shows all-black $Q$-Beads. Notably, the state obtained at $p=0.3$ is already not genuinely tripartite entangled anymore which is directly clear in the generalized BEADS representation.

\begin{figure}[htbp] 
	\centering
	\makebox[\textwidth][c]{
		\begin{minipage}{1\textwidth}
			\begin{subfigure}[t]{0.5\textwidth}
				\centering
				\includegraphics[width=\textwidth, trim={6cm 4cm 5.8cm 5cm},clip]{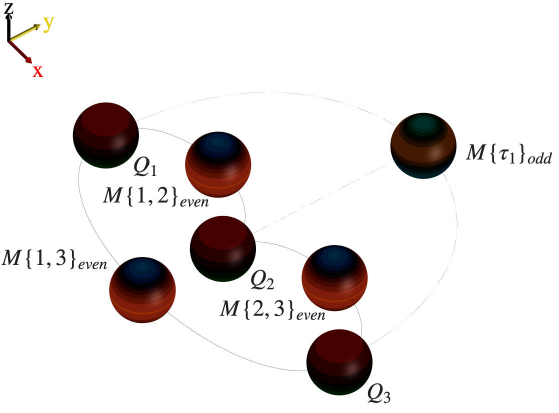}
				\subcaption{$p = 0.1$}
				\label{fig:W_dep0.1_M}
			\end{subfigure}
			\begin{subfigure}[t]{0.5\textwidth}
				\centering
				\includegraphics[width=\textwidth, trim={6cm 4cm 5.8cm 5cm},clip]{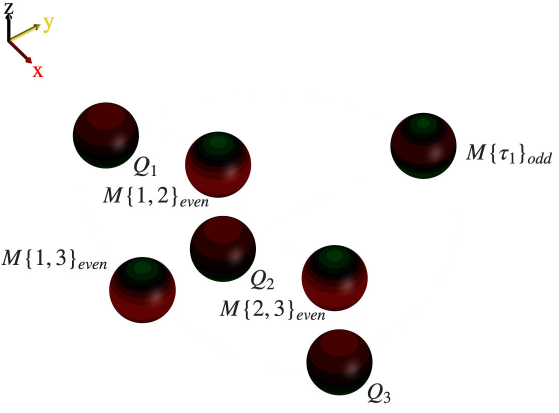}
				\subcaption{$p = 0.2$}
				\label{fig:W_dep0.2_M}
			\end{subfigure}
		\end{minipage}
	}
	
	\vspace{0.3cm}
	
	\makebox[\textwidth][c]{
		\begin{minipage}{1\textwidth}
			\begin{subfigure}[t]{0.5\textwidth}
				\centering
				\includegraphics[width=\textwidth, trim={6cm 4cm 5.8cm 5cm},clip]{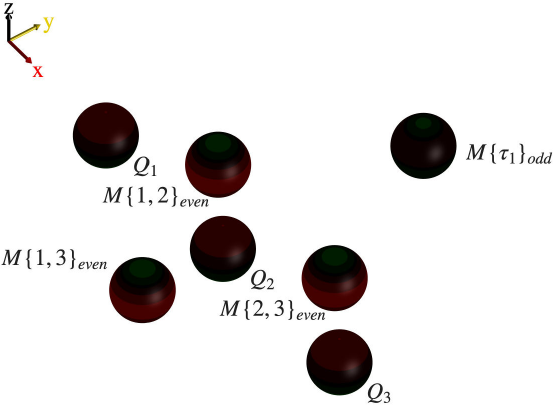}
				\subcaption{$p = 0.3$}
				\label{fig:W_dep0.3_M}
			\end{subfigure}
			\begin{subfigure}[t]{0.5\textwidth}
				\centering
				\includegraphics[width=\textwidth, trim={6cm 4cm 5.8cm 5cm},clip]{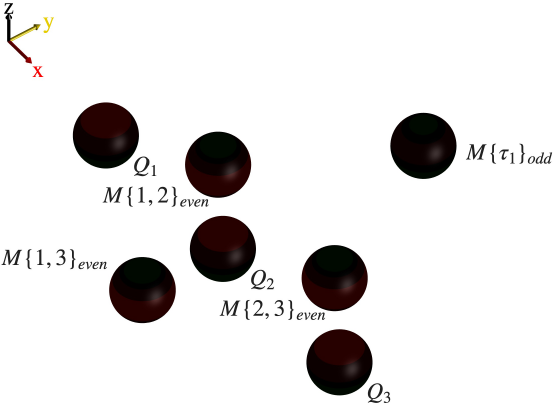}
				\subcaption{$p = 0.4$}
				\label{fig:W_dep0.4_M}
			\end{subfigure}
		\end{minipage}
	}
	
	\vspace{0.3cm}
	
	\makebox[\textwidth][c]{
		\begin{minipage}{1\textwidth}
			\begin{subfigure}[t]{0.5\textwidth}
				\centering
				\includegraphics[width=\textwidth, trim={6cm 4cm 5.8cm 5cm},clip]{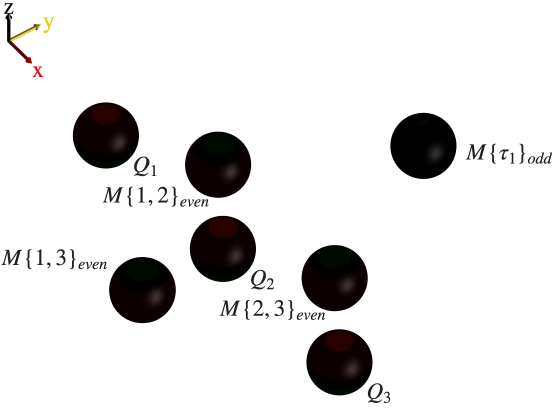}
				\subcaption{$p = 0.5$}
				\label{fig:W_dep0.5_M}
			\end{subfigure}
			\begin{subfigure}[t]{0.5\textwidth}
				\centering
				\includegraphics[width=\textwidth, trim={6cm 4cm 5.8cm 5cm},clip]{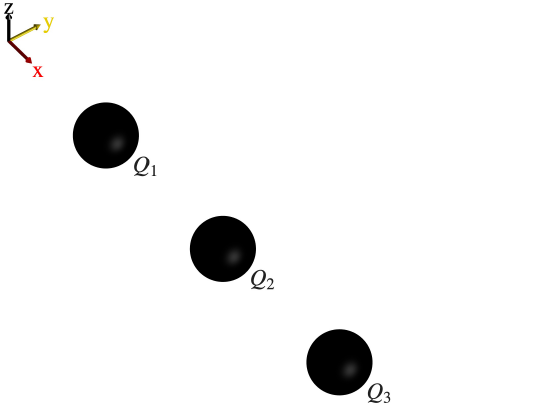}
				\subcaption{$p = 0.75$}
				\label{fig:W_dep0.75_M}
			\end{subfigure}
		\end{minipage}
	}
	\caption{Generalized BEADS representation visualizing the effect of the depolarization channel on the W state (compare \cref{eq:kraus_depolarization_E0}-\cref{eq:kraus_depolarization_E3}). States for different channel parameters $p$ are represented in terms of their total correlation properties using $M$-Beads. Analogous to the GHZ state case, any polarization as well as all genuine correlations vanish with increasing $p$, and genuine tripartite entanglement is effectively destroyed for $p\geq 0.3$. The final state at $p=0.75$ is again the maximally mixed state.}
	\label{fig:w_depolarization}
\end{figure}

A similar observation can be made for the depolarization process of the W state visualized by generalized BEADS representations in \cref{fig:w_depolarization}. Here, the genuine tripartite entanglement is also effectively destroyed at $p = 0.3$ (the residual entanglement at $p=0.2$ is marginally small but nonzero). It is clear that all genuine correlations (bipartite and tripartite) vanish with increasing $p$ (compare \cref{fig:W_dep0.1_M}-\cref{fig:W_dep0.5_M}). The final state at $p=0.7$ is again the maximally mixed state with no remaining correlation function beads and maximally mixed reduced density operators visualized by black $Q$-Beads (see \cref{fig:GHZ_dep0.75_M}).

\subsection{Amplitude damping}
\begin{figure}[htbp] 
	\centering
	\makebox[\textwidth][c]{
		\begin{minipage}{1\textwidth}
			\begin{subfigure}[t]{0.5\textwidth}
				\centering
				\includegraphics[width=\textwidth, trim={6cm 4cm 5.8cm 5cm},clip]{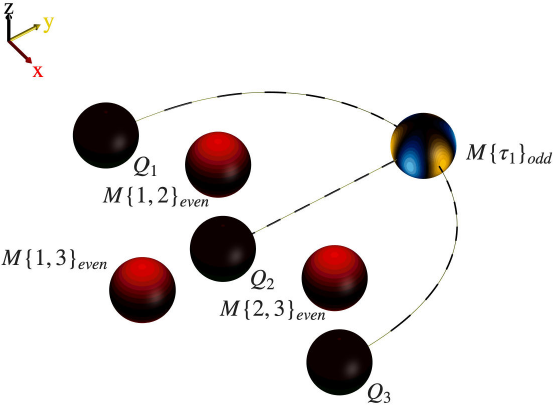}
				\subcaption{$p = 0.1$}
				\label{fig:GHZ_amp0.1_M}
			\end{subfigure}
			\begin{subfigure}[t]{0.5\textwidth}
				\centering
				\includegraphics[width=\textwidth, trim={6cm 4cm 5.8cm 5cm},clip]{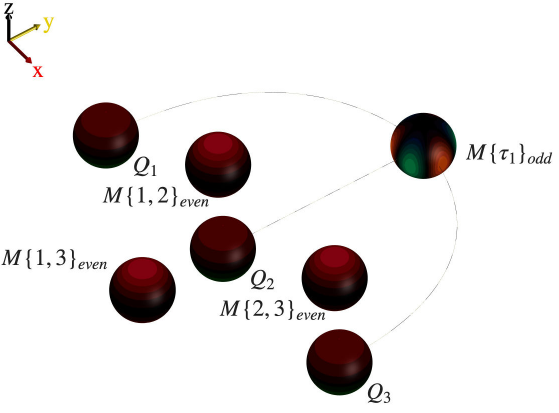}
				\subcaption{$p = 0.3$}
				\label{fig:GHZ_amp0.3_M}
			\end{subfigure}
		\end{minipage}
	}
	
	\vspace{0.3cm}
	
	\makebox[\textwidth][c]{
		\begin{minipage}{1\textwidth}
			\begin{subfigure}[t]{0.5\textwidth}
				\centering
				\includegraphics[width=\textwidth, trim={6cm 4cm 5.8cm 5cm},clip]{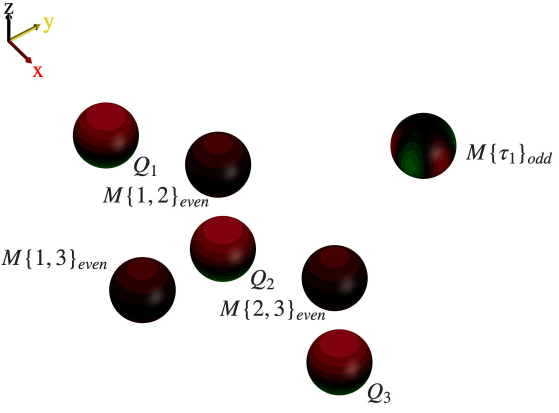}
				\subcaption{$p = 0.5$}
				\label{fig:GHZ_amp0.5_M}
			\end{subfigure}
			\begin{subfigure}[t]{0.5\textwidth}
				\centering
				\includegraphics[width=\textwidth, trim={6cm 4cm 5.8cm 5cm},clip]{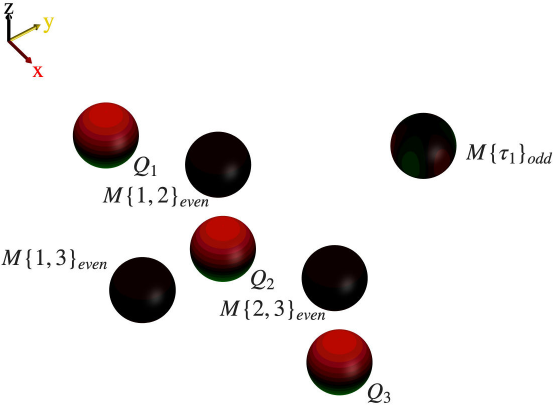}
				\subcaption{$p = 0.7$}
				\label{fig:GHZ_amp0.7_M}
			\end{subfigure}
		\end{minipage}
	}
	
	\vspace{0.3cm}
	
	\makebox[\textwidth][c]{
		\begin{minipage}{1\textwidth}
			\begin{subfigure}[t]{0.5\textwidth}
				\centering
				\includegraphics[width=\textwidth, trim={6cm 4cm 5.8cm 5cm},clip]{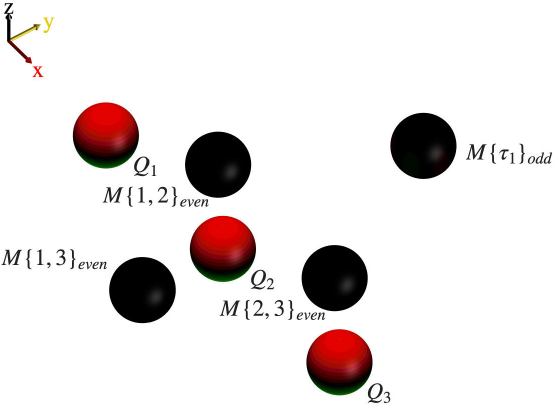}
				\subcaption{$p = 0.9$}
				\label{fig:GHZ_amp0.9_M}
			\end{subfigure}
			\begin{subfigure}[t]{0.5\textwidth}
				\centering
				\includegraphics[width=\textwidth, trim={6cm 4cm 5.8cm 5cm},clip]{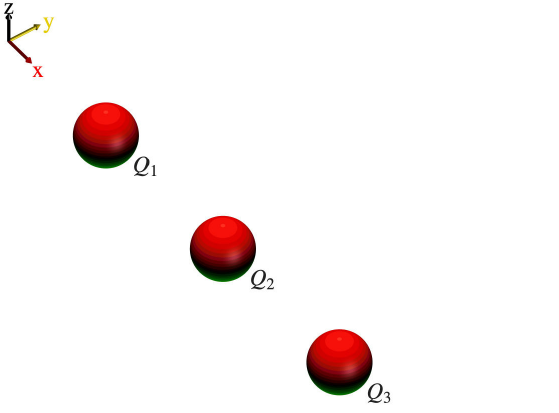}
				\subcaption{$p = 1$}
				\label{fig:GHZ_amp1_M}
			\end{subfigure}
		\end{minipage}
	}
	\caption{Generalized BEADS representation visualizing the effect of the amplitude damping channel on the GHZ state (compare \cref{eq:kraus_amplitude_damping_E0} and \cref{eq:kraus_amplitude_damping_E1}). States for different channel parameters $p$ are represented in terms of their total correlation properties using $M$-Beads. Amplitude damping causes genuine correlations to vanish with increasing $p$ while polarization builds up along the $z$-axis. This is directly reflected in  darkening $M$-Beads, whereas the $Q$-Beads become more brightly colored along the $z$-axis. The final state at $p=1$ is the the pure product state $\ket{000}$. Here, the entire GHZ entanglement is destroyed for  $p\geq0.5$.}
	\label{fig:ghz_amplitude_damping}
\end{figure}

The last type of noise that shall be discussed here is \textit{amplitude damping}. This process can be understood as the transition from the $\ket{1}$ to the $\ket{0}$ level under energy dissipation to the environment. The single-qubit Kraus operators for this type of decoherence process are \cite{nielsen_quantum_2010}
\begin{subequations}
	\begin{align}
		E_0 &= \begin{pmatrix} 1 & 0 \\ 0 & \sqrt{1-p} \end{pmatrix}, \label{eq:kraus_amplitude_damping_E0} \\
		E_1 &= \begin{pmatrix} 0 & \sqrt{p} \\ 0 & 0 \end{pmatrix}.
		\label{eq:kraus_amplitude_damping_E1}
	\end{align}
\end{subequations}
As can be directly seen in the corresponding generalized BEADS representation, the amplitude damping channel applied to the GHZ state (see \cref{fig:ghz_amplitude_damping}) leads to a decrease of the bipartite classical correlations as well as the tripartite entanglement with increasing $p$. Simultaneously, the single-qubit polarization along the $z$-axis increases, which corresponds to the $Q$-Beads increasing in color and brightness, which finally yields the product ground state $\ket{000}$ for $p=1$. Clearly, in the shown visualizations, the genuine tripartite entanglement is destroyed for $p\geq 0.5$.

\begin{figure}[htbp] 
	\centering
	\makebox[\textwidth][c]{
		\begin{minipage}{1\textwidth}
			\begin{subfigure}[t]{0.5\textwidth}
				\centering
				\includegraphics[width=\textwidth, trim={6cm 4cm 5.8cm 5cm},clip]{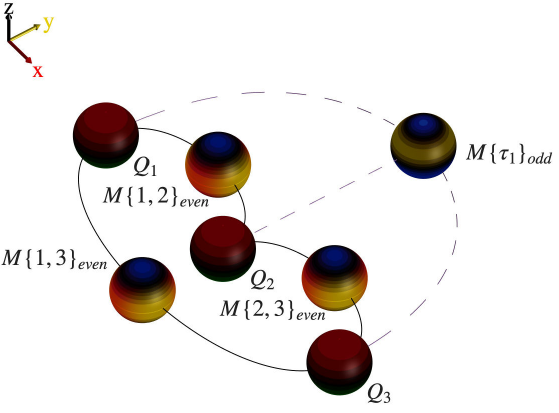}
				\subcaption{$p = 0.1$}
				\label{fig:W_amp0.1_M}
			\end{subfigure}
			\begin{subfigure}[t]{0.5\textwidth}
				\centering
				\includegraphics[width=\textwidth, trim={6cm 4cm 5.8cm 5cm},clip]{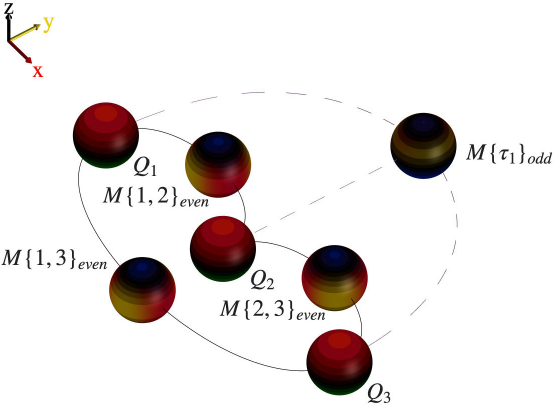}
				\subcaption{$p = 0.3$}
				\label{fig:W_amp0.3_M}
			\end{subfigure}
		\end{minipage}
	}
	
	\vspace{0.3cm}
	
	\makebox[\textwidth][c]{
		\begin{minipage}{1\textwidth}
			\begin{subfigure}[t]{0.5\textwidth}
				\centering
				\includegraphics[width=\textwidth, trim={6cm 4cm 5.8cm 5cm},clip]{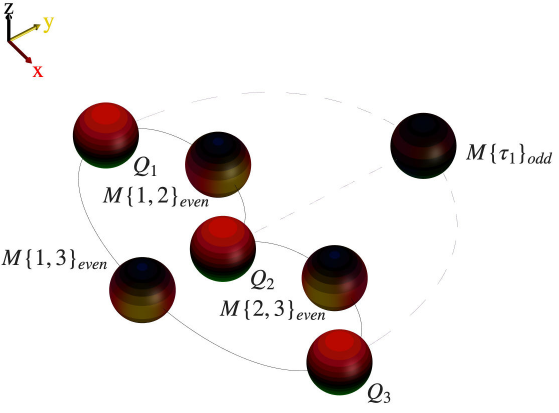}
				\subcaption{$p = 0.5$}
				\label{fig:W_amp0.5_M}
			\end{subfigure}
			\begin{subfigure}[t]{0.5\textwidth}
				\centering
				\includegraphics[width=\textwidth, trim={6cm 4cm 5.8cm 5cm},clip]{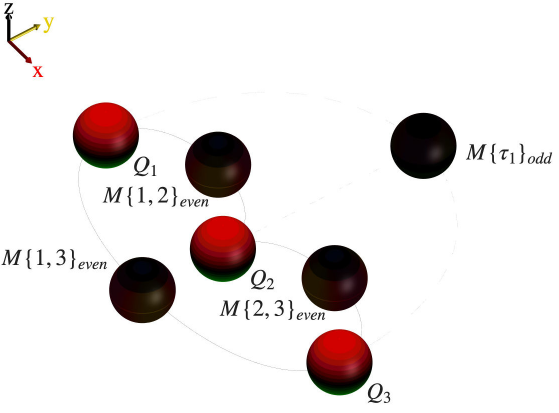}
				\subcaption{$p = 0.7$}
				\label{fig:W_amp0.7_M}
			\end{subfigure}
		\end{minipage}
	}
	
	\vspace{0.3cm}
	
	\makebox[\textwidth][c]{
		\begin{minipage}{1\textwidth}
			\begin{subfigure}[t]{0.5\textwidth}
				\centering
				\includegraphics[width=\textwidth, trim={6cm 4cm 5.8cm 5cm},clip]{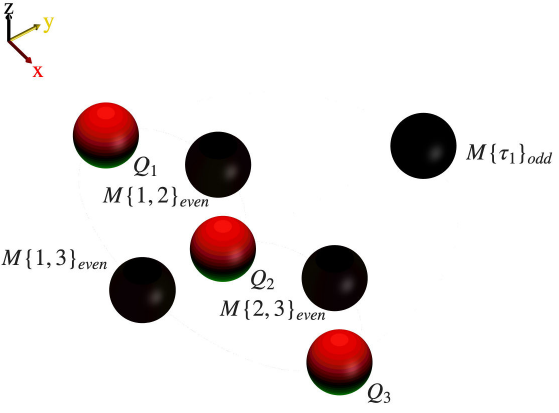}
				\subcaption{$p = 0.9$}
				\label{fig:W_amp0.9_M}
			\end{subfigure}
			\begin{subfigure}[t]{0.5\textwidth}
				\centering
				\includegraphics[width=\textwidth, trim={6cm 4cm 5.8cm 5cm},clip]{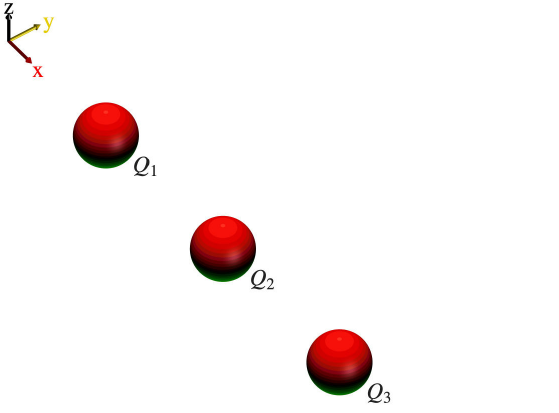}
				\subcaption{$p = 1$}
				\label{fig:W_amp1_M}
			\end{subfigure}
		\end{minipage}
	}
	\caption{Generalized BEADS representation visualizing the effect of the amplitude damping channel on the W state (compare \cref{eq:kraus_amplitude_damping_E0} and \cref{eq:kraus_amplitude_damping_E1}). States for different channel parameters $p$ are represented in terms of their total correlation properties using $M$-Beads. The channel causes the W state to transform analogously to the previously discussed GHZ state. Polarization along the $z$-axis builds up while all genuine correlations vanish. The final state at $p=1$ is again the the pure product state $\ket{000}$. Notably the entanglement of the W state is robust against amplitude damping and persists to a given extent for each $p<1$.}
	\label{fig:w_amplitude_damping}
\end{figure}

In \cref{fig:w_amplitude_damping}, the effect of the amplitude damping channel on the W state is shown using the generalized BEADS representation. Analogously to the GHZ state case, all correlations and especially the characteristic entanglement vanish with increasing $p$ while polarization builds up along the $z$-axis (compare \cref{fig:W_amp0.1_M}-\cref{fig:W_amp0.9_M}). However, contrary to the depolarization channel, the genuine tripartite entanglement persists for any $p<1$ (the entanglement arcs for the case $p=0.9$ are marginally thin though). Hence, W-type entanglement is more robust against amplitude damping but more fragile under depolarization compared to GHZ entanglement. For $p=1$ the final state again becomes the product state $\ket{000}$.

\clearpage

\section{A comparison between sector lengths and the generalized BEADS representation}
\label{sec:sector_length}
Sector lengths $A_k$ as introduced by Aschauer et al. \cite{aschauer_2004} and further explored by Wyderka and Gühne \cite{wyderka_2020} are a useful tool for characterizing the correlation structure of pure and mixed quantum states. A $k$-body sector length can be obtained as 
\begin{equation}
A_k(\rho) = \frac{1}{2^N}\text{Tr}\big[ P_k(\rho)^2 \big],
\end{equation}
where $N$ is the number of qubits in the system and $P_k(\rho)$ is the sum of all $k$-body terms that occur in the Pauli expansion of $\rho$ and where each term is weighted by its expectation value (not its expansion coefficient). Sector lengths are thus very easy to compute. Moreover, as was shown by Wyderka and Gühne \cite{wyderka_2020} they can be used to detect separability or entanglement properties of a given quantum state. For instance, in the three-qubit case it was found that states with $A_3>1$ are not fully separable, whereas $A_3>3$ guarantees the state to be genuinely tripartite entangled. However, it should be stressed that these sector-length-based criteria are sufficient but not necessary, that is, $A_3$ adopting values in the specified ranges guarantees the associated property yet there may exist states, e.g., with $A_3\leq1$ that are not fully separable or even genuinely tripartite entangled. 

In the following, we show a series of exemplary states for which a sector length analysis does mostly not allow to draw conclusions about the underlying correlation structures and complement the analysis with corresponding generalized BEADS representations from which the correlation structure becomes directly clear. It is obvious that $k$-body sector lengths describe the full $k$-body components of the density operator. In the generalized BEADS representation, analyzing these components corresponds to viewing the $T$-Bead picture. First, we shall explore some of the extremal vertices of the tripartite sector length polytope, i.e., the polytope that captures all physical tripartite density operators in terms of their sector lengths. 

\begin{figure}[htbp] 
	\vspace{-.4cm}
	\centering
	\makebox[\linewidth][c]{
		\begin{minipage}{1\textwidth}
			\begin{subfigure}[t]{0.35\textwidth}
				\centering
				\includegraphics[height=7cm]{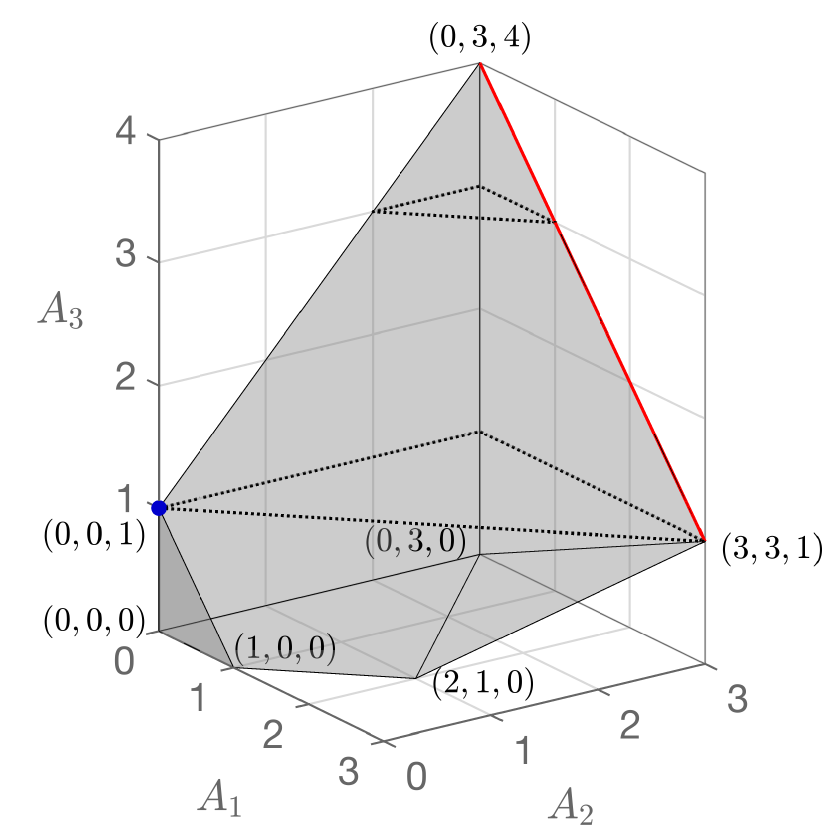}
			\end{subfigure}
			\hspace{1.3cm}
			\begin{subfigure}[t]{0.65\textwidth}
				\centering
				\includegraphics[trim={7cm 5cm 5cm 6cm},
  clip, height=6.8cm]{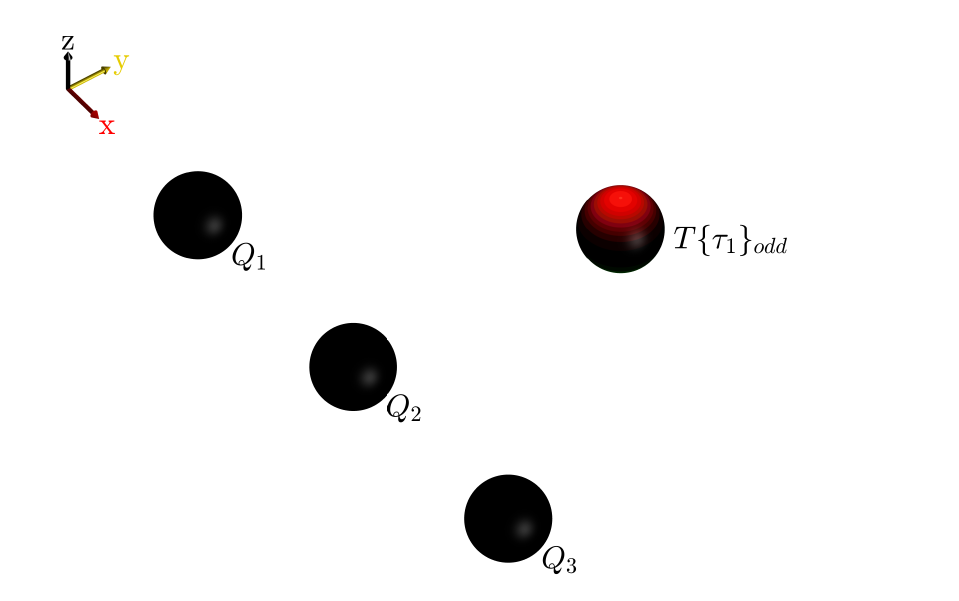}
			\end{subfigure}
		\end{minipage}
	}
	\caption{Left: Sector length polytope of three-qubit states. The red line represents the region of pure states. Dotted lines indicate thresholds above which states are guaranteed to not be fully separable ($A_3>1$) or to be genuinely multipartite entangled ($A_3>3$) \cite{wyderka_2020}. The red line marks the region of pure states.  Here, the mixed state $\rho = 1/4(\ket{000}\bra{000} + \ket{011}\bra{011} +\ket{101}\bra{101} + \ket{110}\bra{110})$ corresponds to the sector length vertex at $A = (0, 0, 1)$ which is indicated by a blue dot.\newline
Right: Generalized BEADS representation of $\rho$. The state only has a nonzero tripartite $T$-Bead and, according to the colors present in its generalized BEADS representation, is fully separable.}
	\label{fig:SL001}
	\vspace{-1.5cm}
\end{figure}

We start with the introductory example $A_1 = 0$, $A_2 = 0$, and $A_3 = 1$. Such a state is easy to construct. The single-body and two-body sector lengths both being zero imply that all bipartite marginals and thus, also all reduced single-qubit density operators, are maximally mixed. In the generalized BEADS representation this corresponds to all-black $Q$-Beads and vanishing bipartite $T$-Beads. The three-body sector length $A_3 = 1$ can then be realized, e.g., in terms of a single three-body Pauli term with coefficient $1/8$ in the expansion of the density operator. This corresponds to a maximally genuine-tripartite classically correlated configuration, for instance $\rho = 1/4(\ket{000}\bra{000} + \ket{011}\bra{011} +\ket{101}\bra{101} + \ket{110}\bra{110}) = 1/8(\mathbbm{1} + \sigma_{1z}\sigma_{2z}\sigma_{3z})$. As the same single-qubit basis is chosen for all qubits, this only gives rise to a fully permutation symmetric tripartite $T(1,2,3\ \tau_1)_{odd}$-Bead with maximum color brightness along the $z$-axis in the BEADS representation (see Fig.~\ref{fig:SL001}). Since the state has $A_3 = 1$, it is not possible to directly infer its separability properties without analyzing the density operator in close detail.

\newpage
However, in the generalized BEADS representation of $\rho$ it is directly possible to tell that the state is fully separable based on the colors of the tripartite $T$-Bead and the vanishing bipartite $T$-Beads.

\begin{figure}[htbp] 
	\centering
	\makebox[\linewidth][c]{
		\begin{minipage}{1\textwidth}
			\begin{subfigure}[t]{0.35\textwidth}
				\centering
				\includegraphics[height=7cm]{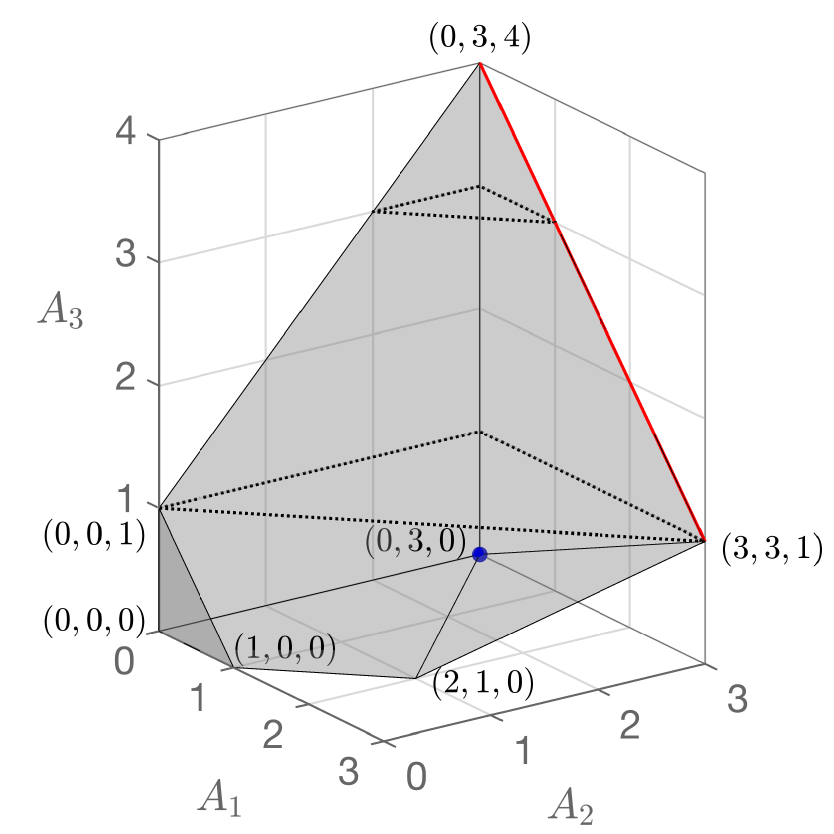}
			\end{subfigure}
			\hspace{1.3cm}
			\begin{subfigure}[t]{0.65\textwidth}
				\centering
				\includegraphics[trim={7cm 5cm 5cm 6cm},
  clip, height=6.8cm]{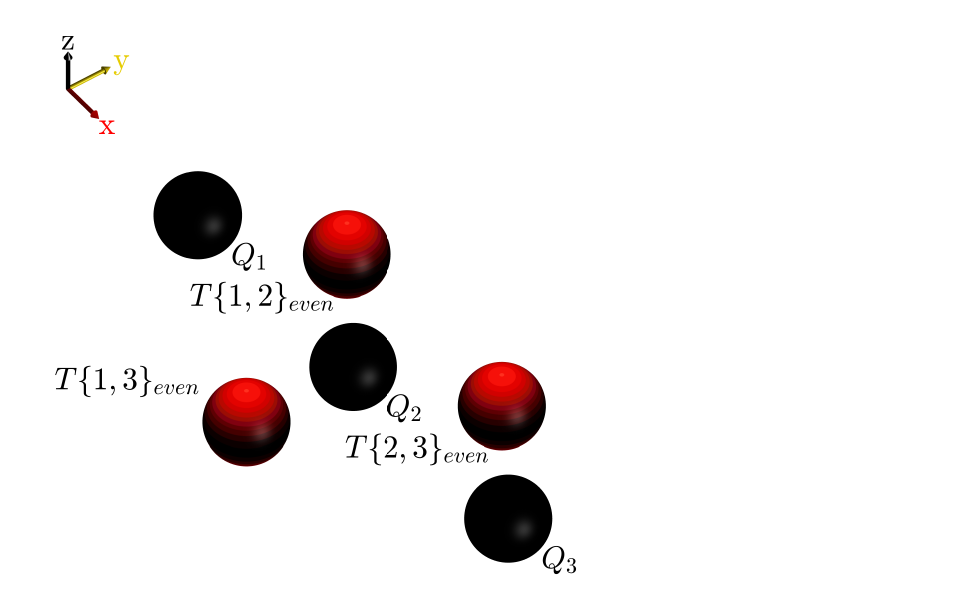}
			\end{subfigure}
		\end{minipage}
	}
	\caption{Left: Sector length polytope of three-qubit states. The red line represents the region of pure states. Dotted lines indicate thresholds above which states are guaranteed to not be fully separable ($A_3>1$) or to be genuinely multipartite entangled ($A_3>3$) \cite{wyderka_2020}. The red line marks the region of pure states.  Here, the mixed state $\rho = 1/2(\ket{000}\bra{000} + \ket{111}\bra{111}) = 1/8(\mathbbm{1} + \sigma_{1z}\sigma_{2z} + \sigma_{1z}\sigma_{3z} + \sigma_{2z}\sigma_{3z})$ corresponds to the sector length vertex at $A = (0, 3, 0)$ which is indicated by a blue dot.\newline
Right: Generalized BEADS representation of $\rho$. The state only has nonzero bipartite $T$-Beads with maximum color brightness along the $z$-axis and, according to the prevalent colors present and missing entanglement arcs in its generalized BEADS representation, is fully separable.}
	\label{fig:SL030_1}
\end{figure}

We now move to a different vertex $A = (0, 3, 0)$ of the sector length polytope. Obviously, this vertex corresponds to the extremal point of the polytope where there are neither single-qubit nor three-body terms contained in the expansion of a corresponding density operator. However, the two-body terms become maximum. Similar to the previous example, a simple case for such a set of sector lengths can be constructed in terms of a density operator that has a single maximum Pauli term for each bipartite marginal. For instance, this corresponds to a classically correlated state $\rho = 1/2(\ket{000}\bra{000} + \ket{111}\bra{111}) = 1/8(\mathbbm{1} + \sigma_{1z}\sigma_{2z} + \sigma_{1z}\sigma_{3z} + \sigma_{2z}\sigma_{3z})$ . Note that here the correlations are genuinely bipartite and identical across all possible bipartite subsystems which is reflected in maximum (coefficient $1/8$ or expectation value 1) two-body Pauli-$ZZ$ terms in the expansion of $\rho$. As can be seen in Fig.~\ref{fig:SL030_1}, the resulting generalized BEADS representation of $\rho$ consequently has all-black (maximally mixed) $Q$-Beads, bipartite $T$-Beads with maximum color brightness along the $z$-axis, and no tripartite $T$-Bead. In this case, it is again only directly possible to identify the state as being fully separable based on its generalized BEADS representation but not based on its sector lengths.

Clearly, the vertex $A = (0, 3, 0)$ lies well below the separability or genuine multipartite entanglement thresholds of the polytope. As stated at the beginning of this chapter, these thresholds do however only induce sufficient conditions, which implies that one may still be able to find a state with sector lengths $A = (0, 3, 0)$ that is not fully separable or even genuinely multipartite entangled (gme).

\begin{figure}[htbp] 
	\vspace{-2.1cm}
	\centering
	\makebox[\linewidth][c]{
		\begin{minipage}{1\textwidth}
			\begin{subfigure}[t]{0.35\textwidth}
				\centering
				\includegraphics[height=7cm]{030}
			\end{subfigure}
			\hspace{1.3cm}
			\begin{subfigure}[t]{0.65\textwidth}
				\centering
				\includegraphics[trim={7cm 5cm 5cm 6cm},
  clip, height=6.8cm]{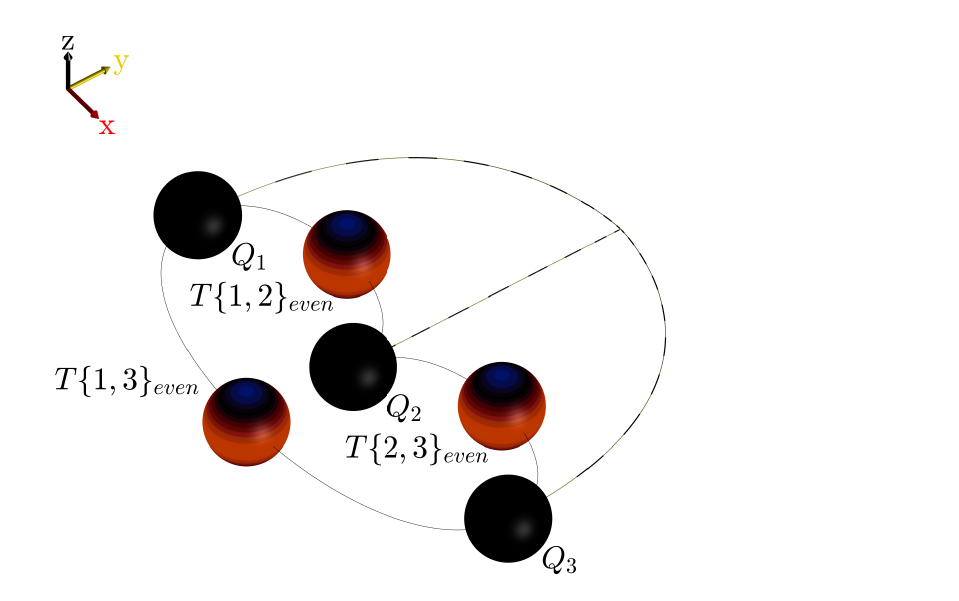}
			\end{subfigure}
		\end{minipage}
	}
	\vspace{-.3cm}
	\caption{Left: Sector length polytope of three-qubit states. The red line represents the region of pure states. Dotted lines indicate thresholds above which states are guaranteed to not be fully separable ($A_3>1$) or to be genuinely multipartite entangled ($A_3>3$) \cite{wyderka_2020}. The red line marks the region of pure states. As for the state shown in Fig.~\ref{fig:SL030_1}, the mixed state $\rho = 1/2(\ket{W}\bra{W} + \ket{\overline{W}}\bra{\overline{W}}) = 1/8(\mathbbm{1} + 1/3[2\sigma_{1x}\sigma_{2x} + 2\sigma_{1y}\sigma_{2y} - \sigma_{1z}\sigma_{2z}] + 1/3[2\sigma_{1x}\sigma_{3x} + 2\sigma_{1y}\sigma_{3y} - \sigma_{1z}\sigma_{3z}] + 1/3[2\sigma_{2x}\sigma_{3x} + 2\sigma_{2y}\sigma_{3y} - \sigma_{2z}\sigma_{3z}])$ also corresponds to a sector length vertex at $A = (0, 3, 0)$ which is indicated by a blue dot. The state is tripartite entangled, but a sector length analysis does not suffice to detect the entanglement.\newline
Right: Generalized BEADS representation of $\rho$. The state only has nonzero bipartite $T$-Beads but is still genuinely tripartite entangled which becomes evident from the entanglement arcs. This state is further discussed in Sec.~\ref{sec:Wmixtures} and references \cite{kaszlikowski_quantum_2008,tran_genuine_2017}.}
	\label{fig:SL030_2}
	\vspace{-.7cm}
\end{figure}

Consider the state $\rho = 1/2(\ket{W}\bra{W} + \ket{\overline{W}}\bra{\overline{W}}) = 1/8(\mathbbm{1} + 1/3[2\sigma_{1x}\sigma_{2x} + 2\sigma_{1y}\sigma_{2y} - \sigma_{1z}\sigma_{2z}] + 1/3[2\sigma_{1x}\sigma_{3x} + 2\sigma_{1y}\sigma_{3y} - \sigma_{1z}\sigma_{3z}] + 1/3[2\sigma_{2x}\sigma_{3x} + 2\sigma_{2y}\sigma_{3y} - \sigma_{2z}\sigma_{3z}])$. This state also corresponds to a set of sector lengths $A = (0, 3, 0)$. However, its correlation structure is vastly different, which becomes evident by looking at the generalized BEADS representation in Fig.~\ref{fig:SL030_2}. It is clear from the entanglement arcs and the $T$-Bead colors that this state has bipartite as well as genuine tripartite entanglement despite having a zero-valued three-body sector length -- properties that are well-achievable in systems with odd numbers of qubits \cite{kaszlikowski_quantum_2008,tran_genuine_2017}. Indeed, it is thus not only possible that genuinely multipartite entangled states can fall below the corresponding sector length threshold (which only depends on $A_3$), but, as shown by this example, to locate an example that resides at the $A_3$-value furthest away from this threshold.

\begin{figure}[htbp] 
	\centering
	\vspace{-.3cm}
	\makebox[\linewidth][c]{
		\begin{minipage}{1\textwidth}
			\begin{subfigure}[t]{0.35\textwidth}
				\centering
				\includegraphics[height=7cm]{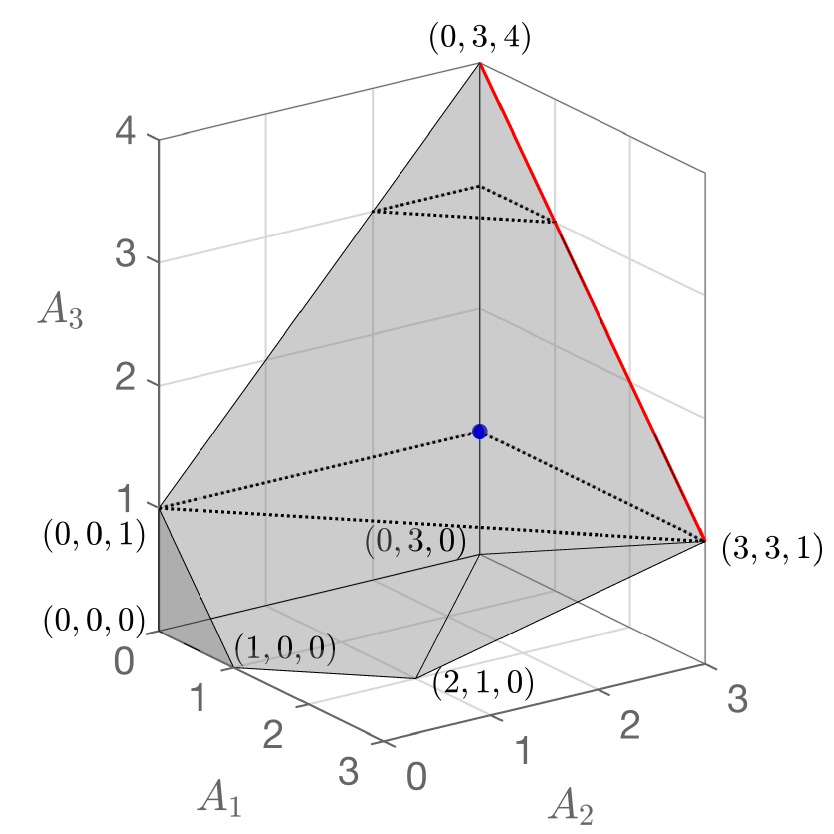}
			\end{subfigure}
			\hspace{1.3cm}
			\begin{subfigure}[t]{0.65\textwidth}
				\centering
				\includegraphics[trim={7cm 5cm 5cm 6cm},
  clip, height=6.8cm]{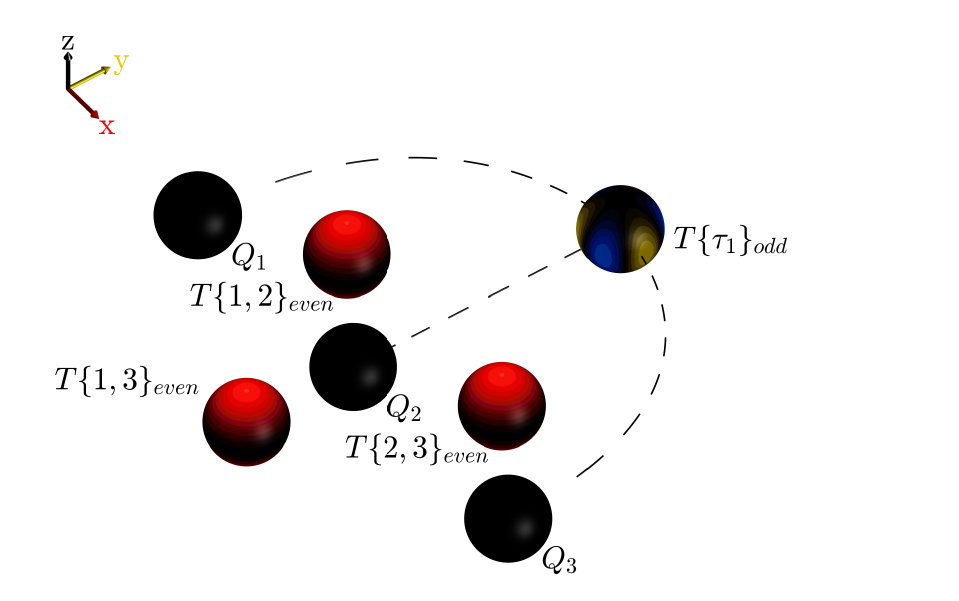}
			\end{subfigure}
		\end{minipage}
	}
	\caption{Left: Sector length polytope of three-qubit states. The red line represents the region of pure states. Dotted lines indicate thresholds above which states are guaranteed to not be fully separable ($A_3>1$) or to be genuinely multipartite entangled ($A_3>3$) \cite{wyderka_2020}. The red line marks the region of pure states. Here, the mixed state $\rho = 1/2(\rho_{\text{GHZ}} + \ \rho_{\text{clas.}})$ with $\rho_{\text{clas.}} = 1/2(\ket{000}\bra{000} + \ket{111}\bra{111})$ corresponds to a sector length vertex at $A = (0, 3, 1)$, which is indicated by a blue dot. The state is genuinely tripartite entangled, but lies below the gme-threshold $A=3$.\newline
Right: Generalized BEADS representation of $\rho$. The state has nonzero bipartite and tripartite $T$-Beads. It is clear that the state is genuinely tripartite (GHZ-type) entangled which becomes evident from the yellow and blue tripartite $T$-Bead color, and the occurring entanglement arcs.}
	\label{fig:SL031}
		\vspace{-2.5cm}
\end{figure}

Starting from $A = (0, 3, 0)$ we now follow along the $A_3$ until we reach the full separability threshold plane at $A_3 = 1$ and specifically the vertex $A = (0, 3, 1)$ in the sector length polytope. It is easy to construct a corresponding state that has genuine tripartite entanglement starting from the classically correlated state $\rho_{\text{clas.}} = 1/2(\ket{000}\bra{000} + \ket{111}\bra{111})$ shown in Fig.~\ref{fig:SL030_1}. Knowing that this state had sector lengths $A = (0, 3, 0)$ it suffices to mix it with another state that has an identical bipartite correlation structure but additional three-body terms associated with genuine tripartite entanglement which remains persistant upon mixing the states. A suitable candidate is given by the GHZ state $\rho_{\text{GHZ}} = 1/2(\ket{000}\bra{000} + \ket{111}\bra{111} + \ket{000}\bra{111} + \ket{111}\bra{000}) = 1/8(\mathbbm{1} + \sigma_{1z}\sigma_{2z} + \sigma_{1z}\sigma_{3z} + \sigma_{2z}\sigma_{3z} + \sigma_{1x}\sigma_{2x}\sigma_{3x} - \sigma_{1x}\sigma_{2y}\sigma_{3y} - \sigma_{1y}\sigma_{2x}\sigma_{3y} - \sigma_{1y}\sigma_{2y}\sigma_{3x})$, and the resulting state $\rho = p\ \rho_{\text{GHZ}} + (1-p)\ \rho_{\text{clas.}}$ is a GHZ-diagonal state which is genuinely tripartite entangled for all $p \neq 0$. This can be easily shown by computing the partial transpose of $\rho$ across any bipartition, which has a negative eigenvalue $-p/2$ for all $p > 0$ \cite{duer_1999, duer_2000} and the biseparability condition $\left|
\langle 000|\rho|111\rangle
\right|
\le
\sqrt{
\langle001|\rho|001\rangle
\langle110|\rho|110\rangle
}
+
\sqrt{
\langle010|\rho|010\rangle
\langle101|\rho|101\rangle
}
+
\sqrt{
\langle011|\rho|011\rangle
\langle100|\rho|100\rangle
}$ \cite{guhne_2010} is violated for all $p > 0$ as the inequality becomes $p/2 \le 0$. Finally, the three-body sector length $A_3 = 1$ can be achieved by choosing $p = 1/2$. The resulting state is shown in Fig.~\ref{fig:SL031}.

\begin{figure}[htbp] 
	\centering
	\vspace{-.3cm}
	\makebox[\linewidth][c]{
		\begin{minipage}{1\textwidth}
			\begin{subfigure}[t]{0.35\textwidth}
				\centering
				\includegraphics[height=7cm]{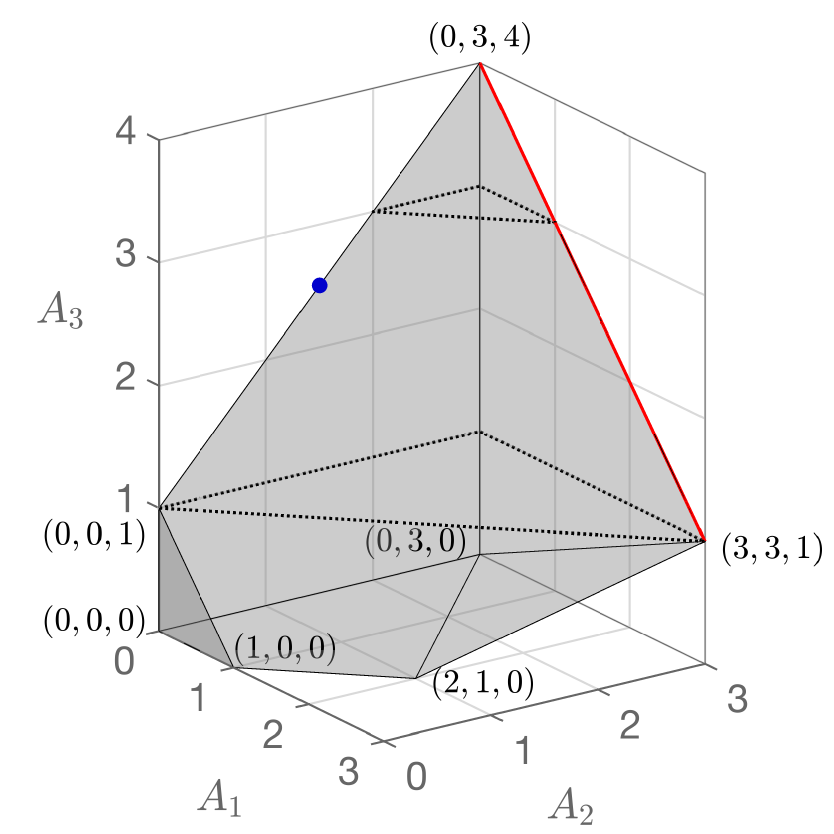}
			\end{subfigure}
			\hspace{1.3cm}
			\begin{subfigure}[t]{0.65\textwidth}
				\centering
				\includegraphics[trim={7cm 5cm 5cm 6cm},
  clip, height=6.8cm]{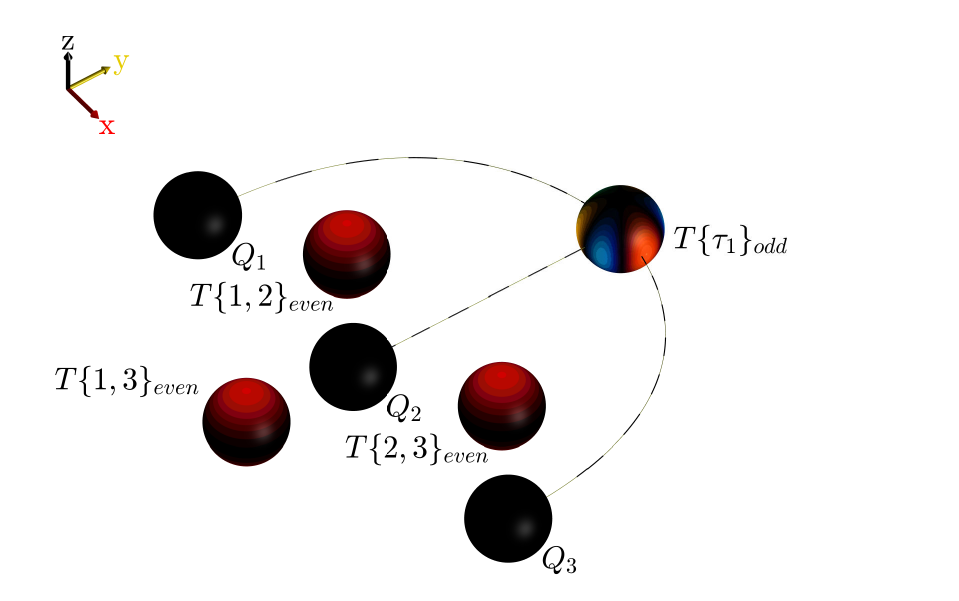}
			\end{subfigure}
		\end{minipage}
	}
	\caption{Left: Sector length polytope of three-qubit states. The red line represents the region of pure states. Dotted lines indicate thresholds above which states are guaranteed to not be fully separable ($A_3>1$) or to be genuinely multipartite entangled ($A_3>3$) \cite{wyderka_2020}. The red line marks the region of pure states. Here, the mixed state $\rho = 1/\sqrt{2} \ket{\text{GHZ}}\bra{\text{GHZ}} + (1-1/\sqrt{2}) \tilde{\rho} = 1/8(\mathbbm{1} + 1/\sqrt{2}[\sigma_{1z}\sigma_{2z} + \sigma_{1z}\sigma_{3z} + \sigma_{2z}\sigma_{3z}] + \sigma_{1x}\sigma_{2x}\sigma_{3x} - 1/\sqrt{2}[\sigma_{1x}\sigma_{2y}\sigma_{3y} + \sigma_{1y}\sigma_{2x}\sigma_{3y} + \sigma_{1y}\sigma_{2y}\sigma_{3x}])$, with $\tilde{\rho} = 1/4(\ket{+++}\bra{+++} + \ket{+--}\bra{+--} +\ket{-+-}\bra{-+-} + \ket{--+}\bra{--+})$, corresponds to the sector length vertex at $A = (0, 1.5, 2.5)$, which is indicated by a blue dot. The state is genuinely tripartite entangled, but lies below the gme-threshold $A=3$ for which a sector length analysis does not suffice to characterize the correlation structure of $\rho$.\newline
Right: Generalized BEADS representation of $\rho$. The state has nonzero bipartite and tripartite $T$-Beads. It is clear that the state is genuinely tripartite entangled which becomes evident from the $T$-Bead colors and the occurring entanglement arcs.}
	\label{fig:SL01525_1}
\end{figure}

\newpage
We can choose a similar approach to modify the state discussed in Fig.~\ref{fig:SL001} to arrive at sector length constellations that lie on the polytope edge connecting vertices $A = (0, 0, 1)$ and $A=(0,3,4)$ (the pure GHZ state). As an example, we specifically consider the vertex $A = (0, 1.5, 2.5)$. First, we take the state $\tilde{\rho} = 1/4(\ket{+++}\bra{+++} + \ket{+--}\bra{+--} +\ket{-+-}\bra{-+-} + \ket{--+}\bra{--+}) = 1/8(\mathbbm{1} + \sigma_{1x}\sigma_{2x}\sigma_{3x})$, which is obtained by applying a global $\sqrt{Y}$-gate to the state shown in Fig.~\ref{fig:SL001}. Analogous to the previous example, we then mix this state incoherently with the GHZ-state, which introduces additional two- and three-body terms, until we reach the desired sector length constellation. Here, the resulting state becomes $\rho = 1/\sqrt{2} \ket{\text{GHZ}}\bra{\text{GHZ}} + (1-1/\sqrt{2}) \tilde{\rho} = 1/8(\mathbbm{1} + 1/\sqrt{2}[\sigma_{1z}\sigma_{2z} + \sigma_{1z}\sigma_{3z} + \sigma_{2z}\sigma_{3z}] + \sigma_{1x}\sigma_{2x}\sigma_{3x} - 1/\sqrt{2}[\sigma_{1x}\sigma_{2y}\sigma_{3y} + \sigma_{1y}\sigma_{2x}\sigma_{3y} + \sigma_{1y}\sigma_{2y}\sigma_{3x}])$. The corresponding generalized BEADS representation is shown in Fig.~\ref{fig:SL01525_1}. It is directly clear from the colors of the $T\{1,2,3\ \tau_1\}_{odd}$-Bead and the dedicated entanglement arc that this state is also genuinely tripartite entangled. Moreover, there is no bipartite entanglement and the bipartite marginals can be identified to be fully separable.

\begin{figure}[htbp] 
	\centering
	\makebox[\linewidth][c]{
		\begin{minipage}{1\textwidth}
			\begin{subfigure}[t]{0.35\textwidth}
				\centering
				\includegraphics[height=7cm]{01525}
			\end{subfigure}
			\hspace{1.3cm}
			\begin{subfigure}[t]{0.65\textwidth}
				\centering
				\includegraphics[trim={7cm 5cm 5cm 6cm},
  clip, height=6.8cm]{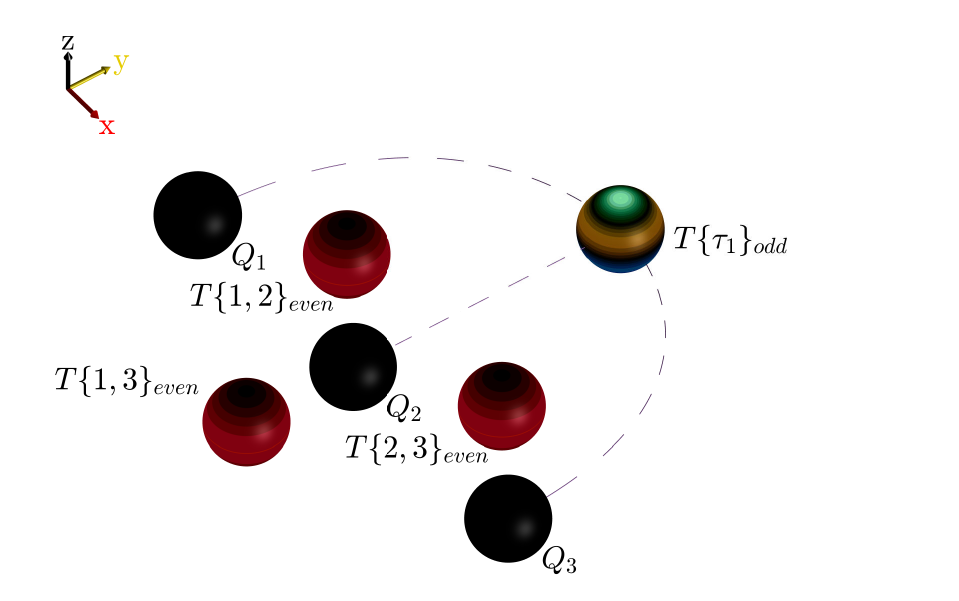}
			\end{subfigure}
		\end{minipage}
	}
	\caption{Left: Sector length polytope of three-qubit states. The red line represents the region of pure states. Dotted lines indicate thresholds above which states are guaranteed to not be fully separable ($A_3>1$) or to be genuinely multipartite entangled ($A_3>3$) \cite{wyderka_2020}. The red line marks the region of pure states. Here, the mixed state $\rho = 3/4\ket{\text{W}}\bra{\text{W}} + 1/4\ket{111}\bra{111} = 1/8(\mathbbm{1} + 1/2[\sigma_{1x}\sigma_{2x} + \sigma_{1y}\sigma_{2y}]+ 1/2[\sigma_{1x}\sigma_{3x} + \sigma_{1y}\sigma_{3y}] + 1/2[\sigma_{2x}\sigma_{3x} + \sigma_{2y}\sigma_{3y}]) + 1/2[\sigma_{1x}\sigma_{2x}\sigma_{3z} + \sigma_{1x}\sigma_{2z}\sigma_{3x} + \sigma_{1z}\sigma_{2x}\sigma_{3x}] + 1/2[\sigma_{1y}\sigma_{2y}\sigma_{3z} + \sigma_{1y}\sigma_{2z}\sigma_{3y} + \sigma_{1z}\sigma_{2y}\sigma_{3y}] - \sigma_{1z}\sigma_{2z}\sigma_{3z})$, also corresponds to the sector length vertex at $A = (0, 1.5, 2.5)$, which is indicated by a blue dot. The state is genuinely tripartite entangled, but lies below the gme-threshold $A=3$ for which a sector length analysis does not suffice to characterize the correlation structure of $\rho$.\newline
Right: Generalized BEADS representation of $\rho$. The state has nonzero bipartite and tripartite $T$-Beads. It is clear that the state is genuinely tripartite entangled which becomes evident from the colors of the $T\{1,2,3\ \tau_1\}_{odd}$-Bead. There is no three-tangle arc but an entanglement-based connected correlation function norm arc for which the state must be only W-type entangled. Furthermore, the bipartite marginals are fully separable, which is clear as only red color and no two-tangle arcs are visible.}
	\label{fig:SL01525_2}
\end{figure}

It is also possible to construct a state corresponding to sector lengths $A = (0, 1.5, 2.5)$ which is strictly $W$-type entangled. This is done by mixing the $W$-state with a single suitable fully separable state -- we choose $\ket{111}$ -- until the desired sector lengths are reached. Here, $A = (0, 1.5, 2.5)$ is obtained for ${\rho = 3/4\ket{\text{W}}\bra{\text{W}} + 1/4\ket{111}\bra{111}} = 1/8(\mathbbm{1} + 1/2[\sigma_{1x}\sigma_{2x} + \sigma_{1y}\sigma_{2y}]+ 1/2[\sigma_{1x}\sigma_{3x} + \sigma_{1y}\sigma_{3y}] + 1/2[\sigma_{2x}\sigma_{3x} + \sigma_{2y}\sigma_{3y}]) + 1/2[\sigma_{1x}\sigma_{2x}\sigma_{3z} + \sigma_{1x}\sigma_{2z}\sigma_{3x} + \sigma_{1z}\sigma_{2x}\sigma_{3x}] + 1/2[\sigma_{1y}\sigma_{2y}\sigma_{3z} + \sigma_{1y}\sigma_{2z}\sigma_{3y} + \sigma_{1z}\sigma_{2y}\sigma_{3y}] - \sigma_{1z}\sigma_{2z}\sigma_{3z})$. Again, this state lies below the gme-threshold of the sector length polytope despite being genuinely tripartite entangled and sector lengths are not applicable for characterizing the correlation structure. On the contrary, the generalized BEADS representation of $\rho$ (see Fig.~\ref{fig:SL01525_2}) directly reveals all correlation properties of the state. First, the colors of $T\{1,2,3\ \tau_1\}_{odd}$ are sufficient to identify this state as being genuinely tripartite entangled. A closer look at the entanglement arcs reveals that the three tangle of $\rho$ is $\tau_3 = 0$ while its entanglement-based connected correlation function norm is not. Hence, the state must be strictly W-type entangled. Moreover, when considering the bipartite subsystems, it is obvious that the state is lacking any bipartite entanglement, which would be commonly encountered for pure W-states or mixed W-states with large enough weights of the W-component. In fact, the partial transposes of the reduced systems of $\rho$, e.g.,

\begin{equation}
\rho_{12}^{\text{T}_2} = \begin{pmatrix}
\frac{1}{4} & 0 & 0 & \frac{1}{4} \\
0 & \frac{1}{4} & 0 & 0 \\
0 & 0 & \frac{1}{4} & 0 \\
\frac{1}{4} & 0 & 0 & \frac{1}{4} \\
\end{pmatrix},
\end{equation} 
have a smallest eigenvalue $\lambda_{\text{min}} = 0$ and it is straightforward to show that choosing a different W-state weight $p < 3/4$ results in $\lambda_{\text{min}} > 0$. Thus, the discussed state represents the limiting case, where the bipartite entanglement vanishes, but the genuine tripartite entanglement persists.

\vspace{-.2cm}

\begin{figure}[htbp] 
	\centering
	\makebox[\linewidth][c]{
		\begin{minipage}{1\textwidth}
			\begin{subfigure}[t]{0.35\textwidth}
				\centering
				\includegraphics[height=7cm]{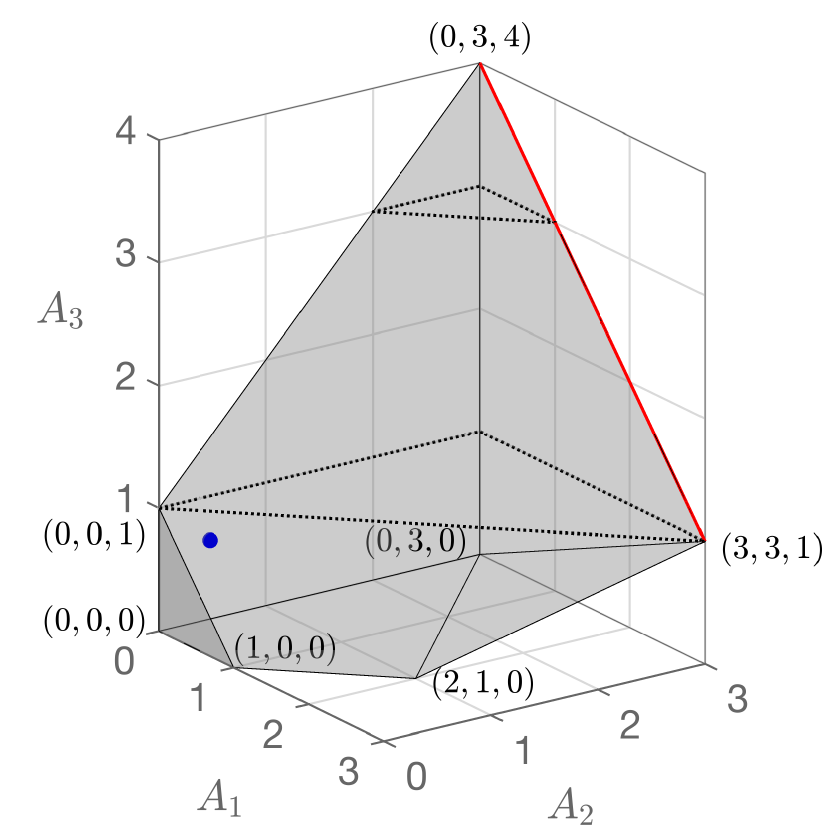}
			\end{subfigure}
			\hspace{1.3cm}
			\begin{subfigure}[t]{0.65\textwidth}
				\centering
				\includegraphics[trim={7cm 5cm 5cm 6cm},
  clip, height=6.8cm]{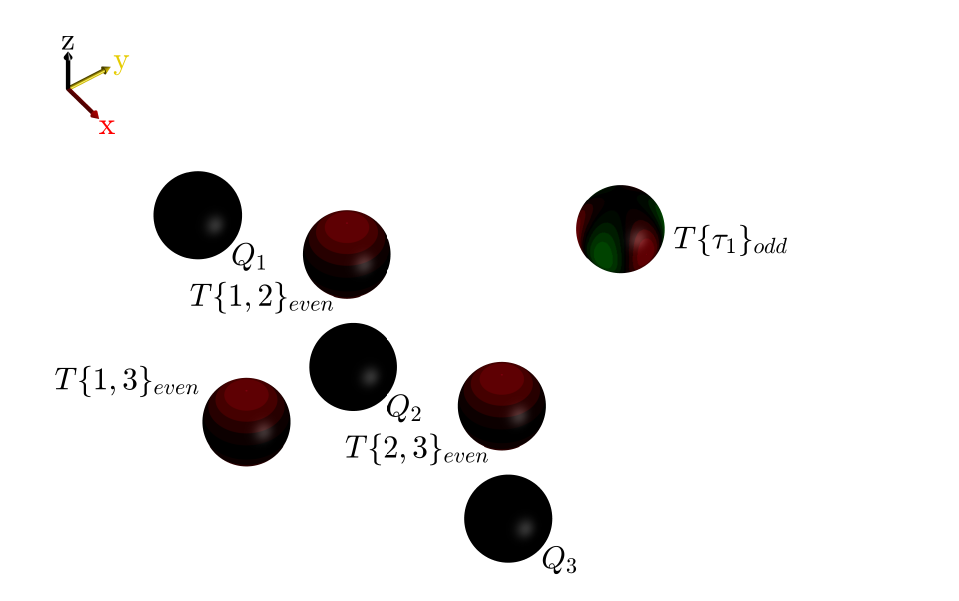}
			\end{subfigure}
		\end{minipage}
	}
	\makebox[\linewidth][c]{
		\begin{minipage}{1\textwidth}
		\centering
		\hspace{1cm}
			\begin{subfigure}[t]{0.65\textwidth}
				\centering
				\includegraphics[trim={7cm 5cm 5cm 5cm},
  clip, height=6.8cm]{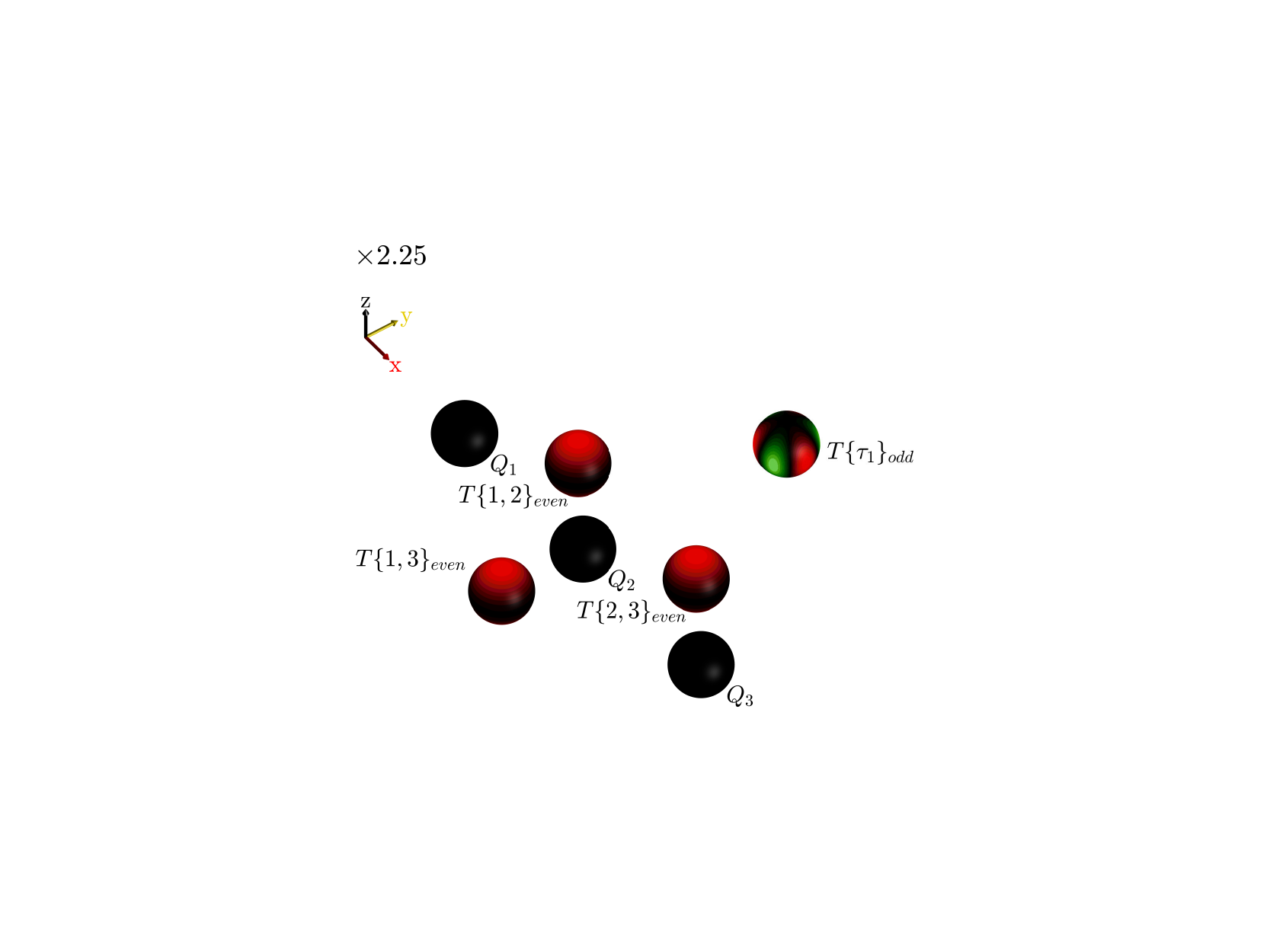}
			\end{subfigure}
		\end{minipage}
	}
	\caption{Top left: Sector length polytope of three-qubit states. The red line represents the region of pure states. Dotted lines indicate thresholds above which states are guaranteed to not be fully separable ($A_3>1$) or to be genuinely multipartite entangled ($A_3>3$) \cite{wyderka_2020}. The red line marks the region of pure states. The mixed state $\rho = 0.4\ket{\text{GHZ}}\bra{\text{GHZ}} + 0.6\frac{\mathbbm{1}}{8} = 1/8(\mathbbm{1} + 2/5[\sigma_{1z}\sigma_{2z} + \sigma_{1z}\sigma_{3z} + \sigma_{2z}\sigma_{3z} + \sigma_{1x}\sigma_{2x}\sigma_{3x} - \sigma_{1x}\sigma_{2y}\sigma_{3y} - \sigma_{1y}\sigma_{2x}\sigma_{3y} - \sigma_{1y}\sigma_{2y}\sigma_{3x}])$ corresponds to the sector length vertex at $A = (0, 0.48, 0.64)$, which is indicated by a blue dot. The state is biseparable, but lies below the separability threshold $A=1$ for which a sector length analysis is unsuitable to characterize the correlation structure of $\rho$.\newline
Top right: Generalized BEADS representation of $\rho$. The state has nonzero bipartite and tripartite $T$-Beads. All $T$-Beads have red or green color. Hence, all marginals are separable and the state is not genuinely tripartite entangled. However, it is not possible to infer the biseparability of the state by considering its mixed state representation.\newline
Bottom: Generalized BEADS representation of $\rho$ with scaled spherical function values for better visibility.}
	\label{fig:SLNoisy04}
	\vspace{-1cm}
\end{figure}

As a last example of this chapter, we consider the state $\rho = 0.4\ket{\text{GHZ}}\bra{\text{GHZ}} + 0.6\frac{\mathbbm{1}}{8} = 1/8(\mathbbm{1} + 2/5[\sigma_{1z}\sigma_{2z} + \sigma_{1z}\sigma_{3z} + \sigma_{2z}\sigma_{3z} + \sigma_{1x}\sigma_{2x}\sigma_{3x} - \sigma_{1x}\sigma_{2y}\sigma_{3y} - \sigma_{1y}\sigma_{2x}\sigma_{3y} - \sigma_{1y}\sigma_{2y}\sigma_{3x}])$, i.e., the GHZ-state with added white noise. Evidently, the added white noise uniformly decreases the $k$-body Pauli terms and thus its sector lengths in the expansion of $\rho$ in magnitude compared to the pure GHZ state. Such states naturally lie on the line connecting vertices $A=(0,0,0)$ (the maximally mixed state) and $A=(0,3,4)$ (the pure GHZ state) of the sector length polytope. The family of noisy GHZ-states is extensively studied in literature, and entanglement as well as separability thresholds are well-known \cite{duer_1999, duer_2000}. Here, the noise weight chosen for $\rho$ gives a biseparable state, as the upper biseparability limit of noisy GHZ states is $3/7$ \cite{duer_1999}. As can be seen in Fig.~\ref{fig:SLNoisy04}, the sector lengths of $\rho$ are $A=(0,0.48, 0.64)$ which lies below the full separability threshold $A_3 = 1$. As a consequence, sector lengths are again unsuitable to identify the separability properties of the state. However, the generalized BEADS representation of $\rho$ does not allow to directly infer that the state is biseparable either. This is due to the fact that the generalized BEADS representation of a mixed state is designed to reveal the correlation properties of each possible marginal of $\rho$. By contrast, biseparability, especially if not fixed to a specific bipartite cut, is a property of the pure state decomposition of the state, and does not generally manifest in the mixed state BEADS representation. Note that it is still true that, as can be directly inferred from the BEADS representation in Fig.~\ref{fig:SLNoisy04}, the state is not gme and the bipartite marginals of $\rho$ are separable.

\begin{figure}[H] 
\vspace{2cm}
\centering
	\includegraphics[width=1.\textwidth]{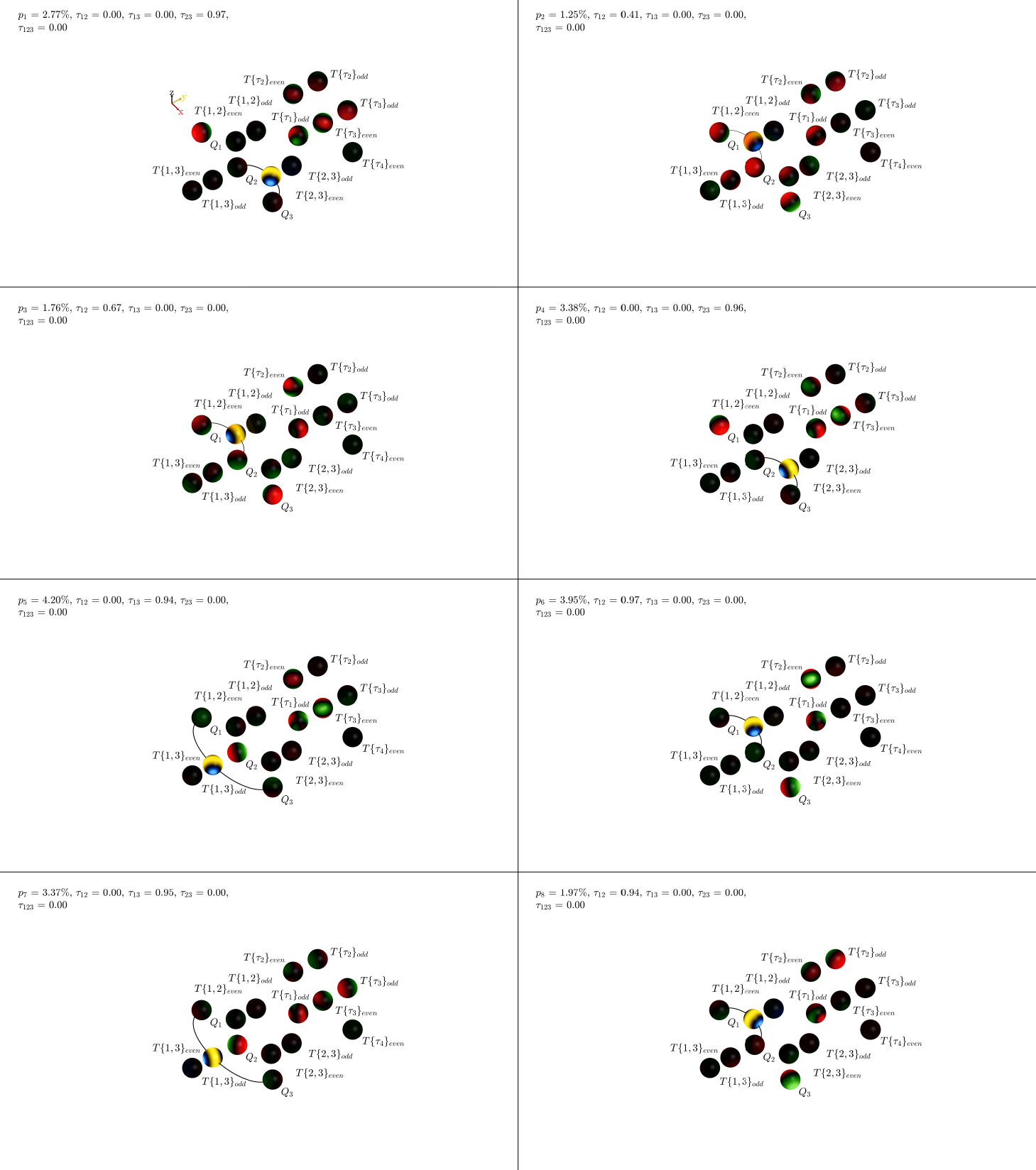}
	\caption{Selection of pure states contributing to an optimized least-entangled decomposition of $\rho = 0.4\ket{\text{GHZ}}\bra{\text{GHZ}} + 0.6\frac{\mathbbm{1}}{8}$ visualized by their generalized BEADS representation. It is immediately clear that all of the pure states are biseparable across some cut although the cut is not fixed among the states. All remaining states contributing in the decomposition of $\rho$ which are not shown here, have the same properties. As a consequence, $\rho$ is a biseparable mixed state.}
	\label{fig:SLNoisy04_pure}
\end{figure}

As generating the generalized BEADS representation involves optimizing for a possible least-entangled pure state decomposition though, it is well-possible to visualize the obtained contributing pure states. As the chosen $\rho$ has full rank and is structurally complicated, the number of pure states contributing to the optimization is close to the Carathéodory bound (here, rank$^2 + 1 = 65$). Hence, in Fig.~\ref{fig:SLNoisy04_pure} we only visualize a selection of eight contributing pure states. It is immediately clear that all of the shown states are biseparable across some but not fixed to a specific bipartite cut. The remaining non-displayed pure states in this decomposition of $\rho$ behave in the same manner.

To summarize this chapter, sector lengths are easy to compute and provide a set of separability and entanglement criteria. However, these conditions are only sufficient. Thus, as for the shown examples, sector lengths are often unsuited to characterize important correlation properties of quantum states. On the other hand, the generalized BEADS representation is computationally more expensive but always allows to uniquely identify the correlation structure even of states where sector lengths are unfit for use.

\newgeometry{left=1cm, right=1cm, top=3cm, bottom=0cm}
\newgeometry{left=2.5cm, right=2.5cm, top=5cm, bottom=0cm}

\clearpage

\clearpage
\restoregeometry
\end{document}